\documentclass[reqno,a4paper,oneside]{amsart}

\usepackage{CJKutf8}
\usepackage{xcolor}
\usepackage{graphicx}
\usepackage[textwidth = 430 pt, textheight = 630 pt]{geometry}
\usepackage{float}
\definecolor{MyDarkBlue}{rgb}{0.15,0.25,0.45}
\definecolor{Cred}{rgb}{0,0,0}
\usepackage{epsfig,rotating}
\usepackage{amsmath,amssymb}
\usepackage{amsfonts}
\usepackage{mathrsfs}
\usepackage{bbm}
\usepackage[normalem]{ulem}
\usepackage{mleftright} 
\usepackage[low-sup]{subdepth} 
\usepackage{latexsym}
\usepackage{amsthm}
\usepackage{tikz}
\usetikzlibrary{matrix,cd,arrows}
\usepackage{lmodern}
\usepackage{microtype}
\usepackage{enumitem} 
\usepackage{setspace} 

\usepackage[all,knot]{xy}
\xyoption{arc}
\newcounter{access}
\usepackage{tikz}
\usetikzlibrary{graphs,decorations.pathmorphing,decorations.markings}
\usepackage{mathtools}
\usepackage[all,knot]{xy}
\xyoption{arc}

\newcommand\mutot{\mu_\caG}
\newcommand\MC{\mu_{{\rm MC}}}

\usepackage[toc,page]{appendix}

\renewcommand{\appendices}{
    \section*{Appendix}\label{appendices}\setcounter{subsection}{0}
    \addcontentsline{toc}{section}{Appendix}
    \setcounter{equation}{0}
    \crefalias{subsection}{appendix}
    \makeatletter
    \renewcommand{\theequation}{\Alph{subsection}.\arabic{equation}}
    \renewcommand{\thesubsection}{\Alph{subsection}}
    \renewcommand{\thethm}{\Alph{subsection}.\arabic{thm}}
    \@addtoreset{equation}{subsection}
    \@addtoreset{thm}{subsection}
    \makeatother
}
\let\oldref\ref
\AtBeginDocument{\renewcommand{\ref}[1]{\cref{#1}}}
\AtBeginDocument{\renewcommand{\eqref}[1]{{\rm (\oldref{#1})}}}

\newcommand{\makecommand}[3]{%
    \foreach \i in #3 {%
        \expandafter\xdef\csname #1\i\endcsname{\noexpand#2{\unexpanded\expandafter{\i}}}%
    }%
}
\newcommand{\latinalphabet}{A,a,B,b,C,c,d,D,E,e,F,f,G,g,H,h,I,i,J,j,K,k,L,l,M,m,N,n,O,o,P,Q,q,R,r,S,s,T,t,U,u,V,v,W,w,X,x,Y,y,Z,z}

\makecommand{fr}{\mathfrak}{\latinalphabet}

\makecommand{fr}{\mathfrak}{{der,gl,sl,so,osp,brst,su,sp,spin,string,Poisson,inn,at}}

\makecommand{sf}{\mathsf}{\latinalphabet}

\makecommand{sf}{\mathsf}{{id,String,Lie,Hom,SU,inv,SO,Map,Diff,HSymp,id,inn,Inn,CE,hom,Gpd,Bibun,pr}}

\makecommand{rm}{\mathrm}{\latinalphabet}

\makecommand{rm}{\mathrm}{{Ham,ad,id,dim,Ad,im,deg}}

\makecommand{ca}{\mathcal}{\latinalphabet}

\makecommand{I}{\mathbbm}{{N,Z,R,C}}

\makecommand{sc}{\mathscr}{\latinalphabet}

\makecommand{tt}{\mathtt}{\latinalphabet}

\newcommand{\reference}[1]{\tcbsubtitle[before skip=\baselineskip]{#1}}
\newcommand{\e}{\ensuremath{\mathrm{e\;\!}}}
\newcommand\fibtimes[2]{\mathbin{_{#1}\times_{#2}}}

\usepackage{hyperref}
\hypersetup{
    hypertexnames=false,
    colorlinks=true,
    citecolor=MyDarkBlue,
    linkcolor=MyDarkBlue,
    urlcolor=MyDarkBlue,
    pdfauthor={Simon-Raphael Fischer},
    pdftitle={Extension of algebroids I},
    pdfsubject={math.DG},
    breaklinks=true
}

\allowdisplaybreaks

\usepackage[many]{tcolorbox} 
\tcbuselibrary{theorems} 
\usepackage[capitalise,nameinlink,noabbrev]{cleveref} 

\theoremstyle{plain}
\newtheorem{theorem}{Theorem}[section]

\newtcbtheorem
  [use counter*=theorem,number within=section,crefname={lemma}{lemmata},crefname={Lemma}{lemmata}]
  {lemmata}
  {Lemma}
  {%
		fontupper=\itshape,
		breakable,
		enhanced,
		sidebyside adapt=both,
		attach boxed title to top center={yshift=-2mm},
    colback=gray!25,
    colframe=gray!70!black,
    fonttitle=\bfseries,
		subtitle style={boxrule=0.3pt,
		colback=gray!25,
		colupper=black,
		fontupper=\normalfont\footnotesize
		}
  }
  {lem}

\newtcbtheorem
  [use counter*=theorem,number within=section,crefname={proposition}{propositions},crefname={Proposition}{propositions}]
  {propositions}
  {Proposition}
  {%
		fontupper=\itshape,
		breakable,
		enhanced,
		sidebyside adapt=both,
		attach boxed title to top center={yshift=-2mm},
    colback=gray!25,
    colframe=gray!70!black,
    fonttitle=\bfseries,
		subtitle style={boxrule=0.3pt,
		colback=gray!25,
		colupper=black,
		fontupper=\normalfont\footnotesize
		}
  }
  {prop}
\newtcbtheorem
  [use counter*=theorem,number within=section,crefname={theorem}{theorems},crefname={Theorem}{theorems}]
  {theorems}
  {Theorem}
  {%
		fontupper=\itshape,
		breakable,
		enhanced,
		sidebyside adapt=both,
		attach boxed title to top center={yshift=-2mm},
    colback=gray!25,
    colframe=gray!70!black,
    fonttitle=\bfseries,
		subtitle style={boxrule=0.3pt,
		colback=gray!25,
		colupper=black,
		fontupper=\normalfont\footnotesize,
		}
  }
  {thm}
\newtcbtheorem
  [use counter*=theorem,number within=section,crefname={corollary}{corollaries},crefname={Corollary}{corollaries}]
  {corollaries}
  {Corollary}
  {%
		fontupper=\itshape,
		breakable,
		enhanced,
		sidebyside adapt=both,
		attach boxed title to top center={yshift=-2mm},
    colback=gray!25,
    colframe=gray!70!black,
    fonttitle=\bfseries,
		subtitle style={boxrule=0.3pt,
		colback=gray!25,
		colupper=black,
		fontupper=\normalfont\footnotesize
		}
  }
  {cor}

\newtcbtheorem
  [use counter*=theorem,number within=section,crefname={conjecture}{conjectures},crefname={Conjecture}{conjectures}]
  {conjectures}
  {Conjecture}
  {%
		fontupper=\itshape,
		breakable,
		enhanced,
		sidebyside adapt=both,
		attach boxed title to top center={yshift=-2mm},
    colback=gray!25,
    colframe=gray!70!black,
    fonttitle=\bfseries,
		subtitle style={boxrule=0.3pt,
		colback=gray!25,
		colupper=black,
		fontupper=\normalfont\footnotesize
		}
  }
  {con}

\theoremstyle{remark}
\newtcbtheorem
  [use counter*=theorem,number within=section,crefname={remark}{remarks},crefname={Remark}{remarks}]
  {remarks}
  {Remark}
  {%
		breakable,
		enhanced,
		sidebyside adapt=both,
		attach boxed title to top center={yshift=-2mm},
    colback=gray!25,
    colframe=gray!70!black,
    fonttitle=\bfseries,
		subtitle style={boxrule=0.3pt,
		colback=gray!25,
		colupper=black,
		fontupper=\normalfont\footnotesize
		}
  }
  {rem}

\newtheorem{remark}[theorem]{Remarks}

\theoremstyle{definition}

\newtcbtheorem
  [use counter*=theorem,number within=section,crefname={definition}{definitions},crefname={Definition}{definitions}]
  {definitions}
  {Definition}
  {%
		breakable,
		enhanced,
		sidebyside adapt=both,
		attach boxed title to top center={yshift=-2mm},
    colback=gray!25,
    colframe=gray!70!black,
    fonttitle=\bfseries,
		subtitle style={boxrule=0.3pt,
		colback=gray!25,
		colupper=black,
		fontupper=\normalfont\footnotesize
		}
  }
  {def}
\newtcbtheorem
  [use counter*=theorem,number within=section,crefname={example}{examples},crefname={Example}{examples}]
  {examples}
  {Example}
  {
		breakable,
		enhanced,
		sidebyside adapt=both,
		attach boxed title to top center={yshift=-2mm},
    colback=gray!25,
    colframe=gray!70!black,
    fonttitle=\bfseries,
		subtitle style={boxrule=0.3pt,
		colback=gray!25,
		colupper=black,
		fontupper=\normalfont\footnotesize
		}
  }
  {ex}

\makeatletter
\renewcommand{\tocsection}[3]{%
  \indentlabel{\@ifnotempty{#2}{\bfseries\ignorespaces#1 #2\quad}}\bfseries#3}
\renewcommand{\tocsubsection}[3]{%
  \indentlabel{\@ifnotempty{#2}{\ignorespaces#1 #2\quad}}#3}

  \def\l@subsection{\@tocline{2}{0pt}{2.5pc}{5pc}{}}
	
	\makeatother 

\newcommand{\ubrace}[3][3]{{
  \mspace{#1mu}%
  \underbrace{\mspace{-#1mu}#2\mspace{-#1mu}}_{#3}%
  \mspace{#1mu}%
}}

\usepackage{tabularx} 
\usepackage{makecell} 
\usepackage{colortbl}

\DeclareFontFamily{U}{mathx}{\hyphenchar\font45}
\DeclareFontShape{U}{mathx}{m}{n}{
      <5> <6> <7> <8> <9> <10>
      <10.95> <12> <14.4> <17.28> <20.74> <24.88>
      mathx10
      }{}
\DeclareSymbolFont{mathx}{U}{mathx}{m}{n}
\DeclareFontSubstitution{U}{mathx}{m}{n}
\DeclareMathAccent{\widecheck}{0}{mathx}{"71}
\DeclareMathAccent{\wideparen}{0}{mathx}{"75}

\definecolor{Gray}{gray}{0.85}

\def\be{\begin{equation}}
\def\ee{\end{equation}}
\def\bs{\begin{subequations}}
\def\es{\end{subequations}}
\def\ba#1\ea{\begin{align}#1\end{align}}
\def\bes{\begin{equation*}}
\def\ees{\end{equation*}}
\def\bas#1\eas{\begin{align*}#1\end{align*}}

\makeatletter
\DeclareFontEncoding{LS1}{}{}
\DeclareFontSubstitution{LS1}{stix}{m}{n}
\DeclareMathAlphabet{\mathKel}{LS1}{stixscr}{m}{n}
\DeclareMathAlphabet{\mathcal}{LS1}{stixscr}{m}{n}
\makeatother
\DeclareMathOperator{\sAut}{\mathKel{A\mkern-5.5mu u\mkern-4mu t\mkern-1.5mu}}
\DeclareMathOperator{\sAd}{\mathKel{A\mkern-5.5mu d}}

\begin{document}
\pagenumbering{Roman}
\renewcommand{\thefootnote}{\fnsymbol{footnote}}

\pagenumbering{Roman}
    \begin{titlepage}
        \vspace*{2cm}
        \begin{center}
            {{\LARGE \bf
                Integrating curved Yang-Mills gauge theories}
								\\[0.2cm]
								\bf Gauge theories related to principal bundles equipped with Lie group bundle actions
            }
            \vskip1cm
            {
						{\Large {Simon-Raphael \textsc{Fischer}\textsuperscript*\textsuperscript{$\dagger$}}}
						\vskip0.5cm
						October 6, 2022: First version \\ May 2, 2025: More examples and classifying setup with simple Lie group bundles \\ \today: Major rewrite, in particular for shortening the paper; no new statements, but an added summary about how to formulate the Lagrangian for principal groupoid bundles
            \vskip0.5cm
            {
                \textsuperscript* 2021–2024: National Center for Theoretical Sciences (\begin{CJK*}{UTF8}{bkai}國家理論科學研究中心\end{CJK*}),
								\\
								Mathematics Division,
								National Taiwan University
								\\
								No.\ 1, Sec.\ 4, Roosevelt Rd., Taipei City 106, Room 407, Cosmology Building, Taiwan
								\ \\
								\begin{CJK*}{UTF8}{bkai}106319 臺北市羅斯福路四段1號　(國立臺灣大學次震宇宙館407室)\end{CJK*}
								\vskip0.5cm
								\textsuperscript{$\dagger$} 2024–today: Mathematical Institute, Georg-August University Göttingen\\
                Bunsenstr. 3--5, 37073 Göttingen, Germany\\[0.3cm]
            {\href{mailto:simon.fischer@mathematik.uni-goettingen.de}{\texttt{simon.fischer@mathematik.uni-goettingen.de}}
						\\
						ORCiD: \href{https://orcid.org/0000-0002-5859-2825}{0000-0002-5859-2825}
						}
                }}
        \end{center}
        \vskip0.8cm
        \begin{center}
            \textbf{Abstract}\footnote{Abbreviations used in this paper: \textbf{LGB(s)} for Lie group bundle(s), \textbf{LAB(s)} for Lie algebra bundle(s).}
        \end{center}
        \begin{quote}
				We construct a gauge theory based on principal bundles $\mathcal{P}$ equipped with a right $\mathcal{G}$-action, where $\mathcal{G}$ is a Lie group bundle instead of a Lie group. Due to the fact that a $\mathcal{G}$-action acts fibre by fibre, pushforwards of tangent vectors via a right-translation act now only on the vertical structure of $\mathcal{P}$. Thus, we generalize pushforwards using a connection on $\mathcal{G}$ which will modify the pushforward. A horizontal distribution on $\mathcal{P}$ invariant under such a modified pushforward will provide a proper notion of Ehresmann connection. For achieving gauge invariance we impose conditions on the connection 1-form $\mu$ on $\mathcal{G}$: $\mu$ has to be a multiplicative form, i.e.\ closed w.r.t.\ a certain simplicial differential $\delta$ on $\mathcal{G}$, and the curvature $R_\mu$ of $\mu$ has to be $\delta$-exact with primitive $\zeta$; $\mu$ will be the generalization of the Maurer-Cartan form of the classical gauge theory, while the $\delta$-exactness of $R_\mu$ will generalize the role of the Maurer-Cartan equation. This introduces the notion of multiplicative Yang-Mills connections, a connection which helped classifying singular foliations and (topological) symmetry breaking. For allowing curved connections on $\mathcal{G}$ in the dynamical theory we will need to generalize the typical definition of the curvature/field strength $F$ on $\mathcal{P}$, too, by adding $\zeta$ to $F$. Concluding with a description of how redo those steps when $\caG$ is a Lie groupoid, and several examples for a gauge theory with a curved $\mu$ will be provided, including the inner group bundle of the Hopf fibration $\mathbb{S}^7 \to \mathbb{S}^4$, and we include a classification for whether these theories admit a classical description. 
        \end{quote}
    \end{titlepage}


\tableofcontents

\pagenumbering{arabic}

\renewcommand{\thefootnote}{\arabic{footnote}}
\setlength{\parindent}{12 pt}


\section{Introduction and summary}

This paper's research concerns curved Yang-Mills-Higgs gauge theories, originally introduced by Alexei Kotov and Thomas Strobl in \cite{CurvedYMH}, where essentially the structural Lie algebra together with its action on the manifold $N$ of values of the Higgs fields is replaced by a general Lie algebroid $E \to N$:

\begin{definitions*}{Lie algebroid}
\tcbsubtitle[before skip=\baselineskip]{e.g.\ \cite[reduced definition of \S 16.1, page 113]{DaSilva}}
Let $E \to N$ be a real vector bundle. Then $E$ is a smooth \textbf{Lie algebroid} if there is a bundle map $\rho: E \to \mathrm{T}N$, called the \textbf{anchor}, and a Lie algebra structure on $\Gamma(E)$ with Lie bracket $\left[ \cdot, \cdot \right]_E$ satisfying\footnote{With $\Gamma(E)$ I denote the space of sections of a vector bundle E, and with $\mathrm{T}N$ the tangent bundle of $N$.}
\begin{align*}
  \left[\mu, f \nu\right]_E = f \left[\mu, \nu\right]_E + \mathcal{L}_{\rho(\mu)}(f) ~ \nu
\end{align*}
for all $f \in C^\infty(N)$ and $\mu, \nu \in \Gamma(E)$, where $\mathcal{L}_{\rho(\mu)}(f)$ is the action of the vector field $\rho(\mu)$ on the function $f$ by derivation.
\end{definitions*}

The idea of replacing Lie algebras with Lie algebroids was proposed by Thomas Strobl in \cite{OriginofCYMH}, with further understanding of the involved gauge transformations in \cite{mayer2009lie}; eventually, this type of infinitesimal gauge theory got summarised and finalised in \cite{CurvedYMH}. My Ph.D.\ thesis was devoted to this type of infinitesimal gauge theory, attempting to find new (physical) examples and understanding the geometry of this infinitesimal gauge theory; see \cite{My1stpaper} and \cite{MyThesis}.

A short summary of this type of generalized infinitesimal gauge theory follows; we have the following ingredients:
\begin{itemize}
	\item $M$ a spacetime;
	\item $N$ a smooth manifold, serves as set for the values of the Higgs field $\Xi: M \to N$;
	\item $E \to N$ a Lie algebroid with anchor $\rho$, replacing the structural Lie algebra $\mathfrak{g}$ and its action $\gamma: \mathfrak{g} \to \mathfrak{X}(N)$ of the classical formulation, where $\mathfrak{X}$ denotes the set of vector fields in this work;
	\item a vector bundle connection $\nabla$ on $E$;
	\item a fibre metric $\kappa$ on $E$, as a substitute of the ad-invariant scalar product on $\mathfrak{g}$;
	\item a Riemannian metric $g$ on $N$, replacing the scalar product on the vector space in which the Higgs field usually has values in and which is invariant under the action of $\gamma$, used for the kinetic term of $\Xi$ which is minimally coupled to the field of gauge bosons $A \in \Omega^1(M; \Xi^*E)$ (1-form on $M$ with values in $\Xi^*E$);
	\item a 2-form on $N$ with values in $E$, $\zeta \in \Omega^2(N;E)$, an additional contribution to the field strength of $A$.
\end{itemize}

Infinitesimal gauge invariance of the Yang-Mills type functional leads to two \textbf{infinitesimal compatibility conditions} and two metric compatibilities to be satisfied between these structures; we will present those compatibility conditions later in this introduction. If the connection on $E$ is flat, the compatibilities imply that the Lie algebroid is locally an action Lie algebroid:

\begin{definitions*}{Action Lie algebroids}
\tcbsubtitle[before skip=\baselineskip]{e.g.\ \cite[\S 16.2, Example 5; page 114]{DaSilva}}
Let $\mleft(\mathfrak{g}, \mleft[\cdot, \cdot \mright]_{\mathfrak{g}}\mright)$ be a Lie algebra equipped with a Lie algebra action $\gamma: \mathfrak{g} \to \mathfrak{X}(N)$ on a smooth manifold $N$. A \textbf{transformation Lie algebroid} or \textbf{action Lie algebroid} is defined as the bundle $E \coloneqq N \times \mathfrak{g}$ over $N$ with anchor
\begin{align*}
\rho(p, v) &\coloneqq \gamma(v)|_p
\end{align*}
for $(p, v) \in E$, and Lie bracket
\begin{align*}
	\mleft.\mleft[\mu, \nu\mright]_E\mright|_p
	&\coloneqq 
	\mleft[\mu_p, \nu_p\mright]_{\mathfrak{g}}
		+ \mleft.\mleft(\mathcal{L}_{\gamma(\mu(p))}(\nu^a) - \mathcal{L}_{\gamma(\nu(p))}(\mu^a) \mright)\mright|_p ~ e_a
\end{align*}
	for all $p \in N$ and $\mu, \nu \in \Gamma(E)$, where one views a section $\mu \in \Gamma(E)$ as a map $\mu: N \to \mathfrak{g}$, and $\mleft( e_a \mright)_a$ is some arbitrary frame of constant sections.
\end{definitions*}

As it is shown in the mentioned references for this type of infinitesimal gauge theory, one gets back to the standard Yang-Mills-Higgs gauge theory if additionally $\zeta \equiv 0$. Thus, the theory represents a covariantized version of gauge theory equipped with an additional 2-form $\zeta$. If $\nabla$ is flat we say in general that we have a \textbf{pre-classical} gauge theory, and if additionally $\zeta \equiv 0$ we have a \textbf{classical} gauge theory. In fact, we will observe in this work that the integration of $\nabla$ is a generalisation of the Maurer-Cartan form, while $\zeta$ sort of reflects the curvature of $\nabla$. Thus, the theory presented here will represent a gauge theory covariantised w.r.t.\ the Maurer-Cartan form; see later.

The 2-form $\zeta$ is needed to allow non-flat $\nabla$, because otherwise only flat $\nabla$ could satisfy the compatibility conditions. But $\zeta$ is not just an auxiliary map, in \cite{MyThesis} I have shown that there is also a class of \textbf{field redefinitions}\footnote{So far, there is no proper name for them.} for the classical formulation of gauge theory. This field redefinition breaks the gauge invariance; so, in order to keep gauge invariance, one needs to add $\zeta$ to the field strength, and at the same time one achieves a richer framework for gauge theories. 
In fact, these field redefinitions are not just any transformation preserving the physics. As pointed out in \cite{MyThesis}, curved Yang-Mills gauge theory is a reformulation of classical gauge theory in such a way that gauge theory is form-invariant w.r.t.\ the field redefinitions, similar to how one reformulates Classical Mechanics so that it is form-invariant w.r.t.\ Galilei or Lorentz transformations. This reformulation precisely introduces those extra terms, $\zeta$ and connection 1-forms of $\nabla$, that is, every classical gauge theory is equivalent to a possibly curved theory with a non-zero $\zeta$ contributing to the field strength. Thus, this gauge theory is not just a ``simple'' generalisation where one simply adds extra terms, and henceforth it is natural to study whether there is an example of a gauge theory where $\zeta$ is non-zero and cannot be transformed to zero by the mentioned field redefinitions; see the \ref{tab:ClassicalCov}, \ref{tab:cYM}, and \ref{tab:cYMH}.

{\renewcommand{\arraystretch}{2}
\begin{table}[H]
	\caption{The idea of covariantisation as for classical mechanics; the classical and covariantised flat theory are equivalent.}
	\label{tab:ClassicalCov}
		\begin{tabularx}{\textwidth}{c c c}
			\rowcolor{gray}
			Classical theory & Covariantised flat theory & Curved Theory \\
			Vector space $V$ & Trivial vector bundle $M \times V$ & Vector bundle $V \to M$ \\
			\rowcolor{Gray}
			$\frac{\partial}{\partial x^i}$ & Canonical flat connection $\nabla^0$ & Vector bundle connection $\nabla$ \\ 
			\multicolumn{1}{X}{Coordinate changes may lead to extra terms} & 
			\multicolumn{2}{c}{Coordinate expressions form-invariant under coordinate changes}
		\end{tabularx}
\end{table}
}
\vspace{-30pt}
\begin{tikzcd}[ampersand replacement=\&]
\phantom{Coordinate changes may lead} \arrow[bend right, equal]{r} \& \phantom{Coordinate}
	\end{tikzcd}
	
	{\renewcommand{\arraystretch}{2}
\begin{table}[H]
\caption{Similarly, one can covariantise gauge theory w.r.t.\ the structural Lie group, leading to the notion of curved Yang-Mills (YM) gauge theory.}
	\label{tab:cYM}
		\begin{tabularx}{\textwidth}{c c c}
			\rowcolor{gray}
			YM theory & Covariantised YM theory & Curved YM theory \\
			Lie group $G$ & Trivial Lie group bundle $M \times G$ & Lie group bundle $G \to M$ \\
			\rowcolor{Gray}
			Maurer-Cartan form & Fibre-wise Maurer-Cartan & Multiplicative YM connection \\ 
			\multicolumn{1}{X}{Field redefinitions lead to extra terms in gauge transformations and field strength} & 
			\multicolumn{2}{c}{\makecell{Expressions form-invariant under field redefinitions, \\ \textbf{but curvature transforms non-trivially}}}
		\end{tabularx}
\end{table}}
\vspace{-30pt}
\begin{tikzcd}[ampersand replacement=\&]
\phantom{Coordinate changes may lead} \arrow[bend right, equal]{r} \& \phantom{Coordinate}
	\end{tikzcd}
	
	{\renewcommand{\arraystretch}{2}
\begin{table}[H]
\caption{The same can be done for Yang-Mills-Higgs theories, leading to the notion of curved Yang-Mills-Higgs (YMH) theory; see also \cite{Fischer:2024vak}.}
	\label{tab:cYMH}
		\begin{tabularx}{\textwidth}{c c c}
			\rowcolor{gray}
			YMH theory & Covariantised YMH theory & Curved YMH theory \\
			Lie group $G$ with right-action on $N$ & Action groupoid $N \times G$ & Lie groupoid $G \rightrightarrows N$ \\
			\rowcolor{Gray}
			Maurer-Cartan form & Fibre-wise Maurer-Cartan & Covariant adjustments \\ 
			\multicolumn{1}{X}{Field redefinitions lead to extra terms in gauge transformations and field strength} & 
			\multicolumn{2}{c}{\makecell{Expressions form-invariant under field redefinitions, \\ \textbf{but curvature transforms non-trivially}}}
		\end{tabularx}
\end{table}}
\vspace{-30pt}
\begin{tikzcd}[ampersand replacement=\&]
\phantom{Coordinate changes may lead} \arrow[bend right, equal]{r} \& \phantom{Expressions form-invariant}
	\end{tikzcd}
	
	In particular, we will study and introduce the notions for curved Yang-Mills gauge theories (\ref{tab:cYM}), and we will provide a full family of examples of new gauge theories which cannot be flattened via field redefinitions. For completeness, gauge theories with a structural groupoid, curved Yang-Mills-Higgs theory, were introduced in \cite{Fischer:2024vak} after this work. In fact, many of the examples provided here were first introduced in \cite{Fischer:2024vak}, and later added to this work here; these examples are a direct consequence of \cite{Fischer:2401.05966}, a first application of the kinematics of curved Yang-Mills gauge theories.
	
	\begin{remarks}{References for connections on principal LGB-bundles}{AnotherPreprintWork}
It is important to mention beforehand that there is another great paper, \cite{OtherPreprintAboutConnection}, which also discusses the notion of connections on principal LGB-bundles. When I wrote this paper, I was not aware of this paper so that there will be some shared results which I found independently of the other authors. This will be obvious due to the fact that our approach is different. Both, this and the other paper, lead to the same sense of Ehresmann connection (based on the same idea of the behaviour of the parallel transport), but are explicitly not directly the same formulation. More like equivalent definitions of the same object, so that it should be worth it to read both research works. Hence, our approach will be different with some shared and some different results: \cite{OtherPreprintAboutConnection} sole purpose was in defining connections on principal bundles with Lie group bundle action (with an application in variational calculus), while we will also introduce and discuss Darboux derivatives, Yang-Mills gauge theory, and related concepts in this context, also clarifying what happens for Yang-Mills-Higgs theories. Furthermore, our field strength on the principal bundle will be different and more general as already pointed out previously; it will consist of the typical curvature measuring the failure of a distribution to be integrable, plus an adjustment term $\zeta$.
\end{remarks}

In \cite{MyThesis} I first focused on Lie algebroids where the anchor map is an isomorphism and, thus, the Lie algebroid is just the tangent bundle of $M$. Locally, such examples can be excluded: The curvature can be always transformed to zero by the field redefinitions. However, globally, this is not always true: I have shown that the tangent bundle of the seven-dimensional sphere $\mathbb{S}^7$ is an example of a curved gauge theory which cannot be transformed to a flat pre-classical gauge theory, otherwise it would be a Lie group which is not possible.

However, I also studied the other edge case of having Lie algebra bundles (LABs) instead of Lie algebroids: \cite{MyThesis} (also in \cite{My1stpaper}) points out that locally, \textit{i.e.}~over a contractible base manifold $N$, there is always a field redefinition transforming the initial gauge theory to a pre-classical one. Hence, we arrive at a similar situation as for tangent bundles. But globally there are examples like the adjoint bundle of the Hopf fibration $\mathbb{S}^7 \to \mathbb{S}^4$, where $\mathbb{S}^n$ ($n \in \mathbb{N}$) denotes the $n$-dimensional sphere.

This is now the starting point of this paper. We want to understand why the adjoint bundle of $\mathbb{S}^7 \to \mathbb{S}^4$ as a structural LAB gives rise to a gauge theory which cannot be described classically, and where it differs from the classical formalism of gauge theory. To truly understand this, and also to possibly get new information about the general case regarding Lie algebroids, we will integrate curved Yang-Mills-Higgs theories in the case of LABs in this paper. Using LABs means that there is no coupling to a Higgs field in the usual sense and therefore we will now just speak instead of a curved Yang-Mills gauge theory. It will turn out, that the Hopf fibration is not just a nice coincidence, but a member of a family of canonical examples: The adjoint bundles of centerless non-flat principal bundles as structural Lie group bundles.

Due to the length of this paper, we will start outlining the paper's main results in a summary; we will focus on an easy presentation in the introduction, especially we will not always restate our assumptions, and the notation will be simpler than the actual formulation.

\subsection{Summary}

The gauge theory presented here will be based on principal bundles $\pi\colon\mathcal{P} \to M$, $M$ a smooth manifold, related to a right action of a Lie group bundle (LGB) $\pi_{\mathcal{G}} \colon \mathcal{G} \to M$. The most important distinction to classical gauge theory is that the action $\Phi$ is a map
\begin{align*}
\mathcal{P} * \mathcal{G} \coloneqq \pi^*\mathcal{G} &\to \mathcal{P},\\
(p, g) &\mapsto p \cdot g,
\end{align*}
where $\pi^*\mathcal{G}$ is the pullback LGB of $\mathcal{G}$ along $\pi$. Observe that $\mathcal{G}_x$ only acts on $\mathcal{P}_x$ as a Lie group, thus an element $g \in \mathcal{G}_x$ cannot act on all of $\mathcal{P}$, where notations like $\mathcal{G}_x$ denote the fibre of $\mathcal{G}$ over an $x \in M$. This leads to certain difficulties for defining gauge theory which we will resolve in this paper.

However, this paper aims to be accessible for all people knowing the basics of ``classical'' gauge theory. Thus, we start with an introduction to LGBs, their actions and their infinitesimal analogues, the LABs, in \ref{LGBSection} to \ref{LGBActionIISection}. We will especially follow \cite{mackenzieGeneralTheory}, but since this and other references are usually introducing these basic notions on (Lie) groupoids, we decided to quickly reintroduce all the needed notions, but without proving those again; see then the references instead. The experienced reader may start directly with \ref{ConnCurvOnPrincLGBBundle}, but there is one partially new result in these preliminary sections: For gauge theory we will need to understand the differential of the $\mathcal{G}$-action $\Phi$ which we will derive in \ref{prop:DiffOfLGBAction} and which will form \textit{the} fundament for many calculations. For this terms like $\mathrm{T}M$ will denote a tangent bundle with fibre $\mathrm{T}_xM$ at $x \in M$; the fibres of other bundles with similar notation are denoted in the same fashion.


\begin{propositions*}{Differential of LGB action $\Phi$}
We have
\begin{equation*}
\mathrm{D}_{(p, g)}\Phi(X, Y)
=
\mathrm{D}_pr_\sigma(X)
	+ \mleft.{\overline{\mleft( \MC\mright)_g \bigl(Y - \mathrm{D}_{x}\sigma (\omega)\bigr)}}\mright|_{p \cdot g}
\end{equation*}
for all $(p, g) \in \mathcal{P} * \mathcal{G}$ and $(X, Y) \in \mathrm{T}_{(p, g)}(\mathcal{P}*\mathcal{G})$, where $x \coloneqq \pi(p)$, $\sigma$ is any (local) section of $\mathcal{G}$ with $\sigma_{x} = g$ and $r_\sigma$ denotes the right-translation of $\sigma$ in $\mathcal{P}$, $\MC$ is the fibre-wise defined Maurer-Cartan form of the fibres, $\overline{\mleft( \MC \mright)_g(\dotsc)}$ denotes its generated fundamental vector field on $\mathcal{P}_x$, and $\omega$ is an element of $\mathrm{T}_xM$ given by
\begin{equation*}
\omega
\coloneqq
\mathrm{D}_p \pi(X)
= 
\mathrm{D}_g\pi_{\mathcal{G}}(Y).
\end{equation*}
\end{propositions*}

In \ref{ConnCurvOnPrincLGBBundle} we will introduce connections and generalized curvatures on principal LGB-bundles: \ref{PrincBundlLGBBased} starts with introducing principal $\mathcal{G}$-bundles $\mathcal{P}$. Such principal bundles were already introduced in \cite[\S 5.7, page 144f.]{GroupoidBasedPrincipalBundles}, but also here we rephrased it in such a way that these bundles' definition has a more familiar shape for readers not so proficient with the study of groupoids. The main difference to Lie group based principal bundles is the previously-mentioned LGB action $\Phi$. Since these bundles were not studied a lot before we will rephrase \cite{GroupoidBasedPrincipalBundles} and introduce basic notions like morphisms and easy examples, for example LGBs $\mathcal{G}$ themselves are principal $\mathcal{G}$-bundles, so that the gauge theory presented in this paper can also be understood as a gauge theory based not only on ``classical'' principal bundles but also on LGBs which were excluded in general in the classical formalism. Furthermore, as in \cite{GroupoidBasedPrincipalBundles}, we generalize certain statements known about classical principal bundles, \textit{e.g.}\ we observe that a section of $\mathcal{P}$ does not trivialize $\mathcal{P}$ in general, however it introduces an isomorphism to $\mathcal{G}$, as will be pointed out in \ref{lem:SectionsNowInduceIsomToLGBsNotNecTriv}.

In \ref{ConnectionSubsection} we finally turn to the notion of connections on $\mathcal{P}$, described as horizontal subbundle $\mathrm{H}\mathcal{P}$ of the tangent bundle $\mathrm{T}\mathcal{P}$, complementary to the vertical bundle $\mathrm{V}\mathcal{P}$, and equivalently we want to describe $\mathrm{H}\mathcal{P}$ as a connection 1-form $A$. As one can already see in the previous proposition about the derivative of the action $\Phi$, due to the fact that right-translations $r_g$ of an element $g \in \mathcal{G}_x$ now only acts on $\mathcal{P}_x$, its derivative $\mathrm{D}r_g$ only acts on the tangent bundle of $\mathcal{P}_x$, the vertical tangent vectors of $\mathcal{P}$ (over $x$). Hence, in order to define the pushforward of non-vertical vectors we make use of auxiliary sections. However, this leads to the problem that there are many different sections with the same value over a fixed base point $x \in M$, all of whose pushforwards of non-vertical vectors are in general different. To get rid of the ambiguity in the choice of section we are going to modify and generalize the pushforward of tangent vectors by right-translations. To do so, we fix a horizontal distribution $\mathrm{H}\mathcal{G}$ on $\mathcal{G}$. The \textbf{(modified) right-pushforward} of $X \in \rmT_p \caP$ ($p \in \caP_x$) by $g \in \caG_x$ is then defined as
\begin{equation*}
\car_{g*}(X) \coloneqq \rmD_{(p, g)}\Phi(X, Y),
\end{equation*}
where $Y \in \rmH_g\caG$ is the unique horizontal vector at $g$ lifting $\rmD_p\pi(X)$.

Then we start to construct the horizontal distribution $\mathrm{H}\mathcal{P}$ initially by its associated parallel transport, shortly denoted now by $\mathrm{PT}^{\mathcal{P}}$; similarly we denote the parallel transport associated to $\mathrm{H}\mathcal{G}$ by $\mathrm{PT}^{\mathcal{G}}$. We define an Ehresmann connection on $\mathcal{P}$ by
\begin{align*}
\mathrm{PT}^{\mathcal{P}}(p \cdot g)
&=
\mathrm{PT}^{\mathcal{P}}(p) \cdot \mathrm{PT}^{\mathcal{G}}(g)
\end{align*}
for all $(p, g) \in \mathcal{P}*\mathcal{G}$. However, in order to define the connection 1-form it is useful to find a definition of such $\mathrm{H}\mathcal{P}$ via a symmetry w.r.t.\ a certain action on $\mathrm{T}\mathcal{P}$: As already high-lighted, this will be a modification of the pushforward via right-translations. Differentiating the previous equations of parallel transports w.r.t.\ the curve parameter achieves an alternative version of this modification, summarised in \ref{prop:IsomorphismRightPushAndDarboux}; we will explain the second summand of the following proposition afterwards.

\begin{propositions*}{The modified right-pushforward}
For $g \in \mathcal{G}_x$ ($x \in M$) we have
\begin{align*}
\mleft.\mathrm{T}\mathcal{P}\mright|_{\mathcal{P}_x} &\to \mleft.\mathrm{T}\mathcal{P}\mright|_{\mathcal{P}_x},\\
X 
&\mapsto 
\mathcal{r}_{g*}(X) 
=
\mathrm{D}_pr_\sigma\mleft( 
	X 
\mright)
	- \mleft.{\overline{
		\mleft. \mleft( \pi^!\Delta\sigma \mright) \mright|_p(X)
	}}\mright|_{p \cdot g}~,
\end{align*}
where $p \in \mathcal{P}_x$, $X \in \mathrm{T}_p \mathcal{P}$, $\pi^!$ denotes the pull-back of forms with $\pi$, and $\sigma$ is any (local) section of $\mathcal{G}$ with $\sigma_x = g$. $\mathcal{r}_{g*}$ is independent of the choice of the local section $\sigma$ by definition, and it is a vector bundle automorphism over the right-translation $r_g$. 

Furthermore, if we have a horizontal distribution $\mathrm{H}\mathcal{P}$ on $\mathcal{P}$, then $\mathrm{H}_p\mathcal{P}$ is isomorphic via $\mathcal{r}_{g*}$ to a complement of $\mathrm{V}_{p \cdot g} \mathcal{P}$ in $\mathrm{T}_{p \cdot g}\mathcal{P}$ (this complement is not necessarily $\mathrm{H}_{p \cdot g}\mathcal{P}$).
\end{propositions*}

This can be generalized to sections $\sigma$ of $\mathcal{G}$ in a straight-forward manner, giving rise to an automorphism $\mathcal{r}_{\sigma*}$ of $\mathrm{T}\mathcal{P}$ over the right-translation $r_\sigma$.
$\Delta\sigma$ is the generalised version of the Darboux derivative, often simply denoted as $\sigma^{-1} \mathrm{d}\sigma$ in the classical formalism, but in some works like \cite[\S 5.1, page 182ff.]{mackenzieGeneralTheory}\ also already written as $\Delta\sigma$ in the classical formalism. As expected, this derivative plays an important role in gauge theory, which is why we will also discuss and introduce the Darboux derivative; usually it only appears in the gauge transformations, but in our case it will already appear now because $\rmH \caG$ may not be trivial. To no suprise, one can introduce the connection 1-form $\mutot$ on $\mathcal{G}$ corresponding to $\rmH \caG$ with values in the LAB $\pi_\caG^*\mathcal{g}$ of $\mathcal{G}$ by defining it as $\MC$ contracted with the projection to the vertical part. The Darboux derivative is then the form-pullback of $\mutot$ (\ref{def:DarbouxDerivativeOnLGBs}) with a section.

\begin{definitions*}{Darboux derivative}
For (local) $\sigma \in \Gamma(\mathcal{G})$ we define the \textbf{Darboux derivative $\Delta \sigma \in \Omega^1(M; \mathcal{g})$} by
\begin{equation*}
\Delta \sigma
=
\sigma^! \mutot~.
\end{equation*}
\end{definitions*}

Using the Darboux derivative, we defined the modified pushforward via right-translations (also called modified right-pushforward) which eventually leads to the definition of an Ehresmann connection on the principal $\mathcal{G}$-bundle $\mathcal{P}$, as pinpointed in \ref{def:FinallyTheConnection}.

\begin{definitions*}{Ehresmann connection on principal LGB-bundles}
We call $\mathrm{H}\mathcal{P}$ an \textbf{Ehresmann connection} or a \textbf{connection on $\mathcal{P}$} if it is \textbf{right-invariant (w.r.t.\ modified right-pushforward)}, \textit{i.e.}\
\begin{align*}
\mathcal{r}_{g*}\mleft( \mathrm{H}_p\mathcal{P} \mright)
&=
\mathrm{H}_{p\cdot g}\mathcal{P}
\end{align*}
for all $p \in \mathcal{P}_x$ and $g \in \mathcal{G}_x$ ($x \coloneqq \pi(p)$).
\end{definitions*}

$\mathrm{H}\mathcal{P}$ will now always be an Ehresmann connection in the following, and one may also say that $\rmH \caP$ is an Ehresmann connection modified by $\rmH\caG$.
The basic idea now is to replace the typical right-pushforward with the modified one in all the involved definitions which is straight-forward. \ref{def:GaugeBosonsOnLGBPrincies} introduces connection 1-forms on principal bundles:

%


\begin{definitions*}{Connection 1-forms on principal LGB-bundles, simplified notation}
A \textbf{connection 1-form} or \textbf{gauge field} on $\mathcal{P}$ is a 1-form $A \in \Omega^1(\mathcal{P}; \pi^*\mathcal{g})$ satisfying:
\begin{itemize}
	\item \textbf{($\mathcal{G}$-equivariance, but w.r.t.\ modified right-pushforward)}
		\begin{equation*} 
			\mathcal{r}_\sigma^! A
			=
			\mathrm{Ad}_{\sigma^{-1}} \circ A
		\end{equation*}
	for all (local) $\sigma \in \Gamma(\mathcal{G})$, where $\mathrm{Ad}$ is the adjoint representation of $\mathcal{G}$ on $\mathcal{g}$.
	\item \textbf{(Identity on $\mathrm{V}\mathcal{P}$)}
	\begin{equation*}
	A\mleft(\widetilde{\nu}\mright)
	=
	\pi^*\nu
	\end{equation*}
	for all (local) $\nu \in \Gamma(\mathcal{g})$, where notations like $\pi^*$ denote the pull-back of sections.
\end{itemize}
\end{definitions*}

Of course we will achieve a 1:1 correspondence to Ehresmann connections in \ref{thm:OurConnectionHasAUniqueoneForm}.

\begin{theorems*}{1:1 correspondence of Ehresmann connections and connection 1-forms}
There is a 1:1 correspondence between Ehresmann connections and connection 1-forms on $\mathcal{P}$:
\begin{itemize}
	\item Let $\mathrm{H}\mathcal{P}$ be an Ehresmann connection on $\mathcal{P}$. Then $\mathrm{H}\mathcal{P}$ defines a connection 1-form $A \in \Omega^1(\mathcal{P}; \pi^*\mathcal{g})$ by
	\begin{equation*}
	A_p\bigl( \overline{v}_p + X_p \bigr)
	=
	(p, v)
	\end{equation*}
	for all $p \in \mathcal{P}_x$ ($x \in M$), $v \in \mathcal{g}_x$ and $X \in \mathrm{H}_p\mathcal{P}$.
	\item Let $A \in \Omega^1(\mathcal{P}; \pi^*\mathcal{g})$ be a connection 1-form on $\mathcal{P}$. Then $A$ defines an Ehresmann connection $\mathrm{H}\mathcal{P}$ on $\mathcal{P}$ via its kernel $\mathrm{Ker}(A)$, that is,
	\begin{equation*}
	\mathrm{H}_p\mathcal{P}
	=
	\mathrm{Ker}(A_p)
	\end{equation*}
	for all $p \in \mathcal{P}$.
\end{itemize}
\end{theorems*}

Not only these results are straight-forward once one has the definition of the modified right-pushforward, also other results will be simple to guess. For example in \ref{GaugeTrafoForA} we turn to the gauge transformations which will have a familiar shape but with the generalized Darboux derivative appearing, see \ref{prop:GaugeTrafoOfGaugeBoson}.

\begin{propositions*}{Gauge transformations of connection 1-forms, simplified notation}
Let $H$ be a (base-preserving) automorphism of $\mathcal{P}$. We then have that $H^!A$ is a connection 1-form on $\mathcal{P}$ and
\begin{equation*}
H^!A
=
{\mathrm{Ad}_{\mleft(\sigma^H\mright)^{-1}}} \circ A 
	+ \mleft(\pi^*\Delta\mright)\sigma^H,
\end{equation*}
where $\sigma^H \in \Gamma(\pi^*\mathcal{G})$ is uniquely defined by
\begin{equation*}
H(p)
=
p \cdot \sigma_p^H
\end{equation*}
for all $p \in \mathcal{P}$, and $\pi^*\Delta$ is the Darboux derivative on $\pi^*\mathcal{G}$ naturally inherited by the pullback of the connection on $\mathcal{G}$.
\end{propositions*}

The 1:1 correspondence between $H$ and $\sigma^H$ arising here leads to similar statements as in classical gauge theory which we will also shortly point out. Afterwards we will show in \ref{prop:LocalGaugeTrafoChangeGauge} that this proposition also has a local version, that is, one looks at the pullback $A_{s_i} \coloneqq s_i^!A$ of $A$ w.r.t.\ a section (also called gauge) $s_i$ of $\mathcal{P}$ defined over an open subset $U_i \subset M$, and then the change of $A$ by changing the gauge $s_i$ to another gauge follows by the previous proposition.

\begin{propositions*}{Gauge transformations as a change of gauge in the local gauge field}
Let $U_i$ and $U_j$ be two open subsets of $M$ so that $U_i \cap U_j \neq \emptyset$, two gauges $s_i \in \Gamma\mleft(\mathcal{P}|_{U_i}\mright)$ and $s_j \in \Gamma\mleft(\mathcal{P}|_{U_j}\mright)$, and the unique $\sigma_{ji} \in \Gamma\mleft( \mleft.\mathcal{G}\mright|_{U_i \cap U_j} \mright)$ with $s_i = s_j \cdot \sigma_{ji}$ on $U_i \cap U_j$. Then we have over $U_i \cap U_j$ that
\begin{equation*}
A_{s_i}
=
\mathrm{Ad}_{\sigma_{ji}^{-1}}\circ A_{s_j}
	+ \Delta\sigma_{ji}.
\end{equation*}
\end{propositions*}

This will finish the discussion about $A$, but of course we also needs its field strength $F$ and its gauge transformations. Up until now $\mathrm{H}\mathcal{G}$ was an arbitrary horizontal distribution, but we will need to fix certain conditions on it to assure gauge invariance later. On one hand, we will argue that the existence of Ehresmann connections is equivalent to the multiplicativity of $\rmH \caG$, that is, viewing $\mathcal{G}$ as the principal $\mathcal{G}$-bundle $\mathcal{P}$ itself whose horizontal distribution $\mathrm{H}\mathcal{P}$ aligns with $\mathrm{H}\mathcal{G}$, it is natural to guess that $\mathrm{H}\mathcal{G}$ should be an Ehresmann connection itself. Ehresmann connections on LGBs modifying themselves were already discussed in works like \cite{LAURENTGENGOUXStienonXuMultiplicativeForms}; there is also \cite{FernandesMarcutMultiplicativeForms} related to a similar subject. We will discuss those in \ref{CurvatureSubsection} in more detail. One of the results of these works is that such Ehresmann connections on LGBs are characterized by the fact that the connection 1-form $\mutot$ on $\caG$ has to be a multiplicative form:

\begin{definitions*}{Multiplicative forms, simplified situation}
\tcbsubtitle[before skip=\baselineskip]{\cite[\S 2.1, special situation of Def.\ 2.1]{crainic2015multiplicative}}
We call an $\omega \in \Omega^1\mleft( \mathcal{G}; \pi_{\mathcal{G}}^*\mathcal{g} \mright)$ a \textbf{multiplicative form} if
\begin{equation*}
\omega_{gq}\mleft( \mathrm{D}_{(g, q)}\Phi(X, Y)  \mright)
=
\mathrm{Ad}_{q^{-1}}\bigl( \omega_{g}(X) \bigr)
	+ \omega_{q}(Y)
\end{equation*}
for all $(g, q) \in \mathcal{G}*\mathcal{G}$ and $(X, Y) \in \mathrm{T}_{(g, q)}(\mathcal{G}*\mathcal{G})$, where $\Phi\colon \mathcal{G} * \mathcal{G} \to \mathcal{G}$ is the multiplication in $\mathcal{G}$.
\end{definitions*}

Assuming that $\mathrm{H}\mathcal{G}$ is multiplicative will also allow us to formulate certain technical identities for the Darboux derivative like a Leibniz rule. In order to calculate the gauge transformation for the curvature/field strength $F$ of $A$ we need to understand the curvature of $\mutot$. Classically, this is described by the Maurer-Cartan equation, however, we generalize this condition, allowing non-flat connections on $\mathcal{G}$. For this we will study a connection on the LAB $\mathcal{g}$ of $\mathcal{G}$, naturally induced by $\mathrm{H}\mathcal{G}$; we will construct this connection in \ref{prop:FinallyTheNablaInduction} and \ref{def:ConnectionOnLAB}, and our construction aligns with \cite[\S 4.5, Prop.\ 4.22]{LAURENTGENGOUXStienonXuMultiplicativeForms}.

\begin{propositions*}{LGB connection induces LAB connection, simplified notation}
The map $\nabla^{\mathcal{G}}\colon \Gamma(\mathcal{g}) \to \Omega^1(M; \mathcal{g})$, $\nu \mapsto \nabla^{\mathcal{G}}\nu$ denoted as an element of $\Omega^1(M; \mathcal{g})$ by $X \mapsto \nabla^{\mathcal{G}}_X \nu$, defined by
\begin{equation*}
\mleft.\nabla^{\mathcal{G}}_X \nu\mright|_x
\coloneqq
\mleft.\frac{\mathrm{d}}{\mathrm{d}t}\mright|_{t=0} \Bigl( \mleft(\Delta \e^{t \nu}\mright)_x (X) \Bigr)
\end{equation*}
for all $x \in M$, $X \in \mathrm{T}_xM$ and $\nu \in \Gamma(\mathcal{g})$, is a vector bundle connection on $\mathcal{g}$, where $t \in \mathbb{R}$. We will call $\nabla^{\mathcal{G}}$ the \textbf{$\mathcal{G}$-connection (on its LAB $\mathcal{g}$)}.
\end{propositions*}

In fact, $\nabla^{\mathcal{G}}$ will play the role of $\nabla$ from the beginning of this introduction. Furthermore, this connection will allow us to define the generalized Maurer-Cartan equation in \ref{thm:GenMCEq} and \ref{def:NowReallyYangMillsConnectio}.

\begin{definitions*}{Multiplicative Yang-Mills connection, simplified notation}
We say that $\mathrm{H}\mathcal{G}$, and thus $\mutot$, is a \textbf{multiplicative Yang-Mills connection (w.r.t.\ a $\zeta \in \Omega^2(M; \mathcal{g})$)}, if it satisfies the \textbf{compatibility conditions}:
\begin{enumerate}
	\item $\mutot$ is multiplicative,
	\item $\mutot$ satisfies the \textbf{generalized Maurer-Cartan equation}, that is,
	\begin{equation*}
	\mleft.\mleft(\mathrm{d}^{\pi_{\mathcal{G}}^*\nabla^{\mathcal{G}}} \mutot
	+ \frac{1}{2} \mleft[ \mutot \stackrel{\wedge}{,} \mutot \mright]_{\pi_{\mathcal{G}}^*\mathcal{g}} \mright)\mright|_g
=
\mathrm{Ad}_{g^{-1}} \circ \mleft.\pi_{\mathcal{G}}^!\zeta\mright|_g
	- \mleft.\pi_{\mathcal{G}}^!\zeta\mright|_g
	\end{equation*}
	for all $g \in \mathcal{G}_x$ ($x \in M$), where $\mleft[\cdot, \cdot\mright]_{\pi_{\mathcal{G}}^*\mathcal{g}}$ is the Lie bracket of the pullback-LAB $\pi_{\mathcal{G}}^*\mathcal{g}$, and the $\stackrel{\wedge}{,}$ denotes the typical graded extension of tensors, that is, the second summand on the left hand side is an element of $\Omega^2(\mathcal{G}; \pi_{\mathcal{G}}^*\mathcal{g})$ given here by
	\begin{equation*}
	\mleft(\frac{1}{2} \mleft[ \mutot \stackrel{\wedge}{,} \mutot \mright]_{\pi_{\mathcal{G}}^*\mathcal{g}} \mright) (X, Y)
	=
	\mleft[ \mutot(X), \mutot(Y) \mright]_{\pi_{\mathcal{G}}^*\mathcal{g}}
	\end{equation*}
	for all $X, Y \in \mathfrak{X}(\mathcal{G})$.
\end{enumerate}
\end{definitions*}

Making a pull-back of the generalized Maurer-Cartan equation w.r.t.\ a section $\sigma \in \Gamma(\mathcal{G})$ is straight-forward and will be provided in \ref{cor:PullbackOfMCSupperEquation}.

\begin{corollaries*}{Pullback of generalized Maurer-Cartan equation}
Let $\mathrm{H}\mathcal{G}$ be a multiplicative Yang-Mills connection w.r.t.\ a $\zeta \in \Omega^2\mleft( M; \mathcal{g} \mright)$. Then
\begin{equation*}
\mathrm{d}^{\nabla^{\mathcal{G}}} \Delta \sigma 
	+ \frac{1}{2} \mleft[ \Delta \sigma \stackrel{\wedge}{,} \Delta\sigma \mright]_{\mathcal{g}} 
	+ \zeta 
=
\mathrm{Ad}_{\sigma^{-1}} \circ \zeta
\end{equation*}
for all $\sigma \in \Gamma(\mathcal{G})$.
\end{corollaries*}

We will then argue that those compatibility conditions are the integrals of the infinitesimal compatibility conditions proposed by Alexei Kotov and Thomas Strobl mentioned earlier (without details): In \cite{LAURENTGENGOUXStienonXuMultiplicativeForms} it was already observed that an Ehresmann connection on an LGB naturally implies that $\nabla^{\mathcal{G}}$ is a Lie bracket derivation.

\begin{lemmata*}{Ehresmann connections induce Lie bracket derivations}
\tcbsubtitle[before skip=\baseskipline]{\cite[\S 4.5, Prop.\ 4.21]{LAURENTGENGOUXStienonXuMultiplicativeForms}}
If $\mathrm{H}\mathcal{G}$ is a multiplicative connection on $\mathcal{G}$, then
\begin{equation*}
\nabla^{\mathcal{G}}\mleft( \mleft[ \mu, \nu \mright]_{\mathcal{g}} \mright)
=
\mleft[ \nabla^{\mathcal{G}} \mu, \nu \mright]_{\mathcal{g}}
	+ \mleft[ \mu, \nabla^{\mathcal{G}} \nu \mright]_{\mathcal{g}}
\end{equation*} 
for all $\mu, \nu \in \Gamma(\mathcal{g})$.
\end{lemmata*}

That $\nabla^{\mathcal{G}}$ is a Lie bracket derivation is one of the infinitesimal compatibility conditions.
Furthermore, we will show in \ref{thm:GenMCEq} that the generalized Maurer-Cartan equation is actually equivalent to its infinitesimal version.

\begin{theorems*}{Generalized Maurer-Cartan equation, changed formulation}
Let $\mathrm{H}\mathcal{G}$ be a multiplicative connection on $\mathcal{G}$ with connected fibres, and $\zeta \in \Omega^2\mleft( M; \mathcal{g} \mright)$. Then $\mathrm{H}\mathcal{G}$ satisfies the generalized Maurer-Cartan equation w.r.t.\ $\zeta$ if and only if 
\begin{equation*}
R_{\nabla^{\mathcal{G}}}(X, Y)\mu
=
\mleft[ \zeta(X, Y), \mu \mright]_{\mathcal{g}}
\end{equation*}
for all $X, Y \in \mathfrak{X}(M)$ and $\nu \in \Gamma(\mathcal{g})$, where $R_{\nabla^{\mathcal{G}}}$ is the curvature of $\nabla^{\mathcal{G}}$.
\end{theorems*}

The condition about the curvature is in fact the second infinitesimal compatibility condition proposed by Alexei Kotov and Thomas Strobl, so that we can conclude that Yang-Mills connections on LGBs serve as an integral of the infinitesimal compatibility conditions which we will point out in \ref{def:YangMillsConnection} and \ref{rem:YangMillsEqualsInfYM}.
%

Additionally we will mention in \ref{rem:SimplicialDifferentialStuff} that multiplicativity of $\mutot$ is in fact a closedness condition w.r.t.\ a certain simplicial differential $\delta$, introduced in \cite[beginning of \S 1.2]{crainic2003differentiable}. The generalized Maurer-Cartan equation is then an exactness-condition of the ``classical'' curvature of $\mutot$, its $\delta$-primitive given by $\zeta$. This is in alignment with the infinitesimal compatibility conditions.

This will finish the discussion about the connection on $\mathcal{G}$, and it will be important for defining the field strength which will be discussed afterwards, starting with \ref{def:NewFieldStrength}. For this we will always assume now that $\mathrm{H}\mathcal{G}$ is a multiplicative Yang-Mills connection w.r.t.\ a $\zeta \in \Omega^2(M; \mathcal{g})$ (while $\mathrm{H}\mathcal{P}$ is an Ehresmann connection modified by such $\rmH\caG$).

\begin{definitions*}{(Generalized) Field strength}
We define the \textbf{(generalized) curvature} or \textbf{(generalized) field strength $F$ (of $A$)} as an element of $\Omega^2\mleft( \mathcal{P}; \pi^*\mathcal{g} \mright)$ by
\begin{equation*}
F
\coloneqq
\mathrm{d}^{\pi^*\nabla^\caG} A \circ \mleft( \pi^{\mathrm{H}\mathcal{P}}, \pi^{\mathrm{H}\mathcal{P}} \mright)
	+ \pi^!\zeta,
\end{equation*}
where $\mathrm{d}^{\pi^*\nabla^\caG}$ is the exterior covariant derivative related to the pullback connection $\pi^*\nabla^\caG$, and $\pi^{\mathrm{H}\mathcal{P}}\colon \mathrm{T}\mathcal{P} \to \mathrm{H}\mathcal{P}$ is the canonical projection onto the associated Ehresmann connection $\mathrm{H}\mathcal{P}$ on $\mathcal{P}$; that is,
\begin{equation*}
F(X, Y)
=
\mathrm{d}^{\pi^*\nabla^\caG} A \mleft( \pi^{\mathrm{H}\mathcal{P}}(X), \pi^{\mathrm{H}\mathcal{P}}(Y) \mright)
	+ \mleft(\pi^*\zeta\mright) \bigl( \mathrm{D}\pi(X), \mathrm{D}\pi(Y) \bigr)
\end{equation*}
for all $X, Y \in \mathfrak{X}(\mathcal{P})$.
\end{definitions*}

By usual arguments, the first summand measures the failure of $\rmH\caP$ being a regular foliation; that is, the gauge theory we will present here not only modifies Ehresmann connections but the modification's curvature additionally modifies the curvature on $\caP$. To avoid confusion with the ordinary curvature (the first summand) one may now need to distinguish between curvatures and field strengths.

With a lengthy calculation evolving around that $\mathrm{H}\mathcal{G}$ is a multiplicative Yang-Mills connection, especially using the generalized Maurer-Cartan equation, we will show in \ref{prop:NewFieldStrengthWithCoolProps} that $F$ transforms in a suitable way when making a pull-back w.r.t.\ the modified right-pushforward, which is the last step needed for defining the physical theory and its associated Lagrangian.

\begin{propositions*}{Properties of the generalized field strength, simplified notation}
We have the following properties of the field strength:
\begin{itemize}
	\item \textbf{(Form of type $\mathrm{Ad}$)}
	\begin{equation*}
	\mathcal{r}_{\sigma}^!F
	=
	\mathrm{Ad}_{\sigma^{-1}} \circ F
	\end{equation*}
	for all (local) $\sigma \in \Gamma(\mathcal{G})$.
	\item \textbf{(Horizontal form)}
	\newline
	For $X, Y \in \mathrm{T}_p\mathcal{P}$ ($p \in \mathcal{P}$) we have
	\begin{equation*}
	F(X, Y)
	=
	0
	\end{equation*}
	if either of $X$ and $Y$ is vertical.
\end{itemize}
\end{propositions*}

We will also derive a structure equation in \ref{prop:StructureEq}.

\begin{propositions*}{Structure equation of the generalized field strength}
We have the \textbf{structure equation}
\begin{equation*}
F
=
\mathrm{d}^{\pi^*\nabla^\caG} A
	+ \frac{1}{2} \mleft[ A \stackrel{\wedge}{,} A \mright]_{\pi^*\mathcal{g}}
	+ \pi^!\zeta.
\end{equation*}
\end{propositions*}

We will argue  that $\zeta$ of course also affects the Bianchi identity, which we have in fact already derived in earlier works.

\begin{lemmata*}{Generalized Bianchi identity}
\tcbsubtitle[before skip=\baseskipline]{\cite[\S 7, Thm.\ 7.3]{My1stpaper} \& \cite[\S 5, Thm.\ 5.1.42]{MyThesis}}
We have the \textbf{(generalized) Bianchi identity}
\begin{equation*}
\mathrm{d}^{\pi^*\nabla^\caG}F
	+ \mleft[ A \stackrel{\wedge}{,} F \mright]_{\pi^*\mathcal{g}}
=
\pi^! \mathrm{d}^{\nabla^\caG} \zeta.
\end{equation*}
\end{lemmata*}

Similarly to the discussion about connections, \ref{CurvatureSubsection} will be concluded with a discussion about the gauge transformation of $F$ in \ref{prop:GaugeTrafoOfCurv}; as for $A$ the gauge transformations of $F$ are given by form-pullbacks with a (base-preserving) principal bundle automorphism $H$ with associated unique $\sigma^H \in \Gamma(\pi^*\mathcal{G})$ given by $H(p) = p \cdot \sigma^H_p$ for all $p \in \mathcal{P}$.

\begin{propositions*}{Gauge transformation of the field strength, simplified notation}
$H^!F$ is the field strength related to $H^!A$ and
\begin{equation*}
H^!F
=
{\mathrm{Ad}_{\mleft(\sigma^H\mright)^{-1}}} \circ F.
\end{equation*}
\end{propositions*}

Again as for $A$, we are interested about its local version regarding pullbacks $F_s \coloneqq s^!F$ w.r.t.\ a gauge $s$ of $\mathcal{P}$ when changing the choice of $s$. This is straight-forward to calculate and will be stated in \ref{prop:LocalGaugeTrafoChangeGaugeFieldStrength}.

\begin{propositions*}{Gauge transformations again as a change of gauge}
Let $U_i$ and $U_j$ be two open subsets of $M$ so that $U_i \cap U_j \neq \emptyset$, two gauges $s_i \in \Gamma\mleft(\mathcal{P}|_{U_i}\mright)$ and $s_j \in \Gamma\mleft(\mathcal{P}|_{U_j}\mright)$, and the unique $\sigma_{ji} \in \Gamma\mleft( \mleft.\mathcal{G}\mright|_{U_i \cap U_j} \mright)$ with $s_i = s_j \cdot \sigma_{ji}$ on $U_i \cap U_j$.

Then we have for the field strength of $A$ over $U_i \cap U_j$ that
\begin{equation*}
F_{s_i}
=
\mathrm{Ad}_{\sigma_{ji}^{-1}}\circ F_{s_j}.
\end{equation*}
\end{propositions*}

Eventually in \ref{CYMSection} we will introduce the physical theory, that is, first of all defining the Lagrangian of a gauge theory based on principal LGB-bundles in \ref{CYMDefGaugeInv}. Due to the fact that we will have pointed out that this theory integrates the infinitesimal curved Yang-Mills-Higgs gauge theory developed by Alexei Kotov and Thomas Strobl (in the case of LABs), we are going to label this theory \textbf{curved Yang-Mills gauge theory} in \ref{def:CYMGTFinally}. To do so, $M$ is now a spacetime, so that we can define the Hodge star operator $*$, and we assume that we have an $\mathrm{Ad}$-invariant fibre metric $\kappa$ on $\mathcal{g}$.

\begin{definitions*}{Curved Yang-Mills gauge theory}
Let $\mleft( U_i \mright)_i$ be an open covering of $M$ so that there are subordinate gauges $s_i \in \Gamma\mleft(\mathcal{P}|_{U_i}\mright)$. Then the top-degree form $\mathfrak{L}_{\mathrm{CYM}}[A] \in \Omega^{\mathrm{dim}(M)}(M; \mathbb{R})$, defined locally by
\begin{equation*}
\mleft.\bigl(\mathfrak{L}_{\mathrm{CYM}}[A]\bigr)\mright|_{U_i}
\coloneqq 
- \frac{1}{2} \kappa \mleft( F_{s_i} \stackrel{\wedge}{,} *F_{s_i} \mright),
\end{equation*}
is called the \textbf{curved Yang-Mills Lagrangian}.
\end{definitions*}

We will have argued in \ref{cor:ContractionOfLocalFIeldWellDefined} that this Lagrangian is well-defined; a trivial consequence of the previously-highlighted gauge transformations. Similarly, we will achieve gauge invariance in \ref{thm:GaugeInvarianceOfLagrangian}.

\begin{theorems*}{Gauge invariance of the curved Yang-Mills Lagrangian}
We have
\begin{equation*}
\mathfrak{L}_{\mathrm{CYM}}\mleft[ H^!A \mright]
=
\mathfrak{L}_{\mathrm{CYM}}[A]
\end{equation*}
for all principal bundle automorphisms $H$.
\end{theorems*}

Finally we will discuss new examples of gauge theory in \ref{CYMExamples}, after pointing out that this gauge theory generalizes the classical formalism. We will also discuss this matter in the context of my previously-mentioned Ph.D.\ thesis, \cite{MyThesis}, and the field redefinitions: Field redefinitions can be thought of as adding multiplicative-exact terms to $\mutot$, \ref{def:FieldRedefsAsDefIntegrated}.

\begin{definitions*}{Field redefinitions on the LGB}
We define \textbf{field redefinitions w.r.t.\ $\lambda \in \Omega^1(M; \mathcal{g})$} of $\mutot$ by
\begin{equation*}
\mleft.\mutot^{\lambda}\mright|_g
\coloneqq
\mleft.\mutot\mright|_g
	+ \sAd_{g^{-1}} \circ \mleft. \pi_{\mathcal{G}}^!\lambda\mright|_g
	- \mleft.\pi_{\mathcal{G}}^!\lambda\mright|_g
\end{equation*}
for all $g \in \caG$
and of $\zeta$ by
\begin{equation*}
\zeta^\lambda
\coloneqq
\zeta
	+ \mathrm{d}^{\nabla^\caG}\lambda + \frac{1}{2} \mleft[ \lambda \stackrel{\wedge}{,} \lambda \mright]_{\mathcal{g}}.
\end{equation*}
\end{definitions*}

It will be easy to see that $\mutot^{\lambda}$ will still be a multiplicative Yang-Mills connection (w.r.t.\ $\zeta^\lambda$) (\ref{prop:YMStaysYM}), and in particular it will lead to the same theory (\ref{thm:EhresmannTransformedEhresmann}):

\begin{theorems*}{Field redefinitions preserving the physics}
We define the \textbf{field redefinitions w.r.t.\ $\lambda \in \Omega^1(M; \mathcal{g})$ of $A$} by
\begin{equation*}
A^\lambda
\coloneqq
A - \pi^!\lambda ~.
\end{equation*}

Then $A^\lambda$ is a connection 1-form w.r.t.\ $\mu^{\lambda}$ and
\begin{equation*}
F^\lambda
\coloneqq
\mathrm{d}^{\pi^*\nabla^{\mathrm{YM},\lambda}} A^\lambda
	+ \frac{1}{2} \mleft[ A^\lambda \stackrel{\wedge}{,} A^\lambda \mright]_{\pi^*\mathcal{g}}
	+ \pi^!\zeta^\lambda
\equiv
F
\end{equation*}
for all $\lambda \in \Omega^1(M; \mathcal{G})$. 
\end{theorems*}

We observe that this is a transformation keeping the Lagrangian invariant such that the dynamics are preserved; $\mathcal{P}$ itself does not change and thus the gauge transformations in form of (local) principal bundle automorphisms also stay the same, and so the kinematics are preserved, too. However, field redefinitions non-trivially transform data like $\mutot$ so that it may become flat, and we may end up at a pre-classical theory describing the same physics. Hence, we want to provide examples which cannot be ``flattened'' in this way: The paper concludes with the reason why this project started, the inner group bundle of the Hopf fibration $\mathbb{S}^7 \to \mathbb{S}^4$ is one of many examples of curved Yang-Mills gauge theory for which $\nabla^\caG$ cannot be flattened; we will present this in \ref{ex:Hopf}.

\begin{examples*}{How to cook new gauge theories}
First an explicit example: Let $P$ be the Hopf bundle
\begin{center}
	\begin{tikzcd}
		\mathrm{SU}(2) \cong \mathbb{S}^3 \arrow{r}	& \mathbb{S}^7 \arrow{d} \\
			& \mathbb{S}^4
	\end{tikzcd}
\end{center}
We define the principal bundle $\mathcal{P}$ and LGB $\mathcal{G}$ as the inner group bundle of $P$
\begin{align*}
\mathcal{P}
\coloneqq
\mathcal{G}
&\coloneqq
c_{\mathrm{SU}(2)}(P)
\coloneqq 
(P\times G) \Big/ G,
\end{align*}
where $(P\times G) / G$ is the associated Lie group bundle, where the action of $G$ on itself is given by its conjugation, $c_g(q) = gqg^{-1}$ for all $g, q \in G$. This principal $c_{\mathrm{SU}(2)}(P)$-bundle admits the structure as curved Yang-Mills gauge theory, which is neither classical nor pre-classical. Furthermore, all possible (curved) Yang-Mills gauge theory structures lead to a curved multiplicative Yang-Mills connection and non-zero $\zeta$, which also implies that this gauge theory cannot be described in the typical formalism of gauge theory.

Similarly, using non-flat principal $G$-bundles $P$ for a Lie group $G$ with no centre leads to curved multiplicative Yang-Mills connections on $c_G(P)$ given by associated connections; those cannot be flattened by field redefinitions. Thus, one quickly achieves curved Yang-Mills gauge theories on other principal $c_G(P)$-bundles like $c_G(P)$ itself, but also $P$ or the Atiyah groupoid of $P$, both of which can also be viewed as principal $c_G(P)$-bundles.
\end{examples*}

The example section will also provide a geometric interpretation using the notion of singular foliations; see \ref{tab:ComparingTheGeometryGaugeTheoryWithSingularFoliations}. This will be based on \cite{Fischer:2401.05966}, which also provides a more general construction than the one in the examples above, making use of outer holonomies; in particular, \cite{Fischer:2401.05966} provides a classification for centerless $G$ with multiplicative Yang-Mills connections which we will reiterate here (\ref{prop:classYMcenterless}). Last, in \ref{conclusions} we will point out how to define associated connections and we summarise how to repeat all those constructions if one replaces $\caG$ by a Lie groupoid over a possible different base manifold; in particular referring to works of us which were uploaded after and motivated by this work.

\subsection{Conclusion}

Since we will argue that a curved Yang-Mills connection which cannot be flattened by field redefinitions only exists globally, one might say that this theory may be just a toy model for physics, but thinking of symplectic geometry, the global aspects may be at least interesting in mathematics; as already mentioned, these connections helped classifying singular foliations and topological symmetry breaking (\cite{Fischer:2401.05966, Fischer:2024hqy}). Alternatively, gauge theory is also a theory about connections, so one could investigate what the sense of connection and curvature given here implies for the multitude of other theories based on connections. More applications will hopefully follow, pointing out the importance of such constructions like the generalised Maurer-Cartan equation.

\subsection{Basic notations and remarks}\label{BasicNotations}

\begin{itemize}
	\item In the main text we usually repeat and reintroduce needed objects for the statements (like "\textbf{Let $M$ be a manifold [...]}" in every statement) (almost) allowing to just read the statements without having read the text introducing it, while the appendix is written as a continuous text which has to be read as a whole in order to understand the essential statements.
	\item As usual, there will be definitions of certain objects depending on other elements, and for keeping notations simple we will not always explicitly denote all dependencies. It will be clear by context on which it is based on, that is for example, if we define an object $A$ using the notion of Lie algebra actions $\gamma$ and we write "Let $X$ be an object $A$", then it will be clear by context which Lie algebra action is going to be used, for example given in a previous sentence writing "Let $\gamma$ be a Lie algebra action".
	\item Throughout this work we always use Einstein's sum convention if suitable.
	\item If not mentioned explicitly, we always assume finite dimensions and morphisms denote base-preserving ones.
	\item With $f^*F$ we denote the pullback/pull-back of the fibre bundles $F \to M$ under a smooth map $f: N \to M$. Similarly we denote the pullbacks of sections of a fibre bundle.
	\item For $V \to M$ a vector bundle over $M$: Do not confuse the pull-back of sections with the pull-back of forms $\omega \in \Omega^l(M; V)$ ($l \in \mathbb{N}_0$), here denoted by $f^!\omega$, which is an element of $\Gamma\left( \mleft(\bigwedge_{m=1}^{l} \mathrm{T}^*N \mright) \otimes f^*V \right) \cong \Omega^l(N; f^*V)$, and not of $\Gamma\left( \mleft(\bigotimes_{m=1}^{l} \mleft(f^*\mathrm{T}M\mright)^*\mright) \otimes f^*V \right)$ like $f^*\omega$. 
	\item Let $F \stackrel{\pi_F}{\to} M$ and $G \stackrel{\pi_G}{\to} N$ be two fibre bundles over smooth manifolds $M$ and $N$, respectively, and let $\phi\colon N \to M$ be a smooth map. Furthermore, let us assume we have a morphism $\Phi\colon G \to F$ of fibre bundles over $\phi$, that is, $\Phi$ is a smooth map such that the following diagram commutes
	\begin{center}
		\begin{tikzcd}
		G \arrow{d}{\pi_G} \arrow{r}{\Phi} & F \arrow{d}{\pi_F}\\
		N \arrow{r}{\phi} & M
		\end{tikzcd}
	\end{center}
especially, $\pi_F \circ \Phi = \phi \circ \pi_G$.
We make often use of that such morphisms have a 1:1 correspondence to \textbf{base-preserving} fibre bundle morphisms $\widetilde{\Phi}\colon G \to \phi^*F$, \textit{i.e.}\ $\widetilde{\Phi}$ is a smooth map with $\phi^*\pi_F \circ \widetilde{\Phi} = \pi_G$. For $p \in N$ the morphism $\widetilde{\Phi}$ has the form
\begin{align*}
\widetilde{\Phi}_p
&=
(p, \Phi_p),
\end{align*}
that is,
\begin{align*}
\widetilde{\Phi}_p(g)
&=
\bigl(p, \Phi_p(g) \bigr)
\end{align*}
for all $g \in G_p$, which is well-defined since $\Phi_p(g) \in F_{\phi(p)}$. The map $\widetilde{\Phi} \mapsto \Phi \coloneqq \mathrm{pr}_2 \circ \widetilde{\Phi}$ is then a bijective map between base-preserving morphisms $G \to \phi^*F$ and morphisms $G \to F$ over $\phi$, where $\mathrm{pr}_2$ is the projection onto the second component. 

In total, $\widetilde{\Phi}$ is a base-preserving morphism if and only if $\Phi$ is a morphism over $\phi$. Observe that $\widetilde{\Phi}$ is an isomorphism (diffeomorphism) if and only if $\Phi$ is a fibre-wise isomorphism (diffeomorphism).

One can extend all of this similarly for more specific types of morphisms like vector bundle-morphisms, and very often we will not mention this 1:1 correspondence explicitly, it should be clear by context. Hence, we will also denote $\widetilde{\Phi}$ by $\Phi$.
%

\item We will often have connections on a bundle, that is, a splitting into a horizontal and vertical bundle of its tangent bundle. When we write "Let $\pi$ be the projection onto the vertical/horizontal bundle", then this will be always w.r.t.\ to the fixed connection even though we are not explicitly mentioning it again.

\item We also need to extend contractions of tensors to graded extensions:
\begin{definitions}{Graded extension of products}{GradingOfProducts}
Let $l \in \mathbb{N}$ and $E_1, \dots E_{l+1} \to M$ be vector bundles over a smooth manifold $M$, and $F \in \Gamma\left( \left(\bigotimes_{m=1}^{l} E_m^*\right) \otimes E_{l+1} \right)$. Then we define the \textbf{graded extension of $F$} as
	\begin{align*}
\Omega^{k_1}(M; E_1) \times \dots \times \Omega^{k_l}(M; E_l)
&\to \Omega^{k}(M; E_{l+1}), \\
(A_1, \dots, A_l)
&\mapsto
F\mleft(A_1\stackrel{\wedge}{,} \dotsc \stackrel{\wedge}{,} A_l\mright),
\end{align*}
where $k := k_1+\dots k_l$ and $k_i \in \mathbb{N}_0$ for all $i\in \{1, \dots, l\}$. $F\mleft(A_1\stackrel{\wedge}{,} \dotsc \stackrel{\wedge}{,} A_l\mright)$ is defined as an element of $\Omega^{k}(M; E_{l+1})$ by
\begin{align*}
&F\mleft(A_1\stackrel{\wedge}{,} \dotsc \stackrel{\wedge}{,} A_l\mright)\mleft(Y_1, \dots, Y_{k}\mright)
\coloneqq \\
&\frac{1}{k_1! \cdot \dots \cdot k_l!} \sum_{\sigma \in S_{k}} \mathrm{sgn}(\sigma) ~ F\left( A_1\left( Y_{\sigma(1)}, \dots, Y_{\sigma(k_1)} \right), \dots, A_l\left( Y_{\sigma(k-k_l+1)}, \dots, Y_{\sigma(k)} \right) \right)
\end{align*}
for all $Y_1, \dots, Y_{k} \in \mathfrak{X}(M)$, where $S_{k}$ is the group of permutations of $\{1, \dots, k\}$ and $\mathrm{sgn}(\sigma)$ the signature of a given permutation $\sigma$. 

$\stackrel{\wedge}{,}$ may be written just as a comma when a zero-form is involved.
%
\end{definitions}

In case of antisymmetric tensors we get of course:

\begin{propositions}{Graded extensions of antisymmetric tensors}{GradedExtensionPlusAntiSymm}
Let $E_1, E_2 \to N$ be real vector bundles of finite rank over a smooth manifold $N$, $F \in \Omega^2(E_1; E_2)$. Then
\begin{equation*}
F \mleft( A \stackrel{\wedge}{,} B \mright)
=
-\mleft( -1 \mright)^{km}
F \mleft( B \stackrel{\wedge}{,} A \mright)
\end{equation*}
for all $A \in \Omega^k(N; E_1)$ and $B \in \Omega^m(N; E_1)$ ($k,m \in \mathbb{N}_0$). Similarly extended to all $F \in \Omega^l(E_1; E_2)$.
\end{propositions}


\begin{proof}
\leavevmode\newline
Trivial, for example by using local coordinates.
\end{proof}

\item References are not only given in the text, the references of referenced statements and definitions are most of the time given below the title of those statements. It will also mention whether the statement as written is a variation or generalization; if it is a strong generalization, then the reference will usually be mentioned in a remark after the statement or its proof.

\end{itemize}

\subsection{Assumed background knowledge}
It is highly recommended to have basic knowledge about differential geometry and gauge theory as presented in \cite[especially Chapter 1 to 5]{Hamilton}, and we will follow the style and labeling as in \cite{Hamilton} when we generalize certain notions; however, sometimes we will still give explicit references to help with more technical details. It can be useful to have knowledge about Lie algebra and Lie group bundles, and even Lie algebroids and Lie groupoids, but we will introduce the basic notions of LGBs and LABs such that it is not necessarily needed to have knowledge about these upfront; but since we will skip the proofs of known statements, it can still be very helpful to have knowledge in those areas.

We often give references about Lie group bundles (LGBs), but the given references are most of the time about Lie groupoids. If the reader has no knowledge about Lie groupoids, then it is important to know that LGBs are a special example of Lie groupoids; Lie groupoids carry ``two projections,'' called \textbf{source} and \textbf{target}. An LGB is a special example of a Lie groupoid whose source equals the target.\footnote{But not every Lie groupoid with equal source and target is an LGB, those are in general bundles of Lie groups which is not completely the same; this nuance will not be important here.} If you look into such a reference, then the source and target are often denoted by $\alpha$ and $\beta$, or by $s$ and $t$, respectively; simply put both to be the same and identify these with our bundle projection which we often denote by $\pi$ or $\pi_{\mathcal{G}}$, where $\mathcal{G}$ is the corresponding LGB. In that way it should be possible to read the references without the need to know Lie groupoids.

\newpage

\section{Needed basics (can be skipped for the experienced reader)}

This section will mostly focus on introducing Lie group and algebra bundles. The experienced reader can skip this section and directly jump to \ref{ConnCurvOnPrincLGBBundle}.

\subsection{Lie group bundles (LGBs)}\label{LGBSection}

\subsubsection{Definition}
\begin{definitions}{Lie group bundle}{LieGroupBundle}
\tcbsubtitle[before skip=\baselineskip]{\cite[\S 1.1, Def.\ 1.1.19; p. 11]{mackenzieGeneralTheory}}
Let $G, \mathcal{G}, M$ be smooth manifolds. A fibre bundle
\begin{center}
	\begin{tikzcd}
		G \arrow{r} & \mathcal{G} \arrow{d}{\pi} \\
		& M
	\end{tikzcd}
\end{center}
is called a \textbf{Lie group bundle (LGB)} if:
\begin{enumerate}
	\item $G$ and each fibre $\mathcal{G}_x \coloneqq \pi^{-1}\mleft( \{x\} \mright)$, $x\in M$, are Lie groups;
	\item there exists a bundle atlas $\mleft\{ \mleft( U_i, \phi_i \mright) \mright\}_{i \in I}$ such that the induced maps
	\begin{align*}
	\phi_{ix}
	&\coloneqq
	\mathrm{pr}_2 \circ \mleft. \phi_i\mright|_{\mathcal{G}_x}: \mathcal{G}_x \to G
	\end{align*}
	are Lie group isomorphisms, where $I$ is an (index) set, $U_i$ are open sets covering $M$, $\phi_i: \mathcal{G}|_U \to U \times G$ subordinate trivializations, and $\mathrm{pr}_2$ the projection onto the second factor. This atlas will be called \textbf{Lie group bundle atlas} or \textbf{LGB atlas}.
\end{enumerate}

We also often say that \textbf{$\mathcal{G}$ is an LGB (over $M$)}, whose structural Lie group is either clear by context or not explicitly needed; and we may also denote LGBs by $G \to \mathcal{G} \stackrel{\pi}{\to} M$ or $\pi \colon \caG \to M$.

The global section $e$ of $\mathcal{G}$ so that $e_p$ is the neutral element of $\mathcal{G}_p$ will be labelled as the \textbf{neutral (element)/identity section (of $\mathcal{G}$)}; we may also denote it by $e^{\mathcal{G}}$.
\end{definitions}

\ifnum\theaccess=1
\begin{remarks}{Principal and Lie group bundles}{LiegroupbundlesNotPrincipalBundles}
Beware, a Lie group bundle is \textbf{not} the same as a principal bundle $P \to M$ with the same fibre type $G$. First of all, the fibres of $P$ are just diffeomorphic to a Lie group, a priori they carry no Lie group structure, while the fibres of $\mathcal{G}$ carry a Lie group structure.
\newline

Second, on $P$ we have a multiplication given as an action of $G$ on $P$
\begin{align*}
P \times G \to P,
\end{align*}
preserving the fibres $P_x$ ($x\in M$) and simply transitive on them. Restricted on $P_x$ we have
\begin{align*}
P_x \times G \to P_x.
\end{align*}
For $\mathcal{G}$ we have canonically a multiplication over $x$ given by
\begin{align*}
\mathcal{G}_x \times \mathcal{G}_x \to \mathcal{G}_x,
\end{align*}
also clearly simply transitive. Observe, the second factor is not "constant", \textit{i.e.}\ we do not have $\mathcal{G}_x \times G \to \mathcal{G}_x$ in general. Hence, there is in general no well-defined product $\mathcal{G} \times \mathcal{G} \to \mathcal{G}$ or $\mathcal{G} \times G \to \mathcal{G}$.
\newline

All of that is also resembled in the existence of sections. The existence of a section of $P$ has a 1:1 correspondence to trivializations of $P$, which is why $P$ in general only admits sections locally; see \textit{e.g.}\ \cite[\S 4.2, Thm.\ 4.2.19; page 219f.]{Hamilton}. $\mathcal{G}$ clearly admits always a global section, even if $\mathcal{G}$ is non-trivial; just take the section which assigns each base point the neutral element of its fibre.
\end{remarks}
\fi

If $M$ is a point we recover the notion of Lie groups, and, as usual, we have the notion of trivial LGBs:

\begin{examples}{Trivial LGB}{TrivialLGBundle}
The \textbf{trivial LGB} is given as the product manifold $M \times G \to M$ with canonical multiplication $(x, g) \cdot (x, q) \coloneqq (x, gq)$.
\end{examples}

In a straightforward manner one defines LGB bundle morphisms:

\begin{definitions}{LGB morphism}{LGB morphism}
\tcbsubtitle[before skip=\baselineskip]{\cite[\S 1.2, special situation of Def.\ 1.2.1 \& 1.2.3, page 12]{mackenzieGeneralTheory}}
Let $\mathcal{H} \stackrel{\pi_{\mathcal{H}}}{\to} N$ and $\pi_{\mathcal{G}} \colon \mathcal{G} \to M$ be two LGBs over two smooth manifolds $N$ and $M$. An \textbf{LGB morphism $F$ over $f$} is a pair of smooth maps $F: \mathcal{H} \to \mathcal{G}$ and $f: N \to M$ such that
\begin{align*}
\pi_{\mathcal{G}} \circ F &= f \circ \pi_{\mathcal{H}},\\
F(gq) &= F(g) ~ F(q)
\end{align*}
for all $g, q \in \mathcal{H}$ with $\pi_{\mathcal{H}}(g) = \pi_{\mathcal{H}}(q)$. We then also say that \textbf{$F$ is an LGB morphism over $f$}. If $N = M$ and $f = \mathrm{id}_M$, then we often omit mentioning $f$ explicitly and just write that \textbf{$F$ is a (base-preserving) LGB morphism}.

We speak of an \textbf{LGB isomorphism (over $f$)} if $F$ is a diffeomorphism.
\end{definitions}

\begin{remarks}{Typical basics for LGB morphisms}{LGBMOrphismRemark}
As usual we have
\begin{itemize}
	\item It is clear that condition 2 in \ref{def:LieGroupBundle} is equivalent to say that $\mathcal{G}$ is locally isomorphic to a trivial LGB; as one may have expected already.

	\item If $F$ is a diffeomorphism, then also $f$, and $F^{-1}$ is an LGB isomorphism, too.
\end{itemize}
\ifnum\theaccess=1
\indent $\bullet$ The right hand side of Eq.\ \eqref{LGBHomomorph} is well-defined because of Eq.\ \eqref{FibreRelationOverf}.

$\bullet$ It is clear that condition 2 in Def.\ \ref{def:LieGroupBundle} is equivalent to say that $\mathcal{G}$ is locally isomorphic to a trivial LGB; as one may have expected already.

$\bullet$ If $F$ is a diffeomorphism, then also $f$: By Eq.\ \eqref{FibreRelationOverf} surjectivity of $f$ is clear; for $y \in M$ just take any $g \in \mathcal{G}_y$, and since $F$ is a bijective, we have a $q \in \mathcal{H}_x$ for some $x\in N$ with $F(q)=g$. By Eq.\ \eqref{FibreRelationOverf} we have $y = \pi_{\mathcal{G}}(F(q)) \stackrel{\eqref{FibreRelationOverf}}{=} f(x)$, thence, surjectivity follows. For injectivity we know by Eq.\ \eqref{LGBHomomorph} and \eqref{FibreRelationOverf} that $F\mleft(e^{\mathcal{H}}_x\mright) = e^{\mathcal{G}}_{f(x)}$, where $e^{\mathcal{H}}_x$ and $e^{\mathcal{G}}_{f(x)}$ denote the unique neutral elements of $\mathcal{H}_x$ and $\mathcal{G}_{f(x)}$, respectively. Assume that there are $x, x^\prime \in N$ with $f(x) = f(x^\prime)$, then we can derive
\begin{align*}
F\mleft(e^{\mathcal{H}}_x\mright)
&=
e^{\mathcal{G}}_{f(x)}
=
e^{\mathcal{G}}_{f(x^\prime)}
=
F\mleft(e^{\mathcal{H}}_{x^\prime}\mright).
\end{align*}
Then we have $e^{\mathcal{H}}_x = e^{\mathcal{H}}_{x^\prime}$ due to that $F$ is bijective, and hence $x = x^\prime$. Therefore $f$ is bijective. Finally, $F^{-1}$ is by assumption also a diffeomorphism, Eq.\ \eqref{LGBHomomorph} clearly carries over, and Eq.\ \eqref{FibreRelationOverf} is w.r.t.\ $f^{-1}$, that is
\begin{align*}
\pi_{\mathcal{H}} \circ F^{-1} &= f^{-1} \circ \pi_{\mathcal{G}}.
\end{align*}
Since $\pi_{\mathcal{H}} \circ F^{-1}$ is smooth and $\pi_{\mathcal{G}}$ is a smooth surjective submersion, it follows that $f^{-1}$ is smooth; this is a well-known fact for right-compositions with surjective submersions, see \textit{e.g.}\ \cite[\S 3.7.2, Lemma 3.7.5, page 153]{Hamilton}. We can conclude that $f$ is a diffeomorphism. Observe that we concluded that $F^{-1}$ is an LGB isomorphism, too.
\fi
\end{remarks}

Similar to the case of Lie groups, \textit{the} example of an LGB are the automorphisms of a vector bundle.

\begin{examples}{Automorphisms of a vector bundle}{AutOfVectorBundleAnLGB}
\tcbsubtitle[before skip=\baselineskip]{\cite[\S 1.1, special situation of Ex.\ 1.1.12, page 8]{mackenzieGeneralTheory}}
Let $V\to M$ be a vector bundle and $\mathrm{Aut}(V) \to M$ its bundle of fibre-wise automorphisms (not to be confused with the sections of $\mathrm{Aut}(V)$ which are the base-preserving automorphisms of $V$). Denote with $W$ the structural vector space of $V$, then $\mathrm{Aut}(V)$ is an LGB with structural Lie group $\mathrm{Aut}(W)$. It is clear that each fibre of $\mathrm{Aut}(V)$ is a Lie group, and the LGB atlas is directly inherited by a vector bundle atlas $\mleft\{ \mleft( U_i, L_i \mright) \mright\}_{i \in I}$ of $V$, where we use a similar notation as for LGB atlases, especially we have vector bundle trivializations $L_{i}: \mleft.V\mright|_{U_i} \to U_i \times W$. Then define an LGB atlas over the same open covering $\mleft( U_i \mright)_i$ by
\begin{align*}
\mleft. \mathrm{Aut}(V) \mright|_{U_i} &\to U_i \times \mathrm{Aut}(W),\\
T &\mapsto \mleft.L_{i} \circ T \circ L_{i}^{-1}\mright|_{\{x\} \times W},
\end{align*}
where $T \in \mleft.\mathrm{Aut}(V)\mright|_x = \mathrm{Aut}(V_x)$,
and $U_i \times \mathrm{Aut}(W)$ acts canonically on $U_i \times W$ in a fibre-wise sense. Then it is trivial to check that these give local trivializations such that $\mathrm{Aut}(V)$ carries the structure as an LGB.
\end{examples}

\subsubsection{Associated Lie group bundles}\label{AssocLGBsSubSection}

For another important example recall that there is the notion of associated fibre bundles and related concepts; following references like \cite[\S1, Construction 1.3.8, page 20]{mackenzieGeneralTheory} and \cite[\S 4.7, page 237ff.; see also Rem.\ 4.7.8, page 242f.]{Hamilton}: Let $P \stackrel{\pi_P}{\to} M$ be a principal bundle with structural Lie group $G$ over a smooth manifold $M$, $N$ another smooth manifold, and $\Psi$ a smooth left $G$-action denoted by
\begin{align*}
G \times N &\to N,\\
(g, v) &\mapsto \Psi(g, v) \coloneqq g \cdot v.
\end{align*}
Then we have a right $G$-action on $P \times N$ given by
\begin{align*}
(P \times N) \times G &\to P \times N,\\
(p,v,g) &\mapsto \mleft( p \cdot g, g^{-1} \cdot v \mright),
\end{align*}
and one can show that the quotient under this action, $P\times_\Psi N \coloneqq ( P \times N) \Big/ G$, yields the structure of a fibre bundle
\begin{center}
	\begin{tikzcd}
		N \arrow{r} & P\times_\Psi N \arrow{d}{\pi_{P\times_\Psi N}} \\
		& M
	\end{tikzcd}
\end{center}
such that the map to the equivalence classes $P \times N \to P \times_\Psi N$, $(p, v) \mapsto [p, v]$, is a smooth surjective submersion,
where the projection $\pi_{P\times_\Psi N}: P\times_\Psi N \to M$ is given by 
\begin{align*}
\pi_{P\times_\Psi N}\mleft( [p, v] \mright)
&\coloneqq
\pi_P(p)
\end{align*}
for all $[p, v] \in P\times_\Psi N$. For $x \in M$, the fibre $\mleft(P\times_\Psi N\mright)_x$ is given by $\mleft( P_x \times N  \mright) \Big/ G = P_x \times_\Psi N$, and the fibre is diffeomorphic to $N$ by $N \ni v \mapsto [p, v] \in \mleft(P\times_\Psi N\mright)_x$ for a fixed $p \in P_x$. We will frequently use this diffeomorphism in the following without further notice.

A very important example are of course associated vector bundles, related to $N$ being a vector space, and $\Psi$ a Lie group representation. Similarly one has associated Lie group bundles: If $G$ acts on another Lie group $H$ by Lie group automorphisms, then $P \times_\psi H$ canonically admits an LGB structure given by
\begin{equation*}
[p, h] \cdot \mleft[p, q\mright]
\coloneqq
\mleft[ p, hq \mright]
\end{equation*}
for all $h, q \in H$ and $p \in P_x$ (that is, $\pi_P(p) = x$). The structural Lie group is given by $H$, and one can check that the identity section is point-wise well-defined by
\begin{equation*}
e_x
=
[p, e]~,
\end{equation*}
where $p \in P_x$ is arbitrary and $e$ is the neutral element of $H$; the inverse of $[p, h] \in \mathcal{H}_x$ is clearly given by
\begin{equation*}
\mleft( [p, h] \mright)^{-1}
=
\mleft[ p, h^{-1} \mright]~.
\end{equation*}

The special situation of $H = G$ is already an important example:

\begin{examples}{Inner group bundle}{InnerLGBs}
\tcbsubtitle[before skip=\baselineskip]{\cite[\S1, paragraph after Def.\ 1.1.19, page 11; comment after Construction 1.3.8, page 20]{mackenzieGeneralTheory}}
The \textbf{inner group bundle} or \textbf{inner LGB} of a principal bundle $P \to M$, denoted by $c_G(P)$, is defined by
\begin{equation*}
c_G(P)
\coloneqq
P \times_{c_G} G,
\end{equation*}
where $c_G: G \times G \to G$ is the left action of $G$ on itself given by \textbf{conjugation}
\begin{equation*}
c_G(g,h)
\coloneqq
c_g (h)
=
\mleft(L_g \circ R_{g^{-1}}\mright)(h)
=
ghg^{-1}
\end{equation*}
for all $g, h \in G$, where we also denote left- and right-multiplications (with $g$) by $L_g$ and $R_g$, respectively. 
%
\end{examples}

\subsubsection{Pullbacks of LGBs}

For this recall that there is the notion of pullbacks of fibre bundles, see \textit{e.g.}\ \cite[\S 4.1.4, page 203ff.; especially Thm.\ 4.1.17, page 204f.]{Hamilton}. That is, if we additionally have a smooth manifold $N$ and a smooth map $f\colon N \to M$, then we have the pullback $f^*\mathcal{G}$ of an LGB $\mathcal{G} \to M$ as a fibre bundle defined as usual by
\begin{equation*}
f^*\mathcal{G}
\coloneqq
N \fibtimes{f}{\pi} \caG
\coloneqq
\left\{
	(x, g) \in N \times \mathcal{G} ~\middle|~
	f(x) = \pi(g)
\right\}.
\end{equation*}
It is an embedded submanifold of $N \times \mathcal{G}$, and the structural fibre is the same Lie group as for $\mathcal{G}$.
That is, the following diagram commutes
\begin{center}
	\begin{tikzcd}
		f^*\mathcal{G} \arrow{d}{\pi_1} \arrow{r}{\pi_2}& \mathcal{G} \arrow{d}{\pi} \\
		N \arrow{r}{f}& M
	\end{tikzcd}
\end{center}
where $\pi_1$ and $\pi_2$ are the projections onto the first and second factor, respectively, of $N \times \mathcal{G}$. Actually, $f^*\mathcal{G}$ carries a natural structure as an LGB.

\begin{corollaries}{Pullbacks of LGBs are LGBs}{PullbackLGB}
\tcbsubtitle[before skip=\baselineskip]{\cite[\S 2.3, simplified situation of the discussion around Prop.\ 2.3.1, page 63ff.]{mackenzieGeneralTheory}}
Let $M, N$ be smooth manifolds, $\mathcal{G} \stackrel{\pi}{\to} M$ an LGB over $M$ and $f\colon N \to M$ a smooth map. Then $f^*\mathcal{G}$ has a unique (up to isomorphisms) LGB structure such that the projection $\pi_2\colon f^*\mathcal{G} \to \mathcal{G}$ onto the second factor is an LGB morphism over $f$ with $\pi_2|_x \colon (f^*\mathcal{G})_x \to \mathcal{G}_{f(x)}$ being a Lie group isomorphism for all $x \in N$.
%
\end{corollaries}

\begin{remarks}{Other references}{referencesPullbackLGB}
The mentioned reference, \cite{mackenzieGeneralTheory}, is rather general, formulated for Lie groupoids. If the reader is only interested into LGBs, then see \textit{e.g.}\ \cite[\S 3, Thm.\ 3.1]{PullbackLGBLAB}.
\end{remarks}

\begin{definitions}{Pullback LGB}{PullbackLGBDef}
Let $M, N$ be smooth manifolds, $\mathcal{G} \to M$ an LGB over $M$ and $f\colon N \to M$ a smooth map. Then we call the LGB structure on $f^*\mathcal{G}$ as given in Cor.\ \ref{cor:PullbackLGB} the \textbf{pullback LGB of $\mathcal{G}$ (under $f$)}. By writing $f^*\mathcal{G}$ we will often refer to this structure without further mention.
\end{definitions}

\subsection{LGB actions, part I}

\subsubsection{Definition}

As for Lie groups, we are interested into their actions: We have an LGB $\mathcal{G} \to M$ over a smooth manifold $M$, and we want to construct an action of $\mathcal{G}$ on another smooth manifold $N$. Each fibre of $\mathcal{G}$ is a Lie group, and we have a notion of Lie groups acting on a manifold $N$. Therefore one could define an LGB action as a collection of Lie group actions, that is, only sections of $\mathcal{G}$ act on $N$; however, one then expects that the general outcome of a product of $\Gamma(\mathcal{G})$ on $N$ would be smooth maps from $M$ to $N$. In order to recover a typical structure of action one could instead introduce a "multiplication rule," i.e.\ each point $p \in N$ can only be multiplied with elements of a specific fibre of $\mathcal{G}$. This "multiplication rule" will be described by a smooth map $f\colon N \to M$ in the sense that only the fibre over $f(p)$ will act on $p$.

\begin{definitions}{Lie group bundle actions}{LiegroupACtion}
\tcbsubtitle[before skip=\baselineskip]{\cite[\S 1.6, special case of Def.\ 1.6.1, page 34]{mackenzieGeneralTheory}}
Let $M, N$ be smooth manifolds, $\pi\colon\mathcal{G} \to M$ an LGB over $M$ and $f\colon N \to M$ a smooth map. Then a \textbf{right-action of $\mathcal{G}$ on $N$} is a smooth map 
\begin{align*}
f^*\mathcal{G} &\to N~,\\
(p, g) &\mapsto p \cdot g~,
\end{align*}
satisfying the following properties:
\begin{align*}
f(p \cdot g) &= \pi(g)~,\\
(p \cdot g) \cdot h &= p \cdot (gh)~,\\
p \cdot e_{f(p)} &= p
\end{align*}
for all $p \in N$ and $g, h \in \mathcal{G}_{f(p)}$, where $e$ is the neutral section of $\mathcal{G}$.
\newline

We analogously define left-actions, and we often write (left or right) \textbf{$\mathcal{G}$-action on $N$}. Furthermore, in order to increase readability as long as the dependency on $f$ is not important, we introduce the notation
\begin{equation*}
N * \mathcal{G}
\coloneqq
f^*\mathcal{G},
\end{equation*}
such that the action's notation has the typical shape $N * \mathcal{G} \to N$; one may also write $N * \mathcal{G} = N \times_M \mathcal{G}$. For left actions similarly $\mathcal{G} * N \to N$; even though $\mathcal{G} * N$ is the same pullback LGB as for right-actions; we also change the order of notation in this case, that is, $(g, p) \in \mathcal{G} * N$ reads $g \in \mathcal{G}_{f(p)}$ and $p \in N$. 
\end{definitions}

\begin{remarks}{Relation to the structure of the canonical pullback $f^*\caG$}{ActionAndPullbackLGBs}
Observe that by the definition of $f^*\mathcal{G}$ we can also write
\begin{align*}
f(p \cdot g) &= f(p),
\end{align*}
so, the $\mathcal{G}$-action is defined in such a way that $f$ is invariant under it. If $f$ is the projection of $N$ as a bundle over $M$, then this means that an LGB action is fibre-preserving. Moreover, the fibre-wise group structure on $\mathcal{G}$ naturally defines a $\mathcal{G}$-action on $\mathcal{G}$; in this situation $f$ would be $\pi$ itself, see also \ref{ex:LGBActingOnItself} later. Furthermore, having $M = \{*\}$ recovers the notion of a Lie group action.
\newline

The other conditions are the typical conditions for actions, especially such that we get a $\mathcal{G}$-action on $f^*\mathcal{G}$ by
\begin{equation*}
(p, g) \cdot q
\coloneqq
\mleft( p \cdot q, q^{-1} g \mright)
\end{equation*}
for all $p \in N$ and $g, q \in \mathcal{G}_{f(p)}$.\footnote{In alignment to \ref{def:LiegroupACtion}, this action is a map $(f \circ \pi_1)^*\mathcal{G} \to f^*\mathcal{G}$, where $\pi_1$ is the projection onto the first factor in $f^*\mathcal{G}$.}
As usual, this gives rise to an equivalence relation, whose set of equivalence classes denoted by $f^*\mathcal{G} \Big/ \mathcal{G}$ is isomorphic to $N$ (as a set) by $[p, g] \mapsto p \cdot g$, where we denote equivalence classes of $(p, g) \in f^*\mathcal{G}$ by $[p, g]$.
\end{remarks}

\begin{remarks}{Localizing LGB actions}{LocalLGBAction}
We can localize the LGB action in the following sense: Fix any open neighbourhood $U$ of $M$. Then $f^{-1}(U)$ is an open neighbourhood of $N$, and we can restrict the action to $f^{-1}(U)$, resulting into a map
\begin{align*}
\mleft.(N*\mathcal{G})\mright|_{f^{-1}(U)} \to f^{-1}(U),
\end{align*}
because of $f(p \cdot g) = \pi(g) = f(p)$, that is, if $(p, g) \in \mleft.(N*\mathcal{G})\mright|_{f^{-1}(U)}$, then $p \in f^{-1}(U)$ and so $p \cdot g \in f^{-1}(U)$. In fact, by the definition of $N*\mathcal{G}$ as $f^*\mathcal{G}$, this describes a $\mathcal{G}|_{U}$-action on $f^{-1}(U)$
\begin{align*}
f^{-1}(U) * \mleft.\mathcal{G}\mright|_{U} \coloneqq \mleft( f|_{f^{-1}(U)}\mright)^*(\mathcal{G}|_U) = \mleft.(N*\mathcal{G})\mright|_{f^{-1}(U)} &\to f^{-1}(U).
\end{align*}
If $x \in M$ is a regular value of $f$, then $f^{-1}(\{x\})$ is an embedded submanifold of $N$ due to the regular value theorem. In that case one could apply the same arguments to restrict the action on $f^{-1}(\{x\})$. Since $\mathcal{G}_x$ is a Lie group this describes a Lie group action on $f^{-1}(\{x\})$.
\end{remarks}


One can probably see that it is straightforward to extend a lot of the typical notions of Lie group actions to LGB actions; hence, we mainly focus on the definitions and properties which we need in this paper. 

\begin{definitions}{Left and right translations}{LRTranslations}
\tcbsubtitle[before skip=\baselineskip]{\cite[\S 1.4, special situation of Def.\ 1.4.1 and its discussion, page 22]{mackenzieGeneralTheory}}
Let $M, N$ be smooth manifolds, $\mathcal{G} \stackrel{\pi}{\to} M$ an LGB over $M$ and $f \colon N \to M$ a smooth map. Furthermore assume that we have a right action $N * \mathcal{G} \to N$. We define the \textbf{right translation} over $x \in M$ with $g \in \mathcal{G}_x$ as a map $r_g$ defined by
\begin{align*}
f^{-1}(\{x\}) &\to f^{-1}(\{x\}),\\
p &\mapsto p \cdot g,
\end{align*}
and we define the \textbf{orbit map} through $p \in f^{-1}(\{x\})$ as a map $\Phi_p$ given by
\begin{align*}
\mathcal{G}_x &\to N,\\
g&\mapsto p \cdot g.
\end{align*}
For $\sigma \in \Gamma(\mathcal{G})$ we define the \textbf{right translation} on $N$ as a map $r_\sigma$ by
\begin{align*}
N &\to N,\\
p &\mapsto p \cdot \sigma_{f(p)}.
\end{align*}

If one has a section of $f$, \textit{i.e.}\ a smooth map $\tau: M \to N$, $x \mapsto \tau_x$, with $f \circ \tau = \mathrm{id}_M$, then we can define the \textbf{orbit map} through $\tau$ as a map $\Phi_\tau$ given by
\begin{align*}
\mathcal{G} &\to N,\\
g &\mapsto \tau_{\pi(g)} \cdot g.
\end{align*}
\end{definitions}

\begin{remarks}{Left action and translation}{LeftTranslation}
Similarly we define left translations for left actions, which we similarly denote by $l_g$ and $l_\sigma$. By \ref{rem:LocalLGBAction} we can define $r_\sigma$ (and $l_\sigma$) also for local sections $\sigma \in \Gamma(\mathcal{G}|_U)$ by restricting $N$ onto $f^{-1}(U)$, where $U$ is some open subset of $M$; then $r_\sigma, l_\sigma: f^{-1}(U) \to f^{-1}(U)$. In the same manner one achieves a restriction for $\Phi_\tau \colon \mathcal{G}|_U \to f^{-1}(U)$, if $\tau: U \to f^{-1}(U)$.

In case of $N$ being $\mathcal{G}$ itself we will introduce right (and left) translations later via capital letters.
\end{remarks}

\begin{remarks}{Group action on sections}{SectionMultiplication}
Assume that $N$ is a fibre bundle over $M$ with $f$ as its projection. Also observe that $\Gamma(\mathcal{G})$ is clearly a group and it may be possible to endow it with an infinite-dimensional Lie group structure (but this is not important here now). In the same fashion as before, we can define a $\Gamma(\mathcal{G})$-action on $\Gamma(N)$ given by
\begin{align*}
\mleft(\tau \cdot \sigma\mright)_x
&\coloneqq
\tau_x \cdot \sigma_x
\end{align*}
for all $\tau \in \Gamma(N)$, $\sigma \in \Gamma(\mathcal{G})$, and $x \in M$. It is straightforward to show that this is well-defined.
\end{remarks}

\begin{remarks}{Smoothness of translations and orbit maps}{SmoothnessOfACtionTranslations}
It is a straightforward exercise to show that $\Phi_p$ and $\Phi_\tau$ are smooth, and that $r_\sigma$ and $l_\sigma$ are diffeomorphisms with inverse $r_{\sigma^{-1}}$ and $l_{\sigma^{-1}}$, respectively; the same holds for $r_g$ and $l_g$ if $x$ is a regular value of $f$, or more general, if $f^{-1}(\{x\})$ is an embedded submanifold.
\end{remarks}

Motivated by the previous remark, it might be useful to require that $f$ is a submersion, or that $N$ is actually some bundle over $M$ and $f$ its projection. In fact, this will be the case later.

%
%

\subsubsection{Examples of LGB actions}

If $M$ is a point or $f$ a constant map, then we recover the typical notion of a Lie group action acting on $N$.
Additionally, we have the following examples. The notation will be as in \ref{def:LRTranslations}.

\begin{examples}{LGB acting on itself}{LGBActingOnItself}
Each LGB $\mathcal{G} \stackrel{\pi}{\to} M$ acts on itself from the left and right, having $N \coloneqq \mathcal{G}$ and $f \coloneqq \pi$,
\begin{align*}
\mathcal{G} * \mathcal{G} \coloneqq \pi^*\mathcal{G} &\to \mathcal{G},\\
(g,h) &\mapsto gh.
\end{align*}
That this satisfies all properties for an LGB action is clear by definition of LGBs; and using the localisation of this action and the local trivialisation of $\caG$ it is clearly smooth.
We will also call the $\mathcal{G}$-action on itself the \textbf{multiplication in $\mathcal{G}$}. 

Similarly one can argue that the \textbf{inverse map} $\mathcal{G} \to \mathcal{G}$, $g \mapsto g^{-1}$, is smooth, too.
\end{examples}

\begin{examples}{Trivial action}{TrivialAction}
\tcbsubtitle[before skip=\baselineskip]{\cite[\S 1.6, special situation of Ex.\ 1.6.3, page 35]{mackenzieGeneralTheory}}
The projection $\pi_1$ onto the first factor of $f^*\mathcal{G} \stackrel{\pi_1}{\to} N$ satisfies the properties of a right $\mathcal{G}$-action on $N$, that is, the action is given by
\begin{align*}
N * \mathcal{G} &\to N,\\
(p, g) &\mapsto p \cdot g \coloneqq p.
\end{align*}
That this action satisfies the properties of an action for all $f$ is trivial, hence we call it the \textbf{trivial action}.
\end{examples}

\begin{examples}{Actions of trivial LGBs}{TrivialLGBAction}
Assume that $\mathcal{G}$ is trivial, that is $\mathcal{G} \cong M \times G$ as LGBs, where $G$ is the structural Lie group of $\mathcal{G}$. In that case, for $p \in N$, the product $p \cdot q$ is only defined if $q \in \mathcal{G}$ is of the form $(f(p), g)$, where $g \in G$. We also have
\begin{align*}
f^*\mathcal{G}
&\cong
N \times G,
\end{align*}
the trivial $G$-LGB over N. Hence, let us define a map by
\begin{align*}
N \times G &\to N,\\
(p, g) &\mapsto p \cdot g \coloneqq p \cdot (f(p), g),
\end{align*}
which is clearly a smooth map since it is a composition of the $\mathcal{G}$-action on $N$ and
\begin{align*}
N \times G &\to f^*\mathcal{G},\\
(p, g) &\mapsto (p, f(p), g).
\end{align*}
The latter is smooth because $(p, g) \mapsto (p, f(p), g)$ is a smooth map $N \times G \to N \times M \times G$ and $f^*\mathcal{G}$ is an embedded submanifold of $N \times M \times G$. Using \ref{def:LiegroupACtion}, it is trivial to see that $p \cdot e = p$, and
\begin{align*}
p \cdot (gh)
&=
p \cdot \underbrace{(f(p), gh)}_{\mathclap{= (f(p), g) \cdot (f(p), h)}}
=
\bigl( p \cdot (f(p), g) \bigr) \cdot (f(p), h)
=
(p \cdot g) \cdot h
\end{align*}
for all $g,h \in G$. Hence, we have a $G$-action on $N$, and by construction it is equivalent to the $\mathcal{G}$-action on $N$. 

Observe, that one can therefore recover the notion of Lie group actions not only via $M = \{*\}$, the point manifold, (or equivalently via constant maps $f$) but also via trivial LGBs. Translations with constant sections of $M \times G$ w.r.t.\ the action $N*\mathcal{G}\to N$ are trivially to be seen as translations with an element of $G$ w.r.t.\ the action $N \times G \to N$.

Regarding general $\caG$, not just trivial ones: Due to the discussion in \ref{rem:LocalLGBAction} we can thence conclude that every $\mathcal{G}$-action is locally a typical Lie group action. If $f$ is a submersion, then the action is also a $G$-action on each fibre of $f$.
\end{examples}

Hence, if one wants something "truly" new regarding LGB actions, then one has to look at global structures of non-trivial LGBs and their actions. In fact, the following type of example can provide such a new structure, which we will understand later once we have introduced the physical theory.

\begin{examples}{Inner group bundle acting on associated fibre bundles}{AssocLGACtingOnAssocVec}
\tcbsubtitle[before skip=\baselineskip]{\cite[\S 1.6, simplified version of Ex.\ 1.6.4, page 35]{mackenzieGeneralTheory}}
Let $P \stackrel{\pi_P}{\to} M$ be a principal $G$-bundle with structural Lie group $G$ over a smooth manifold $M$, and recall \ref{ex:InnerLGBs}. Furthermore, let $F$ be another smooth manifold, equipped with a smooth left $G$-action $\Psi \colon G \times F \to F$. In total we have two associated bundles over $M$: 
\begin{center}
	\begin{tikzcd}
		G \arrow{r} & c_G(P) \arrow{d}{\pi_{c_G(P)}} && F \arrow{r} & \mathcal{F} \coloneqq P \times_\Psi F \arrow{d}{\pi_{\mathcal{F}}} \\
		& M & && M
	\end{tikzcd}
\end{center}
the inner group bundle of $P$ and an associated $F$-bundle, respectively. Then we have a left $c_G(P)$-action on $\mathcal{F}$ given by
\begin{align*}
c_G(P) * \mathcal{F}
\coloneqq
\pi_{\mathcal{F}}^*c_G(P) &\to \mathcal{F},\\
\bigl( \mleft[ p, g \mright], \mleft[ p, v \mright] \bigr)
&\mapsto
\mleft[ p, \Psi \mleft(g, v\mright) \mright]
=
\mleft[ p \cdot g, v \mright]
\end{align*}
for all $p \in P_x$ ($x \in M$), $g \in G$ and $v \in F$. Recall that the group of sections of $c_G(P)$ is isomorphic to the group of principal bundle automorphisms of $P$ as a group. In other words, this action is coming from the well-known action of gauge transformations on associated bundles; see for example \cite[\S 5.3]{Hamilton}.
\end{examples}

\subsection{Lie algebra bundles (LABs)}

\subsubsection{Definition}

Lie algebras are the infinitesimal version of Lie groups, hence, we expect something similar for LGBs, the Lie algebra bundles:

\begin{definitions}{Lie algebra bundle (LAB}{LAB}
\tcbsubtitle[before skip=\baselineskip]{\cite[\S 3.3, Definition 3.3.8, page 104]{mackenzieGeneralTheory}}
Let $\mathfrak{g}$ be a Lie algebra, and $\mathcal{g}, M$ be smooth manifolds. A vector bundle
\begin{center}
	\begin{tikzcd}
	\mathfrak{g} \arrow{r} & \mathcal{g} \arrow{d}{\pi} \\
	& M
	\end{tikzcd}
\end{center}
is called a \textbf{Lie algebra bundle (LAB)} if:
\begin{enumerate}
	\item $\mathfrak{g}$ and each fibre $\mathcal{g}_x$, $x \in M$, are Lie algebras;
	\item there exists a bundle atlas $\mleft\{ \mleft( U_i, \phi_i \mright) \mright\}_{i \in I}$ such that the induced maps
	\begin{align*}
	\phi_{ix}
	&\coloneqq
	\mathrm{pr}_2 \circ \mleft. \phi_i\mright|_{\mathcal{g}_x}: \mathcal{g}_x \to \mathfrak{g}
	\end{align*}
	are Lie algebra isomorphisms, where $I$ is an (index) set, $U_i$ are open sets covering $M$, $\phi_i: \mathcal{g}|_U \to U \times \mathfrak{g}$ subordinate trivializations, and $\mathrm{pr}_2$ the projection onto the second factor. This atlas will be called \textbf{Lie algebra bundle atlas} or \textbf{LAB atlas}.
\end{enumerate}
We often say that \textbf{$\mathcal{g}$ is an LAB (over $M$)}, whose structural Lie algebra is either clear by context or not explicitly needed; and we may also denote LABs by $\mathfrak{g} \to \mathcal{g} \stackrel{\pi}{\to} M$.
\end{definitions}

Of course, we have the typical trivial examples:

\begin{examples}{Trivial examples}{TrivialLABs}
We recover the notion of a Lie algebra, if $M$ consist of just one point. Moreover, the \textbf{trivial LAB} is given as the product manifold $\mathcal{g} \coloneqq M \times \mathfrak{g} \to M$. We have obviously a canonical smooth field of Lie brackets on this bundle $\mleft[ \cdot, \cdot \mright]_{\mathcal{g}}: \Gamma(\mathcal{g}) \times \Gamma(\mathcal{g}) \to \Gamma(\mathcal{g})$, \textit{i.e.}\ $\mleft[ \cdot, \cdot \mright]_{\mathcal{g}} \in \Gamma\mleft(\bigwedge^2 \mathcal{g}^* \otimes \mathcal{g} \mright)$ which restricts to the Lie algebra bracket $\mleft[ \cdot, \cdot \mright]_{\mathfrak{g}}$ of $\mathfrak{g}$ on each fibre. The bracket is given by
\begin{align*}
\mleft[ (x, X), (x, Y) \mright]_{\mathcal{g}}
&\coloneqq
\mleft( x, \mleft[ X, Y \mright]_{\mathfrak{g}} \mright)
\end{align*}
for all $(x, X), (x, Y) \in M \times \mathfrak{g}$. Smoothness is an immediate consequence.
\end{examples}

The definition of LAB morphisms is straight-forward:

\begin{definitions}{LAB morphism}{LABmorphism}
\tcbsubtitle[before skip=\baselineskip]{\cite[\S. 4.3, simplified version of Def.\ 4.3.1, page 158]{mackenzieGeneralTheory}}
Let $\pi_{\mathcal{G}} \colon \mathcal{G} \to M$ and $\mathcal{h} \stackrel{\pi_{\mathcal{h}}}{\to} N$ be two LGBs over two smooth manifolds $M$ and $N$. An \textbf{LAB morphism} is a pair of smooth maps $F \colon \mathcal{h} \to \mathcal{g}$ and $f \colon N \to M$ such that
\begin{align*}
\pi_{\mathcal{g}} \circ F &= f \circ \pi_{\mathcal{h}},\\
F &\text{ linear},\\
F\mleft(\mleft[g, q\mright]_{\mathcal{h}_p}\mright) &= \mleft[F(g), F(q)\mright]_{\mathcal{g}_{f(p)}}
\end{align*}
for all $g, q \in \mathcal{h}_p$ ($p \in N$), where $\mleft[\cdot, \cdot\mright]_{\mathcal{h}_p}$ and $\mleft[\cdot, \cdot\mright]_{\mathcal{g}_{f(p)}}$ are Lie brackets of $\mathcal{h}_p$ and $g_{f(p)}$, respectively. We also say that \textbf{$F$ is an LAB morphism over $f$}. If $N = M$ and $f = \mathrm{id}_M$, then we often omit mentioning $f$ explicitly and just write that \textbf{$F$ is a (base-preserving) LAB morphism}. We speak of an \textbf{LAB isomorphism (over $f$)} if $F$ is a diffeomorphism.
\end{definitions}

\begin{remarks}{Smooth field of Lie brackets}{FieldOfLieBrackets}
We have similar remarks as in \ref{rem:LGBMOrphismRemark}. Additionally, we have locally a canonical smooth field of Lie brackets which restricts to a Lie bracket on each fibre because every LAB is locally isomorphic to a trivial LAB as in \ref{ex:TrivialLABs}. Define a field of Lie brackets $\mleft[ \cdot, \cdot \mright]_{\mathcal{g}}: \Gamma(\mathcal{g}) \times \Gamma(\mathcal{g}) \to \Gamma(\mathcal{g})$, \textit{i.e.}\ $\mleft[ \cdot, \cdot \mright]_{\mathcal{g}} \in \Gamma\mleft(\bigwedge^2 \mathcal{g}^* \otimes \mathcal{g} \mright)$, by
\begin{equation*}
\mleft[ X, Y \mright]_{\mathcal{g}}
\coloneqq
\mleft[ X, Y \mright]_{\mathcal{g}_x}
\end{equation*}
for all $X, Y \in \mathcal{g}_x$ ($x \in M$). Using a local trivialization, this bracket is locally of the form as in \ref{rem:LGBMOrphismRemark} such that smoothness follows. In fact, as also argued in \cite[\S 16.2, Example 2, page 114]{DaSilva}, every vector bundle equipped with a smooth field of Lie brackets is an LAB.
\end{remarks}

Endomorphisms of a vector bundle are of course another important example of LABs.

\begin{examples}{Endomorphisms of a vector bundle}{EndVAnLAB}
\tcbsubtitle[before skip=\baselineskip]{\cite[\S 3.3, part of Ex.\ 3.3.4]{mackenzieGeneralTheory}}
Let $V \to M$ be a vector bundle, and denote with $\mathrm{End}(V) \to M$ its bundle of fibre-wise endomorphisms (its sections are the base-preserving bundle endomorphisms of $V$). This is clearly an LAB whose field of Lie brackets is given by the commutator.
\end{examples}

As we have seen for LGBs, the pullback of LABs is again an LAB.

\begin{corollaries}{Pullbacks of LABs are LABs}{PullBackLABIsLAB}
\tcbsubtitle[before skip=\baselineskip]{\cite[\S 3, Thm.\ 3.2]{PullbackLGBLAB}}
Let $M, N$ be smooth manifolds, $\mathcal{g} \stackrel{\pi}{\to} M$ an LAB over $M$ and $f \colon N \to M$ a smooth map. Then the pullback vector bundle $f^*\mathcal{g}$ has a unique (up to isomorphisms) LAB structure such that the projection $\pi_2\colon f^*\mathcal{g} \to \mathcal{g}$ onto the second factor is an LAB morphism over $f$ with $\pi_2|_x\colon (f^*\mathcal{g})_x \to \mathcal{g}_{f(x)}$ being a Lie algebra isomorphism for all $x \in N$.
%
\end{corollaries}


\begin{definitions}{Pullback LAB}{PullbackLABDef}
Let $M, N$ be smooth manifolds, $\mathcal{g} \to M$ an LGB over $M$ and $f: N \to M$ a smooth map. Then we call the LAB structure on $f^*\mathcal{g}$ as given in \ref{cor:PullBackLABIsLAB} the \textbf{pullback LAB of $\mathcal{g}$ (under $f$)}.
\end{definitions}

By writing $f^*\mathcal{g}$ we will often refer to this structure without further mention.

\subsubsection{From LGBs to LABs}

Let us now quickly discuss how LGBs and LABs are related; it is very similar to the relation of Lie groups and algebras, now somewhat fibre-wise. We will follow the style of \cite[\S 1.5.2, page 40ff.]{Hamilton}\ and \cite[\S 3.5, page 119ff.]{mackenzieGeneralTheory}; our approach will be using left-invariant vector fields but the mentioned latter reference actually uses right-invariant vector fields.

Let us start with introducing the basic notations; in contrast to the discussion around \ref{def:LRTranslations} we will denote the following maps with capital letters when referring to $\caG$ acting on itself.

\begin{definitions}{Left and right translation and conjugation}{LeftRightTranslationConjugation}
Let $\mathcal{G} \to M$ be an LGB over a smooth manifold $M$. For $g \in \mathcal{G}_x$ ($x \in M$) we define the following maps:
\begin{itemize}
	\item \textbf{Left translation} given by
		\begin{align*}
			L_g: \mathcal{G}_x &\to \mathcal{G}_x,\\
			h &\mapsto g h.
		\end{align*}
	\item \textbf{Right translation} given by
		\begin{align*}
			R_g: \mathcal{G}_x &\to \mathcal{G}_x,\\
			h &\mapsto h g.
		\end{align*}
	\item \textbf{Conjugation} given by
		\begin{align*}
			c_g: \mathcal{G}_x &\to \mathcal{G}_x,\\
			h &\mapsto g h g^{-1}.
		\end{align*}
\end{itemize}
\end{definitions}

\begin{remarks}{Conjugation by sections}{conjugSect}
\tcbsubtitle[before skip=\baselineskip]{\cite[\S 1.4, Def.\ 1.4.6 and its discussion afterwards, page 24f.]{mackenzieGeneralTheory}}
By definition of $\mathcal{G}$, all these maps are smooth. Furthermore, they clearly satisfy the typical properties as known for these maps since $\mathcal{G}_x$ is a Lie group for all $x \in M$ In particular, while the conjugation $c_g$ is a Lie group automorphism of $\mathcal{G}_x$, it describes an LGB automorphism of $\mathcal{G}$ if extended to sections. That is for $\sigma \in \Gamma(\mathcal{G})$ we define the conjugation $c_\sigma$ as a smooth map by
\begin{align*}
\mathcal{G} &\to \mathcal{G},\\
q &\mapsto c_\sigma(q) \coloneqq \mleft(L_\sigma \circ R_{\sigma^{-1}}\mright)(q) = \mleft( R_{\sigma^{-1}} \circ L_\sigma \mright)(q) = \sigma_{\pi(q)} \cdot q \cdot \sigma_{\pi(q)}^{-1}.
\end{align*}
It is clear that $c_\sigma(gq) = c_\sigma(g) \cdot c_\sigma(q)$ for all $g, q \in \mathcal{G}$ with $\pi(g) = \pi(q)$, and that a smooth inverse is given by $c_{\sigma^{-1}}$; thence, $c_\sigma$ is an LGB isomorphism of $\mathcal{G}$ on itself, an automorphism, in sense of \ref{def:LGB morphism}. It is also trivial to check that we have $c_{\sigma \cdot \tau} = c_\sigma \circ c_\tau$, where $\tau$ is another section of $\mathcal{G}$. Here we defined $R_\sigma$ as $r_\sigma$ as in \ref{def:LRTranslations}. Similarly for left translations.
\end{remarks}

Since these are diffeomorphism of the fibres, it makes sense to say that a left-invariant vector field of $\mathcal{G}$ has to be a vertical vector field, that is, it is in the kernel of $\mathrm{D}\pi$, the total differential/tangent map of the projection of $\mathcal{G} \stackrel{\pi}{\to} M$. For this recall that there is the notion of a \textbf{vertical bundle} for fibre bundles $F \stackrel{\varpi}{\to} M$, which is defined as a subbundle $\mathrm{VF} \to F$ of the tangent bundle $\mathrm{T}F \to F$ given as the kernel of $\mathrm{D}\varpi \colon \mathrm{T}F \to \mathrm{T}M$. The fibres $\mathrm{V}_v F$ of $\mathrm{V}F$ at $v \in F$ are then given by 
\begin{align*}
\mathrm{V}_v F
&=
\mathrm{T}_v F_x,
\end{align*}
where $x \coloneqq \varpi(v) \in M$ and $F_x$ is the fibre of $F$ at $x$. $F_x$ is an embedded submanifold of $F$, thence, by definition a section $X \in \Gamma(\mathrm{V}F)$ restricts to a vector field on the fibres, that is, 
\begin{align*}
X|_{F_x} \in \mathfrak{X}(F_x).
\end{align*}

In our case $F = \mathcal{G}$, hence $F_x = \mathcal{G}_x$ is a Lie group, so, the vertical bundle just consists of the tangent bundles of Lie groups of all fibres. All of these are generated by their Lie algebra at $e_x$, the identity element of $\mathcal{G}_x$. Hence, it is natural to guess that the LAB for $\mathcal{G}$ will be $\mathrm{V}\mathcal{G}|_{e_M}$, where $e_M$ is the image of $M$ under the identity/neutral section of $\mathcal{G}$, thus, an embedding of $M$ into $\mathcal{G}$. $\mathrm{V}\mathcal{G}|_{e_M}$ is a fibre bundle, and clearly a vector bundle. Equivalently, since the identity section $e$ is an embedding, we think of $\mathrm{V}\mathcal{G}|_{e_M}$ as the pullback vector bundle $e^*\mathrm{V}\mathcal{G}$, which is conveniently a vector bundle over $M$. That is, the construction of an LAB out of an LGB is precisely the same as in the Lie group setting: Restricting left-invariant vector fields to the identity. For all $\nu_x \in \rmV_{e_x}\caG$ ($x \in M$) we define its induced left-invariant vector field $X_\nu$ as a tangent vector of $\caG_x$ given by
\begin{equation*}
\mleft.X_\nu\mright|_g
\coloneqq
\mathrm{D}_{e_x}L_g \mleft( \nu_x \mright)
\end{equation*}
for all $g \in \mathcal{G}_x$. Similarly for sections of $e^*\rmV\caG$, giving rise to a left-invariant vertical vector field on $\caG$.

\begin{definitions}{The LAB of an LGB}{LABOfAnLGB}
\tcbsubtitle[before skip=\baselineskip]{\cite[\S 3.5, special situation of Def.\ 3.5.1, page 120]{mackenzieGeneralTheory}}
Let $\mathcal{G} \to M$ be an LGB over a smooth manifold $M$, and denote with $e$ the identity section of $\mathcal{G}$. Then we define the \textbf{LAB $\mathcal{g}$ of $\mathcal{G}$} as the vector bundle $e^*\mathrm{V}\mathcal{G}$ with bracket 
\begin{equation*}
\mleft[ \mu_x, \nu_x \mright]_{\cag}
\coloneqq
\mleft.\mleft[ X_\mu, X_\nu \mright]\mright|_{e_x}
\end{equation*}
for all $x \in M$ and $\mu_x, \nu_x \in \rmV_{e_x}\caG$.
\end{definitions}

In the following $\mathcal{g}$ will usually denote the LAB of $\mathcal{G}$.

\begin{examples}{Endomorphisms as LAB of fibre-wise automorphisms}{EndosAreLABOfAutos}
\tcbsubtitle[before skip=\baselineskip]{E.g.\ canonically extend the proof for why the Lie algebra of automorphisms are endomorphisms of a vector space as in \cite[\S 1.5.4, page 45ff.]{Hamilton}}
Recall \ref{ex:AutOfVectorBundleAnLGB} and \ref{ex:EndVAnLAB}. For a vector bundle $V \to M$ one can show that the LAB of $\mathrm{Aut}(V)$ is given by $\mathrm{End}(V)$. 
\end{examples}

\begin{examples}{Associated LAB as LAB of associated LGB}{AssociatedLABsFromAssocLGBs}
\tcbsubtitle[before skip=\baselineskip]{Similar to \cite[\S 3.1, Prop.\ 3.1.4, page 90]{mackenzieGeneralTheory}}
Recall associated LGBs of \ref{AssocLGBsSubSection}. So, let $G, H$ be Lie groups, $P \to M$ a principal $G$-bundle over a smooth manifold $M$, and $\psi$ a $G$-action on $H$ by Lie group isomorphisms. Then we have the associated LGB $\mathcal{H} \coloneqq P \times_\psi H$.

$\psi_* \coloneqq \mleft[ g \mapsto \mathrm{D}_{e_G}\psi_g \mright]$ is the induced $G$-representation on $\mathfrak{h}$ by Lie algebra isomorphisms, the Lie algebra of $H$, where $e_G$ is the neutral element of $G$. Hence, we also have the associated bundle $\mathcal{h} \coloneqq P \times_{\psi_*} \mathfrak{h}$ which is naturally the LAB of $\mathcal{H}$.
%
%
As a special example, the LAB of the inner group bundle $c_G(P)$, \ref{ex:InnerLGBs}, is the \textbf{adjoint bundle $\mathrm{Ad}(P)$ of $P$}, $\mathrm{Ad}(P) \coloneqq P \times_{\mathrm{Ad}} \mathfrak{g}$, where $\mathrm{Ad}$ is the adjoint representation of $G$.
\end{examples}

\begin{examples}{LAB of pullback LGB is pullback LAB}{LABOfPullbackLGBIsPullbackLAB}
\tcbsubtitle[before skip=\baselineskip]{\cite[\S 3, Thm.\ 3.5, page 21]{PullbackLGBLAB}}
Let $\mathcal{G} \to M$ be an LGB over a smooth manifold $M$, and let $f\colon N \to M$ be a smooth map defined on another smooth manifold $N$. Then the LAB of $f^*\mathcal{G}$ is isomorphic to the pullback LAB $f^*\mathcal{g}$.
\end{examples}

Either directly or with the previous example it is straightforward to show the following:

\begin{corollaries}{Differentials of LGB morphisms are LAB morphisms}{LGBToLABHomomorphis}
\reference{\cite[\S 3.5, section about morphisms, page 124f.]{mackenzieGeneralTheory}}
Let $\mathcal{H} \to N$ and $\mathcal{G} \to M$ be two LGBs over two smooth manifolds $N$ and $M$, and we denote with $\mathcal{h}$ and $\mathcal{g}$ the LABs of $\mathcal{H}$ and $\mathcal{G}$, respectively. Furthermore, assume that we have an LGB morphism $F: \mathcal{H} \to \mathcal{G}$ over a smooth map $f: N \to M$. Then
\begin{align*}
\mleft.\mathrm{D} F\mright|_{\mathcal{h}}: \mathcal{h} &\to \mathcal{g}
\end{align*}
is an LAB morphism over $f$.
\end{corollaries}

\subsubsection{Vertical Maurer-Cartan form of LGBs}

Naturally, we have the canonical Maurer-Cartan form on each fibre; as a 1-form on the total space of $\caG$ it is only defined on the vertical bundle $\rmV\caG$ and it is a straight-forward exercise to show smoothness of the following:

\begin{definitions}{Vertical Maurer-Cartan form of LGBs}{MCFormOnLGBs}
\tcbsubtitle[before skip=\baselineskip]{Straight-forward generalisation of the Maurer-Cartan form.}
Let $\mathcal{G} \stackrel{\pi}{\to} M$ be an LGB over a smooth manifold $M$. We define the \textbf{vertical Maurer-Cartan form $\MC$} as an element of $\Gamma\mleft( \mathrm{V}^*\mathcal{G} \otimes \pi^*\mathcal{g} \mright)$ given by
\begin{align*}
\mleft(\MC\mright)_g(v)
&\coloneqq
\mleft( \mathrm{D}_g L_{g^{-1}} \mright)(v)
\end{align*}
for all $g \in \mathcal{G}$ and $v \in V_g\mathcal{G}$,
where $\mathrm{V}^*\mathcal{G}$ is the dual bundle of $\mathrm{V}\mathcal{G}$.
\end{definitions}

Recall \ref{BasicNotations}, we have a 1:1 correspondence of $\MC$ to the following commuting diagram
\begin{center}
	\begin{tikzcd}
		\mathrm{V}\mathcal{G} \arrow{d} \arrow{r}{\MC} & \mathcal{g} \arrow{d}\\
		\mathcal{G} \arrow{r}{\pi} & M
	\end{tikzcd}
\end{center}
which is the same diagram as in \cite[\S 3.5, special situation of Prop.\ 3.5.3, page 121]{mackenzieGeneralTheory}.
Since $\MC$ restricts to the Maurer-Cartan form of the fibres, this induces a canonical isomorphism:

\begin{corollaries}{Vertical tangent space of $\mathcal{G}$}{TLGBAsLGB}
\tcbsubtitle[before skip=\baselineskip]{\cite[\S 3.5, a reformulation of Proposition 3.5.3, page 121]{mackenzieGeneralTheory}}
Let $\mathcal{G} \stackrel{\pi}{\to} M$ be an LGB over a smooth manifold $M$. Then we have an isomorphism of vector bundles
\begin{align*}
\mathrm{V}\mathcal{G} &\cong \pi^*\mathcal{g}.
\end{align*}
\end{corollaries}

Observe that we know by \ref{cor:PullBackLABIsLAB} that $\mathrm{V}\mathcal{G}$ admits a unique LAB structure such that $\mathrm{V}\mathcal{G} \cong \pi^*\mathcal{g}$ is an isomorphism of LABs. By $e^*\rmV\caG = \cag$, this implies that $\mathrm{V}\mathcal{G}$ is trivial if and only if $\mathcal{g}$ is trivial; as also argued in \cite[\S 3.5, discussion after Cor.\ 3.5.4, page 121]{mackenzieGeneralTheory}.
Last but not least, sections of $\mathrm{V}\mathcal{G}$ are therefore generated by sections of $\mathcal{g}$, the left-invariant vector fields.


\subsubsection{Exponential map of LGBs}\label{ExponentialMapSubsection}


Similarly there is a natural exponential map, extending the one defined in each fibre.

\begin{definitions}{Exponential map}{ExpOfLGB}
\tcbsubtitle[before skip=\baselineskip]{\cite[\S 3.6, second part of Example 3.6.2, page 133f.]{mackenzieGeneralTheory}}
Let $\mathcal{G} \to M$ be an LGB over a smooth manifold $M$, and $\mathcal{g} \to M$ its LAB. Then we define the \textbf{exponential map $\exp \colon \mathcal{g} \to \mathcal{G}$} by
\begin{align*}
\exp(X) &\coloneqq \e^X \coloneqq \exp_{\mathcal{G}_x}(X)
\end{align*}
for all $x \in M$ and $X \in \mathcal{g}_x$, where $\exp_{\mathcal{G}_x}$ is the exponential map of the Lie group $\mathcal{G}_x$. Its extension to sections, also denoted by $\exp\colon \Gamma(\mathcal{g}) \to \Gamma(\mathcal{G})$, is canonically given by
\begin{align*}
\mleft.\exp(\nu)\mright|_x
&\coloneqq
\exp(\nu_x)
\end{align*}
for all $x \in M$ and $\nu \in \Gamma(\mathcal{g})$. Usually, the LGB is given by context, otherwise we will denote the exponential map by $\exp_{\mathcal{G}}$ instead.
\end{definitions}

The exponential map is smooth because it describes the flow of left-invariant vector fields as it also happens for Lie groups. Recall that we denote left-invariant vector fields by $X_\nu$ where $\nu \in \Gamma(\mathcal{g})$.

\begin{corollaries}{The exponential map as flow}{ExpAsFlow}
\tcbsubtitle[before skip=\baselineskip]{\cite[discussion at the beginning of \S 3.6, Prop.\ 3.6.1 and its discussion afterwards; page 132f.]{mackenzieGeneralTheory}}
Let $\mathcal{G} \stackrel{\pi}{\to} M$ be an LGB over a smooth manifold $M$, and $\mathcal{g} \to M$ its LAB. Then the flow of a left-invariant vector field $X_\nu \in L(\mathcal{G})$ ($\nu \in \Gamma(\mathcal{g})$) is a complete flow $\phi: \mathbb{R} \times \mathcal{G} \to \mathcal{G}$ given by
\begin{equation*}
\phi(t, g)
=
g \cdot \e^{t \nu_{\pi(g)}}
\end{equation*}
for all $t \in \mathbb{R}$ and $g \in \mathcal{G}$. 
Especially, the map
\begin{align*}
\mathbb{R} \times M &\to \mathcal{G},\\
(t, x) &\mapsto \e^{t\nu_x}
\end{align*}
is smooth.
\end{corollaries}
%

\begin{remarks}{Simplifying notation related to the exponential map}{SimplyExpNotation}
Due to \ref{def:ExpOfLGB} we recover a lot of the typical properties of the exponential map, if we understand these properties point-wise. 
As in \cite[\S 3.6, discussion after Prop.\ 3.6.1, page 133]{mackenzieGeneralTheory}, we say that $\mathbb{R} \ni t \mapsto \e^{t\nu}$ is smooth in the sense of the second statement of \ref{cor:ExpAsFlow}, and by construction we have
\begin{equation*}
\nu_x
=
\mleft.\frac{\mathrm{d}}{\mathrm{d}t}\mright|_{t=0} \e^{t\nu_x}
\end{equation*}
for all $x\in M$, so that we also write
\begin{equation*}
\nu
=
\mleft.\frac{\mathrm{d}}{\mathrm{d}t}\mright|_{t=0} \e^{t\nu}.
\end{equation*}
\end{remarks}

\subsection{LGB actions, part II}\label{LGBActionIISection}

Finally, we come to the last part of the \textit{basics} for LGBs and their notions which we require.

\subsubsection{LGB and LAB representations}

As usual, representations are a special type of group action, with an infinitesimal analogue. 

\begin{definitions}{LGB representations}{LGBRep}
\reference{\cite[\S 1.7, special situation of the remark before Def.\ 1.7.1, page 43]{mackenzieGeneralTheory}}
Let $\mathcal{G} \stackrel{\pi}{\to} M$ be an LGB over a smooth manifold $M$, $V \stackrel{p}{\to} M$ be a vector bundle, and assume that we have a left $\mathcal{G}$-action on $V$ denoted by $\Psi\colon \mathcal{G}*V \coloneqq p^*\mathcal{G}\to V$. Then we say that $\Psi$ is a \textbf{$\mathcal{G}$-representation on $V$} if it is linear, that is,
\begin{align*}
\Psi(g, \alpha v )
&=
\alpha \Psi(g, v)~,\\
\Psi(g, v + w)
&=
\Psi(g, v) + \Psi(g, w )
\end{align*}
for all $\alpha \in \mathbb{R}$, and $(g, v), (g, w) \in \mathcal{G}*V$. In alignment with previous notations we may also write $\Psi(g, v) = \Psi_g(v)$, or $\Psi(g, v) = g \cdot v$.
\end{definitions}


By fixing a base point $x \in M$ we clearly have the typical notion of a $\mathcal{G}_x$-representation on $V_x$. Equivalently, we could therefore obviously define LGB representations as a base-preserving LGB morphism $\Psi: \mathcal{G} \to \mathrm{Aut}(V)$ as also in \cite[\S 1.7, Def.\ 1.7.1, page 43]{mackenzieGeneralTheory}.

%
%

\begin{examples}{Representations of trivial LGBs}{LGRepsOnVecs}
Similarly as to \ref{ex:TrivialLGBAction}, if $\mathcal{G} = M \times G$ is a trivial LGB over $M$, $G$ its structural Lie group, and $V = M \times W$ a trivial vector bundle with structural vector space $W$, then $\Psi$ is equivalent to a $G$-representation on $W$. Vice versa, every Lie group representation gives a representation of trivial LGBs on trivial vector bundles.
\end{examples}

Also recall Ex.\ \ref{ex:AutOfVectorBundleAnLGB}: We can similarly construct another LGB needed for the adjoint representation.

\begin{examples}{Automorphisms of LABs as LGB}{AutosOfLABsAsLGB}
\reference{\cite[\S 1.7, special situation of Ex.\ 1.7.12, page 46]{mackenzieGeneralTheory}}
Let $\mathcal{g}$ be an LAB over the smooth manifold $M$. We denote with $\mathrm{Aut}(\mathcal{g}) \to M$ the bundle of fibre-wise Lie algebra automorphisms of $\mathcal{g}$, where the sections of $\mathrm{Aut}(\mathcal{g})$ are the base-preserving LAB automorphisms of $\mathcal{g}$. As in \ref{ex:AutOfVectorBundleAnLGB} one can show that $\mathrm{Aut}(\mathcal{g})$ is an LGB by using LAB trivializations of $\mathcal{g}$ instead of vector bundle trivializations.

The notation of $\mathrm{Aut}(\mathcal{g})$ may be confusing with the notation for vector bundle automorphisms. We usually refer to LAB automorphism when speaking of LABs, and state it explicitly if we just mean the vector bundle version.
\end{examples}

Recall the notation introduced in \ref{def:LeftRightTranslationConjugation}:

\begin{examples}{Adjoint LGB representation}{LGBAdjointRep}
\reference{\cite[\S 3.5, special situation of Prop.\ 3.5.20, page 131]{mackenzieGeneralTheory}}
For $g \in \mathcal{G}_x$ ($x \in M$) we have the conjugation $c_g$ which is a $\mathcal{G}_x$-automorphism. In the usual way we define the \textbf{adjoint representation of $\mathcal{G}$} as a base-preserving LGB morphism $\mathcal{G} \to \mathrm{Aut}(\mathcal{g})$ given by
\begin{align*}
\mathrm{Ad}_g &\coloneqq \mathrm{D}_{e_x} c_g
\end{align*}
for all $g \in \mathcal{G}_x$ ($x \in M$), which restrict fibre-wise to the typical adjoint representation (\ref{ex:LGRepsOnVecs}).
\end{examples}

Infinitesimally, we have LAB actions and representations; recall Cor.\ \ref{cor:PullBackLABIsLAB}.

\begin{definitions}{LAB actions}{LABACtions}
\reference{\cite[\S 4.1, reformulated version for LABs of Def.\ 4.1.1, page 149]{mackenzieGeneralTheory}}
Let $M, N$ be smooth manifolds, $\mathcal{g} \stackrel{\pi}{\to} M$ an LAB, and $f\colon N \to M$ a smooth map. Then a \textbf{$\mathcal{g}$-action on $N$} is a base-preserving vector bundle morphism 
\begin{align*}
f^*\mathcal{g} &\to \mathrm{T}N~,\\
\nu &\mapsto \rho(\nu)~,
\end{align*}
satisfying
\begin{equation*}
\mathrm{D}f \circ \rho
=
0
\end{equation*}
and such that the induced map
\begin{align*}
\Gamma(\mathcal{g}) &\to \Gamma(f^*\mathcal{g}) \to \mathfrak{X}(N)~,\\
\mu &\mapsto f^*\mu \mapsto \rho\mleft( f^*\mu \mright)
\end{align*}
is a homomorphism of Lie algebras, that is,
\begin{equation*}
\rho\biggl( f^*\mleft(\mleft[ \mu, \nu \mright]_{\mathcal{g}}\mright) \biggr)
=
\bigl[ \rho\mleft(f^*\mu\mright), \rho\mleft(f^*\nu\mright) \bigr]
\end{equation*}
for all $\mu, \eta \in \Gamma(\mathcal{g})$.
\end{definitions}


As for Lie group and algebra actions, an LGB action induces an LAB action.

\begin{lemmata}{LGB actions induce LAB actions}{LGBInduceLABAction}
\reference{\cite[\S 4.1, special situation of Thm.\ 4.1.6, page 152]{mackenzieGeneralTheory}}
Let $M, N$ be smooth manifolds, $\mathcal{G} \stackrel{\pi}{\to} M$ an LGB over $M$ and $f\colon N \to M$ a smooth map. Then any right $\mathcal{G}$-action $\Phi\colon N * \mathcal{G} = f^*\mathcal{G} \to N$ on $N$ induces a $\mathcal{g}$-action $\rho$ on $N$ by 
\begin{align*}
\rho
&\coloneqq
\mleft.\mathrm{D}\Phi\mright|_{f^*\mathcal{g}}~,
\end{align*}
i.e.\
\begin{align*}
\rho(\eta)
&=
\mathrm{D}_{\mleft(p, e_{f(p)}\mright)}\Phi(\eta)
\end{align*}
for all $\eta \in (f^*\mathcal{g})_p$ ($p\in N$).
\end{lemmata}

\begin{remarks}{Induced LAB action by left LGB actions}{LeftActionsAndTheirSignProblem}
Similarly to the discussion as in \cite[\S 3.4, page 141ff.]{Hamilton}, one can show the same for left actions, but one has to use the multiplication with the inverse, that is, 
\begin{equation*}
\rho
\coloneqq
\mleft.\mathrm{D}\Phi^\prime\mright|_{f^*\mathcal{g}}~,
\end{equation*}
where
\begin{align*}
\Phi^\prime: \mathcal{G} * N &\to N~,\\
(g, p) &\mapsto g^{-1} \cdot p~.
\end{align*}
\end{remarks}

\begin{remarks}{LAB actions as fibrewise actions}{LABHomomorpInActionVariants}
If $f$ is a submersion, then one can think of the LAB action as a ``collection'' of Lie algebra actions on the fibres $f^{-1}(\{x\})$ ($x \in M$).
\end{remarks}


Of course, by restricting to linear vector fields one achieves representations.

\begin{definitions}{LAB representations}{LABReps}
\reference{\cite[\S 3.3, Def.\ 3.3.13, page 107]{mackenzieGeneralTheory}}
Let $\mathcal{g} \to M$ be an LAB over a smooth manifold $M$, and $V \to M$ be a vector bundle. A \textbf{$\mathcal{g}$-representation on $V$} is an LAB morphism $\psi \colon \mathcal{g} \to \mathrm{End}(V)$.
\end{definitions}

\begin{corollaries}{LAB representations are specific LAB actions}{LABRepsAreLABActions}
\reference{\cite[\S 4.1, special consequence of Prop.\ 4.1.7, page 153]{mackenzieGeneralTheory}}
Let $\mathcal{g} \to M$ be an LAB over a smooth manifold $M$, $V \stackrel{p}{\to} M$ be a vector bundle, and $\psi \colon \mathcal{g} \to \mathrm{End}(V)$ a $\mathcal{g}$-representation on $V$. Then $\psi$ defines an LAB action $\overline{\psi} \colon p^*\mathcal{g} \to \mathrm{T}V$ by
\begin{equation*}
\overline{\psi}(v, \nu)
\coloneqq
-\psi(\nu)(v)
\end{equation*}
for all $(v, \nu) \in p^*\mathcal{g}$, where one makes use of the identification $\mathrm{T}_vV_x = \mathrm{V}_vV \cong V_x$ ($x \coloneqq p(v)$).
\end{corollaries}
%
%

This is a known relationship in the case of Lie algebra representations on vector spaces and their associated actions; usually this is proven by assuming a Lie group representation, however, one can show it without assuming integrability of the Lie algebra action. See for example \cite[\S 2.1, proof of Prop.\ 2.1.16, page 22]{MyThesis} and extend this to LABs.

Of course, the adjoint representation is an important example inherited by an LGB representation.

\begin{examples}{Adjoint LAB representations}{LABAdjointRep}
\reference{\cite[\S 3.3, special situation of Ex.\ 3.3.15, page 108]{mackenzieGeneralTheory}}
Let us assume the same situation as in \ref{ex:LGBAdjointRep} for the adjoint representation of the LGB $\mathcal{G}\to M$, $\mathrm{Ad}\colon \mathcal{G} \to \mathrm{Aut}(\mathcal{g})$. As discussed earlier, its infinitesimal version is a $\mathcal{g}$-representation on itself, the \textbf{adjoint representation $\mathrm{ad}$ of $\mathcal{g}$}. By construction,
\begin{align*}
\mathrm{ad}_\nu(\mu)
&\coloneqq
\mathrm{ad}(\nu)(\mu)
=
\mleft[ \nu, \mu \mright]_{\mathcal{g}_x}
\end{align*}
for all $\nu, \mu \in \mathcal{g}_x$ ($x\in M$).
\end{examples}

\subsubsection{Fundamental vector fields}

The LAB actions inherited by LGB actions are also called fundamental vector fields; we are following \cite[\S 3.4, generalization of Def.\ 3.4.1, page 143]{Hamilton} for the labelling:

\begin{definitions}{Fundamental vector fields}{FundVecs}
Let $M$ and $N$ be two smooth manifolds, $\mathcal{G} \to M$ an LGB over $M$, $f\colon N \to M$ a smooth map, and assume we have a right $\mathcal{G}$-action on $N$, denoted as $\Phi\colon N * \mathcal{G} \to N$. For $\nu \in \Gamma(\mathcal{g})$ we define its induced \textbf{fundamental vector field $\overline{\nu}$} as an element of $\mathfrak{X}(N)$ by
\begin{equation*}
\overline{\nu}_p
\coloneqq
\mathrm{D}_{e_{f(p)}}\Phi_p\mleft(\nu_{f(p)}\mright)
\end{equation*}
for all $p \in N$, where $\Phi_p$ is the orbit map defined in \ref{def:LRTranslations}. For left actions we define fundamental vector fields similarly by
\begin{equation*}
\overline{\nu}_p
\coloneqq
\mathrm{D}_{e_{f(p)}}\Phi^\prime_p\mleft(\nu_{f(p)}\mright)
\end{equation*}
for all $p \in N$,
where $\Phi^\prime_p$ is a slightly adjusted orbit map given by
\begin{align*}
\mathcal{G}_{f(p)} &\to N~,\\
g &\mapsto g^{-1} \cdot p~.
\end{align*}
\end{definitions}

\begin{remarks}{Fundamental vectors pointwise}{FundVecsNotations}
Again, point-wise over $x \coloneqq f(p)$ this is just the typical definition of a fundamental vector field with respect to $\nu_x \in \mathcal{g}_x$ (except that $f^{-1}(\{x\})$ may not be a manifold). Hence, one has also a point-wise definition which we will also denote similarly by $\overline{\nu_x}$. If $f^{-1}(\{x\})$ is not a manifold, then $\mleft.\overline{\nu_x}\mright|_p$ is just a formal notation, and it only defines an element of $\mathrm{T}_pN$ which may not be related to a vector field tangent to $f^{-1}(\{x\})$, not even to a vector field on $N$ if $\nu_x$ does not formally come from a fixed section of $\mathcal{g}$.

However, if $f$ is \textit{e.g.}\ the projection of a bundle, then we have a $\mathcal{G}$-action on the fibre $f^{-1}(\{x\})$ as manifold, and therefore $\overline{\nu}|_{f^{-1}(\{x\})}$ and $\overline{\nu_x}$ give rise to a vector field on $f^{-1}(\{x\})$. In other words, fundamental vector fields are vertical vector fields as expected and are fibre-wise fundamental vector fields coming from a Lie group action.
%
\end{remarks}

\begin{remarks}{Map to fundamental vector fields an LAB action}{FundVecsAreLABActions}
As anticipated, this is related to the LAB action coming from the right $\mathcal{G}$-action $\Phi: N*\mathcal{G} \to N$. The following can also be shown for left actions in a similar manner by recalling \ref{rem:LeftActionsAndTheirSignProblem}.

By \ref{lem:LGBInduceLABAction} we have an LAB action $\rho\colon f^*\mathcal{g} \to \mathrm{T}N$ given by $\mleft.\rho \coloneqq \mathrm{D}\Phi\mright|_{f^*\mathcal{g}}$. We have
\begin{align*}
\rho(p, \nu)
&=
\mleft. \frac{\mathrm{d}}{\mathrm{d}t} \mright|_{t=0} \Phi\mleft( p, \e^{t\nu} \mright)
=
\mleft. \frac{\mathrm{d}}{\mathrm{d}t} \mright|_{t=0} \mleft( p \cdot \e^{t\nu} \mright)
=
\mathrm{D}_{e_{f(p)}}\Phi_p(\nu)
=
\overline{\nu}_p
\end{align*}
for all $(p, \nu) \in f^*\mathcal{g}$, where $t \in \mathbb{R}$, and $\e$ is the exponential map of $\mathcal{G}$. Thus, the map to fundamental vector fields
\begin{align*}
\Gamma(\mathcal{g}) &\mapsto \mathfrak{X}(N),\\
\nu &\mapsto \overline{\nu},
\end{align*}
is equivalent to the map
\begin{align*}
\Gamma(\mathcal{g}) &\to \mathfrak{X}(N),\\
\nu &\mapsto \rho\mleft( f^*\nu \mright)
\end{align*}
induced by $\rho$ as in \ref{def:LABACtions}, which also implies that the map to the fundamental vector fields is a homomorphism of Lie algebras. Last but not least, by \ref{def:LABACtions} it follows that fundamental vector fields are in the kernel of $\mathrm{D}f$.
\end{remarks}

\subsubsection{Differential of smooth LGB actions}

We will often need to differentiate LGB actions, and thus we have to understand the tangent space of pullback manifolds:

\begin{lemmata}{Tangent bundle of pullback fibre bundles}{PullbackFibreBundleItsTangentSp}
Let $M$ and $N$ be two smooth manifolds, $F \stackrel{\pi}{\to} M$ a fibre bundle over $M$, and $f\colon N \to M$ a smooth map. Then we have for its tangent spaces
\begin{align*}
\mathrm{T}_{(p, v)}\mleft( f^*F \mright)
&=
\bigl\{
	(X, Y)
	~\big|~
	X \in \mathrm{T}_pN, Y\in \mathrm{T}_vF \text{ with } \mathrm{D}_pf(X) = \mathrm{D}_v\pi(Y)
\bigr\}
=
\rmT N \fibtimes{\rmD f}{\rmD \pi} \rmT F
\end{align*}
for all $(p, v) \in f^*F$.
\end{lemmata}

This is a straight-forward exercise to show by making use of the fact that $f$ and $\pi$ are transverse to each other.
%
%
With this we can finally show the following proposition; also recall the notations introduced in \ref{def:LRTranslations}, \ref{def:LeftRightTranslationConjugation}, \ref{def:MCFormOnLGBs} and \ref{def:FundVecs}.

\begin{propositions}{Differential of smooth LGB actions}{DiffOfLGBAction}
Let $M$ and $N$ be smooth manifolds, $\pi_\scG \colon \mathscr{G} \to M$ an LGB over $M$, $\Phi \colon N \to M$ a smooth map, and assume we have a right $\mathscr{G}$-action on $N$, $\Psi\colon N * \scG \to N$. Then we have
\begin{align*}
\mathrm{D}_{(p, g)}\Psi(X, Y)
&=
\mathrm{D}_pr_\sigma(X)
	+ \mleft.{\overline{\mleft( \MC\mright)_g \bigl(Y - \mathrm{D}_{x}\sigma (\omega)\bigr)}}\mright|_{p \cdot g}
\end{align*}
for all $(p, g) \in N \fibtimes\Phi{\pi_\scG} \scG$ and $(X, Y) \in \rmT N \fibtimes{\rmT\Phi}{\rmT\pi_\scG} \rmT \scG$, where $x \coloneqq \Phi(p) = \pi_\scG(g)$, $\sigma$ is any (local) section of $\mathscr{G}$ with $\sigma_{x} = g$, and $\omega$ is an element of $\mathrm{T}_xM$ given by
\begin{align*}
\omega
&\coloneqq
\mathrm{D}_p \Phi(X)
= 
\mathrm{D}_g\pi_\scG(Y)~.
\end{align*}
\end{propositions}


\begin{proof}
Due to $N*\mathcal{G} = f^*\mathcal{G}$ we are going to use \ref{lem:PullbackFibreBundleItsTangentSp}. That is, fix $(p, g) \in N*\mathcal{G}$ and $X \in \mathrm{T}_pN$, $Y \in \mathrm{T}_g \mathcal{G}$ with
\begin{align*}
\mathrm{D}_p f(X) &= \mathrm{D}_g\pi(Y) \eqqcolon \omega \in \mathrm{T}_{f(p)}M.
\end{align*}
Recall that we can localize LGB actions in sense of \ref{rem:LocalLGBAction}; so, let $x \coloneqq f(p) = \pi(g)$, and fix a trivialization of $\mathcal{G}$ around $x$. Then it is clear that there is a (local)\footnote{For simplicity of notation we omit the notation of restricting on some open subset of $M$.} section $\sigma \in \Gamma(\mathcal{G})$ with $\sigma_x = g$. Observe that we can write
\begin{align*}
Y
&=
Y 
	- \mathrm{D}_x \sigma (\omega)
	+ \mathrm{D}_x \sigma (\omega).
\end{align*}
The two first summands result into a vertical tangent vector due to
\begin{align*}
\mathrm{D}_g\pi\bigl( Y - \mathrm{D}_x\sigma(\omega) \bigr)
&=
\mathrm{D}_g\pi( Y ) - \mathrm{D}_x\underbrace{(\pi \circ \sigma)}_{= \mathrm{id}_M}(\omega)
=
\mathrm{D}_g \pi (Y) - \omega
=
0.
\end{align*}
So, we write 
\begin{align*}
Y^v
&\coloneqq
Y - \mathrm{D}_x \sigma(\omega)
\end{align*}
We have proven that $Y^v \in \mathrm{V}_g\mathcal{G}$, and thus $(0_p, Y^v) \in \mathrm{T}_{(p, g)}(f^*\mathcal{G})$ by \ref{lem:PullbackFibreBundleItsTangentSp}, where $0_p$ is the zero tangent vector of $\mathrm{T}_pN$. In the same fashion we also have
\begin{align*}
\mleft(X,  \mathrm{D}_x \sigma (\omega) \mright)
&\in \mathrm{T}_{(p, g)}(f^*\mathcal{G})
\end{align*}
because of
\begin{align*}
\mathrm{D}_g\pi \Bigl( \mathrm{D}_x \sigma (\omega)  \Bigr)
&=
\omega
=
\mathrm{D}_g \pi (Y)
=
\mathrm{D}_p f (X),
\end{align*}
thence we can write in $\mathrm{T}_{(p, g)}(f^*\mathcal{G})$
\begin{align*}
(X, Y)
&=
\mleft(X, \mathrm{D}_x \sigma (\omega) + Y^v \mright)
=
\mleft(X, \mathrm{D}_x \sigma (\omega) \mright)
	+ (0_p, Y^v).
\end{align*}
Hence also
\begin{align}\label{DifferentialOfActionSplitFirst}
\mathrm{D}_{(p, g)}\Phi(X, Y)
&=
\mathrm{D}_{(p, g)}\Phi\mleft(X, \mathrm{D}_x \sigma (\omega) \bigr) \mright)
	+ \mathrm{D}_{(p, g)}\Phi(0_p, Y^v)
\end{align}
The second summand is quickly calculated as
\begin{align*}
\mathrm{D}_{(p, g)}\Phi(0_p, Y^v)
&=
\mleft.\frac{\mathrm{d}}{\mathrm{d}t}\mright|_{t=0} \Phi(p, \gamma)
=
\mleft.\frac{\mathrm{d}}{\mathrm{d}t}\mright|_{t=0} \underbrace{(p \cdot \gamma)}
	_{\Phi_p(\gamma)}
=
\mathrm{D}_{g}\Phi_p (Y^v)
\end{align*}
where $\gamma: I \to \mathcal{G}_x$ ($I$ open interval of $\mathbb{R}$ containing 0) is a curve with $\gamma(0) = g$ and $\mathrm{d}/\mathrm{d}t|_{t=0} \gamma = Y^v$; this is due to the verticality of $Y^v$, and so $\mathrm{D}_g\Phi_p(Y^v)$ is also well-defined since $\Phi_p$ is a map $\mathcal{G}_x \to N$. Now recall \ref{def:MCFormOnLGBs} and \ref{def:FundVecs}, and then observe that we can write
\begin{align}\label{ClassicalWayToWriteLeibnizRuleWithLeftPushForwardInsteadOfMCForm}
\mathrm{D}_{g}\Phi_p (Y^v)
&=
\mleft(\mathrm{D}_{g}\Phi_p \circ \mathrm{D}_{e_x}L_g \circ \mathrm{D}_g L_{g^{-1}} \mright)(Y^v)
=
\mathrm{D}_{e_x}\underbrace{\mleft( \Phi_p \circ L_g \mright)}
	_{\mathclap{ \mathcal{G} \ni q \mapsto p \cdot gq = \Phi_{p \cdot g}(q) }}
	\mleft(\mleft( \MC\mright)_g (Y^v) \mright)
=
\mleft.{\overline{\mleft( \MC \mright)_g (Y^v)}}\mright|_{p \cdot g}
\end{align}
making use of the verticality of $Y^v$ such that operators like $\mathrm{D}L_g$ can act on $Y^v$. For the first summand in Eq.\ \eqref{DifferentialOfActionSplitFirst} we use
\begin{align*}
\mathrm{D}_{(p, g)}\Phi\bigl(X, \mathrm{D}_x \sigma (\omega) \bigr)
&=
\mathrm{D}_{(p, g)}\Phi\mleft(X, \mathrm{D}_x \sigma \mleft( \rmD_pf(X) \mright) \mright)
\\
&=
	\rmD_p \ubrace[0]{\mleft( \Phi \circ \mleft( {{\rm id}}_N, \sigma_f \mright) \mright)}{p \mapsto p \cdot \sigma_{f(p)} = r_\sigma(p)}(X)
\\
&=
\mathrm{D}_pr_\sigma(X)~.
\end{align*}
So, we get in total
\begin{align*}
\mathrm{D}_{(p, g)}\Phi(X, Y)
&=
\mathrm{D}_pr_\sigma(X)
	+ \mleft.{\overline{\mleft( \MC\mright)_g (Y^v)}}\mright|_{p \cdot g}~.
\end{align*}
\end{proof}

\section{Connections and curvature on principal LGB-bundles}\label{ConnCurvOnPrincLGBBundle}

\subsection{Principal bundles with structural LGB}\label{PrincBundlLGBBased}

\subsubsection{Definition}

The principal bundle $\mathcal{P}$ we are interested into is still a fibre bundle related to the same Lie group as the one behind the LGB $\mathcal{G}$, especially also $\mathrm{dim}(\mathcal{P}) = \mathrm{dim}(\mathcal{G})$, but it is equipped with an LGB action; we will now follow \cite[\S 5.7, page 144f.]{GroupoidBasedPrincipalBundles}.

\begin{definitions}{Principal bundles with structural LGB}{PrinciBdleWithStruLGB}
\reference{\cite[simplification of the beginning of \S 5.7, page 144f.]{GroupoidBasedPrincipalBundles}}
Let $G$ be a Lie group, $M$ a smooth manifold, and an LGB $\mathcal{G} \to M$ which acts on the right on another fibre bundle $\mathcal{P} \to M$ with structural fibre $G$
\begin{center}
	\begin{tikzcd}
	G \arrow{r} & \mathcal{P} \arrow{d}{\pi_{\mathcal{P}}} & \arrow[bend right]{l} \mathcal{G} \arrow{d}{\pi_{\mathcal{G}}} & \arrow{l} G \\
	& M & M
	\end{tikzcd}
\end{center}
where the right-action is defined on $\mathcal{P} * \mathcal{G}$ given by $\pi_{\mathcal{P}}^*\mathcal{G}$.
Then we call $\mathcal{P}$ a \textbf{principal $\mathcal{G}$-bundle} if 
\begin{enumerate}
	\item The right $\mathcal{G}$-action on $\mathcal{P}$ \textbf{is simply transitive on the fibres}, that is, the restriction of $\mathcal{P}*\mathcal{G} \to \mathcal{P}$ to $x\in M$
	\begin{equation*}
	\mathcal{P}_x \times \mathcal{G}_x \to \mathcal{P}_x
	\end{equation*}
	induces bijective orbit maps
	\begin{align*}
	\mathcal{G}_x &\to \mathcal{P}_x,
	\\
	g &\mapsto p \cdot g
	\end{align*}
	for $p \in \mathcal{P}_x$.
	\item There exist base-preserving \textbf{$\mathcal{G}$-equivariant diffeomorphisms $\varphi_i: \mathcal{P}|_{U_i} \to \mathcal{G}|_{U_i}$}	subordinate to an open covering $\mleft( U_i \mright)_i$ of $M$, that is,
	\begin{align*}
	\pi_{\mathcal{G}} \circ \varphi_i &= \pi_{\mathcal{P}},
	\\
	\varphi_i(p \cdot g)
	&=
	\varphi_i(p) \cdot g,
	\end{align*}
	where the multiplication on the right hand side is inherited by $\mathcal{G} * \mathcal{G} \to \mathcal{G}$, the multiplication on $\mathcal{G}$.
	%
\end{enumerate}

The LGB $\mathcal{G}$ is the \textbf{structural LGB} of the principal bundle $\mathcal{P}$.
\end{definitions}

\begin{remarks}{Discussion about the definition of $\mathcal{G}$-principal bundles}{LGBPrincDefDiscussion}
There are several things in need to be discussed:
\begin{enumerate}
	\item The mentioned reference, \cite{GroupoidBasedPrincipalBundles}, introduces these principal bundles in a different manner; we will come back later to this in \ref{rem:AlternativePrincBdlDef}.
	\item By \ref{rem:ActionAndPullbackLGBs} the $\mathcal{G}$-action on $\mathcal{P}$ is fibre-preserving, thus, the restriction of $\mathcal{P}*\mathcal{G} \to \mathcal{P}$ to $x \in M$ is indeed a map $\mathcal{P}_x \times \mathcal{G}_x \to \mathcal{P}_x$.
	\item The $\mathcal{G}$-equivariant diffeomorphisms $\varphi_i$ give rise to a \textbf{bundle atlas of $\mathcal{G}$-equivariant bundle charts $\xi_i\colon \mathcal{P}|_{U_i} \to U_i \times G$}. Assume w.l.o.g.\ that $U_i$ is small enough so that there is an LGB chart $\phi_i\colon \mathcal{G}|_{U_i} \to U_i \times G$ of $\mathcal{G}$; recall \ref{def:LieGroupBundle}. Then $\xi_i \coloneqq \phi_i \circ \varphi_i$, which clearly satisfies
	\begin{align*}
	\xi_i(p \cdot g)
	&=
	\xi_i(p) \cdot \phi_i(g),
	\end{align*}
	where the multiplication on the right hand side is the canonical one for $U_i \times G$ as a trivial LGB (recall \ref{ex:TrivialLGBundle}),
	and this can be viewed as $\mathcal{G}$-equivariance (under the trivialization induced by $\phi_i$). We usually then speak of \textbf{$\mathcal{G}$-equivariance w.r.t.\ the LGB morphism $\phi_i$}.
	
	We will call such an atlas and its charts a \textbf{principal bundle atlas} and \textbf{principal bundle charts} for $\mathcal{P}$, respectively. For simplicity we may also refer to $\varphi_i$ as principal bundle chart giving rise to the principal bundle atlas.
	\item Due to the existence of $\varphi_i$ one does not need to claim upfront that $\mathcal{P}$ is a $G$-fibre bundle. However, for readability, we decided to structure the definition like this. We kept a similar ``lecture-type-style'' as for ``ordinary'' principal bundles as provided in \cite[\S 4.2, Def.\ 4.2.1, page 207f.]{Hamilton}.
	\item Since $\pi_{\mathcal{P}}$ is a surjective submersion we know by \ref{rem:SmoothnessOfACtionTranslations} that right-translations $r_g$ ($g \in \mathcal{G}_x$) are diffeomorphisms on $\mathcal{P}_{x}$. Furthermore, following \cite[\S 4.2, discussion after Def.\ 4.2.1, page 208f.]{Hamilton}, by definition we have a simply transitive $\mathcal{G}_x$-action (as a Lie group) on $\mathcal{P}_x$, and the isotropy group for each $p \in \mathcal{P}_x$ is trivial. Henceforth, the orbit map $\Phi_p$ gives rise to a $\mathcal{G}_x$-equivariant diffeomorphism
	\begin{align*}
	\mathcal{G}_x &\to \mathcal{P}_x,
	\end{align*}
	where the action on $\mathcal{G}_x$ is the right-action on itself regarding the $\mathcal{G}_x$-equivariance; see for example \cite[\S 3.8, Thm.\ 3.8.8, page 165]{Hamilton}.
	\item Similarly by definition, for each $x \in U_i$ we know that $\mleft.\mleft(\varphi_i\mright)\mright|_x: \mathcal{P}_x \to \mathcal{G}_x$ is a $\mathcal{G}_x$-equivariant diffeomorphism. In fact, together with $\pi_{\mathcal{G}} \circ \varphi_i = \pi_{\mathcal{P}}$ this clearly gives an equivalent definition of $\varphi_i$.
\end{enumerate}
\end{remarks}

\subsubsection{Examples}

Let us provide examples of such principal bundles.

\begin{examples}{The ``classical'' principal bundle}{TheCLassicalPrincAsEx}
We recover principal $G$-bundles as principal $\mathcal{G} \coloneqq M \times G$-bundles. Recall \ref{ex:TrivialLGBAction}, the LGB action of a trivial LGB $\mathcal{G} \cong M \times G$ is equivalent to a $G$-action, the right-translation with an element $g \in G$ is then the right-translation of the corresponding constant section in $\mathcal{G}$; this action is clearly simply transitive. The principal $G$-bundle atlas is then naturally inherited by the existing principal $\mathcal{G}$-bundle atlas. In fact, reverting this argument proves that principal $G$-bundles are equivalent to principal $\mathcal{G} \coloneqq M \times G$-bundles.

We often refer to principal bundles related to trivial LGBs as \textbf{typical} or \textbf{classical principal bundles}, or, as usual, principal $G$-bundle. 
\end{examples}

\begin{examples}{The ``trivial'' principal $\mathcal{G}$-bundle}{TrivialPrincAsLGB}
\reference{\cite[\S 5.7, Remark 5.34, page 145f.]{GroupoidBasedPrincipalBundles}}
$\mathcal{G}$ itself is a principal $\mathcal{G}$-bundle, equipped with its canonical right-action inherited by its multiplication $\mathcal{G}*\mathcal{G} \to \mathcal{G}$. The principal bundle atlas then just consists of the identity map $\mathrm{id}_{\mathcal{G}}$.

By \ref{ex:TheCLassicalPrincAsEx}, it may be natural to call this the \textbf{trivial $\mathcal{G}$-principal bundle} even if $\mathcal{G}$ itself might not be trivial. It will be clearer later why one may choose to do so, but in particular one essential property is that this example shows that we are going to define a gauge theory for which the structural LGBs themselves are allowed as principal bundles.
\end{examples}

In order to explain the "triviality" of \ref{ex:TrivialPrincAsLGB} we need to introduce morphisms of principal bundles.

\subsubsection{Morphism of principal LGB-bundles}

Let us define morphisms of principal LGB-bundles

\begin{definitions}{Morphism of principal bundles with structural LGB}{MorphOfPrincBundles}
\reference{\cite[\S 5.7, page 144f.]{GroupoidBasedPrincipalBundles}}
Let $M$ and $N$ be smooth manifolds, $\mathcal{H} \stackrel{\pi_{\mathcal{H}}}{\to} N$ and $\pi_{\mathcal{G}} \colon \mathcal{G} \to M$ LGBs, and $\mathcal{P}^\prime \stackrel{\pi^\prime}{\to} N$ and $\pi\colon\mathcal{P} \to M$ principal $\mathcal{H}$- and $\mathcal{G}$-bundles, respectively. A \textbf{principal bundle morphism} between $\mathcal{P}^\prime$ and $\mathcal{P}$ is a triple of smooth maps $F: \mathcal{H} \to \mathcal{G}$, $f: N \to M$ and $H: \mathcal{P}^\prime \to \mathcal{P}$ such that the pair $(F, f)$ is an LGB morphism as in \ref{def:LGB morphism} and
\begin{align*}
\pi \circ H &= f \circ \pi^\prime~, \\
H(p \cdot h) &= H(p) \cdot F(h)
\end{align*}
for all $(p, h) \in \mathcal{P}^\prime * \mathcal{H} = \mleft(\pi^\prime\mright)^*\mathcal{H}$. We also speak of a \textbf{principal bundle morphism over $f$ w.r.t.\ the LGB morphism $F$}.

We speak of a \textbf{principal bundle isomorphism (over $f$, w.r.t.\ $F$)} if $H$ is a diffeomorphism. 

If $H$ is a base-preserving isomorphism $\mathcal{P} \to \mathcal{P}$ w.r.t.\ $F = \mathrm{id}_{\mathcal{G}}$, then we say that $H$ is a \textbf{(global) gauge transformation} or \textbf{principal bundle automorphism}, and the set of all such automorphisms is denoted by $\sAut(\mathcal{P})$. If $H$ is only defined on an open subset of $M$, then we may also speak of a \textbf{local gauge transformation}.
\end{definitions}

\begin{remarks}{Notions of local triviality}{PrincipalMorphBasics}
\reference{\cite[\S 5.7, in particular Remark 5.34, page 144f.]{GroupoidBasedPrincipalBundles}}
\indent $\bullet$ 
It is straightforward to check that the equations are well-defined.


$\bullet$ Similar to \ref{rem:LGBMOrphismRemark}: If $H$ is a diffeomorphism, then $F$ is an LGB isomorphism, and $H^{-1}$ is a principal bundle isomorphism, too.

$\bullet$ Finally, observe that we have a local isomorphism of $\mathcal{P}$ to $\mathcal{G}$: Fix a principal bundle chart $\varphi: \mathcal{P}|_U \to \mathcal{G}|_U$ ($U$ some open subset of $M$). Then take $H \coloneqq \varphi$, $F \coloneqq \mathrm{id}_{\mathcal{G}|_U}$ and $f \coloneqq \mathrm{id}_U$; by the definition of a principal bundle chart this gives a principal bundle isomorphism. Therefore one could say that every principal bundle is locally "trivial" in the sense of being isomorphis to an LGB as principal bundle; recall \ref{ex:TrivialPrincAsLGB}. 

Additionally using \ref{rem:LGBPrincDefDiscussion} we have a local isomorphism of $\mathcal{P}$ to a trivial LGB; keeping the same notation as in \ref{rem:LGBPrincDefDiscussion}, choose $H \coloneqq \xi_i$, $F \coloneqq \phi_i$ and $f \coloneqq \mathrm{id}_{U_i}$. Hence, locally every principal bundle is also classical in the sense of \ref{ex:TheCLassicalPrincAsEx}.
\end{remarks}

Equivalent to the last remark, there is a natural isomorphism induced by local sections of $\mathcal{P}$, which are, however, no trivializations in general. 

\begin{lemmata}{Local sections of principal bundles induce isomorphisms to $\caG$}{SectionsNowInduceIsomToLGBsNotNecTriv}
\reference{\cite[\S 5.7, Remark 5.34, page 145f.]{GroupoidBasedPrincipalBundles}}
Let $\pi_{\mathcal{G}} \colon \mathcal{G} \to M$ be an LGB over a smooth manifold $M$, and $\mathcal{P} \to M$ a principal $\mathcal{G}$-bundle. Let $s: U \to \mathcal{P}$ be a smooth local section of $\mathcal{P}$ over an open subset $U$ of $M$. Then the orbit map $\Phi_s$ through $s$, given as in \ref{def:LRTranslations} by
\begin{align*}
\mathcal{G}|_U &\to \mathcal{P}|_U,\\
g &\mapsto s_{\pi_{\mathcal{G}}(g)} \cdot g,
\end{align*}
is a base-preserving principal bundle isomorphism w.r.t.\ $\mathrm{id}_{\mathcal{G}|_U}$.
\end{lemmata}

This fits the expectation that the choice of a section of $\mathcal{P}$ is actually the choice of a designated neutral element, naturally inducing a Lie group structure in each fibre and thus an LGB structure, which may not be trivial; aligning the more general definition of principal bundles with the definition of LGBs. In contrast, LGBs always carry a global section, for examples the neutral section $e$.

Let us now introduce general pullbacks of principal bundles; recall \ref{cor:PullbackLGB}.

\begin{corollaries}{Pullbacks of principal LGB-bundles are principal LGB-bundles}{PullBacksArePrincToo}
\reference{\cite[\S 5.7, Remark 5.34, page 145f.]{GroupoidBasedPrincipalBundles}}
Let $M, N$ be smooth manifolds, $f: N \to M$ a smooth map, $\mathcal{G} \to M$ an LGB, and $\mathcal{P} \to M$ a principal $\mathcal{G}$-bundle. Then there is a unique (up to isomorphisms) principal $f^*\mathcal{G}$-bundle structure on $f^*\mathcal{P}$, such that the projection $\pi_2: f^*\mathcal{P} \to \mathcal{P}$ onto the second factor is a principal bundle morphism (over $f$) w.r.t.\ the projection $\pi_2^{\mathcal{G}}: f^*\mathcal{G} \to \mathcal{G}$ onto the second factor as LGB morphism, and such that $\pi_2|_x: (f^*\mathcal{P})_x \to \mathcal{P}_{f(x)}$ is a $\mathcal{G}_x$-equivariant diffeomorphism w.r.t.\ the LGB isomorphism $\mleft.\pi_2^{\mathcal{G}}\mright|_x$.
%
\end{corollaries}

\begin{definitions}{Pullback principal bundle}{PullbackPrincBundleDef}
Let $M, N$ be smooth manifolds, $f: N \to M$ a smooth map, $\mathcal{G} \to M$ an LGB, and $\mathcal{P} \to M$ a principal $\mathcal{G}$-bundle. Then we call the principal $f^*\mathcal{G}$-bundle structure on $f^*\mathcal{P}$ as given in \ref{cor:PullBacksArePrincToo} the \textbf{pullback principal bundle of $\mathcal{P}$ (under $f$)}.

We will refer to this structure often without further mention.
\end{definitions}

We can actually show that $\mathcal{P}*\mathcal{G}$ is not only $\pi^*\mathcal{G}$ as pullback LGB by definition but it is also isomorphic to $\pi^*\mathcal{P}$ as principal bundle. For this recall \ref{ex:TrivialPrincAsLGB}, i.e.\ LGBs are "trivially" also principal bundles, hence we know that $\mathcal{P} * \mathcal{G}$ is a principal $\pi^*\mathcal{G}$-bundle.

\begin{corollaries}{$\mathcal{P}*\mathcal{G}$ is the pullback of $\mathcal{P}$ along its projection}{ProductSpaceIsPItself}
Let $\mathcal{G} \to M$ be an LGB over a smooth manifold $M$ and $\pi\colon\mathcal{P} \to M$ a principal $\mathcal{G}$-bundle. Then we have a base-preserving principal bundle isomorphism 
\begin{align*}
\mathcal{P}*\mathcal{G} &\cong \pi^*\mathcal{P}
\end{align*}
w.r.t.\ the LGB isomorphism given as the identity map on $\mathcal{P} * \mathcal{G} = \pi^*\mathcal{G}$.
\end{corollaries}

\begin{proof}
\leavevmode\newline
We have a global section of the pullback principal bundle $\pi^*\mathcal{P} \to \mathcal{P}$ given by
\begin{align*}
\mathcal{P} &\to \pi^*\mathcal{P},\\
p &\mapsto (p, p).
\end{align*}
This is clearly a well-defined global section of $\pi^*\mathcal{P}$, so that by \ref{lem:SectionsNowInduceIsomToLGBsNotNecTriv} we achieve the desired isomorphism $\pi^*\mathcal{P} \cong \pi^*\mathcal{G} = \mathcal{P} * \mathcal{G}$.
\end{proof}

\begin{remarks}{Alternative definition of principal bundles}{AlternativePrincBdlDef}
By \ref{lem:SectionsNowInduceIsomToLGBsNotNecTriv} this isomorphism is explicitly given by
\begin{align*}
\mathcal{P}*\mathcal{G} &\to \pi^*\mathcal{P},\\
(p, g) &\mapsto (p,p) \cdot (p, g) = (p, p \cdot g).
\end{align*}
In fact, the cited reference of \ref{def:PrinciBdleWithStruLGB}, \cite[simplification of the beginning of \S 5.7, page 144f.]{GroupoidBasedPrincipalBundles}, takes the existence of such a diffeomorphism as the essential sole part of defining principal bundles. Such an approach essentially avoids the second part of \ref{def:PrinciBdleWithStruLGB}.
\end{remarks}

\begin{remarks}{Why "principal"?}{ItIsAPrincipalAction}
The last result and remark also outline why we are speaking of a \textit{principal} bundle; this is basically due to similar arguments as for ordinary principal bundles, in particular one can construct a canonical principal $\caG$-bundle structure on $\caP \to \caP\Big/\caG$; see \cite[\S 5.7, Lemma 5.35, page 146.]{GroupoidBasedPrincipalBundles}.
\end{remarks}

Let us conclude our discussion about principal bundles by introducing a typical label appearing in physics.

\begin{definitions}{Gauges of a principal bundle}{GaugesOfPrincipalBundles}
Let $\mathcal{G} \to M$ be an LGB over a smooth manifold $M$, and $\mathcal{P} \to M$ a principal $\mathcal{G}$-bundle. A \textbf{gauge} of $\mathcal{P}$ is a section of $\mathcal{P}$. If the section is globally defined on $M$, then we speak of a \textbf{global gauge}, otherwise we may just say \textbf{local gauge} or just gauge.
\end{definitions}
%

\subsection{Generalized distributions and connections}\label{ConnectionSubsection}

Finally, let us now define the gauge theory, starting with horizontal distributions. 
We expect a basic understanding of horizontal distributions and their relationship to what we call connections (on principal and vector bundles). The bare-bones start with horizontal distributions.

\begin{definitions}{Horizontal distribution}{EhresmannConnectionBasics}
Let $F \to M$ be a fibre bundle over a smooth manifold $M$. Then a \textbf{horizontal distribution/bundle of $F$} is a smooth subbundle $\mathrm{H}F$ of $\mathrm{T}F$ with
\begin{align*}
\mathrm{T}F
&=
\mathrm{H}F \oplus \mathrm{V}F.
\end{align*}
For $p \in M$ the fibre is denoted by $\mathrm{H}_p F$, which we may call a \textbf{horizontal tangent space}.
\end{definitions}

As usual for gauge theory we will understand connections as horizontal distributions with a certain symmetry along the fibres, induced by the LGB action. For this let us first understand the vertical structure; recall \ref{def:FundVecs} and \ref{rem:FundVecsNotations}. The following is a straightforward generalization of what one knows in the typical formulation of gauge theory, see e.g.\ \cite[\S 5.1, part 2 to 4 of Prop.\ 5.1.3, page 258f]{Hamilton}.

\begin{corollaries}{The natural invariance of the vertical bundle of $\mathcal{P}$}{VerticalBundleOfPrincIsNearlyAsUsual}
Let $\mathcal{G} \to M$ be an LGB over a smooth manifold $M$, $\mathcal{g}$ its LAB, and $\mathcal{P}\stackrel{\pi}{\to} M$ a principal $\mathcal{G}$-bundle. Then
\begin{align*}
\mathrm{D}_p r_g\mleft( \mathrm{V}_p \mathcal{P} \mright)
&=
\mathrm{V}_{p \cdot g} \mathcal{P}
\end{align*}
for all $(p, g) \in \mathcal{P} * \mathcal{G}$, and we have an isomorphism of vector bundles
\begin{align*}
\mathrm{V}\mathcal{P}
&\cong
\pi^*\mathcal{g}
\end{align*}
given by
\begin{align*}
\pi^*\mathcal{g} &\to \mathrm{V}\mathcal{P},\nonumber\\
(p, \nu) &\mapsto \overline{\nu}_p.
\end{align*}
\end{corollaries}

\begin{remarks}{Extending the notation of fundamental vector fields}{FundVecNotationOnPullbackBundle}
For $\mu \coloneqq (p, \nu)$ we may also write
\begin{align*}
\overline{\mu}_p
&\coloneqq
\overline{\nu}_p,
\end{align*}
which simplifies the notation in certain circumstances.
\end{remarks}

\begin{proof}[Proof of \ref{cor:VerticalBundleOfPrincIsNearlyAsUsual}]
\leavevmode\newline
Recall \ref{rem:LGBPrincDefDiscussion} for this proof. By definition it is clear that each fibre $\mathcal{P}_x$ ($x \in M$) is a principal $\mathcal{G}_x$-bundle over $\{x\}$ whose $\mathcal{G}_x$-action is the $\mathcal{G}$-action restricted to $x$, and thus we know
\begin{equation*}
\mathrm{D}_pr_g\mleft(\mathrm{V}_p\mathcal{P}_x\mright)
=
\mathrm{V}_{p\cdot g} \mathcal{P}_x
\end{equation*}
for all $p \in \mathcal{P}_x$ and $g \in \mathcal{G}_x$
Due to the fact that $\mathcal{P}_x$ is a bundle over a point, we have $\mathrm{V}\mathcal{P}_x = \mathrm{T}\mathcal{P}_x = \mleft.\mathrm{V}\mathcal{P}\mright|_{\mathcal{P}_x}$. Thus,
\begin{equation*}
\mathrm{D}_pr_g\mleft(\mathrm{V}_p\mathcal{P}\mright)
=
\mathrm{V}_{p\cdot g} \mathcal{P}.
\end{equation*}
Due to the fact that the $\mathcal{G}$-action is simply transitive we can derive that its induced $\mathcal{g}$-action $\rho$ is a vector bundle isomorphism: By \ref{rem:FundVecsAreLABActions} this LAB action $\rho$ is precisely given by fundamental vector fields as proposed and $\rho$ has values in $\mathrm{V}\mathcal{P}$, thence, a vector bundle morphism $\pi^*\cag \to \rmV\caP$. We know that the orbit maps $\Phi_p\colon \mathcal{G}_x \to \mathcal{P}_x$ through $p \in \mathcal{P}_x$ are $\mathcal{G}_x$-equivariant diffeomorphisms, so that $\mathrm{D}_{e_x}\Phi_p\colon \mathcal{g}_x \to \mathrm{V}_p\mathcal{P}$ is a vector space isomorphism. But we also have
\begin{equation*}
\rho(p, \nu)
=
\mathrm{D}_{e_x}\Phi_p(\nu)
\end{equation*}
for all $\nu\in \mathcal{g}_x$, and therefore $\rho$ is fibre-wise an isomorphism of vector spaces such that it is an isomorphism of vector bundles.
\end{proof}

This result was highly expected, however, the right-invariance for horizontal distributions is not as straight-forward and will require a choice to make:

\subsubsection{Idea and motivation}\label{TheBigMotivationBehindEverything}

\begin{center}
\begin{tikzpicture}[scale=0.80]
\draw (0,-3.5) to[out=10,in=170] (4.25,-3.5) to[out=350,in=190] (8.5,-3.5);
\draw (1.25,2.5) to[out=280,in=90](1.5,0) to[out=270,in=80] (1.25,-2.5);
\filldraw [fill=gray!20!white, draw=white] (3.5,2.5) to[out=280,in=90](3.75,0) to[out=270,in=80] (3.5,-2.5) -- (5,-2.5)  to[out=80,in=270](5.25,0) to[out=90,in=280] (5,2.5) -- cycle;
\draw (4.25,2.5) to[out=280,in=95](4.41,1) to[out=275,in=85] (4.41,-1) to[out=265,in=80] (4.25,-2.5); 

\draw (4,-1) node {\textcolor[rgb]{1,0,0}{$p$}};
\draw [<-, draw=blue] (5,1) to[out=275,in=85] (5,-1);
\draw (5.5,0) node {\textcolor[rgb]{0,0,1}{$\cdot g$}};
\filldraw [fill=red, draw=red] (4.41,1) circle (1pt);
\filldraw [fill=red, draw=red] (4.41,-1) circle (1pt);
\draw (5.75,2.5) to[out=280,in=90](6,0) to[out=270,in=80] (5.75,-2.5);
\draw (7.25,2.5) to[out=280,in=90](7.5,0) to[out=270,in=80] (7.25,-2.5);
\draw (2.75,2.475) to[out=280,in=90](3,0) to[out=270,in=80] (2.75,-2.475);

\draw (9.5,-3.5) node {$M$};
\draw (5,2) node {$\mathcal{P}_U$};
\draw (3.5,-3.4) node {$($};
\filldraw [fill=red, draw=red] (4.25,-3.5) circle (1pt);
\draw (4.25,-3.9) node {\textcolor[rgb]{1,0,0}{$x$}};
\draw (5,-3.6) node {$)$};
\draw (4.25, -3) node {$U$};
\draw (0,0) node {$\mathcal{P}$};
\draw [->] (0,-0.4) -- (0,-3.1);
\draw (0.4,-1.75) node {$\pi$};
\draw (3.7,1) node {\textcolor[rgb]{1,0,0}{$p \cdot g$}};
\draw (10.5,-3.5) to[out=10,in=170] (14.75,-3.5) to[out=350,in=190] (19,-3.5); 
\filldraw [fill=gray!20!white, draw=white] (14,2.5) to[out=280,in=90](14.25,0) to[out=270,in=80] (14,-2.5) -- (15.5,-2.5)  to[out=80,in=270](15.75,0) to[out=90,in=280] (15.5,2.5) -- cycle; 
\draw (11.75,2.5) to[out=280,in=90](12,0) to[out=270,in=80] (11.75,-2.5); 
\draw (14.75,2.5) to[out=280,in=90](15,0) to[out=270,in=80] (14.75,-2.5); 
\draw (16.25,2.5) to[out=280,in=90](16.5,0) to[out=270,in=80] (16.25,-2.5); 
\draw (17.75,2.5) to[out=280,in=90](18,0) to[out=270,in=80] (17.75,-2.5); 

\draw (13.25,2.475) to[out=280,in=90](13.5,0) to[out=270,in=80] (13.25,-2.475); 
\filldraw [fill=blue, draw=blue] (15,0) circle (1pt);
\draw (15.3,0) node {\textcolor[rgb]{0,0,1}{$g$}};

\draw (19,0) node {$\mathcal{G}$}; 
\draw [->] (19,-0.4) -- (19,-3.1);
\draw (18.5,-1.75) node {$\pi_{\mathcal{G}}$};

\draw (15.5,2) node {$\mathcal{G}_U$};
\draw (14,-3.4) node {$($};
\filldraw [fill=red, draw=red] (14.75,-3.5) circle (1pt);
\draw (14.75,-3.9) node {\textcolor[rgb]{1,0,0}{$x$}};
\draw (15.5,-3.6) node {$)$};
\draw (14.75, -3) node {$U$};

\path[<-] (8.5,0) edge [bend left] (11,0);
\end{tikzpicture}
\end{center}

For $\pi_\caG\colon\mathcal{G} \to M$ an LGB over a smooth manifold $M$ and $\pi\colon\mathcal{P} \to M$ a principal $\mathcal{G}$-bundle we fix a point $p \in \mathcal{P}_x$ ($x \in M$) and can multiply that with an element $g \in \mathcal{G}_x$. Infinitesimally, we are interested into how this multiplication by $g$ affects tangent vectors, especially non-vertical ones.

However, as we have seen in \ref{def:LRTranslations} and \ref{prop:DiffOfLGBAction} (and its proof) the push-forward of horizontal vectors is not well-defined anymore on non-vertical vectors if one uses a fixed element of an LGB; $r_g$ is just a map $\mathcal{P}_x \to \mathcal{P}_x$. In order to study push-forwards of non-vertical vectors, we need information of the $\mathcal{G}$-action in an open neighbourhood $U$ around the fibres over $x$. Hence we want to use a section $\sigma \in \Gamma(\mathcal{G}|_U)$ with $\sigma_x= g$ instead.

\begin{center}
\begin{tikzpicture}[scale=0.80]
\draw (0,-3.5) to[out=10,in=170] (4.25,-3.5) to[out=350,in=190] (8.5,-3.5);
\draw (1.25,2.5) to[out=280,in=90](1.5,0) to[out=270,in=80] (1.25,-2.5);
\filldraw [fill=gray!20!white, draw=white] (3.5,2.5) to[out=280,in=90](3.75,0) to[out=270,in=80] (3.5,-2.5) -- (5,-2.5)  to[out=80,in=270](5.25,0) to[out=90,in=280] (5,2.5) -- cycle;
\draw (4.25,2.5) to[out=280,in=95](4.41,1) to[out=275,in=85] (4.41,-1) to[out=265,in=80] (4.25,-2.5); 

\draw (4,-1) node {\textcolor[rgb]{1,0,0}{$p$}};
\draw [<-, draw=blue] (5,1) to[out=275,in=85] (5,-1);
\draw [<-, draw=blue] (4.8,1) to[out=275,in=85] (4.8,-1);
\draw [<-, draw=blue] (4.6,1) to[out=275,in=85] (4.6,-1);
\draw (5.5,0) node {\textcolor[rgb]{0,0,1}{$\cdot \sigma$}};
\filldraw [fill=red, draw=red] (4.41,1) circle (1pt);
\filldraw [fill=red, draw=red] (4.41,-1) circle (1pt);
\draw (5.75,2.5) to[out=280,in=90](6,0) to[out=270,in=80] (5.75,-2.5);
\draw (7.25,2.5) to[out=280,in=90](7.5,0) to[out=270,in=80] (7.25,-2.5);
\draw (2.75,2.475) to[out=280,in=90](3,0) to[out=270,in=80] (2.75,-2.475);

\draw (9.5,-3.5) node {$M$};
\draw (5,2) node {$\mathcal{P}_U$};
\draw (3.5,-3.4) node {$($};
\filldraw [fill=red, draw=red] (4.25,-3.5) circle (1pt);
\draw (4.25,-3.9) node {\textcolor[rgb]{1,0,0}{$x$}};
\draw (5,-3.6) node {$)$};
\draw (4.25, -3) node {$U$};
\draw (0,0) node {$\mathcal{P}$};
\draw [->] (0,-0.4) -- (0,-3.1);
\draw (0.4,-1.75) node {$\pi$};
\draw (3.7,1) node {\textcolor[rgb]{1,0,0}{$p \cdot \sigma_x$}};
\draw (10.5,-3.5) to[out=10,in=170] (14.75,-3.5) to[out=350,in=190] (19,-3.5); 
\filldraw [fill=gray!20!white, draw=white] (14,2.5) to[out=280,in=90](14.25,0) to[out=270,in=80] (14,-2.5) -- (15.5,-2.5)  to[out=80,in=270](15.75,0) to[out=90,in=280] (15.5,2.5) -- cycle; 
\draw (11.75,2.5) to[out=280,in=90](12,0) to[out=270,in=80] (11.75,-2.5); 
\draw (14.75,2.5) to[out=280,in=90](15,0) to[out=270,in=80] (14.75,-2.5); 
\draw (16.25,2.5) to[out=280,in=90](16.5,0) to[out=270,in=80] (16.25,-2.5); 
\draw (17.75,2.5) to[out=280,in=90](18,0) to[out=270,in=80] (17.75,-2.5); 

\draw (13.25,2.475) to[out=280,in=90](13.5,0) to[out=270,in=80] (13.25,-2.475); 
\draw [draw=blue] (14.25,0) to[out=10,in=170] (15,0) to[out=350,in=190] (15.75,0);
\draw (13.9,0) node {\textcolor[rgb]{0,0,1}{$\sigma$}};

\draw (19,0) node {$\mathcal{G}$}; 
\draw [->] (19,-0.4) -- (19,-3.1);
\draw (18.5,-1.75) node {$\pi_{\mathcal{G}}$};

\draw (15.5,2) node {$\mathcal{G}_U$};
\draw (14,-3.4) node {$($};
\filldraw [fill=red, draw=red] (14.75,-3.5) circle (1pt);
\draw (14.75,-3.9) node {\textcolor[rgb]{1,0,0}{$x$}};
\draw (15.5,-3.6) node {$)$};
\draw (14.75, -3) node {$U$};

\path[<-] (8.5,0) edge [bend left] (11,0);
\end{tikzpicture}
\end{center}

Of course, such a section with $\sigma_x= g$ is in general not unique, especially the tangential behaviour of the section's image (as an embedding of the base) may contribute to the push-forward; given a horizontal distribution, a horizontal vector may be still horizontal after a push-forward with one LGB section but not with respect to another LGB section. A similar problem may arise if we would work with local trivializations in order to use \ref{ex:TrivialLGBAction} but this requires the choice of a trivialisation. Thus, we need to adjust the typical definition of connections on principal bundles to account for making such a choice.

In order to understand what has to be changed, let us revisit the "typical" situation, that is, let $\mathcal{P}$ be a classical principal bundle as in \ref{ex:TheCLassicalPrincAsEx}, i.e.\ $\mathcal{G} = M \times G$ is trivial with Lie group $G$. The $\mathcal{G}$-action is equivalent to a $G$-action on $\mathcal{P}$ by \ref{ex:TrivialLGBAction}, a push-forward with $g \in G$ w.r.t.\ the latter action is equivalent to the push-forward with a constant section in $\mathcal{G}$ for which we may still simply write $g$.

Equip $\mathcal{P}$ with a "typical" Ehresmann connection, that is, a horizontal distribution $\mathrm{H}\mathcal{P}$ of $\mathcal{P}$ with
\begin{equation}\label{ClassicalSymmetryOfHorizontalDistr}
\mathrm{D}_pr_g\mleft(\mathrm{H}\mathcal{P}_p\mright) = \mathrm{H}\mathcal{P}_{p \cdot g}
\end{equation}
for all $p \in \mathcal{P}$ and $g \in G$ as constant section. Recall that a connection is 1:1 to its parallel transport. This means, corresponding to a piece-wise smooth base curve $\alpha\colon [0, t] \to M$ ($t > 0$) with $\alpha(0) = x$, we have a \textbf{parallel transport along $\alpha$} as a map $\mathrm{PT}_\alpha^{\mathcal{P}}\colon \mathcal{P}_x \to \Gamma(\alpha^*\mathcal{P})$ satisfying
\begin{align}\label{PTProp1}
\mleft.\mathrm{PT}^{\mathcal{P}}_{\alpha*\alpha^\prime}\mright|_t
&=
\mleft.\mathrm{PT}^{\mathcal{P}}_{\alpha^\prime}\mright|_t
	\circ \mleft.\mathrm{PT}_\alpha^{\mathcal{P}}\mright|_t,\\
\mleft.\mathrm{PT}_{\alpha^-}^{\mathcal{P}}\mright|_t
&=
\mleft.\mleft( \mathrm{PT}_\alpha^{\mathcal{P}} \mright)^{-1}\mright|_t,\label{PTProp2}
\end{align}
where $\alpha^\prime$ is just another similarly defined base curve with $\alpha^\prime(0)= \alpha(t)$, $\alpha*\alpha^\prime$ their concatenation ($\alpha$ coming first and with a suitable parametrization such that $\alpha*\alpha^\prime$ is a map $[0, t] \to M$), and $\alpha^-$ denotes $\alpha$ traversed backwards; in particular, the parallel transport is a diffeomorphism between the fibres. Moreover, Eq.\ \eqref{ClassicalSymmetryOfHorizontalDistr} integrates and is equivalent to
\begin{equation}\label{ClassicalSymmetryofPTs}
\mleft.\mathrm{PT}_\alpha^{\mathcal{P}}(p \cdot g)\mright|_t
=
\mleft.\mathrm{PT}_\alpha^{\mathcal{P}}(p)\mright|_t \cdot g.
\end{equation}
Thinking of the associated $\mathcal{G}$-action, $g$ is an element of $\mathcal{G}_x$ such that the right hand side is in general not well-defined
anymore due to the fact that $\mleft.\mathrm{PT}_\alpha^{\mathcal{P}}(p)\mright|_t \in \mathcal{P}_{\alpha(t)}$; it is well-defined if interpreting $g$ as a constant section of $\mathcal{G} = M \times G$, denoted now by $\overline{g} \in \Gamma(\mathcal{G})$ for bookkeeping reasons. The left hand side uses $\overline{g}_{\alpha(0)} = \overline{g}_{x} = (x, g)$, the right hand side $\overline{g}_{\alpha(t)} = (\alpha(t), g)$, and by \ref{ex:TrivialLGBAction} we rewrite the $G$-action to a $\mathcal{G}$-action:
\begin{align*}
p \cdot g
&=
p \cdot (x, g)
=
p \cdot \overline{g}_x,&
\mathrm{PT}_\alpha^{\mathcal{P}}(p) \cdot g
&=
\mathrm{PT}_\alpha^{\mathcal{P}}(p) \cdot (\alpha, g)
=
\mathrm{PT}_\alpha^{\mathcal{P}}(p) \cdot \alpha^*\overline{g},
\end{align*}
where we recall that $\alpha^*\mathcal{P}$ is a principal $\alpha^*\mathcal{G}$-bundle in sense of \ref{def:PullbackPrincBundleDef}. 

Let us equip $\pi_{\mathcal{G}} \colon \mathcal{G} \to M$ with its \textbf{canonical flat connection} $\mathrm{H}\mathcal{G} \coloneqq \pi^*_{\mathcal{G}}\mathrm{T}M$ which induces a parallel transport $\mathrm{PT}_\alpha^{\mathcal{G}} \colon \mathcal{G}_{x} \to \Gamma\mleft(\alpha^*\mathcal{G}\mright)$ with $\mleft.\mathrm{PT}_\alpha^{\mathcal{G}}\mright|_t(x, g) = ( \alpha(t), g)$, especially $\mathrm{PT}_\alpha^{\mathcal{G}}\mleft( \overline{g}_x \mright) = \alpha^*\overline{g}$. In total, we can rewrite Eq.\ \eqref{ClassicalSymmetryofPTs} to
\begin{align*}
\mathrm{PT}_\alpha^{\mathcal{P}}\mleft(p \cdot \overline{g}_x\mright)
&=
\mathrm{PT}_\alpha^{\mathcal{P}}(p) \cdot \alpha^*\overline{g}
=
\mathrm{PT}_\alpha^{\mathcal{P}}(p) \cdot \mathrm{PT}_\alpha^{\mathcal{G}}\mleft( \overline{g}_x \mright).
\end{align*}
This opens a gateway to define Eq.\ \eqref{ClassicalSymmetryofPTs} on general principal $\mathcal{G}$-bundles. That is, now let $\mathcal{P}$ be again a general principal $\mathcal{G}$-bundle. Fix any horizontal distribution $\mathrm{H}\mathcal{G}$ on $\mathcal{G}$ inducing a parallel transport $\mathrm{PT}_\alpha^{\mathcal{G}}$, then an \textbf{Ehresmann connection} on $\mathcal{P}$ is a parallel transport $\mathrm{PT}_\alpha^{\mathcal{P}}$ on $\mathcal{P}$ satisfying Eq.\ \eqref{PTProp1}, \eqref{PTProp2} and
\begin{equation}\label{PTHomomNEw}
\mathrm{PT}_\alpha^{\mathcal{P}}(p \cdot g)
=
\mathrm{PT}_\alpha^{\mathcal{P}}(p) \cdot \mathrm{PT}_\alpha^{\mathcal{G}}(g)
\end{equation}
for all $p \in \mathcal{P}_x$ and $g \in \mathcal{G}_x$. The right hand side is now well-defined since both, $\mathrm{PT}_\alpha^{\mathcal{P}}(p)$ and $\mathrm{PT}_\alpha^{\mathcal{G}}(g)$, are elements of $\Gamma(\alpha^*\mathcal{P})$ and $\Gamma\mleft(\alpha^*\mathcal{G}\mright)$, respectively. We now want to derive its infinitesimal analogue similar to Eq.\ \eqref{ClassicalSymmetryOfHorizontalDistr}. Recall that we can view the parallel transports like $\mathrm{PT}_\alpha^{\mathcal{P}}(p)$ also as a map $[0, t] \to \mathcal{P}$ with $\pi \circ \mathrm{PT}_\alpha^{\mathcal{P}}(p) = \alpha$ (recall \ref{BasicNotations}; alternatively use \ref{lem:PullbackFibreBundleItsTangentSp} and project onto the second component for the following derivatives). Then by construction
\begin{equation*}
Y
\coloneqq
\mleft.\frac{\mathrm{d}}{\mathrm{d}t}\mright|_{t=0} \mathrm{PT}_\alpha^{\mathcal{G}}(g)
\in
\mathrm{H}_g\mathcal{G}
\end{equation*}
for all $g \in \mathcal{G}$, that is, it is horizontal in $\mathcal{G}$; similarly for the parallel transport on $\mathcal{P}$,
\begin{equation*}
X
\coloneqq
\mleft.\frac{\mathrm{d}}{\mathrm{d}t}\mright|_{t=0} \mathrm{PT}_\alpha^{\mathcal{P}}(p)
\in
\mathrm{H}_p\mathcal{P}.
\end{equation*}

We differentiate both sides of Eq.\ \eqref{PTHomomNEw} with $\mathrm{d}/\mathrm{d}t|_{t=0}$; the left hand side implies that the right hand side is an element of $\mathrm{H}_{p\cdot g}\mathcal{P}$, while the right hand side gives
\begin{equation*}
\mleft.\frac{\mathrm{d}}{\mathrm{d}t}\mright|_{t=0} \mleft(
	\mathrm{PT}_\alpha^{\mathcal{P}}(p) \cdot \mathrm{PT}_\alpha^{\mathcal{G}}(g)
\mright)
=
\mathrm{D}_{(p, g)}\Phi\mleft(
	X, 
	Y
\mright).
\end{equation*}
We can now already derive an infinitesimal definition of $\rmH\caP$, that is, an Ehresmann connection $\rmH\caP$, given $\rmH \caG$, is precisely such a horizontal distribution such that $\rmD \Phi$ restricts to the horizontal distributions:
\begin{equation*}
\rmD \Phi|_{\rmH\caP \fibtimes{\rmD\pi}{\rmD\pi_\caG} \rmH\caG} \colon 
\rmH\caP \fibtimes{\rmD\pi}{\rmD\pi_\caG} \rmH\caG \to \rmH \caP,
\end{equation*}
where $\Phi\colon \mathcal{P}* \mathcal{G} \to \mathcal{P}$ denotes the $\mathcal{G}$-action on $\mathcal{P}$;
one could say that ``horizontal vectors are a closed algebra.''

\begin{figure}
\begin{minipage}[]{0.5\textwidth} 
\centering
\begin{tikzpicture}[scale=1.2]
		\filldraw [fill=gray!20!white, draw=white] (3.5,2.5) to[out=280,in=95](3.66,1) to[out=275,in=85] (3.66,-1) to[out=265,in=80] (3.5,-2.5) -- (5,-2.5) to[out=80, in=265] (5.16, -1) to[out=85, in=275] (5.16, 1) to[out=95,in=280] (5,2.5) -- cycle;
		\draw (4.25,2.5) to[out=280,in=95](4.41,1) to[out=275,in=85] (4.41,-1) to[out=265,in=80] (4.25,-2.5);
		\definecolor{darkgreen}{rgb}{0,0.58,0}
		\draw [draw=darkgreen] (3.66,-1) to[out=280,in=200] (4.41,-1) to[out=20,in=100] (5.16, -1);
		\draw [draw=darkgreen] (3.66,1) to[out=280,in=200] (4.41,1) to[out=20,in=100] (5.16,1);
		\filldraw [fill=red, draw=red] (4.41,1) circle (1pt);
		\filldraw [fill=red, draw=red] (4.41,-1) circle (1pt);
		\draw [<-, draw=blue] (5.16,1) to[out=275,in=85] (5.16,-1);
		\draw [->, thick] (4.41,-1) -- (4.71,-0.88);
		\draw [->, thick] (4.41,1) -- (4.71,1.12);
		\draw (3.1,1) node {\textcolor[rgb]{0,0.58,0}{$\mathrm{H}_{p \cdot g}\mathcal{P}$}};
		\draw (3.2,-1) node {\textcolor[rgb]{0,0.58,0}{$\mathrm{H}_p\mathcal{P}$}};
		\draw (5.5,0) node {\textcolor[rgb]{0,0,1}{$\cdot g$}};
		\draw (3,2.2) node {$\mathcal{P}_U$};
		\draw (4.71,-0.6) node {$X$};
		\draw (5,1.4) node {$\mathrm{D}_pr_g(X)$};
	\end{tikzpicture}
	\end{minipage}\hfill
	\begin{minipage}[]{0.5\textwidth}
	\begin{center}
	\begin{tikzpicture}[scale=1.2]
		\filldraw [fill=gray!20!white, draw=white] (3.5,2.5) to[out=280,in=95](3.66,1) to[out=275,in=85] (3.66,-1) to[out=265,in=80] (3.5,-2.5) -- (5,-2.5) to[out=80, in=265] (5.16, -1) to[out=85, in=275] (5.16, 1) to[out=95,in=280] (5,2.5) -- cycle;
		\draw (4.25,2.5) to[out=280,in=95](4.41,1) to[out=275,in=85] (4.41,-1) to[out=265,in=80] (4.25,-2.5);
		\definecolor{darkgreen}{rgb}{0,0.58,0}
		\draw [draw=darkgreen] (3.66,-1) to[out=280,in=200] (4.41,-1) to[out=20,in=100] (5.16, -1);
		\draw [draw=darkgreen] (3.66,1) to[out=280,in=200] (4.41,1) to[out=20,in=100] (5.16,1);
		\filldraw [fill=red, draw=red] (4.41,1) circle (1pt);
		\filldraw [fill=red, draw=red] (4.41,-1) circle (1pt);
		\draw [<-, draw=blue] (5.16,1) to[out=275,in=85] (5.16,-1);
		\draw [->, thick] (4.41,-1) -- (4.71,-0.88);
		\draw [->, thick] (4.41,1) -- (4.63,2);
		\draw [->,dotted,thick] (4.41,1) -- (4.71,1.12);
		\draw [<-,dotted,thick] (4.63,2) -- (4.71,1.12);
		\draw (3,1) node {\textcolor[rgb]{0,0.58,0}{$\mathrm{H}_{p \cdot \sigma_x}\mathcal{P}$}};
		\draw (3.2,-1) node {\textcolor[rgb]{0,0.58,0}{$\mathrm{H}_p\mathcal{P}$}};
		\draw (5.5,0) node {\textcolor[rgb]{0,0,1}{$\cdot \sigma$}};
		\draw (3,2.2) node {$\mathcal{P}_U$};
		\draw (4.71,-0.6) node {$X$};
		\draw (5,2.25) node {$\mathrm{D}_pr_\sigma(X)$};
		\draw (5.9,1.56) node {$\thicksim \mathrm{D}_p(\sigma\circ\pi)(X)$};
	\end{tikzpicture}
	\end{center}
	\end{minipage}
	\caption{Push-forward of a horizontal tangent vector $X$ with constant section (left) and general section (right), where $\mathcal{P}$ is a classical principal bundle as in \ref{ex:TheCLassicalPrincAsEx} equipped with a "typical" connection $\mathrm{H}\mathcal{P}$ of principal $G$-bundles ($G$ the structural Lie group).}
	\label{fig:Pushforwardwandwithoutparallelsection}
\end{figure}
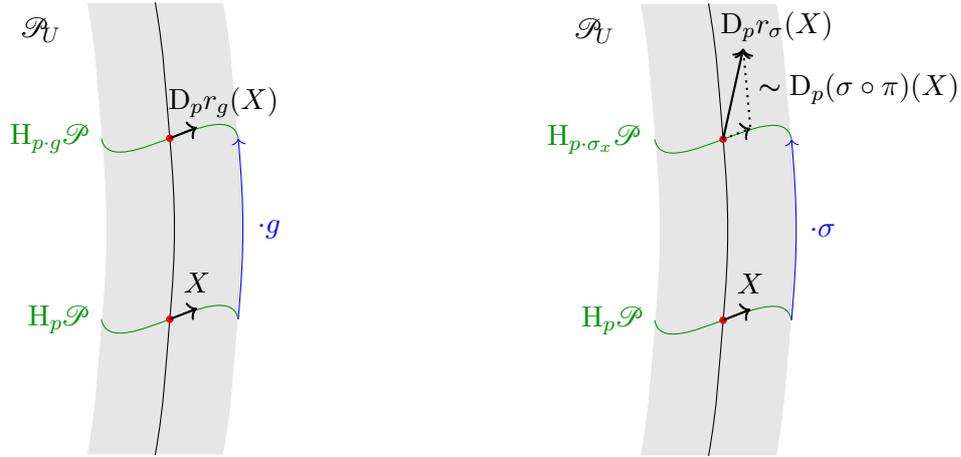

Using \ref{prop:DiffOfLGBAction}:
\begin{align*}
\mathrm{D}_{(p, g)}\Phi\mleft(
	X, 
	Y
\mright)
&=
\mathrm{D}_pr_\sigma\mleft( 
	X 
\mright)
	+ \mleft.{\overline{\mleft( \MC\mright)_g \Bigl(
		Y
		- \mathrm{D}_x \sigma \bigl( \dot{\alpha}(0) \bigr)
	\Bigr)}}\mright|_{p \cdot g}
\end{align*}
where $\sigma$ is any (local) section of $\mathcal{G}$ with $\sigma_x = g$, and
\begin{align*}
\dot{\alpha}(0)
&=
\mleft.\frac{\mathrm{d}}{\mathrm{d}t}\mright|_{t=0} \alpha
=
\mathrm{D}_p\pi (X)
\end{align*}
because of $\pi \circ \mathrm{PT}_\alpha^{\mathcal{P}}(p) = \alpha$ (similarly w.r.t.\ $\mathrm{PT}_\alpha^{\mathcal{G}}$ so that it follows that $(X, Y)$ is a tangent vector of $\mathcal{P}*\mathcal{G}$). As we already have shown several times, for example recall the beginning of the proof of \ref{prop:DiffOfLGBAction}, we have
\begin{align*}
Y - \mathrm{D}_x \sigma \bigl( \dot{\alpha}(0) \bigr)
&\in
\mathrm{V}_g\mathcal{G},
\end{align*}
so that the canonical projection $\pi^{\mathrm{vert}, \mathcal{G}}: \mathrm{T}\mathcal{G} \to \mathrm{V}\mathcal{G}$ onto the vertical structure of $\mathcal{G}$ acts as identity on it. By making use of the horizontality of $Y$, we can write
\begin{align*}
\mleft( \MC\mright)_g \Bigl(
		Y - \mathrm{D}_x \sigma \bigl( \dot{\alpha}(0) \bigr)
	\Bigr)
&=
\mleft( \MC \circ \pi^{\mathrm{vert}, \mathcal{G}}\mright)_g \bigl(
		Y
		- \mathrm{D}_p (\sigma \circ \pi)(X)
	\bigr)
\\
&=
-
\mleft( \MC \circ \pi^{\mathrm{vert}, \mathcal{G}}\mright)_g \bigl(
		\mathrm{D}_p (\sigma \circ \pi)(X)
	\bigr).
\end{align*}
It may not surprise that $\MC \circ \pi^{\mathrm{vert}, \mathcal{G}}$ is by construction actually \textit{the} connection 1-form on $\mathcal{G}$ corresponding to $\mathrm{H}\mathcal{G}$, therefore we will denote $\MC \circ \pi^{\mathrm{vert}, \mathcal{G}}$ as the \textbf{connection 1-form $\mutot$ of $\mathcal{G}$}. We get in total (see also \ref{fig:Pushforwardwandwithoutparallelsection})
\begin{equation}\label{OiTHatIsHowWeFormulateHorizSymmetry}
\mathrm{D}_pr_\sigma\mleft( 
	X 
\mright)
	- \mleft.{\overline{
		\mleft( \mutot\mright)_g \bigl(
		\mathrm{D}_p (\sigma \circ \pi)(X)
	\bigr)
	}}\mright|_{p \cdot g}
\in
\mathrm{H}_{p \cdot g}\mathcal{P}
\end{equation}
for all $X \in \mathrm{H}_p\mathcal{P}$, and we would like to take this as the starting point of defining a connection on $\mathcal{P}$, denoting this by $\car_{g*}(X)$, the \textbf{modified (or covariantised) right-pushforward}. 

By construction, $\car_{g*}(X)$ is independent of the chosen section $\sigma$ because we have seen that it is equivalent to $\mathrm{D}_{(p, g)}\Phi\mleft(X, Y \mright)$, where $Y$ is the unique element of $\rmH_g\caG$ lifting $\rmD_p\pi(X)$. Last but not least, if $X$ is a vertical vector, then
\begin{equation*}
\mathrm{D}_p\pi(X) = 0~,
\end{equation*}
and henceforth
\begin{equation*}
\car_{g*}(X)
=
\mathrm{D}_p r_\sigma (X)
=
\mathrm{D}_p r_g (X)~.
\end{equation*}
Thence, the symmetry behind Eq.\ \eqref{OiTHatIsHowWeFormulateHorizSymmetry} will be compatible with the one of \ref{cor:VerticalBundleOfPrincIsNearlyAsUsual} on the vertical bundle. Let us first study the second summand in Eq.\ \eqref{OiTHatIsHowWeFormulateHorizSymmetry} which defines the \textit{Darboux derivative}.

\subsubsection{Darboux derivative on LGBs}\label{DiscussingDarbouxDerivativeGeneral}

By \ref{cor:VerticalBundleOfPrincIsNearlyAsUsual} we know that $\mathrm{V}\mathcal{P}$ is the pullback of $\mathcal{g}$. Since connections are projections onto the vertical bundle we are interested into the pullback situation, and therefore we will put some remarks after some definitions in the following in order to discuss the situation of pullback LGBs. Hence let us start with a general remark about the pullback situation to which we will later refer in the other remarks regarding pullbacks.

\begin{remarks}{Pullback LGBs and their connections}{GeneralPullBackLGBConnSituation}
Let $\mathcal{G} \stackrel{\pi_\mathcal{G}}{\to} M$ be an LGB over a smooth manifold $M$, and $\mathrm{H}\mathcal{G}$ be a horizontal distribution of $\mathcal{G}$, where we denote with $\pi^{\mathrm{vert}} \colon \mathrm{T}\mathcal{G} \to \mathrm{V}\mathcal{G}$ the associated projection onto its vertical bundle; we will view $\pi^{\mathrm{vert}}$ as an element of $\Omega^1(\mathcal{G}; \mathrm{V}\mathcal{G})$. Furthermore, let $f \colon N \to M$ be a smooth map defined on another smooth manifold $N$. By \ref{lem:PullbackFibreBundleItsTangentSp} we know that $\mathrm{T}(f^*\mathcal{G})$ consists of pairs of tangent vectors $(X, Y) \in \mathrm{T}_{(p, g)}(f^*\mathcal{G})$ ($(p, g) \in f^*\mathcal{G}$) with $X \in \mathrm{T}_pN$, $Y \in \mathrm{T}_g\mathcal{G}$ satisfying
\begin{equation*}
\mathrm{D}_pf(X) = \mathrm{D}_g\pi_{\mathcal{G}}(Y)~.
\end{equation*}
In the following we denote with $\mathrm{pr}_i$ ($i \in \{1,2\}$) the projections onto the $i$-th component of $f^*\mathcal{G}$.

The projection of $f^*\mathcal{G} \to N$ is $\mathrm{pr}_1$, thus $\mathrm{Dpr}_1(X, Y) = X$. The vertical bundle $\mathrm{V}(f^*\mathcal{G})$ as the kernel of $\mathrm{Dpr}_1$ then consists of pairs $(X = 0, Y)$, and therefore $\mathrm{D}_g\pi_{\mathcal{G}}(Y) = 0$ which implies $Y \in \mathrm{V}_g\mathcal{G}$. Thence,
\begin{equation*}
\mathrm{V}\mleft( f^*\mathcal{G} \mright)
\cong
\mathrm{pr}_2^*\mleft( \mathrm{V}\mathcal{G} \mright)~.
\end{equation*}
It is then trivial to check that
\begin{equation*}
\mathrm{pr}_2^!\pi^{\mathrm{vert}}
\in
\Omega^1\bigl(f^*\mathcal{G}; \mathrm{pr}_2^*\mleft( \mathrm{V}\mathcal{G} \mright)\bigr)
\cong
\Omega^1\bigl(f^*\mathcal{G}; \mathrm{V}\mleft( f^*\mathcal{G} \mright)\bigr)
\end{equation*}
is a projection onto $\mathrm{V}\mleft( f^*\mathcal{G} \mright)$, and gives therefore rise to a horizontal distribution on $f^*\mathcal{G}$, the \textbf{pullback connection on $f^*\mathcal{G}$}. Especially we have
\begin{equation*}
\mleft(\mathrm{pr}_2^!\pi^{\mathrm{vert}}\mright)_{(p,g)}(X, Y)
=
\mleft( 0, \pi^{\mathrm{vert}}_g(Y) \mright)~.
\end{equation*}
\end{remarks}

As we have previously seen, we have a connection 1-form on $\caG$ with values in $\cag$.

\begin{definitions}{The connection 1-form on $\caG$}{TotMCFormOnLGB}
Let $\pi_{\mathcal{G}}\colon\mathcal{G} \to M$ be an LGB over a smooth manifold $M$, and $\mathrm{H}\mathcal{G}$ be a horizontal distribution of $\mathcal{G}$, where we denote with $\pi^{\mathrm{vert}} \colon \mathrm{T}\mathcal{G} \to \mathrm{V}\mathcal{G}$ the corresponding projection onto its vertical bundle. Then we define its associated \textbf{connection 1-form $\mutot \in \Omega^1\mleft(\mathcal{G}; \pi_{\mathcal{G}}^*\mathcal{g}\mright)$} by
\begin{align*}
\mleft( \mutot \mright)_g(Y)
&\coloneqq
\mleft.\mleft(\MC \circ \pi^{\mathrm{vert}}\mright)\mright|_g(Y)
=
\mleft( \mathrm{D}_g L_{g^{-1}} \mright)\mleft(\pi^{\mathrm{vert}}(Y)\mright)
\end{align*}
for all $g \in \mathcal{G}$ and $Y \in \mathrm{T}_g\mathcal{G}$.
\end{definitions}
%


\begin{remarks}{Pullback situation: Part I}{PullbackTotMCForm}
Given the situation as in \ref{rem:GeneralPullBackLGBConnSituation}, then we have
\begin{equation*}
L_{(p, g)}\bigl( (p, q) \bigr)
=
\bigl(p, L_g(q)\bigr)
\end{equation*}
for all $(p, g), (p, q) \in f^*\mathcal{G}$, where the left-translation on the left and right hand side are the ones of $f^*\mathcal{G}$ and $\mathcal{G}$, respectively. Thence,
\begin{equation*}
L_{(p, g)}
=
\mleft(
	\mathrm{pr}_1, L_{\mathrm{pr_2}(p, g)} \circ \mathrm{pr}_2
\mright)~,
\end{equation*}
and hence,
\begin{align*}
\mathrm{D}_{(p, g)} L_{(p, g)^{-1}}
&=
\mleft.
\mleft(
	\mathrm{D}_{(p,g)}\mathrm{pr}_1, \mathrm{D}_{\mathrm{pr}_2(p, g)}L_{\mathrm{pr}_2\mleft(p, g^{-1}\mright)} \circ \mathrm{D}_{(p,g)}\mathrm{pr}_2
\mright)
\mright|_{\mathrm{V}_{(p, g)}\mleft( f^*\mathcal{G} \mright)}
\\
&=
\mleft.
\mleft(
	0, 
	\mleft( \MC \mright)_{\mathrm{pr}_2(p, g)} \circ \mathrm{D}_{(p,g)}\mathrm{pr}_2
\mright)\mright|_{\mathrm{V}_{(p, g)}\mleft( f^*\mathcal{G} \mright)}~,
\end{align*}
using that 
$L_{(p, g)}\colon \mleft(f^*\mathcal{G}\mright)_p \to \mleft(f^*\mathcal{G}\mright)_p$ so that $\mathrm{D}_{(p, g)}L_{(p, g)^{-1}}\colon\mathrm{V}_{(p, g)}\mleft( f^*\mathcal{G} \mright) \to \mathrm{V}_{(p, e_x)}\mleft( f^*\mathcal{G} \mright) = \mleft(f^*\mathcal{g}\mright)_{p}$, where $x \coloneqq f(p)$ and $e_x$ is the neutral element of $\mathcal{G}_x$.
In total:
\begin{align*}
\mleft(\mu_{f^*\mathcal{G}}\mright)_{(p, g)}
&=
\mathrm{D}_{(p, g)} L_{(p, g)^{-1}} \circ \mleft.\mathrm{pr}_2^!\pi^{\mathrm{vert}}\mright|_{(p, g)}
\\
&=
\mleft(
	0, 
	\mleft( \MC \mright)_{\mathrm{pr}_2(p, g)} \circ \mathrm{D}_{(p,g)}\mathrm{pr}_2
\mright)
	\circ \mleft.\mathrm{pr}_2^!\pi^{\mathrm{vert}}\mright|_{(p, g)}
\\
&=
\mleft(
	0, 
	\mleft.\mleft( \MC \circ \pi^{\mathrm{vert}}\mright)\mright|_{\mathrm{pr}_2(p, g)} \circ \mathrm{D}_{(p,g)}\mathrm{pr}_2
\mright)
\\
&=
\mleft(
	0, 
	\mleft( \mutot\mright)_{\mathrm{pr}_2(p, g)} \circ \mathrm{D}_{(p,g)}\mathrm{pr}_2
\mright)
\\
&=
\mleft.\mleft(\mathrm{pr}_2^!\mutot\mright)\mright|_{(p, g)}~.
\end{align*}
\end{remarks}

\begin{remarks}{$\mutot$ as Maurer-Cartan form on trivial LGBs}{TrivialLGBsAndTheirMCForm}
Let $G$ be a Lie group with Lie algebra $\mathfrak{g}$, and $\mathcal{G} \coloneqq M \times G \stackrel{\mathrm{pr}_1}{\to} M$ be the trivial LGB equipped with its canonical flat connection $\mathrm{H}\mathcal{G} \coloneqq \mathrm{pr}_1^*\mathrm{T}M$, where $\mathrm{pr}_1$ is the projection onto the first component in $M \times G$. Its LAB $\mathcal{g}$ is also a trivial bundle, $M \times \mathfrak{g}$, and we have several identities (recall \ref{cor:TLGBAsLGB})
\begin{equation*}
\mathrm{T}\mathcal{G}
\cong
\mathrm{pr}_1^*\mathrm{T}M \oplus \mathrm{pr}_2^*\mathrm{T}G
\cong
\mathrm{pr}_1^*\mathrm{T}M \oplus \mathfrak{g}
\cong
\mathrm{pr}_1^*\mathrm{T}M \oplus \mathrm{pr}_1^*\mathcal{g}
=
\mathrm{H}\mathcal{G} \oplus \mathrm{V}\mathcal{G},
\end{equation*}
where $\mathrm{pr}_2\colon M \times G \to G$ is the projection onto the second component,
and the projection onto the vertical bundle is then equivalent to $\mathrm{Dpr}_2 \in \Omega^1\mleft(\mathcal{G}; \mathrm{pr}_2^*\mathrm{T}G\mright)$. 

Now let us view $G$ as an LGB over a point $\{*\}$, then $\mathcal{G} = f^*G$, where $f: M \to \{*\}$. Making use of the uniqueness of $\pi^{\mathrm{vert}} = \mathrm{id}_{\mathrm{T}G}$ for $G \to \{*\}$, we have for the pullback connection $\mathrm{pr}_2^!\pi^{\mathrm{vert}} = \mathrm{Dpr}_2$ which is precisely the projection for $\mathcal{G}$, and therefore we can use \ref{rem:PullbackTotMCForm} to derive
\begin{equation*}
\mutot
=
\mathrm{pr}_2^!\mu_G,
\end{equation*}
where $\mu_G$ is the Maurer-Cartan form of $G$.
\end{remarks}

The Maurer-Cartan form is important for gauge transformations because it induces the \textbf{Darboux derivative $\Delta$} as for example given in \cite[\S 5.1, page 182ff.]{mackenzieGeneralTheory}. We can canonically extend this to an LGB setup:

\begin{definitions}{Generalised Darboux derivative}{DarbouxDerivativeOnLGBs}
Let $\pi_{\mathcal{G}} \colon \mathcal{G} \to M$ be an LGB over a smooth manifold $M$, and $\mathrm{H}\mathcal{G}$ be a horizontal distribution of $\mathcal{G}$.
For $\sigma \in \Gamma(\mathcal{G})$ we define the \textbf{Darboux derivative $\Delta \sigma \in \Omega^1(M; \mathcal{g})$}
\begin{equation*}
\Delta \sigma
=
\sigma^! \mutot
=
\mleft(\sigma^* \mutot \mright) \circ \mathrm{D}\sigma.
\end{equation*}
We may also write $\Delta^{\mathcal{G}}$ instead of $\Delta$ in order to accentuate the LGB.
\end{definitions}

\begin{remarks}{Several notations of the Darboux derivative}{RemarkABoutDarbouxNotationWRTPullback}
The notation with the pullback $\sigma^*$ is only needed if one wants to view $\Delta \sigma$ as a $C^\infty(M)$-linear map $\mathfrak{X}(M) \to \Gamma(\mathcal{g})$; in this case it is a composition of base-preserving vector bundle morphisms $\mathrm{D}\sigma\colon \mathrm{T}M \to \sigma^*\mathrm{T}\mathcal{G}$ and $\sigma^*\mutot\colon \sigma^*\mathrm{T}\mathcal{G} \to \sigma^*\pi_{\mathcal{G}}^*\mathcal{g} \cong \mathcal{g}$ which can be extended to sections, where $\pi$ is the projection of $\mathcal{G}$.

Of course one can also view $\Delta \sigma$ as a map $\mathrm{T}M \to \mathcal{g}$ which one can write as the composition of $\mathrm{D}\sigma: \mathrm{T}M \to \mathrm{T}\mathcal{G}$ and $\mutot: \mathrm{T}\mathcal{G} \to \mathcal{g}$, that is,
\begin{equation*}
\Delta \sigma
=
\mutot \circ \mathrm{D}\sigma.
\end{equation*}
If one wants to emphasize base points of the involved pullback bundles, that is, if one views $\mutot$ as a map $\mathrm{T}\mathcal{G} \to \pi_{\mathcal{G}}^*\mathcal{g}$, then one can write point-wise
\begin{equation*}
\mleft(\mutot\mright)_{\sigma_x} \circ \mathrm{D}_x\sigma
=
\bigl(
	\sigma_x, \mleft.\mleft(\Delta \sigma\mright)\mright|_x
\bigr)
\end{equation*}
for all $x \in M$.
\end{remarks}

As a derivative we expect a Leibniz rule as for the classical Darboux derivative (see \cite[\S 5.1, Eq.\ 2, page 182]{mackenzieGeneralTheory}). However, we cannot make a general statement about that yet, as long as we do not fix a more specific type of connection on $\mathcal{G}$. We will come back to this later.

Let us now discuss the pullback situation for the Darboux derivative.

\begin{remarks}{Pullback situation: Part II}{PullBackDarboux}
Following \ref{rem:PullbackTotMCForm} and its notation and results (see \ref{rem:GeneralPullBackLGBConnSituation} for the initial setup), we can calculate $\Delta^{f^*\mathcal{G}}\mleft( f^*\sigma \mright) \in \Omega^1(N; f^*\mathcal{g})$ for $\sigma \in \mathcal{G}$; first we write
\begin{equation*}
\mleft.f^*\sigma\mright|_p
=
\mleft( p, \sigma_{f(p)} \mright)
\end{equation*}
for all $p \in N$, and thus
\begin{align}\label{PullBackDarbouxOnPullbackSections}
\Delta^{f^*\mathcal{G}}\mleft( f^*\sigma \mright)
&=
\mleft( f^*\sigma \mright)^! \mu_{f^*\mathcal{G}}
\nonumber
\\
&=
\mleft( f^*\sigma \mright)^! \mathrm{pr}_2^! \mutot
\nonumber
\\
&=
\mleft( \mathrm{pr}_2\circ f^*\sigma \mright)^! \mutot
\nonumber
\\
&=
\mleft( \sigma \circ f \mright)^!\mutot
\nonumber
\\
&=
\mleft(\mleft( \sigma \circ f \mright)^*\mutot \mright)
	\circ \mathrm{D}(\sigma\circ f)
\nonumber
\\
&=
\underbrace{\mleft( f^* \sigma^*\mutot \mright) \circ f^*\mleft(\mathrm{D}\sigma\mright)}_{= f^*\mleft( \mleft(\sigma^*\mutot \mright) \circ \mathrm{D}\sigma \mright)} \circ ~\mathrm{D}f
\nonumber
\\
&=
f^!\mleft(\mleft( \sigma^*\mutot \mright) \circ \mathrm{D}\sigma\mright)
\nonumber
\\
&=
f^!\mleft( \Delta^{\mathcal{G}} \sigma \mright),
\end{align}
where we rewrote the chain rule
\begin{equation*}
\mathrm{D}(\sigma \circ f)
=
f^*\mleft(\mathrm{D}\sigma\mright) \circ \mathrm{D}f,
\end{equation*}
but as in \ref{rem:RemarkABoutDarbouxNotationWRTPullback} one can decide to omit this notation. Such Darboux derivatives related to the pullback connection on $f^*\mathcal{G}$ we call the \textbf{pullback Darboux derivative}, and similar to the notation of pullback vector bundle connections we write
\begin{equation*}
f^*\Delta^{\mathcal{G}}
\coloneqq
\Delta^{f^*\mathcal{G}}.
\end{equation*}
\end{remarks}

\begin{remarks}{Canonical flat Darboux derivative}{DarbouxOnCanonFlat}
Given the situation as in \ref{rem:TrivialLGBsAndTheirMCForm} we can use \ref{rem:PullBackDarboux} to derive
\begin{equation*}
\Delta^{\mathcal{G}}
=
f^*\Delta^0,
\end{equation*}
where $\Delta^0$ on the right hand side is the Darboux derivative of the Lie group $G$ as a bundle over point. Trivially, $\Delta^0 g \equiv 0$ for all $g \in G$, and thus
\begin{equation*}
\Delta^{\mathcal{G}}(f^*g)
=
f^!\mleft( \Delta^0 g\mright)
=
0
\end{equation*}
for all $g \in G$ viewed as a section of $G \to \{0\}$, so that $f^*g$ are the constant sections of $\mathcal{G}$. In other words such \textbf{$\sigma$ are parallel w.r.t.\ $\Delta$}. Since constant sections generate all sections of $\mathcal{G}$, we then speak of the \textbf{canonical flat Darboux derivative}, and constant sections are its parallel sections.

However, just because $\Delta^0$ is a zero map does not mean that also $\Delta^{\mathcal{G}}$ is zero. On one hand because of a Leibniz rule which we will discuss later in more detail, and on the other hand, as mentioned in \ref{BasicNotations}, we can view general sections $\sigma \in \Gamma(\mathcal{G})$ (not necessarily constant) equivalently as a smooth map $\mathrm{pr}_2 \circ \sigma\colon M \to G$, denoted by $\overline{\sigma}$ for bookkeeping reasons. By \ref{rem:TrivialLGBsAndTheirMCForm} we then get
\begin{equation*}
\Delta^{\mathcal{G}} \sigma
=
\sigma^! \mathrm{pr}_2^!\mu_G
=
\mleft( \mathrm{pr}_2 \circ \sigma \mright)^!\mu_G
=
\Delta^G \overline{\sigma}
\end{equation*}
for all $\sigma \in \Gamma(\mathcal{G})$, where $\Delta^G$ is the ``classical'' Darboux derivative related to $\mu_G$. This emphasizes why we can speak of a canonical flat derivative due to the fact that $\mu_G$ is flat (Maurer-Cartan equation), and we may simply write $\Delta^{\mathcal{G}} = \Delta^G$.
\end{remarks}

We can actually rewrite the infinitesimal action; recall the notation introduced in \ref{rem:FundVecNotationOnPullbackBundle}.

\begin{remarks}{Darboux derivative in the infinitesimal LGB action}{DarbouxInActionInfinit}
Recall \ref{prop:DiffOfLGBAction}; keeping the same notation as in this theorem but denoting the projection of $\mathcal{P}$ by $\pi$, we checked several times that $Y - \mathrm{D}_{x}\sigma (\omega)$ is vertical, and thus we can now write
\begin{align*}
\mathrm{D}_{(p, g)}\Phi(X, Y)
&=
\mathrm{D}_pr_\sigma(X)
	+ \mleft.{\overline{\mleft( \MC\mright)_{g} \bigl(Y - \mathrm{D}_{x}\sigma (\omega)\bigr)}}\mright|_{p \cdot g}
\\
&=
\mathrm{D}_pr_\sigma(X)
	+ \mleft.{\overline{\mleft( \MC \circ \pi^{\mathrm{vert}}\mright)_{g} \bigl(Y - \mathrm{D}_{x}\sigma (\omega)\bigr)}}\mright|_{p \cdot g}
\\
&=
\mathrm{D}_pr_\sigma(X)
	+ \mleft.{\overline{\mleft( \MC \circ \pi^{\mathrm{vert}}\mright)_{g} (Y)}}\mright|_{p \cdot g}
	- \mleft.{\overline{\mleft( \MC \circ \pi^{\mathrm{vert}}\mright)_{\sigma_x} \bigl( \mathrm{D}_{x}\sigma (\omega)\bigr)}}\mright|_{p \cdot g}
\\
&=
\mathrm{D}_pr_\sigma(X)
	+ \mleft.{\overline{\mleft( \mutot\mright)_{g} (Y)}}\mright|_{p \cdot g}
	- \mleft.{\overline{ \mleft.(\Delta \sigma)\mright|_x (\omega)}}\mright|_{p \cdot g}
\\
&=
\mathrm{D}_pr_\sigma(X)
	- \mleft.{\overline{ \mleft.\mleft(\pi^!\Delta \sigma\mright)\mright|_p (X)}}\mright|_{p \cdot g}
	+ \mleft.{\overline{\mleft( \mutot\mright)_{g} (Y)}}\mright|_{p \cdot g}
\end{align*}
If $\mathcal{G}$ is a trivial LGB, then this emphasizes again that we recover the typical Leibniz rule by choosing a constant section $\sigma$, using \ref{rem:TrivialLGBsAndTheirMCForm} and \ref{rem:DarbouxOnCanonFlat}.
\end{remarks}

\begin{remarks}{Idea of Ehresmann connections using the Darboux derivative}{DefOfConnectionIdeaWithDarboux}
Recall the notation and discussion around the terms in Eq.\ \eqref{OiTHatIsHowWeFormulateHorizSymmetry} which will be important for our definition of a connection on principal LGB-bundles. The terms in Eq.\ \eqref{OiTHatIsHowWeFormulateHorizSymmetry} can be similarly rewritten as in
\begin{align*}
\mathrm{D}_pr_\sigma\mleft( 
	X 
\mright)
	- \mleft.{\overline{
		\mleft( \mutot\mright)_g \bigl(
		\mathrm{D}_p (\sigma \circ \pi)(X)
	\bigr)
	}}\mright|_{p \cdot g}
&=
\mathrm{D}_pr_\sigma\mleft( 
	X 
\mright)
	- \mleft.{\overline{
		\bigl(\mleft(\Delta\sigma\mright)_x \circ \mathrm{D}_p\pi\bigr) (X)
	}}\mright|_{p \cdot g}
\\
&=
\mathrm{D}_pr_\sigma\mleft( 
	X 
\mright)
	- \mleft.{\overline{
		\mleft. \mleft( \pi^!\Delta\sigma \mright) \mright|_p(X)
	}}\mright|_{p \cdot g}~.
\end{align*}
We will use this form for the definition of the principal bundle connection. As in \ref{rem:DarbouxInActionInfinit}, if $\mathcal{G}$ is trivial and $\sigma$ a constant section corresponding to a Lie group element $g$, then these terms are just
\begin{equation*}
\mathrm{D}_p r_g (X)~,
\end{equation*}
making use of \ref{ex:TrivialLGBAction}.
\end{remarks}

In fact, the Darboux derivative naturally induces a connection on $\mathcal{g}$ as LAB of $\mathcal{G}$. One may have expected that since $\mathrm{H}\mathcal{G}$ should infinitesimally induce a horizontal distribution on $\mathcal{g}$; recall the exponential map introduced in \ref{ExponentialMapSubsection}. Also recall that by definition we know that the vertical bundle $\mathrm{V}\mathcal{g}$ of the LAB $\pi_{\mathcal{g}}\colon\mathcal{g} \to M$ is isomorphic to $\pi_{\mathcal{g}}^*\mathcal{g}$; in the following we will use the natural projection onto the second component of $\pi_{\mathcal{g}}^*\mathcal{g}$ but defined as a morphism on $\mathrm{V}\mathcal{g}$, denoted by $\mathrm{pr}_2\colon \mathrm{V}\mathcal{g} \to \mathcal{g}$.

\begin{propositions}{LGB connection induces LAB connection}{FinallyTheNablaInduction}
Let $\pi_{\mathcal{G}}\colon\mathcal{G} \to M$ be an LGB over a smooth manifold $M$, and $\mathrm{H}\mathcal{G}$ be a horizontal distribution of $\mathcal{G}$. Then the map $\nabla^{\mathcal{G}}\colon \Gamma(\mathcal{g}) \to \Omega^1(M; \mathcal{g})$, $\nu \mapsto \nabla^{\mathcal{G}}\nu$ denoted as an element of $\Omega^1(M; \mathcal{g})$ by $X \mapsto \nabla^{\mathcal{G}}_X \nu$, defined by
\begin{equation*}
\mleft.\nabla^{\mathcal{G}}_X \nu\mright|_x
\coloneqq
\mathrm{pr}_2 \mleft(
\mleft.\frac{\mathrm{d}}{\mathrm{d}t}\mright|_{t=0} \Bigl( \mleft(\Delta \e^{t \nu}\mright)_x (X) \Bigr)
\mright)
\end{equation*}
for all $x \in M$, $X \in \mathrm{T}_xM$ and $\nu \in \Gamma(\mathcal{g})$, is a well-defined vector bundle connection on $\mathcal{g}$, where $t \in \mathbb{R}$, and $\mathrm{pr}_2: \mathrm{V}\mathcal{g} \to \mathcal{g}$ is the projection onto the second component of $\mathrm{V}\mathcal{g}$ naturally viewed as pullback bundle.
\end{propositions}

\begin{remarks}{Simplified notation}{RemarkAboutPrTwoInDarbouxDerivative}
The notation of $\mathrm{pr}_2$ is usually omitted in such constructions due to the fact that 
\begin{equation*}
t \mapsto \mleft(\Delta \e^{t \nu}\mright)_x (X)
\end{equation*}
is a curve with values in $\mathcal{g}_x$, a vector space, so that one canonically uses the identification of tangent spaces of vector spaces with itself to show that $\mleft.\nabla^{\mathcal{G}}_X \nu\mright|_x \in \mathcal{g}_x$.
\end{remarks}


\begin{proof}[Proof of \ref{prop:FinallyTheNablaInduction}]
\leavevmode\newline
We will not prove this statement in full generality: If a reader is interested into the proof, then please check out an earlier version of this paper on arXiv. However, the general statement was rather tedious to show in a rigorous manner, while we will only study $\nabla^\caG$ for \textit{multiplicative} $\rmH\caG$, that is, a parallel transport on $\caG$ by Lie group automorphism (see later). This case is well-understood, and in the multiplicative situation \ref{prop:FinallyTheNablaInduction} is shown in \cite[Proposition 4.22]{LAURENTGENGOUXStienonXuMultiplicativeForms}.
\end{proof}

\begin{definitions}{LGB connection on its LAB}{ConnectionOnLAB}
Let $\pi_{\mathcal{G}}\colon\mathcal{G} \to M$ be an LGB over a smooth manifold $M$, and $\mathrm{H}\mathcal{G}$ be a horizontal distribution of $\mathcal{G}$. Then we call the vector bundle connection $\nabla^{\mathcal{G}}$ on $\mathcal{g}$ of \ref{prop:FinallyTheNablaInduction} the \textbf{$\mathcal{G}$-connection (on the LAB $\mathcal{g}$)}.
\end{definitions}


\begin{examples}{Canonical flat $\mathcal{G}$-connection}{CanonicalFlatGConnection}
If we again focus on trivial LGBs with their canonical flat connection as in Remark \ref{rem:TrivialLGBsAndTheirMCForm} and \ref{rem:DarbouxOnCanonFlat}, we can quickly derive that $\nabla^{\mathcal{G}}$ is then the canonical flat connection on the trivial LAB $\mathcal{g} = M \times \mathfrak{g}$ of $\mathcal{G} = M \times G$, $\mathfrak{g}$ the Lie algebra of the Lie group $G$. Let $\nu$ be a constant section of $\mathcal{g}$, then $\e^{t\nu}$ is a constant section of $\mathcal{G}$ for all $t \in \mathbb{R}$, so that by \ref{rem:DarbouxOnCanonFlat}
\begin{equation*}
\Delta^{\mathcal{G}} \e^{t\nu}
=
0,
\end{equation*}
and thus
\begin{equation*}
\nabla^{\mathcal{G}} \nu
=
0
\end{equation*}
for all constant sections $\nu \in \Gamma(\mathcal{g})$. By the uniqueness of the canonical flat connection (w.r.t.\ a trivialisation) we conclude that $\nabla^{\mathcal{G}}$ is the canonical flat connection on $\mathcal{g} = M \times \mathfrak{g}$.
\end{examples}

\begin{remarks}{Pullback situation: Part III}{PullBackConnectionOfNablaG}
Following \ref{rem:PullBackDarboux} (see \ref{rem:GeneralPullBackLGBConnSituation} for the initial setup) and all the involved notation we can again derive something related to the pullback LGB $f^*\mathcal{G}$. First of all, let us denote the exponential of $f^*\mathcal{G}$ and $\mathcal{G}$ by $\e_{f^*\mathcal{G}}$ and $\e_{\mathcal{G}}$, respectively, then we clearly have
\begin{equation*}
\e_{f^*\mathcal{G}}^{(p, \nu_x)}
=
\exp_{\mleft(f^*\mathcal{G}\mright)_p}(p, \nu_x)
=
\mleft( p, \exp_{\mathcal{G}_x}(\nu_x) \mright)
=
\mleft( p, \e_{\mathcal{G}}^{\nu_x} \mright)
\end{equation*}
for all $(p, \nu_x) \in \mleft(f^*\mathcal{g}\mright)_p$, where $x \coloneqq f(p)$ and so $\nu_x \in \mathcal{g}_x$, and where we made use of that $\mleft(f^*\mathcal{G}\mright)_p \cong \mathcal{G}_x$ as Lie groups via the projection $\mathrm{pr}_2$ onto the second component (recall \ref{cor:PullbackLGB}), that is, we have used the well-known relation
\begin{equation*}
\mathrm{pr}_2\mleft( \exp_{\mleft(f^*\mathcal{G}\mright)_p}(p, \nu_x) \mright)
=
\mleft(\exp_{\mathcal{G}_x} \circ~ \mathrm{D}_{(p, e_x)}\mathrm{pr}_2\mright)(p, \nu_x)
=
\exp_{\mathcal{G}_x}(\nu_x),
\end{equation*}
see \textit{e.g.}\ \cite[\S 1.7, Thm.\ 1.7.16, page 59]{Hamilton} for the general formula; then use that $\mathrm{pr}_2|_{\mleft( f^*\mathcal{G} \mright)_p}$ is bijective. Hence, we get
\begin{equation*}
\e_{f^*\mathcal{G}}^{f^*\nu}
=
f^*\mleft( \e_{\mathcal{G}}^\nu \mright)
\end{equation*}
for all $\nu \in \Gamma(\mathcal{g})$. Due to clatch of notation we relabel the projection $\mathrm{V}\mathcal{g} \to \mathcal{g}$ in \ref{prop:FinallyTheNablaInduction} to $\pi_2$, and then observe, by using Eq.\ \eqref{PullBackDarbouxOnPullbackSections},
\begin{align*}
\mleft.\nabla^{f^*\mathcal{G}}_Y (f^*\nu)\mright|_p
&=
\pi_2 \mleft(
\mleft.\frac{\mathrm{d}}{\mathrm{d}t}\mright|_{t=0} \Bigl( \mleft(\Delta^{f^*\mathcal{G}} \e_{f^*\mathcal{G}}^{t \cdot f^*\nu}\mright)_p (Y) \Bigr)
\mright)
\\
&=
\pi_2 \mleft(
\mleft.\frac{\mathrm{d}}{\mathrm{d}t}\mright|_{t=0} \Bigl( \mleft(\Delta^{f^*\mathcal{G}} \bigl( f^*\mleft( \e_{\mathcal{G}}^{t\nu} \mright) \bigr) \mright)_p (Y) \Bigr)
\mright)
\\
&=
\pi_2 \mleft(
\mleft.\frac{\mathrm{d}}{\mathrm{d}t}\mright|_{t=0} \Bigl( \mleft( f^!\mleft( \Delta^{\mathcal{G}} \e_{\mathcal{G}}^{t\nu} \mright) \mright)_p (Y) \Bigr)
\mright)
\\
&=
\pi_2 \mleft(
\mleft.\frac{\mathrm{d}}{\mathrm{d}t}\mright|_{t=0} \biggl( p, 
\mleft( \Delta^{\mathcal{G}} \e_{\mathcal{G}}^{t\nu} \mright)_{f(p)} \bigl( \mathrm{D}_pf(Y) \bigr) 
\biggr)
\mright)
\\
&=
\pi_2 \mleft( 0,
\mleft.\frac{\mathrm{d}}{\mathrm{d}t}\mright|_{t=0} \biggl( 
\mleft( \Delta^{\mathcal{G}} \e_{\mathcal{G}}^{t\nu} \mright)_{f(p)} \bigl( \mathrm{D}_pf(Y) \bigr) 
\biggr)
\mright)
\\
&=
\pi_2 \mleft(
\mleft.\frac{\mathrm{d}}{\mathrm{d}t}\mright|_{t=0} \biggl( 
\mleft( \Delta^{\mathcal{G}} \e_{\mathcal{G}}^{t\nu} \mright)_{f(p)} \bigl( \mathrm{D}_pf(Y) \bigr) 
\biggr)
\mright)
\\
&=
\mleft.\nabla^{\mathcal{G}}_{\mathrm{D}_pf(Y)} \nu\mright|_{f(p)}
\end{align*}
for all $\nu \in \Gamma(\mathcal{g})$, $p \in N$ and $Y \in \mathrm{T}_p N$. By the uniqueness of pullback vector bundle connections we therefore get
\begin{equation*}
\nabla^{f^*\mathcal{G}}
=
f^*\nabla^{\mathcal{G}}.
\end{equation*}
\end{remarks}

\subsubsection{First step towards towards associated bundles}

In order to provide a concise definition of connection 1-forms on principal bundles, we will need to introduce some canonical form of action which will be obviously related to the action needed for an analogue to the notion of associated bundles related to typical principal bundles. However, we will neither discuss nor introduce a more general notion of associated bundles in this paper; so, there will be just the "first step". 

\begin{propositions}{Canonical $\caG$-action on pullbacks of principal bundles}{CanonicalLGBActionForAssociatedBundles}
Let $\pi_{\mathcal{G}}\colon \mathcal{G} \to M$ be an LGB over a smooth manifold $M$ and $\pi\colon\mathcal{P} \to M$ a principal $\mathcal{G}$-bundle. Furthermore let $N$ be another smooth manifold and $f\colon N \to M$ a smooth map on which $\mathcal{G}$ acts on the left as in \ref{def:LiegroupACtion}. Then on the pullback manifold $\mathcal{P} \times_M N \coloneqq f^*\mathcal{P}$, whose pairs of points are now reordered as in
\begin{align*}
\mathcal{P} \times_M N
&\coloneqq
\left\{
	(p, x) \in \mathcal{P} \times N
	~\middle|~
	\pi(p) = f(x)
\right\},
\end{align*}
we have a right $\mathcal{G}$-action given by
\begin{align*}
\mleft(\mathcal{P} \times_M N\mright) * \mathcal{G} &\to \mathcal{P} \times_M N,\\
(p, x, g) &\mapsto (p, x) \cdot g \coloneqq \mleft( p \cdot g, g^{-1} \cdot x \mright)
\end{align*}
where $\mleft(\mathcal{P} \times_M N\mright) * \mathcal{G}$ is given as pullback of $\mathcal{G}$ w.r.t.\ the map $\overline{\pi}: \mathcal{P} \times_M N \to M$, $(p,x) \mapsto \pi(p)$.
\end{propositions}

\begin{proof}
\leavevmode\newline
By \ref{def:LiegroupACtion} it is clear that this action is well-defined, due to $f(x)= \pi(p) = \pi_\caG(g) = \pi_\caG\mleft(g^{-1}\mright)$, and
\begin{equation*}
\pi(p\cdot g) = \pi_{\mathcal{G}}(g) = f\mleft( g^{-1} \cdot x \mright)
\end{equation*}
for all $(p,x,g) \in \mleft(\mathcal{P} \times_M N\mright) * \mathcal{G}$,
and this also implies
\begin{equation*}
\overline{\pi}\bigl( (p, x) \cdot g \bigr)
=
\pi(p\cdot g)
=
\pi_{\mathcal{G}}(g)~.
\end{equation*}
By construction we also have a smooth action, since it is the composition of maps
\begin{align*}
\mleft(\mathcal{P} \times_M N\mright) * \mathcal{G} &\to (\mathcal{P} * \mathcal{G}) \times (\mathcal{G} * N) \to \mathcal{P} \times_M N,\\
(p, x, g) &\mapsto \bigl( (p, g), (g, x) \bigr) \mapsto \mleft( p \cdot g, g^{-1} \cdot x \mright).
\end{align*}
The first arrow is clearly an embedding, and the second arrow is just the restriction of the smooth diagonal action,
\begin{align*}
(\mathcal{P} * \mathcal{G}) \times (\mathcal{G} * N) &\to \mathcal{P} \times N,\\
\bigl( (p, g), (q, p) \bigr) &\mapsto \mleft( p \cdot g, q^{-1} \cdot p \mright),
\end{align*}
onto an embedded submanifold, and for its image make use of that $\mathcal{P} \times_M N$ is an embedded submanifold of $\mathcal{P} \times N$. Associativity follows simply by
\begin{equation*}
\bigl((p, x) \cdot g \bigr) \cdot q
=
\mleft( p \cdot g, g^{-1} \cdot x \mright) \cdot q
=
\mleft( p \cdot gq, (gq)^{-1} \cdot x \mright)
=
\mleft( p, x \mright) \cdot (gq)
\end{equation*}
for all $g, q \in \mathcal{G}_y$ ($y \coloneqq \pi(p) = f(x)$); it is trivial to check that $\mleft( p, x \mright) \cdot e_y = (p, x)$.
\end{proof}

As mentioned in \ref{rem:ItIsAPrincipalAction}, constructions of quotients related to the LGB action are possible here. Thus, using the last proposition, one should be able to construct associated bundles in this more general setting, but in the same fashion as usual. We mainly need this proposition for the following examples.

\begin{examples}{Adjoint action on the vertical bundle of $\mathcal{P}$}{AdjointACtionOnVP}
Recall the adjoint representation of $\mathcal{G}$ on its LAB $\pi_\cag\colon\mathcal{g} \to M$, \ref{ex:LGBAdjointRep}. Using the notation of \ref{prop:CanonicalLGBActionForAssociatedBundles} we have a right $\mathcal{G}$-action on $\mathcal{P} \times_M \mathcal{g} = \{(p, v) \in \mathcal{P} \times \mathcal{g} ~ | ~ \pi(p) = \pi_{\mathcal{g}}(v) \}$ given by
\begin{equation*}
(p, v) \cdot g
\coloneqq
\mleft( p \cdot g, \mathrm{Ad}_{g^{-1}}(v) \mright)
\end{equation*}
for all $p \in \mathcal{P}_x$ ($x \in M$), $v \in \mathcal{g}_x$ and $g \in \mathcal{G}_x$. Observe that $\mathcal{P} \times_M \mathcal{g} = \pi^*\mathcal{g}$ which is isomorphic to $\mathrm{V}\mathcal{P}$ by \ref{cor:VerticalBundleOfPrincIsNearlyAsUsual}. We will denote this action shortly by
\begin{equation*}
\sAd_{g^{-1}}(p, v)
\coloneqq
(p,v) \cdot g,
\end{equation*}
the \textbf{adjoint representation of $\mathcal{G}$ on $\mathrm{V}\mathcal{P}$}; not to be confused with the adjoint representation of $\pi^*\mathcal{G}$ on $\pi^*\mathcal{g}$. In fact, it is trivial to check that $\sAd\colon \mathcal{G} \to \mathrm{Aut}(\pi^*\mathcal{g})$, $g \mapsto \sAd_{g}$, is a $\mathcal{G}$-representation on $\mathcal{P} \times_M \mathcal{g} = \pi^*\mathcal{g} \cong \mathrm{V}\mathcal{P}$.

The quotient of $\mathcal{P} \times_M \mathcal{g}$ w.r.t.\ this group action should lead to a structure which is the generalization of the adjoint bundle in typical formulations of gauge theory.
\end{examples}

Similarly, we can define the conjugation action on its integrated bundle; recall the conjugation defined in \ref{def:LeftRightTranslationConjugation}. Especially observe that the conjugation defines a left $\mathcal{G}$-action on itself by
\begin{align*}
\mathcal{G} * \mathcal{G} &\to \mathcal{G},\\
(g,q) &\mapsto c_g(q) = g q g^{-1}.
\end{align*}

\begin{examples}{Conjugation action over $\mathcal{P}$}{ConjugationActionForTheGeneralInnerGroupBundle}
Let us look at $\mathcal{P} \times_M \mathcal{G} \cong \pi^*\mathcal{G}$. Then we have a right $\mathcal{G}$-action on $\mathcal{P} \times_M \mathcal{G}$ given by
\begin{align*}
(p, q) \cdot g
&\coloneqq
\mleft( p \cdot g, c_{g^{-1}}(q) \mright)
=
\mleft( p \cdot g, g^{-1} q g \mright)
\end{align*}
for all $p \in \mathcal{P}_x$ ($x \in M$), and $q,g \in \mathcal{G}_x$.
We will denote this also by
\begin{align*}
\mathcal{c}_{g^{-1}}(p, q)
\coloneqq
(p, q) \cdot g,
\end{align*}
the \textbf{conjugation of $\mathcal{G}$ on the integral of $\mathrm{V}\mathcal{P}$}. Taking a quotient of $\mathcal{P} \times_M \mathcal{G}$ over this action should lead to a generalization of inner group bundles as introduced in \ref{ex:InnerLGBs}.
\end{examples}

Let us finally define connections on principal bundles.

\subsubsection{Generalized connection 1-forms on principal bundles}

As usual, for a given horizontal distribution $\mathrm{H}\mathcal{P}$ of a principal $\mathcal{G}$-bundle $\pi\colon\mathcal{P} \to M$ over a smooth manifold $M$ we have by construction that
\begin{align*}
\mleft.\mathrm{D}_p \pi\mright|_{\mathrm{H}_p\mathcal{P}}\colon \mathrm{H}_p\mathcal{P} \to \mathrm{T}_xM
\end{align*}
is a vector space isomorphism for all $x \in M$ and $p \in \mathcal{P}_x$; similarly for $\mathcal{G}$ itself. We want to provide an identification of horizontal spaces in the sense of Equation \eqref{OiTHatIsHowWeFormulateHorizSymmetry}, also recall \ref{rem:DefOfConnectionIdeaWithDarboux}; especially recall the Darboux derivative introduced in \ref{DiscussingDarbouxDerivativeGeneral}. The following proposition and definition will be needed.

\begin{propositions}{The modified right-pushforward as well-defined isomorphism}{IsomorphismRightPushAndDarboux}
Let $\pi_{\mathcal{G}} \colon \mathcal{G} \to M$ be an LGB over a smooth manifold $M$ and $\pi\colon\mathcal{P} \to M$ a principal $\mathcal{G}$-bundle, and we have horizontal a distribution $\mathrm{H}\mathcal{G}$ on $\mathcal{G}$. Furthermore, for $g \in \mathcal{G}_x$ ($x \in M$) define the map\footnote{The font of $\mathcal{r}$ is a calligraphic r.}
\begin{align*}
\mleft.\mathrm{T}\mathcal{P}\mright|_{\mathcal{P}_x} &\to \mleft.\mathrm{T}\mathcal{P}\mright|_{\mathcal{P}_x},\\
X 
&\mapsto 
\mathcal{r}_{g*}(X) 
\coloneqq
\mathrm{D}_pr_\sigma\mleft( 
	X 
\mright)
	- \mleft.{\overline{
		\mleft. \mleft( \pi^!\Delta\sigma \mright) \mright|_p(X)
	}}\mright|_{p \cdot g},
\end{align*}
where $p \in \mathcal{P}_x$, $X \in \mathrm{T}_p \mathcal{P}$, and $\sigma$ is any (local) section of $\mathcal{G}$ with $\sigma_x = g$. Then $\mathcal{r}_{g*}$ is independent of the choice of the local section $\sigma$, and it is a vector bundle automorphism over the right-translation $r_g$. 

Furthermore, if we also have a horizontal distribution $\mathrm{H}\mathcal{P}$ on $\mathcal{P}$, then $\mathrm{H}_p\mathcal{P}$ is isomorphic via $\mathcal{r}_{g*}$ to a complement of $\mathrm{V}_{p \cdot g} \mathcal{P}$ in $\mathrm{T}_{p \cdot g}\mathcal{P}$ (this complement is not necessarily $\mathrm{H}_{p \cdot g}\mathcal{P}$).
\end{propositions}

\begin{proof}
\leavevmode\newline
In the following we have $(p, g) \in \mathcal{P} * \mathcal{G}$, $X \in \mathrm{T}_p \mathcal{P}$, and $\sigma$ is any (local) section of $\mathcal{G}$ with $\sigma_x = g$ ($x \coloneqq \pi(p)$).

$\bullet$ The well-definedness is just a simple consequence of the discussion around Equation \eqref{OiTHatIsHowWeFormulateHorizSymmetry}. As mentioned previously there is a unique $Y \in \mathrm{H}_g\mathcal{G}$ with
\begin{align*}
\mathrm{D}_g\pi_{\mathcal{G}}(Y)
&=
\mathrm{D}_p\pi(X)
\in
\mathrm{T}_x M.
\end{align*}
By \ref{rem:DarbouxInActionInfinit} we can derive
\begin{align*}
\mathrm{D}_{(p, g)}\Phi(X,Y)
&=
\mathcal{r}_{g*}(X),
\end{align*}
where we made use of that $Y$ is horizontal in $\mathcal{G}$ so that $\mleft(\mutot\mright)_g(Y) = 0$, and where $\Phi$ denotes the right $\mathcal{G}$-action on $\mathcal{P}$ as a map $\mathcal{P}*\mathcal{G} \to \mathcal{P}$. Therefore the independence of $\mathcal{r}_{g*}$ w.r.t.\ the choice of $\sigma$ follows, and this also shows that $\car_{g*}$ is a smooth map since it is a composition of smooth maps (interpreting $Y$ as the image of a smooth horizontal lift).

$\bullet$ $\mathcal{r}_{g*}$ is clearly a vector bundle morphism over $r_g$ by construction. Let us now show fibre-wise that $\mathcal{r}_{g*}$ is injective, then it is also bijective by dimensional reasons and thus an automorphism: First of all, given $p$ and $X$ choose a curve $\gamma\colon [0,1] \to M$ with $\gamma(0) = \pi(p) = x$ with velocity $\dot{\gamma}(0) = \rmD_p\pi(X)$. By usual arguments, there is a parallel section $\sigma$, that is, $\sigma_x=g$ and $\mleft.\Delta \sigma\mright|_x(\dot{\gamma}(0)) = 0$. Then 
\begin{equation*}
\car_{g*}(X) = \rmD_pr_\sigma(X)~,
\end{equation*}
so that it is clear that $\car_{g*}$ is injective (\ref{rem:SmoothnessOfACtionTranslations}). It also follows that $\rmD_{p}\pi(\car_{g*}(X)) = \rmD_{p}\pi(X)$, therefore altogether, also by dimensional reasons: The image $\mathrm{Im}\mleft( \mleft.\car_{g*}\mright|_{\mathrm{H}_p\mathcal{P}} \mright)$ of $\mleft.\car_{g*}\mright|_{\mathrm{H}_p\mathcal{P}}$ has to be a complement subspace of $\mathrm{V}_{p \cdot g}\mathcal{P}$ in $\mathrm{T}_{p \cdot g}\mathcal{P}$. $\mathrm{Im}\mleft( \mleft.\car_{g*}\mright|_{\mathrm{H}_p\mathcal{P}} \mright)$ is not necessarily equal to $\mathrm{H}_{p \cdot g}\mathcal{P}$ in general but of the same dimension, especially also of the same dimension as $\mathrm{H}_p \mathcal{P}$.
\end{proof}

Hence, we formally define:

\begin{definitions}{Modified pushforward via right-translation}{WeModTheRIghtPush}
Let $\mathcal{G} \to M$ be an LGB over a smooth manifold $M$ and $\pi\colon\mathcal{P} \to M$ a principal $\mathcal{G}$-bundle, and we have a horizontal distribution $\mathrm{H}\mathcal{G}$ on $\mathcal{G}$. Then we define the \textbf{modified right-pushforward $\mathcal{r}_{g*}$ (with $g \in \mathcal{G}_x$, $x \in M$)} as the vector bundle isomorphism $\mleft.\mathrm{T}\mathcal{P}\mright|_{\mathcal{P}_x} \to \mleft.\mathrm{T}\mathcal{P}\mright|_{\mathcal{P}_x}$ over $r_g$ as given in \ref{prop:IsomorphismRightPushAndDarboux} by
\begin{equation*}
\mathcal{r}_{g*}(X)
\coloneqq
\mathrm{D}_pr_\sigma\mleft( X \mright)
	- \mleft.{\overline{
		\mleft. \mleft( \pi^!\Delta\sigma \mright) \mright|_p(X)
	}}\mright|_{p \cdot g}
\end{equation*}
for all $p \in \mathcal{P}_x$ and $X \in \mathrm{T}_p \mathcal{P}$, where $\sigma$ is any (local) section of $\mathcal{G}$ with $\sigma_x = g$.

Similarly, for a (local) section $\sigma$ of $\mathcal{G}$ we define the \textbf{modified right-pushforward $\mathcal{r}_{\sigma*}$ with $\sigma$} as a (local) vector bundle isomorphism $\mathrm{T}\mathcal{P} \to \mathrm{T}\mathcal{P}$ by
\begin{equation*}
\mathcal{r}_{\sigma*}(X)
\coloneqq
\mathcal{r}_{\sigma_x*}(X)
=
\mathrm{D}_pr_\sigma\mleft( X \mright)
	- \mleft.{\overline{
		\mleft. \mleft( \pi^!\Delta\sigma \mright) \mright|_p(X)
	}}\mright|_{p \cdot \sigma_{x}}
\end{equation*}
for all $X \in \mathrm{T}_p \mathcal{P}$ ($p \in \mathcal{P}_x$).
\end{definitions}

\begin{remarks}{Alternative versions of the modified pushforward}{AlternModPush}
As we saw in the discussion around Equation \eqref{OiTHatIsHowWeFormulateHorizSymmetry} and as we also have seen in the proof of \ref{prop:IsomorphismRightPushAndDarboux}, there are several alternative identities of the modified right-pushforward: 
\begin{itemize}
	\item There is a unique $Y \in \mathrm{H}_g\mathcal{G}$ with
	\begin{equation*}
	\mathcal{r}_{g*}(X)
	=
	\mathrm{D}_{(p, g)}\Phi(X,Y)
	\end{equation*}
	where $\Phi$ denotes the right $\mathcal{G}$-action on $\mathcal{P}$, which the reader may prefer over the formulation with the Darboux derivative. However, we will see in this work that the Darboux derivative allows to extend the ``classical'' arguments in a straight-forward way.
	\item If $\sigma$ is parallel along $\omega \coloneqq \mathrm{D}_p\pi(X)$, that is $\mleft.\Delta \sigma\mright|_x(\omega) = 0$, then
	\begin{equation*}
	\mathcal{r}_{g*}(X)
	=
	\rmD_pr_\sigma(X)~.
	\end{equation*}
\end{itemize}
\end{remarks}

\begin{remarks}{Restriction onto vertical bundle gives typical right-pushforward}{ModRightPushOnVertic}
It is trivial to check that we have
\begin{equation*}
\mathcal{r}_{g*}(X)
=
\mathrm{D}_pr_g(X)
\eqqcolon
r_{g*}(X)
\end{equation*}
for all $X \in \mathrm{V}_p\mathcal{P}$; we actually have proven this directly after Eq.\ \eqref{OiTHatIsHowWeFormulateHorizSymmetry}. As also pinpointed in \ref{rem:DefOfConnectionIdeaWithDarboux}, if $\mathcal{G}$ is a trivial LGB equipped with its canonical flat connection and $\sigma$ a constant section, then it is easy to see that $\mathcal{r}_{\sigma*} = \mathcal{r}_{g*} = r_{g*}$, where $r_g$ has to be understood as the right-translation of the canonical Lie group action inherited by $\mathcal{G}$ in the sense of \ref{ex:TrivialLGBAction}.
\end{remarks}

\begin{remarks}{Modified pushforward as 1-form}{RSigmaAnAuto}
\ref{prop:IsomorphismRightPushAndDarboux} trivially extends to $\mathcal{r}_{\sigma*}$, that is, $\mathcal{r}_{\sigma*}$ is a (local) automorphism of $\mathrm{T}\mathcal{P}$ over $r_\sigma$. Thus, we can view $\mathcal{r}_{\sigma*}$ as an element of $\Omega^1\mleft(\mathcal{P}; r_\sigma^*\mathrm{T}\mathcal{P}\mright)$.
\end{remarks}

Hence, the following definition makes sense, especially if thinking about what we discussed for Eq.\ \eqref{OiTHatIsHowWeFormulateHorizSymmetry}, also recall \ref{rem:DefOfConnectionIdeaWithDarboux}.

\begin{definitions}{Ehresmann connection on principal LGB-bundles}{FinallyTheConnection}
Let $\mathcal{G} \to M$ be an LGB over a smooth manifold $M$, $\pi\colon\mathcal{P} \to M$ a principal $\mathcal{G}$-bundle, and $\mathrm{H}\mathcal{G}$ and $\mathrm{H}\mathcal{P}$ horizontal distributions on $\mathcal{G}$ and $\mathcal{P}$, respectively. We call $\mathrm{H}\mathcal{P}$ an \textbf{Ehresmann connection connection on $\mathcal{P}$ (modified by $\rmH\caG$)} if it is \textbf{right-invariant (w.r.t.\ modified right-pushforward)}, \textit{i.e.}\
\begin{align*}
\mathcal{r}_{g*}\mleft( \mathrm{H}_p\mathcal{P} \mright)
&=
\mathrm{H}_{p\cdot g}\mathcal{P}
\end{align*}
for all $p \in \mathcal{P}_x$ and $g \in \mathcal{G}_x$ ($x \coloneqq \pi(p)$).
\end{definitions}

Most of the time we may just write \textit{connection} instead of \textit{Ehresmann connection}.

\begin{remarks}{Moerdijk-Mrčun condition}{StienonsDef}
There is also a definition of connections on such and more general principal bundles in \cite[\S 5.7, paragraph before Prop.\ 5.38, page 148]{GroupoidBasedPrincipalBundles} and a corresponding gauge theory in \cite{Signori:2008jd}. However, this reference provides a different type of definition; translated to our situation, it is based on assuming that $\mathcal{G}$ is defined over (in general) another base manifold $N$. In order to define the LGB action on $\mathcal{P}$ this reference introduces a moment map $\mu\colon \mathcal{P} \to N$ so that the action of an element $\mathcal{G}_y$ ($y \in N$) is defined on $\mu^{-1}(\{y\})$, especially the infinitesimal action $r_g$ of a fixed LGB element $g$ acts on tangent vectors of $\mu^{-1}(\{y\})$, not necessarily on the vertical structure of $\mathcal{P}$. Hence, in order to circumvent the problem we discussed in \ref{TheBigMotivationBehindEverything}, the reference's definition of a connection is then based on assuming that the horizontal distribution is tangent to the ``fibres'' $\mu^{-1}(\{y\})$, so the infinitesimal action $r_g$ now acts well-defined on the horizontal structure. Henceforth, the reference does not need to choose a connection on $\caG$.

However, this is not a suitable definition for us, because we want to allow the projection of $\mathcal{P}$ itself as moment map such that the fibres of $\mu$ are the fibres of $\caP$ so that the horizontal distributions cannot be tangent to the fibres of $\mu$. The reference's definition is rather restrictive for us, while our definition works for all principal $\mathcal{G}$-bundles, but one has to choose a connection on $\mathcal{G}$.
\end{remarks}

Now recall that the whole motivation behind \ref{def:FinallyTheConnection} comes from \eqref{OiTHatIsHowWeFormulateHorizSymmetry} (also recall \ref{rem:DefOfConnectionIdeaWithDarboux}) which itself stems from \eqref{PTHomomNEw}. We see that that Eq.\ \eqref{PTHomomNEw} is satisfied here by usual arguments:

\begin{corollaries}{Parallel transport and horizontal lift}{OtherDefsForEhresmann}
Assume the situation as in \ref{def:FinallyTheConnection}, then the notion of Ehresmann connection is equivalent to the following definitions:
\begin{enumerate}
	\item An Ehresmann connection on $\caP$ is a \textbf{parallel transport ${\rm PT}^\caP$} satisfying
	\begin{equation*}
		\mathrm{PT}_\alpha^{\mathcal{P}}(p \cdot g)
		=
		\mathrm{PT}_\alpha^{\mathcal{P}}(p) \cdot \mathrm{PT}_\alpha^{\mathcal{G}}(g)
	\end{equation*}
	for all $p \in \caP_x$, $g \in \caG_x$ and curves $\alpha$ starting at $x \in M$, where ${\rm PT}^\caG$ is the parallel transport corresponding to $\rm H \caG$.
	\item An Ehresmann connection on $\caP$ is a \textbf{horizontal lift} defined as a vector bundle morphism $\chi\colon \pi^*\rmT M \to \rmT \caP$ satisfying
	\begin{equation*}
	\chi(p\cdot g, X) = \car_{g*}\bigl( \chi(p, X) \bigr)
	\end{equation*}
	for all $p \in \caP_x$ ($x\in M$), $g \in \caG_x$ and $X \in \rmT_xM$.
\end{enumerate}
\end{corollaries}

As in ordinary principal bundle settings one can show that parallel transports are complete here. Before we jump to the examples let us quickly discuss the existence of Ehresmann connections: 

\begin{corollaries}{Existence of Ehresmann connections}{ExOfEhresmann}
Assume the situation as in \ref{def:FinallyTheConnection}, then an Ehresmann connection on $\caP$ exists if and only if the connection on $\caG$ is \textbf{multiplicative}, that is, its associated parallel transport ${\rm PT}^\caG$ acts by Lie group automorphisms.
\end{corollaries}

\begin{proof}
\leavevmode\newline
We will argue with the notion of parallel transport as in \ref{cor:OtherDefsForEhresmann}.
The proof of existence of the Ehresmann connection on $\caP$ given a multiplicative ${\rm PT}^\caG$ is similar to ordinary existence arguments because multiplicative connections are indeed Ehresmann connections on $\caG$ viewed as principal $\caG$-bundle, where $\rmH \caG$ modifies itself; then make use of \ref{lem:SectionsNowInduceIsomToLGBsNotNecTriv} and usual gluing arguments. For the other direction:
\begin{equation*}
\mathrm{PT}_\alpha^{\mathcal{P}}(p \cdot g \cdot q)
=
\mathrm{PT}_\alpha^{\mathcal{P}}(p) \cdot \mathrm{PT}_\alpha^{\mathcal{G}}(gq)
\end{equation*}
for all $(p, g, q) \in \mathcal{P}* \mathcal{G} * \mathcal{G}$ and a suitable curve $\alpha$, but also
\begin{equation*}
\mathrm{PT}_\alpha^{\mathcal{P}}(p \cdot g \cdot q)
=
\mathrm{PT}_\alpha^{\mathcal{P}}(p) \cdot \mathrm{PT}_\alpha^{\mathcal{G}}(g) \cdot \mathrm{PT}_\alpha^{\mathcal{G}}(q)~.
\end{equation*}
Due to the fact that we have a simply transitive $\mathcal{G}$-action on $\mathcal{P}$, it follows that the connection on $\mathcal{G}$ is multiplicative. 
\end{proof}

Multiplicative connections will be introduced later in more detail when we discuss the field strength.
Let us now introduce several examples.

\begin{examples}{Recovering the classical definition}{OurConnectionIsReallyMoreGeneral}
As we already discussed several times, especially recall \ref{rem:ModRightPushOnVertic} and \ref{ex:TheCLassicalPrincAsEx}, but ultimately by the discussion for Equation \eqref{OiTHatIsHowWeFormulateHorizSymmetry}: We recover the typical definition of a connection on a principal bundle if $\mathcal{P}$ is a classical principal bundle and if the trivial LGB $\mathcal{G}$ is equipped with its canonical flat connection $\mathrm{H}\mathcal{G}$. We call such a connection a \textbf{classical connection}.
\end{examples}

\begin{examples}{Associated LGBs}{AssociatedLGBsAndTheirCanonicalConnection}
Recall that LGBs themselves are principal bundles as in \ref{ex:TrivialPrincAsLGB}. Let $G, H$ be Lie groups, $P \to M$ a principal $G$-bundle over a smooth manifold $M$, and $\psi$ a $G$-representation on $H$. Then we have the LGB associated to the principal bundle $P$ and the representation $\psi$ on $H$
\begin{center}
	\begin{tikzcd}
	H \arrow{r}& \mathcal{H} \coloneqq P \times_\psi H \arrow{d} \\
	& M
	\end{tikzcd}
\end{center}
Fix a typical classical connection on $P$; as also introduced in \ref{TheBigMotivationBehindEverything} we have an associated parallel transport $\mathrm{PT}_\alpha^P\colon P_x \to \Gamma(\alpha^*P)$ along a curve $\alpha\colon [0, t] \to M$ ($t > 0$), where $\alpha(0) \eqqcolon x$. We have a canonical well-defined parallel transport $\mathrm{PT}_\alpha^\mathcal{H}\colon \mathcal{H}_x \to \Gamma(\alpha^*\mathcal{H})$ given by the \textbf{associated connection}, that is,
\begin{align*}
\mathrm{PT}_\alpha^\mathcal{H}\bigl( [p, h] \bigr)
&=
\mleft[ \mathrm{PT}_\alpha^P(p), h \mright]
\end{align*}
for all $[p, h] \in \mathcal{H}_x$. 
Observe that we have
\begin{align*}
\mathrm{PT}_\alpha^\mathcal{H}\Bigl( [p, h] \cdot \mleft[ p, h^\prime \mright] \Bigr)
&=
\mathrm{PT}_\alpha^\mathcal{H}\Bigl( \mleft[ p, h h^\prime \mright] \Bigr)
\\
&=
\mleft[ \mathrm{PT}_\alpha^P(p), h h^\prime \mright]
\\
&=
\mleft[ \mathrm{PT}_\alpha^P(p), h \mright]
	\cdot \mleft[ \mathrm{PT}_\alpha^P(p), h^\prime \mright]
\\
&=
\mathrm{PT}_\alpha^\mathcal{H}\bigl( [p, h] \bigr)
	\cdot \mathrm{PT}_\alpha^\mathcal{H}\Bigl( \mleft[p, h^\prime\mright] \Bigr)
\end{align*}
for all $[p, h], \mleft[ p, h^\prime \mright] \in \mathcal{H}_x$. Thus, associated connections define multiplicative and in particular Ehresmann connections on $\caH$(whose horizontal distribution modifies itself). Observe however, that not every Ehresmann connection is multiplicative; we will come back to this in \ref{CYMExamples}.
%
\end{examples}

Last but not least, we have a corresponding connection 1-form. For this we need to formally define the pullback of 1-forms with respect to the modified right-pushforward.

\begin{definitions}{The pullback of forms via modified right-pushforward}{PullbackOfFormsViaModRight}
Let $\mathcal{G} \to M$ be an LGB over a smooth manifold $M$, $\pi\colon\mathcal{P} \to M$ a principal $\mathcal{G}$-bundle, and $\mathrm{H}\mathcal{G}$ a horizontal distribution on $\mathcal{G}$. For $\omega \in \Omega^k(\mathcal{P}; \pi^*\mathcal{g})$ ($k \in \mathbb{N}_0$) we define the \textbf{pullback via the modified right-pushforward $\mathcal{r}_g^!\mleft(\mleft.\omega\mright|_{\mathcal{P}_x}\mright)$ (with $g\in\mathcal{G}_x$, over $\mathcal{P}_x$, $x \in M$)} as an element of $\Gamma \mleft( \Lambda^k\mleft.\mathrm{T}^*\mathcal{P}\mright|_{\mathcal{P}_x} \otimes \mathcal{g}_x \mright)$ by
\begin{align*}
\mleft.\mleft(\mathcal{r}_g^!\mleft(\mleft.\omega\mright|_{\mathcal{P}_x}\mright)\mright)\mright|_p \mleft( Y_1, \dotsc, Y_k\mright)
&\coloneqq
\omega_{p \cdot g} \bigl( \mathcal{r}_{g*}(Y_1), \dotsc, \mathcal{r}_{g*}(Y_k)\bigr)
\end{align*}
for all $p \in \mathcal{P}_x$ and $Y_1, \dotsc, Y_k \in \mathrm{T}_p\mathcal{P}$. Similarly we define the \textbf{pullback via the modified right-pushforward $\mathcal{r}^!_{\sigma}\omega$ with a (local) section $\sigma \in \Gamma(\mathcal{G})$} as an element of $\Omega^k(\mathcal{P}; \pi^*\mathcal{g})$ by 
\begin{align*}
\mleft.\mleft(\mathcal{r}_\sigma^!\omega\mright)\mright|_p \mleft( Y_1, \dotsc, Y_k\mright)
&\coloneqq
\mleft.\mleft(\mathcal{r}_{\sigma_x}^!\mleft(\mleft.\omega\mright|_{\mathcal{P}_x}\mright)\mright)\mright|_p \mleft( Y_1, \dotsc, Y_k\mright)
=
\mleft.\mleft(r_\sigma^*\omega\mright)\mright|_{p} \bigl( \mathcal{r}_{\sigma*}(Y_1), \dotsc, \mathcal{r}_{\sigma*}(Y_k)\bigr)
\end{align*}
for all $p \in \mathcal{P}_x$ and $Y_1, \dotsc, Y_k \in \mathrm{T}_p\mathcal{P}$.
\end{definitions}

\begin{remarks}{Extending to vector fields}{SectionPullbackModif}
By \ref{prop:IsomorphismRightPushAndDarboux} (and \ref{rem:RSigmaAnAuto}) these definitions are well-defined. A short note about the notation on the very right hand side of the second definition: This notation allows us to extend it to vector fields, that is,
\begin{align*}
\mleft(\mathcal{r}_\sigma^!\omega\mright) \mleft( Y_1, \dotsc, Y_k\mright)
=
\mleft(r_\sigma^*\omega\mright)\bigl( \mathcal{r}_{\sigma*}(Y_1), \dotsc, \mathcal{r}_{\sigma*}(Y_k)\bigr)
\end{align*}
for all $Y_1, \dotsc, Y_k \in \mathfrak{X}(\mathcal{P})$. Observe that we have $\mathcal{r}_{\sigma*}(Y_l) \in \Gamma\mleft(r_\sigma^*\mathrm{T}\mathcal{P}\mright)$ ($l \in \{1, \dotsc, k\}$) and $r_\sigma^*\omega \in \Gamma\mleft( \Lambda^k r_\sigma^*\mathrm{T}\mathcal{P} \otimes \pi^*\mathcal{g} \mright)$, using $r_\sigma^*\pi^*\mathcal{g} \cong (\pi \circ r_\sigma)^*\mathcal{g} = \pi^*\mathcal{g}$, such that the right hand side is well-defined.
\end{remarks}

For the following definition recall the adjoint $\mathcal{G}$-representation on $\mathrm{V}\mathcal{P}$, \ref{ex:AdjointACtionOnVP}, and the isomorphism for $\mathrm{V}\mathcal{P}$ in \ref{cor:VerticalBundleOfPrincIsNearlyAsUsual}.

\begin{definitions}{Connection 1-forms on principal LGB-bundles}{GaugeBosonsOnLGBPrincies}
Let $\mathcal{G} \to M$ be an LGB over a smooth manifold $M$, $\pi\colon\mathcal{P} \to M$ a principal $\mathcal{G}$-bundle, and $\mathrm{H}\mathcal{G}$ a horizontal distribution on $\mathcal{G}$. A \textbf{connection 1-form} or \textbf{gauge field} on $\mathcal{P}$ is a 1-form $A \in \Omega^1(\mathcal{P}; \pi^*\mathcal{g})$ satisfying:
\begin{itemize}
	\item \textbf{($\mathcal{G}$-equivariance, but w.r.t.\ modified right-pushforward)}
		\begin{align*} 
			\mathcal{r}_\sigma^! A
			&=
			\sAd_{\sigma^{-1}} \circ A
		\end{align*}
	for all (local) $\sigma \in \Gamma(\mathcal{G})$.
	\item \textbf{(Identity on $\mathrm{V}\mathcal{P}$)}
	\begin{align*}
	A\mleft(\overline{\nu}\mright)
	&=
	\pi^*\nu
	\end{align*}
	for all (local) $\nu \in \Gamma(\mathcal{g})$.
\end{itemize}
\end{definitions}

\begin{remarks}{Pointwise notation}{PointwiseNotationOfConnectioNOneForms}
The $\mathcal{G}$-equivariance reads point-wise for $g \coloneqq \sigma_x$ ($x\in M$)
\begin{equation*}
\mleft.\mleft(\mathcal{r}_g^! \mleft(A|_{\mathcal{P}_x}\mright)\mright)\mright|_{p}(X)
=
\mleft(
	p \cdot g, 
	\mathrm{Ad}_{g^{-1}}\mleft( \hat{A}_p(X) \mright)
\mright)
\end{equation*}
for all $p \in \mathcal{P}_x$ and $X \in \mathrm{T}_p\mathcal{P}$,
where we wrote $A_p = \mleft(p, \hat{A}_p\mright)$ with $\hat{A}_p \in \Gamma\mleft( \mathrm{T}^*_p\mathcal{P} \otimes \mathcal{g}_x \mright)$, and $\mathrm{Ad}$ is the adjoint representation of $\mathcal{G}$. In the typical formulation of gauge theory the base point component is usually omitted due to that $\mathcal{g}$ is then trivial and so also $\mathrm{V}\mathcal{P} (\cong \pi^*\mathcal{g})$.

The identity on $\mathrm{V}\mathcal{P}$ reads pointwise
\begin{equation*}
A_p(\overline{v_p})
=
(p, v_p)
\end{equation*}
for all $v_p \in \mathcal{g}_x$. For readability we may also omit the basepoint information and just write
\begin{equation*}
A_p(\overline{v_p})
\equiv
v_p.
\end{equation*}
Furthermore, recall the notation introduced in \ref{rem:FundVecNotationOnPullbackBundle}, then for $\mu \coloneqq (p, v_p)$ we can therefore write
\begin{equation*}
A_p\mleft( \overline{\mu}_p \mright)
=
\mu.
\end{equation*}
\end{remarks}

Finally, we identify (Ehresmann) connections and connection 1-forms on principal LGB-bundles in the typical way; for this recall Cor.\ \ref{cor:VerticalBundleOfPrincIsNearlyAsUsual}.

\begin{theorems}{Ehresmann connections and connection 1-forms are 1:1}{OurConnectionHasAUniqueoneForm}
Let $\mathcal{G} \to M$ be an LGB over a smooth manifold $M$ and $\pi\colon\mathcal{P} \to M$ a principal $\mathcal{G}$-bundle, and let $\mathrm{H}\mathcal{G}$ be a horizontal distribution on $\mathcal{G}$. Then there is a 1:1 correspondence between Ehresmann connections and connection 1-forms on $\mathcal{P}$:
\begin{itemize}
	\item Let $\mathrm{H}\mathcal{P}$ be an Ehresmann connection on $\mathcal{P}$. Then $\mathrm{H}\mathcal{P}$ defines a connection 1-form $A \in \Omega^1(\mathcal{P}; \pi^*\mathcal{g})$ by
	\begin{equation*}
	A_p\bigl( \overline{v}_p + X_p \bigr)
	=
	(p, v)
	\end{equation*}
	for all $p \in \mathcal{P}_x$ ($x \in M$), $v \in \mathcal{g}_x$ and $X \in \mathrm{H}_p\mathcal{P}$.
	\item Let $A \in \Omega^1(\mathcal{P}; \pi^*\mathcal{g})$ be a connection 1-form on $\mathcal{P}$. Then $A$ defines an Ehresmann connection $\mathrm{H}\mathcal{P}$ on $\mathcal{P}$ via its kernel $\mathrm{Ker}(A)$, that is,
	\begin{equation*}
	\mathrm{H}_p\mathcal{P}
	=
	\mathrm{Ker}(A_p)
	\end{equation*}
	for all $p \in \mathcal{P}$.
\end{itemize}
\end{theorems}

\begin{proof}
\leavevmode\newline
The proof is very similar to the proof of "typical connections", as \textit{e.g.}\ provided in \cite[\S 5.2, Thm.\ 5.2.2, page 262]{Hamilton}.

$\bullet$ For the first bullet point we need to show that \ref{def:GaugeBosonsOnLGBPrincies} is satisfied, and the identity behaviour on $\mathrm{V}\mathcal{P}$ quickly follows by definition: We have
\begin{align*}
\mleft.A\mleft( \overline{\nu} \mright)\mright|_{p}
&=
A_p\mleft( \overline{\nu}_p \mright)
=
(p, \nu_x)
=
\mleft.\mleft(\pi^*\nu\mright)\mright|_p
\end{align*}
for all $\nu \in \Gamma(\mathcal{g})$ and $p \in \mathcal{P}_x$ ($x \in M$). Therefore it is only left to show the $\mathcal{G}$-equivariance. As mentioned in \ref{rem:FundVecsNotations}, for $v\in \mathcal{g}_x$ the vector field $\overline{v}$ on $\mathcal{P}_x$ is a fundamental vector field coming from the $\mathcal{G}_x$-action on $\mathcal{P}_x$. Hence, we know, e.g.\ by \cite[\S 3.4, Prop.\ 3.4.6, page 145f.]{Hamilton}, that
\begin{align*}
\mathcal{r}_{g*}\mleft( \overline{v} \mright)
&=
r_{g*}\mleft( \overline{v} \mright)
=
\overline{\mathrm{Ad}_{g^{-1}}(v)}
\end{align*}
for all $g \in \mathcal{G}_x$, also using \ref{rem:ModRightPushOnVertic}.\footnote{This is very straightforward to prove.} Thus,
\begin{align*}
\mleft.\mleft(\mathcal{r}^!_\sigma A\mright)\mright|_p\bigl( \overline{v}_p + X_p \bigr)
&=
A_{p \cdot \sigma_x}\bigl( 
	\underbrace{\mathcal{r}_{\sigma_x*}\mleft(\overline{v}_p\mright)}
		_{= \mleft. \mathcal{r}_{\sigma_x*}\mleft( \overline{v} \mright) \mright|_{p \cdot  \sigma_x}}
	+ \underbrace{\mathcal{r}_{\sigma_x*}(X_p) }_{\in \mathrm{H}_{p \cdot \sigma_x}\mathcal{P}}
\bigr)
\\
&=
\mleft( p \cdot \sigma_x, \mathrm{Ad}_{\sigma_x^{-1}}(v) \mright)
\\
&=
\sAd_{\sigma_x^{-1}}(p, v)
\\
&=
\sAd_{\sigma_x^{-1}}\mleft(A_p\bigl( \overline{v}_p + X_p \bigr)\mright)
\\
&=
\mleft.\mleft(\sAd_{\sigma^{-1}} \circ A\mright)\mright|_p \bigl( \overline{v}_p + X_p \bigr)
\end{align*}
for all $p \in \mathcal{P}_x$, $\sigma \in \Gamma(\mathcal{G})$, $v \in \mathcal{g}_x$ and $X_p \in \mathrm{H}_p\mathcal{P}$. This finishes the proof for the first bullet point.

$\bullet$ For the second bullet point we make use of \ref{cor:VerticalBundleOfPrincIsNearlyAsUsual} and
\begin{align*}
A\mleft( \overline{\nu} \mright)
&=
\pi^*\nu
\end{align*}
for all (local) sections $\nu \in \Gamma(\mathcal{g})$. This implies that $A$ has not only values in $\mathrm{V}\mathcal{P}$, but acts also as identity on $\mathrm{V}\mathcal{P}$ and has thus constant rank. Thus, $\mathrm{H}\mathcal{P} = \mathrm{Ker}(A)$ is a distribution complementary to $\mathrm{V}\mathcal{P}$.
It is only left to show the right-invariance of $\mathrm{H}\mathcal{P}$. So let $X \in \mathrm{H}_p\mathcal{P}$ for $p \in \mathcal{P}$. Then
\begin{align*}
A_{p \cdot \sigma_x}\bigl( \mathcal{r}_{\sigma*}(X) \bigr)
&=
\mleft(\mathcal{r}^!_\sigma(A)\mright)_p(X)
=
\sAd_{\sigma_x^{-1}}\bigl( A_p(X) \bigr)
=
0
\end{align*}
for all (local) $\sigma \in \Gamma(\mathcal{G})$, where $x \coloneqq \pi(p)$. Thence, $\mathcal{r}_{\sigma*}(X) = \mathcal{r}_{\sigma_x*}(X) \in \mathrm{H}_{p \cdot \sigma_x}\mathcal{P}$, and therefore we derive by \ref{prop:IsomorphismRightPushAndDarboux} and dimensional reasons that $\mathcal{r}_{\sigma_x*}\mleft(\mathrm{H}_p \mathcal{P}\mright) = \mathrm{H}_{p \cdot \sigma_x}\mathcal{P}$. Henceforth $\mathrm{H}\mathcal{P}$ is an Ehresmann connection.
\end{proof}

We finish this subsection with the following technical and useful corollary.

\begin{corollaries}{Commutation of modified push-forward and projections}{ModifiedRightPushyCommutesWithProj}
Let $\mathcal{G} \to M$ be an LGB over a smooth manifold $M$ and $\mathcal{P} \to M$ a principal $\mathcal{G}$-bundle, also let $\mathrm{H}\mathcal{G}$ be a horizontal distribution on $\mathcal{G}$ and $\mathrm{H}\mathcal{P}$ an Ehresmann connection on $\mathcal{P}$; denote with $\pi_h$ and $\pi_v$ the associated projections on the horizontal and vertical bundle of $\mathcal{P}$, respectively. Then we have
\begin{align*}
\mathcal{r}_{g*} \circ \pi_h
&=
\pi_h \circ \mathcal{r}_{g*},\\
\mathcal{r}_{g*} \circ \pi_v
&=
\pi_v \circ \mathcal{r}_{g*}
\end{align*}
for all $g \in \mathcal{G}$.
\end{corollaries}

\begin{remarks}{Extending to sections}{RemOohThesePullbacksConfusOrNotToConfus}
This extends of course to (local) sections $\sigma \in \Gamma(\mathcal{G})$, and as discussed in \ref{rem:RemarkABoutDarbouxNotationWRTPullback} we can extend these equations to vector fields on $\mathcal{P}$ via pullbacks, that is,
\begin{align*}
\mathcal{r}_{\sigma*} \circ \pi_h
&=
r_\sigma^*\pi_h \circ \mathcal{r}_{\sigma*},\\
\mathcal{r}_{\sigma*} \circ \pi_v
&=
r_\sigma^*\pi_v \circ \mathcal{r}_{\sigma*},
\end{align*}
making use of that $\mathcal{r}_{\sigma*}$ is an isomorphism over $r_{\sigma*}$, recall \ref{prop:IsomorphismRightPushAndDarboux}.
\end{remarks}

\begin{proof}[Proof of \ref{cor:ModifiedRightPushyCommutesWithProj}]
\leavevmode\newline
By \ref{cor:VerticalBundleOfPrincIsNearlyAsUsual} (also recall \ref{rem:ModRightPushOnVertic}) and \ref{def:FinallyTheConnection} we have
\begin{align*}
\mathcal{r}_{g*}\mleft(\mathrm{V}_p\mathcal{P}\mright)
&=
\mathrm{V}_{p\cdot g}\mathcal{P},\\
\mathcal{r}_{g*}\mleft( \mathrm{H}_p\mathcal{P} \mright)
&=
\mathrm{H}_{p\cdot g}\mathcal{P},
\end{align*}
thus, $\mathcal{r}_{g*}$ preserves the splitting $\mathrm{T}\mathcal{P} = \mathrm{H}\mathcal{P} \oplus \mathrm{V}\mathcal{P}$. This concludes the proof.
\end{proof}

\subsection{Gauge transformations}\label{GaugeTrafoForA}

Let us now study gauge transformations of $A$; for this recall the definition of gauge transformations in \ref{def:MorphOfPrincBundles}. As in ordinary gauge theory we expect a relationship between gauge transformations and certain LGB valued maps. For the following also recall \ref{ex:ConjugationActionForTheGeneralInnerGroupBundle}.

\begin{definitions}{LGB-valued conjugation maps}{GaugeTrafosLGBMaps}
Let $\mathcal{G} \to M$ be an LGB over a smooth manifold $M$, and $\pi\colon\mathcal{P} \to M$ a principal $\mathcal{G}$-bundle. Then we define the group \textbf{$C^\infty(\mathcal{P}; \mathcal{G})^{\mathcal{G}}$ of $\mathcal{G}$-valued conjugation maps} as a set by
\begin{align*}
C^\infty(\mathcal{P}; \mathcal{G})^{\mathcal{G}}
&=
\bigl\{
	\sigma \in \Gamma(\pi^*\mathcal{G})
	~\big|~
	\sigma_{p \cdot g}
	=
	\mathcal{c}_{g^{-1}}( \sigma_p )
	\text{ for all } (p, g) \in \mathcal{P}*\mathcal{G}
\bigr\}.
\end{align*}
Its group structure is inherited by the point-wise group structure of $\Gamma(\pi^*\mathcal{G})$; recall \ref{cor:PullbackLGB}.
\end{definitions}

\begin{remarks}{Group structure on $C^\infty(\mathcal{P}; \mathcal{G})^{\mathcal{G}}$}{GroupStructureOnCInftPGG}
It is trivial to check that $C^\infty(\mathcal{P}; \mathcal{G})^{\mathcal{G}}$ is indeed a subgroup of $\Gamma(\pi^*\mathcal{G})$.
\end{remarks}

$C^\infty(\mathcal{P}; \mathcal{G})^{\mathcal{G}}$ canonically acts on $\mathcal{P}$ on the right via the given $\mathcal{G}$-action: Let $\sigma \in C^\infty(\mathcal{P}; \mathcal{G})^{\mathcal{G}}$, then we define
\begin{equation}\label{MultiWithPulli}
p \cdot \sigma_p
\coloneqq
p \cdot \mathrm{pr}_2\mleft( \sigma_p \mright)
\end{equation}
for all $p \in \mathcal{P}$, where $\mathrm{pr}_2\colon \pi^*\mathcal{G} \to \mathcal{G}$ is the projection onto the second component. In essence, we drop the notation of $\mathrm{pr}_2$, as if we view $\sigma$ as a map $\mathcal{P} \to \mathcal{G}$ of fibre bundles over $\pi$; recall \ref{BasicNotations}. If we denote the right $\mathcal{G}$-action on $\mathcal{P}$ by $\Phi$, then observe by $\sigma_p = (p, \mathrm{pr}_2(\sigma_p))$ that we could also write
\begin{equation*}
p \cdot \sigma_p
=
\Phi \bigl( p, \mathrm{pr}_2\mleft( \sigma_p \mright) \bigr)
=
\Phi\mleft( \sigma_p \mright)~.
\end{equation*}

As in the case of typical/classical principal bundles, we have the following statement.

\begin{propositions}{Gauge transformations as $\mathcal{G}$-valued conjugation maps}{GaugeTrafoAsBundleIsomIsASectionOfConjugationMaps}
Let $\mathcal{G} \to M$ be an LGB over a smooth manifold $M$, and $\pi\colon\mathcal{P} \to M$ a principal $\mathcal{G}$-bundle. Then there is a well-defined group isomorphism of gauge transformations and $\mathcal{G}$-valued conjugation maps given by
\begin{align*}
\sAut(\mathcal{P}) &\to C^\infty(\mathcal{P}; \mathcal{G})^{\mathcal{G}},\\
H &\mapsto \sigma^H,
\end{align*}
where $\sigma^H \in C^\infty(\mathcal{P}; \mathcal{G})^{\mathcal{G}}$ is defined by
\begin{equation*}
H(p) = p \cdot \sigma^H_p
\end{equation*}
for all $p \in \mathcal{P}$.
\end{propositions}

\begin{proof}
\leavevmode\newline
This result can be proven similarly as for Lie group based principal bundles; however smoothness needs to be discussed shortly. 

$\bullet$ First of all, due to the fact that $H$ is base-preserving we know that $H(p)$ is in the same fibre as $p$ ($p \in \mathcal{P}$), and due to that the $\mathcal{G}$-action on $\mathcal{P}$ is simply transitive there is a unique element of $\pi^*\mathcal{G}|_p$, denoted by $\sigma^H_p$, such that
\begin{equation*}
H(p)
=
p \cdot \sigma_p^H
=
p \cdot \mathrm{pr}_2\mleft(\sigma_p^H\mright),
\end{equation*}
where $\mathrm{pr}_2\colon \mathcal{P}*\mathcal{G} = \pi^*\mathcal{G} \to \mathcal{G}$ is the smooth projection onto the second component.
Let us first show smoothness of $p \mapsto \sigma_p^H$: Recall \ref{rem:PrincipalMorphBasics}, every principal $\caG$-bundle is locally a classical principal $G$-bundle and $\caG$ is trivial in that case. Since $\sAut(\caP)$ consist of base-preserving automorphisms, we can restrict those locally to those ``classical'' setups, in particular, $\sAut(\caP)$ are then ordinary principal bundle automorphisms for which this proposition is known, and we can conclude that $\sigma^H$ is smooth since it aligns with precisely such a section of a trivial $G$-bundle over $\caP$ which is associated to an ordinary principal bundle automorphism.

$\bullet$ That $\sigma^H$ is an element of $C^\infty(\mathcal{P};\mathcal{G})^{\mathcal{G}}$ follows either similarly or one shows it directly:
\begin{equation*}
(p \cdot g) \cdot \mathrm{pr}_2\mleft(\sigma^H_{p \cdot g}\mright)
=
H(p \cdot g)
=
H(p) \cdot g
=
p \cdot \mleft(\mathrm{pr}_2\mleft(\sigma^H_p\mright) ~ g \mright)
\end{equation*}
for all $(p, g) \in \mathcal{P}*\mathcal{G}$,
so that, by the simply transitivity of the action,
\begin{equation*}
g ~ \mathrm{pr}_2\mleft(\sigma^H_{p \cdot g}\mright)
=
\mathrm{pr}_2\mleft(\sigma^H_p\mright) ~ g~,
\end{equation*}
and thus
\begin{equation*}
\sigma^H_{p \cdot g}
=
\Bigl( p \cdot g, \mathrm{pr}_2\mleft(\sigma^H_{p \cdot g}\mright) \Bigr)
=
\Bigl( p \cdot g, g^{-1} ~ \mathrm{pr}_2\mleft(\sigma^H_p\mright) ~ g \Bigr)
=
\mathcal{c}_{g^{-1}}\mleft( \sigma^H_p \mright)~.
\end{equation*}
Thus, $\sigma^H \in C^\infty(\mathcal{P}; \mathcal{G})^{\mathcal{G}}$.

$\bullet$ The inverse of $H \mapsto \sigma^H$ is clearly given by
\begin{align*}
C^\infty(\mathcal{P}; \mathcal{G})^{\mathcal{G}} &\to \sAut(\mathcal{P})~,\\
\sigma &\mapsto H^\sigma~,
\end{align*}
where 
\begin{equation*}
H^\sigma(p)
\coloneqq
p \cdot \sigma_p
\end{equation*}
for all $p \in \mathcal{P}$. Smoothness here is now obvious, it is also clearly base-preserving, and we have 
\begin{equation*}
H^\sigma(p \cdot g)
=
p \cdot g ~ \underbrace{\sigma_{p \cdot g}}_{\mathclap{ = \mathcal{c}_{g^{-1}}\mleft(\sigma_p\mright) }}
=
p \cdot g ~ \underbrace{\mathrm{pr}_2\mleft( \mathcal{c}_{g^{-1}}\mleft(\sigma_p\mright) \mright)}_{\mathclap{ = c_{g^{-1}}\mleft( \mathrm{pr}_2(\sigma_p) \mright) }}
=
p \cdot \mathrm{pr}_2\mleft( \sigma_p \mright) ~ g
=
p \cdot \sigma_p ~ g
=
H^\sigma(p) \cdot g
\end{equation*}
for all $(p, g) \in \mathcal{P} * \mathcal{G}$. Thus, $H^\sigma \in \sAut(\mathcal{P})$, and so this finishes the proof.
%
\end{proof}

Using this, we can finally formulate the gauge transformations of connection 1-forms; for this recall \ref{cor:PullBacksArePrincToo} and the Darboux derivative, \ref{def:DarbouxDerivativeOnLGBs}. Also recall the pullback Darboux derivative, \ref{rem:PullBackDarboux}.

\begin{propositions}{Gauge transformations of connection 1-forms}{GaugeTrafoOfGaugeBoson}
Let $\mathcal{G} \to M$ be an LGB over a smooth manifold $M$ and $\pi\colon\mathcal{P} \to M$ a principal $\mathcal{G}$-bundle, also let $\mathrm{H}\mathcal{G}$ be a horizontal distribution on $\mathcal{G}$ and $A \in \Omega^1(\mathcal{P}; \pi^*\mathcal{g})$ be a connection 1-form on $\mathcal{P}$. Furthermore, let $H \in \sAut(\mathcal{P})$. We then have that $H^!A$ is a connection 1-form on $\mathcal{P}$ and
\begin{equation*}
H^!A
=
{\sAd_{\mathrm{pr}_2\circ\mleft(\sigma^H\mright)^{-1}}} \circ A 
	+ \mleft(\pi^*\Delta\mright)\sigma^H~,
\end{equation*}
where $\sigma^H \in C^\infty(\mathcal{P}; \mathcal{G})^{\mathcal{G}}$ is defined as in \ref{prop:GaugeTrafoAsBundleIsomIsASectionOfConjugationMaps} and $\mathrm{pr}_2\colon \pi^*\mathcal{G} \to \mathcal{G}$ is the projection onto the second component.

Similar to \ref{MultiWithPulli} we may shortly just write 
\begin{equation*}
H^!A
=
{\sAd_{\mleft(\sigma^H\mright)^{-1}}} \circ A
	+ \mleft(\pi^*\Delta\mright)\sigma^H~.
\end{equation*}
\end{propositions}

\begin{proof}
\leavevmode\newline
First of all observe that $H^!A \in \Omega^1(\mathcal{P}; H^*\pi^*\mathcal{g})$, and trivially $H^*\pi^*\mathcal{g} \cong (\pi\circ H)^*\mathcal{g} = \pi^*\mathcal{g}$. We will make use of this and similar isomorphisms in the following without further mentioning it. We will often not bookkeep the basepoint component in the following pairs and triples; so, everything has to be read in such a way that we have again values in the correct space. It would just be cumbersome to keep track of all these natural isomorphisms, and we decided for readability to avoid this bookkeeping most of the time.

$\bullet$ We then have
\begin{align*}
\mleft(\mathcal{r}_\sigma^!H^!A\mright)_p(X)
&=
\mleft( H^!A \mright)_{p \cdot \sigma_x}\bigl(
	\mathcal{r}_{\sigma*}(X)
\bigr)
\\
&=
\mleft( H^!A \mright)_{p \cdot \sigma_x}\mleft(
	\mathrm{D}_pr_\sigma\mleft( 
	X 
\mright)
	- \mleft.{\overline{
		\mleft. \mleft( \pi^!\Delta\sigma \mright) \mright|_p(X)
	}}\mright|_{p \cdot \sigma_{x}}
\mright)
\\
&=
A_{H\mleft(p \cdot \sigma_x\mright)}\mleft( \mathrm{D}_{p \cdot \sigma_x}H\mleft(
	\mathrm{D}_pr_\sigma\mleft( 
	X 
\mright)
	- \mleft.{\overline{
		\mleft. \mleft( \pi^!\Delta\sigma \mright) \mright|_p(X)
	}}\mright|_{p \cdot \sigma_{x}}
\mright)\mright)
\end{align*}
for all (local) $\sigma \in \Gamma(\mathcal{G})$, $p \in \mathcal{P}_x$ ($x \in M$) and $X \in \mathrm{T}_p\mathcal{P}$. We also get by definition of $\sAut(\mathcal{P})$
\begin{equation*}
H(p \cdot \sigma_x)
=
H(p) \cdot \sigma_x~,
\end{equation*}
that is,
\begin{equation*}
H\circ r_\sigma
=
r_\sigma \circ H
\end{equation*}
and thus
\begin{equation*}
\mathrm{D}_{p\cdot \sigma_x}H \circ \mathrm{D}_p r_\sigma
=
\mathrm{D}_p \mleft( H \circ r_\sigma \mright)
=
\mathrm{D}_p \mleft( r_\sigma \circ H \mright)
=
\mathrm{D}_{H(p)} r_\sigma \circ \mathrm{D}_pH.
\end{equation*}
Now also observe that
\begin{equation}\label{GaugeTrafoDiffOnFUndVect}
\mathrm{D}_p H \mleft( \overline{\nu}_p \mright)
=
\mleft(\mathrm{D}_p H \circ \mathrm{D}_{e_x}\Phi_p\mright) (\nu)
=
\mathrm{D}_{e_x}\underbrace{\mleft(H \circ\Phi_p \mright)}_{\mathclap{ \mathcal{G}_x \ni g \mapsto H(p \cdot g) = H(p) \cdot g }} (\nu)
=
\mathrm{D}_{e_x} \Phi_{H(p)} (\nu)
=
\overline{\nu}_{H(p)}
\end{equation}
for all $\nu \in \mathcal{g}_x$, where $\Phi_p$ is the orbit map through $p$. Due to that $H$ is base-preserving we derive
\begin{equation*}
\mleft. \mleft( \pi^!\Delta\sigma \mright) \mright|_{H(p)}\bigl(\mathrm{D}_pH(X)\bigr)
=
\mleft( H^! \pi^! \Delta \sigma \mright)_p (X)
=
\mleft( (\pi \circ H)^! \Delta \sigma \mright)_p (X)
=
\mleft( \pi^! \Delta \sigma \mright)_p (X).
\end{equation*}
In total we get
\begin{align*}
\mleft(\mathcal{r}_\sigma^!H^!A\mright)_p(X)
&=
A_{H(p) \cdot \sigma_x}\mleft( 
	\mathrm{D}_{H(p)}r_\sigma \bigl(\mathrm{D}_pH( X )\bigr)
	- \mleft.{\overline{
		\mleft. \mleft( \pi^!\Delta\sigma \mright) \mright|_{H(p)}\bigl( \mathrm{D}_pH(X) \bigr)
	}}\mright|_{H(p) \cdot \sigma_{x}}
\mright)
\\
&=
A_{H(p) \cdot \sigma_x}\mleft(
	\mathcal{r}_{\sigma*}\bigl(\mathrm{D}_pH( X )\bigr)
\mright)
\\
&=
\mleft( \mathcal{r}_\sigma^!A \mright)_{H(p)}\bigl(\mathrm{D}_pH( X )\bigr)
\\
&=
\mleft( \sAd_{\sigma^{-1}} \circ A \mright)_{H(p)}\bigl(\mathrm{D}_pH( X )\bigr)
\\
&=
\mleft( H(p) \cdot \sigma_x, ~\mathrm{Ad}_{\sigma_x^{-1}}\mleft( \widehat{A}_{H(p)}\bigl(\mathrm{D}_pH( X )\bigr) \mright) \mright)
\\
&=
\sAd_{\sigma^{-1}}
\mleft( H(p), ~\widehat{A}_{H(p)}\bigl(\mathrm{D}_pH( X )\bigr) \mright)
\\
&=
\sAd_{\sigma^{-1}}
\mleft( \mleft(H^!A\mright)_p(X) \mright),
\end{align*}
where we wrote $A = \mleft( p, \widehat{A}_p \mright)$ with $\widehat{A} \coloneqq \pi_2 \circ A$ ($\pi_2: \pi^*\mathcal{g} \to \mathcal{g}$ the projection onto the second component), and thus
\begin{align*}
\mathcal{r}_\sigma^!H^!A
&=
\sAd_{\sigma^{-1}}\circ~ H^!A.
\end{align*}
By Eq.\ \eqref{GaugeTrafoDiffOnFUndVect} we also get $\mathrm{D}H(\overline{\nu}) = H^*\overline{\nu}$ for all (local) $\nu \in \Gamma(\mathcal{g})$ and thus
\begin{equation*}
\mleft(H^!A\mright)\mleft( \overline{\nu} \mright)
=
\mleft(H^*A\mright)\bigl( \mathrm{D}H\mleft( \overline{\nu} \mright) \bigr)
=
\mleft(H^*A\mright)\mleft( H^*\overline{\nu} \mright)
=
H^*\bigl( A\mleft( \overline{\nu}\mright) \bigr)
=
H^*\pi^*\nu
=
(\pi \circ H)^*\nu
=
\pi^*\nu.
\end{equation*}
Hence, $H^!A$ is a connection 1-form on $\mathcal{P}$.

$\bullet$ For the last part recall \ref{prop:GaugeTrafoAsBundleIsomIsASectionOfConjugationMaps}, especially we have a unique $\sigma^H \in C^\infty(\mathcal{P}; \mathcal{G})^{\mathcal{G}}$ such that
\begin{equation*}
H(p)
=
p \cdot \sigma^H_p
=
p \cdot \mathrm{pr}_2\mleft(\sigma^H_p\mright)
=
\Phi\mleft(p, \mathrm{pr}_2\mleft(\sigma^H_p \mright)\mright)
=
\Bigl( \Phi \circ \mleft( \mathrm{id}_{\mathcal{P}}, \mathrm{pr}_2\circ\sigma^H \mright) \Bigr)(p)
\end{equation*}
for all $p \in \mathcal{P}_x$ ($x\in M$), where $\Phi$ is the right $\mathcal{G}$-action on $\mathcal{P}$, also recall \ref{MultiWithPulli}, that is, $\mathrm{pr}_2\colon \pi^*\mathcal{G} \to \mathcal{G}$ is the projection onto the second component. Define $\overline{\sigma} \coloneqq \mathrm{pr}_2 \circ \sigma^H$, $\mathcal{P} \ni p \mapsto \overline{\sigma}_p \in \mathcal{G}$; by \ref{rem:PullbackTotMCForm}, \ref{rem:PullBackDarboux} and \ref{prop:DiffOfLGBAction}, especially \ref{rem:DarbouxInActionInfinit}, we can calculate
\begin{align}\label{DiffOfPrincAutom}
\mathrm{D}_p H(X)
&=
\mathrm{D}_{\mleft(p, \overline{\sigma}_p\mright)}\Phi \bigl( X, \mathrm{D}_p\overline{\sigma}(X) \bigr)
\nonumber
\\
&=
\mathrm{D}_pr_\sigma(X)
	- \mleft.{\overline{ \mleft.\mleft(\pi^!\Delta \sigma\mright)\mright|_p (X)}}\mright|_{p \cdot \overline{\sigma}_p}
	+ \mleft.{\overline{\mleft( \mutot\mright)_{\overline{\sigma}_p} \bigl(\mathrm{D}_p\overline{\sigma}(X) \bigr)}}\mright|_{p \cdot \overline{\sigma}_p}
\nonumber
\\
&=
\mathcal{r}_{\overline{\sigma}_p*}(X)
	+ \mleft.{\overline{\mleft( \mutot\mright)_{\mathrm{pr}_2\mleft(\sigma^H_p\mright)} \mleft(\mleft(\mathrm{D}_{\sigma^H_p}\mathrm{pr}_2 \circ \mathrm{D}_p\sigma^H\mright)(X) \mright)}}\mright|_{H(p)}
\nonumber
\\
&=
\mathcal{r}_{\overline{\sigma}_p*}(X)
	+ \mleft.{\overline{\mleft( \mathrm{pr}_2^!\mutot\mright)_{\sigma^H_p} \mleft(\mathrm{D}_p\sigma^H(X) \mright)}}\mright|_{H(p)}
\nonumber
\\
&=
\mathcal{r}_{\overline{\sigma}_p*}(X)
	+ \mleft.{\overline{\mleft( \mu_{\pi^*\mathcal{G}}\mright)_{\sigma^H_p} \mleft(\mathrm{D}_p\sigma^H(X) \mright)}}\mright|_{H(p)}
\nonumber
\\
&=
\mathcal{r}_{\overline{\sigma}_p*}(X)
	+ \mleft.{\overline{ \mleft(\Delta^{\pi^*\mathcal{G}}\sigma^H\mright)_p(X) }}\mright|_{H(p)}
\nonumber
\\
&=
\mathcal{r}_{\overline{\sigma}_p*}(X)
	+ \mleft.{\overline{ \mleft(\mleft(\pi^*\Delta\mright)\sigma^H\mright)_p(X) }}\mright|_{H(p)}
\end{align}
for all $X \in \mathrm{T}_p\mathcal{P}$, where $\sigma$ is a (local) section of $\mathcal{G}$ so that $\sigma_x = \overline{\sigma}_p = \mathrm{pr}_2\mleft( \sigma^H_p \mright)$.

Then
\begin{align*}
\mleft(H^!A\mright)_p(X)
&=
A_{H(p)}\bigl( \mathrm{D}_pH(X) \bigr)
\\
&=
A_{H(p)}\mleft( 
	\mathcal{r}_{\overline{\sigma}_p*}(X)
	+ \mleft.{\overline{ \mleft(\mleft(\pi^*\Delta\mright)\sigma^H\mright)_p(X) }}\mright|_{H(p)}
\mright)
\\
&=
A_{p \cdot \overline{\sigma}_p}\mleft( 
	\mathcal{r}_{\overline{\sigma}_p*}(X)
\mright)
	+ \mleft(\mleft(\pi^*\Delta\mright)\sigma^H\mright)_p(X)
\\
&=
\mleft(\mathcal{r}_{\overline{\sigma}_p}^!A\mright)_p(X)
	+ \mleft(\mleft(\pi^*\Delta\mright)\sigma^H\mright)_p(X)
\\
&=
\sAd_{\overline{\sigma}_p^{-1}}\bigl(A_p(X)\bigr)
	+ \mleft(\mleft(\pi^*\Delta\mright)\sigma^H\mright)_p(X)
\\
&=
\sAd_{\mathrm{pr}_2\mleft( \mleft(\sigma_p^H\mright)^{-1} \mright)}\bigl(A_p(X)\bigr)
	+ \mleft(\mleft(\pi^*\Delta\mright)\sigma^H\mright)_p(X),
\end{align*}
which finishes the proof, where we used the trivial relation (alternatively recall \ref{cor:PullbackLGB})
\begin{align*}
\mleft(\mathrm{pr}_2\mleft( \sigma_p^H \mright)\mright)^{-1}
&=
\mathrm{pr}_2\mleft( \mleft(\sigma_p^H\mright)^{-1} \mright)~.
\end{align*}
\end{proof}

Of course there is also the sense of gauge transformation with respect to gauges as in typical gauge theory, recall \ref{def:GaugesOfPrincipalBundles}.

\begin{definitions}{Local gauge field}{LocalGaugeField}
Let $\mathcal{G} \to M$ be an LGB over a smooth manifold $M$ and $\pi\colon\mathcal{P} \to M$ a principal $\mathcal{G}$-bundle, also let $\mathrm{H}\mathcal{G}$ be a horizontal distribution on $\mathcal{G}$ and $A \in \Omega^1(\mathcal{P}; \pi^*\mathcal{g})$ be a connection 1-form on $\mathcal{P}$. Furthermore, let $s \in \Gamma(\mathcal{P}|_U)$ be a (local) gauge over an open subset $U \subset M$. Then we define the \textbf{local connection 1-form} of \textbf{local gauge field $A_s \in \Omega^1\mleft(U; \mleft.\mathcal{g}\mright|_U\mright)$ (w.r.t.\ $s$)} by
\begin{align*}
A_s
&\coloneqq
s^!A.
\end{align*}
\end{definitions}

\begin{remarks}{Well-definedness}{PullBackGaugeFieldRemark}
$A$ has values in $\pi^*\mathcal{g}$, and thus $A_s$ has values in $s^*\pi^*\mathcal{g} \cong (\pi \circ s)^*\mathcal{g} = \mathrm{id}_U^*\mathcal{g} \cong \mleft.\mathcal{g}\mright|_U$.
\end{remarks}

Gauge transformations now naturally arise in a change of the gauge $s$. So, let $U_i$ and $U_j$ be two open subsets of $M$ so that $U_i \cap U_j \neq \emptyset$. For two gauges $s_i \in \Gamma\mleft(\mathcal{P}|_{U_i}\mright)$ and $s_j \in \Gamma\mleft(\mathcal{P}|_{U_j}\mright)$ there is then a unique $\sigma_{ji} \in \Gamma\mleft( \mleft.\mathcal{G}\mright|_{U_i \cap U_j} \mright)$ such that
\begin{equation*}
s_i
=
s_j \cdot \sigma_{ji}
=
\Phi_{s_j} \circ \sigma_{ji}
\end{equation*}
on $U_i \cap U_j$,
where $\Phi_{s_j}$ is the orbit through $s_j$, especially also recall \ref{lem:SectionsNowInduceIsomToLGBsNotNecTriv}.
The unique existence is clear by the definition of principal bundles, while the smoothness follows by \ref{lem:SectionsNowInduceIsomToLGBsNotNecTriv} so that we can write $\sigma_{ji}$ as the composition of smooth maps
\begin{equation*}
\sigma_{ji}
=
\Phi^{-1}_{s_j} \circ s_i~.
\end{equation*}
In the following we also introduce the notation
\begin{equation*}
\mleft(\Phi^{-1}_{s_j}\mright)^{-1}(p)
\coloneqq
\mleft(\Phi^{-1}_{s_j}(p)\mright)^{-1}
\end{equation*}
for all $p \in \mathcal{P}_{U_j}$, that is, $\Phi^{-1}_{s_j}$ is the inverse of the orbit map $\Phi_{s_j}$, but the additional $(\cdot)^{-1}$ is the inverse of $\Phi^{-1}_{s_j}(p)$ as an element of $\mathcal{G}_{U_j}$.

Instead of calculating directly how $A_{s_i}$ and $A_{s_j}$ are related we want to use \ref{prop:GaugeTrafoOfGaugeBoson}. For this it will be useful to understand that gauge transformations $\sAut(\mathcal{P})$ are locally isomorphic to the group of sections $\Gamma(\mathcal{G})$ via a gauge, analogously to the typical formulation of gauge theory. While the isomorphism of $\sAut(\mathcal{P})$ to $C^\infty(\mathcal{P};\mathcal{G})^{\mathcal{G}} \subset \Gamma(\pi^*\mathcal{G})$ is global, the following isomorphism is in general only local.
%

\begin{propositions}{Gauge transformations and sections of the structural LGB}{GaugeTrafosAsLGBSectionsLocal}
Let $\pi_{\mathcal{G}}\colon\mathcal{G} \to M$ be an LGB over a smooth manifold $M$, and $\pi\colon\mathcal{P} \to M$ a principal $\mathcal{G}$-bundle. Also let $s \in \Gamma\mleft(\mleft.\mathcal{P}\mright|_U\mright)$ be a gauge defined over some open subset $U$ of $M$. Then we have a group isomorphism given by
\begin{align*}
C^\infty \mleft( \mleft.\mathcal{P}\mright|_U; \mleft.\mathcal{G}\mright|_U \mright)^{\mleft.\mathcal{G}\mright|_U} 
&\to 
\Gamma\mleft(\mleft.\mathcal{G}\mright|_U\mright),
\\
\sigma &\mapsto \mathrm{pr}_2 \circ \sigma \circ s,
\end{align*}
where $\mathrm{pr}_2\colon \pi^*\mathcal{G} \to \mathcal{G}$ is the projection onto the second component,
with inverse
\begin{align*}
\Gamma\mleft(\mleft.\mathcal{G}\mright|_U\mright)
&\to
C^\infty \mleft( \mleft.\mathcal{P}\mright|_U; \mleft.\mathcal{G}\mright|_U \mright)^{\mleft.\mathcal{G}\mright|_U},
\\ 
\tau &\mapsto \sigma^\tau,
\end{align*}
where 
\begin{equation*}
\sigma^\tau_p
\coloneqq
\mleft(p, c_{\mleft(\Phi^{-1}_{s}(p)\mright)^{-1}} \mleft( \tau_{\pi(p)} \mright) \mright)
=
\mleft(p, \mleft(\Phi^{-1}_{s}(p)\mright)^{-1} ~ \tau_{\pi(p)} ~ \Phi^{-1}_{s}(p) \mright)
\end{equation*}
for all $p \in \mathcal{P}_U$,
with $\Phi_s$ being the orbit map through $s$.
\end{propositions}

\begin{remarks}{Other notation}{OtherNotationForLGBSectionsAsGaugeTrafo}
Due to the fact that $s^*\pi^*\mathcal{G} \cong (\pi \circ s)^*\mathcal{G} = \mathrm{id}_U^*\mathcal{G} \cong \mleft.\mathcal{G}\mright|_U$, and due to what we discussed in \ref{BasicNotations} about pullback sections, we could rewrite the first map to
\begin{align*}
C^\infty \mleft( \mleft.\mathcal{P}\mright|_U; \mleft.\mathcal{G}\mright|_U \mright)^{\mleft.\mathcal{G}\mright|_U} 
&\to 
\Gamma\mleft(\mleft.\mathcal{G}\mright|_U\mright),
\\
\sigma &\mapsto s^*\sigma.
\end{align*}
\end{remarks}

\begin{proof}[Proof of \ref{prop:GaugeTrafosAsLGBSectionsLocal}]
\leavevmode\newline
For $\sigma \in C^\infty \mleft( \mleft.\mathcal{P}\mright|_U; \mleft.\mathcal{G}\mright|_U \mright)^{\mleft.\mathcal{G}\mright|_U} \subset \Gamma\mleft( \pi^*\mleft.\mathcal{G}\mright|_U \mright)$ observe that $\mathrm{pr}_2 \circ \sigma \circ s$ is by construction a section of $\mleft.\mathcal{G}\mright|_U$ due to 
\begin{equation*}
\pi_{\mathcal{G}} \circ \mathrm{pr}_2 \circ \sigma \circ s
=
\pi \circ \mathrm{pr}_1 \circ \sigma \circ s
=
\pi \circ s
=
\mathrm{id}_U,
\end{equation*}
where $\mathrm{pr}_1\colon \pi^*\mathcal{G} \to \mathcal{P}$ is the projection onto the first component.
For the supposed inverse we have similarly that $\sigma^\tau$ is a section of $\Gamma(\pi^*\mleft.\mathcal{G}\mright|_U)$ by construction; then observe, using \ref{lem:SectionsNowInduceIsomToLGBsNotNecTriv},
\begin{align*}
\sigma_{p \cdot g}^\tau
&=
\mleft(p \cdot g, c_{\mleft(\Phi^{-1}_{s}(p \cdot g)\mright)^{-1}} \mleft( \tau_{\pi(p \cdot g)} \mright) \mright)
\\
&=
\mleft(p \cdot g, c_{g^{-1} ~ \mleft(\Phi^{-1}_{s}(p)\mright)^{-1}} \mleft( \tau_{\pi(p)} \mright) \mright)
\\
&=
\mleft(p \cdot g, \mleft(c_{g^{-1}} \circ c_{ \mleft(\Phi^{-1}_{s}(p)\mright)^{-1}} \mright) \mleft( \tau_{\pi(p)} \mright) \mright)
\\
&=
\mathcal{c}_{g^{-1}}
\mleft(p, c_{ \mleft(\Phi^{-1}_{s}(p)\mright)^{-1}} \mleft( \tau_{\pi(p)} \mright) \mright)
\\
&=
\mathcal{c}_{g^{-1}}
\mleft( \sigma^\tau_p \mright)
\end{align*}
for all $(p, g) \in \mleft.\mathcal{P}\mright|_U * \mleft.\mathcal{G}\mright|_U$, thus, $\sigma^\tau \in C^\infty \mleft( \mleft.\mathcal{P}\mright|_U; \mleft.\mathcal{G}\mright|_U \mright)^{\mleft.\mathcal{G}\mright|_U}$. Recall
\begin{equation*}
p = s_{\pi(p)} \cdot \Phi^{-1}_s(p)
\end{equation*} 
for all $p \in \mleft.\mathcal{P}\mright|_U$,
so that
\begin{equation*}
\Phi^{-1}_{s} \circ s
=
e|_U,
\end{equation*}
where $e$ is the neutral section of $\mathcal{G}$. On one hand
\begin{equation*}
\mleft(\mathrm{pr_2} \circ \sigma^\tau\mright)(s_x)
=
\mathrm{pr}_2\mleft(\sigma^\tau_{s_x}\mright)
=
c_{\mleft(\Phi^{-1}_{s}(s_x)\mright)^{-1}} \mleft( \tau_{\pi(s_x)} \mright)
=
c_{e_x}\mleft( \tau_x \mright)
=
\tau_x
\end{equation*}
for all $x \in M$ and $\tau \in \Gamma(\mleft.\mathcal{G}\mright|_U)$, and by \ref{def:GaugeTrafosLGBMaps} and \ref{ex:ConjugationActionForTheGeneralInnerGroupBundle} we have one the other hand
\begin{align*}
\sigma^{\mathrm{pr}_2 \circ \sigma \circ s}_p
&=
\Biggl(p, \mleft(c_{\mleft(\Phi^{-1}_{s}(p)\mright)^{-1}} \circ \mathrm{pr}_2 \mright) \mleft(\sigma_{s_{\pi(p)}}\mright) \Biggr)
\\
&=
\Biggl(p, \mleft(c_{\mleft(\Phi^{-1}_{s}(p)\mright)^{-1}} \circ \mathrm{pr}_2 \mright) \mleft(\sigma_{p \cdot \mleft( \Phi^{-1}_s(p) \mright)^{-1}}\mright) \Biggr)
\\
&=
\Biggl(p, \mleft(c_{\mleft(\Phi^{-1}_{s}(p)\mright)^{-1}} \circ \mathrm{pr}_2 \circ \mathcal{c}_{\Phi^{-1}_{s}(p)} \mright) \mleft(\sigma_{p}\mright) \Biggr)
\\
&=
\Biggl(p, \mleft(c_{\mleft(\Phi^{-1}_{s}(p)\mright)^{-1}} \circ c_{\Phi^{-1}_{s}(p)} \circ \mathrm{pr}_2 \mright) \mleft(\sigma_{p}\mright) \Biggr)
\\
&=
\bigl( p, \mathrm{pr}_2 (\sigma_p) \bigr)
\\
&=
\sigma_p
\end{align*}
for all $p \in \mleft.\mathcal{P}\mright|_U$ and $\sigma \in C^\infty \mleft( \mleft.\mathcal{P}\mright|_U; \mleft.\mathcal{G}\mright|_U \mright)^{\mleft.\mathcal{G}\mright|_U}$. Thus, $\sigma \mapsto \mathrm{pr}_2 \circ \sigma \circ s$ is bijective with inverse $\tau \mapsto \sigma^\tau$, and so it is only left to show that $\sigma \mapsto \mathrm{pr}_2 \circ \sigma \circ s$ is a group homomorphism. For $\sigma^\prime \in C^\infty \mleft( \mleft.\mathcal{P}\mright|_U; \mleft.\mathcal{G}\mright|_U \mright)^{\mleft.\mathcal{G}\mright|_U}$ we have
\begin{equation*}
\mleft.\sigma \sigma^\prime\mright|_p
=
\mleft( p, \mathrm{pr}_2( \sigma_p) ~ \mathrm{pr}_2\mleft( \sigma^\prime_p \mright) \mright),
\end{equation*}
thus,
\begin{equation*}
\mleft(\mathrm{pr}_2\circ \mleft( \sigma \sigma^\prime \mright) \circ s\mright)_x
=
\mathrm{pr}_2\mleft( \sigma_{s_x}\mright) ~ \mathrm{pr}_2\mleft( \sigma^\prime_{s_x} \mright)
=
\mleft( \mleft( \mathrm{pr}_2 \circ \sigma \circ s \mright) \cdot \mleft( \mathrm{pr}_2 \circ \sigma^\prime \circ s \mright) \mright)_x
\end{equation*}
for all $x \in M$. This finishes the proof.
\end{proof}

\begin{propositions}{Change of gauge as a local bundle automorphism}{GaugeChangeAsBundleAutomorph}
Let $\mathcal{G} \to M$ be an LGB over a smooth manifold $M$, and $\pi\colon\mathcal{P} \to M$ a principal $\mathcal{G}$-bundle. Also let $U_i$ and $U_j$ be two open subsets of $M$ so that $U_i \cap U_j \neq \emptyset$, two gauges $s_i \in \Gamma\mleft(\mathcal{P}|_{U_i}\mright)$ and $s_j \in \Gamma\mleft(\mathcal{P}|_{U_j}\mright)$, and the unique $\sigma_{ji} \in \Gamma\mleft( \mleft.\mathcal{G}\mright|_{U_i \cap U_j} \mright)$ with $s_i = s_j \cdot \sigma_{ji}$ on $U_i \cap U_j$.

Then the map $H_{ji}$ defined by
\begin{align*}
\mleft.\mathcal{P}\mright|_{U_i \cap U_j} &\to \mleft.\mathcal{P}\mright|_{U_i \cap U_j},\\
p &\mapsto \mleft.s_i\mright|_{\pi(p)} \cdot \Phi_{s_j}^{-1}(p),
\end{align*}
is an element of $\sAut\mleft( \mleft.\mathcal{P}\mright|_{U_i \cap U_j} \mright)$,
where $\Phi_{s_j}$ is the orbit map through $s_j$.

Using the notation of \ref{prop:GaugeTrafoAsBundleIsomIsASectionOfConjugationMaps}, we also get that $\sigma^{H_{ji}}$ is related to $\sigma_{ji}$ via $s_j$ in sense of \ref{prop:GaugeTrafosAsLGBSectionsLocal}, that is,
\begin{equation*}
\sigma^{H_{ji}}_p
=
\mleft(p, c_{\mleft(\Phi^{-1}_{s_j}(p)\mright)^{-1}} \mleft( \mleft.\sigma_{ji}\mright|_{\pi(p)} \mright) \mright)
\end{equation*}
for all $p \in \mleft.\mathcal{P}\mright|_{U_i \cap U_j}$, and
\begin{equation*}
\sigma_{ji}
=
\mathrm{pr}_2 \circ \sigma^{H_{ji}} \circ s_j,
\end{equation*}
where $\mathrm{pr}_2\colon \pi^*\mathcal{G} \to \mathcal{G}$ is the projection onto the second component.
\end{propositions}

\begin{proof}
\leavevmode\newline
Recall \ref{lem:SectionsNowInduceIsomToLGBsNotNecTriv} for the following proof.

We have $\mleft.s_i\mright|_{\pi(p)} \in \mathcal{P}_{\pi(p)}$ and $\Phi_{s_j}$ is base-preserving, and thus $\Phi_{s_j}^{-1}(p) \in \mathcal{G}_{\pi(p)}$, so that $H_{ji}$ is well-defined and also base-preserving because the right $\mathcal{G}$-action on $\mathcal{P}$ preserves the fibres.
$H_{ji}$ is clearly smooth by construction. Finally, 
\begin{equation*}
H_{ji}(p \cdot g)
=
\mleft.s_i\mright|_{\pi(p\cdot g)} \cdot \Phi_{s_j}^{-1}(p \cdot g)
=
\mleft.s_i\mright|_{\pi(p)} \cdot \Phi_{s_j}^{-1}(p) \cdot g
=
H_{ji}(p) \cdot g
\end{equation*}
for all $(p, g) \in \mleft.\mleft( \mathcal{P} * \mathcal{G} \mright)\mright|_{U_i \cap U_j}$, thus, $H_{ji}$ is a principal bundle morphism. The inverse $H^{-1}_{ji}$ is given by
\begin{equation*}
H^{-1}_{ji}
=
H_{ij}
\end{equation*}
due to
\begin{equation*}
\mleft(H_{ij} \circ H_{ji}\mright)(p)
=
\mleft.s_j\mright|_x \cdot \Phi^{-1}_{s_i}\mleft( \mleft.s_i\mright|_{x} \cdot \Phi_{s_j}^{-1}(p) \mright)
=
\mleft.s_j\mright|_x \cdot {\underbrace{\Phi^{-1}_{s_i}\mleft( \mleft.s_i\mright|_{x} \mright)}_{= e_x}} \cdot \Phi_{s_j}^{-1}(p)
=
\mleft.s_j\mright|_x \cdot \Phi_{s_j}^{-1}(p)
=
p
\end{equation*}
for all $p \in \mathcal{P}_{U_i \cap U_j}$, where $x \coloneqq \pi(p)$; by symmetry of course similarly when swapping $j$ and $i$. Thus, $H_{ji}$ is bijective and its inverse smooth, so that $H_{ji} \in \sAut\mleft( \mleft.\mathcal{P}\mright|_{U_i \cap U_j} \mright)$.

Now write
\begin{align*}
H_{ji}(p)
&=
\mleft.s_i\mright|_{x} \cdot \Phi_{s_j}^{-1}(p)
\\
&=
\mleft.s_j\mright|_{x} \cdot \mleft.\sigma_{ji}\mright|_x \cdot \Phi_{s_j}^{-1}(p)
\\
&=
\mleft.s_j\mright|_{x} \cdot \Phi_{s_j}^{-1}(p) \cdot \mleft( \Phi_{s_j}^{-1}(p) \mright)^{-1} \cdot \mleft.\sigma_{ji}\mright|_x \cdot \Phi_{s_j}^{-1}(p)
\\
&=
p \cdot c_{\mleft(\Phi^{-1}_{s_j}(p)\mright)^{-1}} \mleft( \mleft.\sigma_{ji}\mright|_{x} \mright),
\end{align*}
and due to $H_{ji}(p) = p \cdot \mathrm{pr}_2\mleft(\sigma^{H_{ji}}_p\mright)$ we can immediately derive
\begin{equation*}
\sigma^{H_{ji}}_p
=
\mleft(p, c_{\mleft(\Phi^{-1}_{s_j}(p)\mright)^{-1}} \mleft( \mleft.\sigma_{ji}\mright|_{\pi(p)} \mright) \mright),
\end{equation*}
thus, $\sigma^{H_{ji}}$ is related to $\sigma_{ji}$ w.r.t.\ $s_j$ in sense of \ref{prop:GaugeTrafosAsLGBSectionsLocal} and so the remaining part of the proposition's statement follows by \ref{prop:GaugeTrafosAsLGBSectionsLocal}.
\end{proof}

Using this (local) gauge transformation we can now relate $A_{s_i}$ and $A_{s_j}$.

\begin{propositions}{Gauge transformations as a change of gauge}{LocalGaugeTrafoChangeGauge}
Let $\mathcal{G} \to M$ be an LGB over a smooth manifold $M$ and $\pi\colon\mathcal{P} \to M$ a principal $\mathcal{G}$-bundle, also let $\mathrm{H}\mathcal{G}$ be a horizontal distribution on $\mathcal{G}$ and $A \in \Omega^1(\mathcal{P}; \pi^*\mathcal{g})$ be a connection 1-form on $\mathcal{P}$. Also let $U_i$ and $U_j$ be two open subsets of $M$ so that $U_i \cap U_j \neq \emptyset$, two gauges $s_i \in \Gamma\mleft(\mathcal{P}|_{U_i}\mright)$ and $s_j \in \Gamma\mleft(\mathcal{P}|_{U_j}\mright)$, and the unique $\sigma_{ji} \in \Gamma\mleft( \mleft.\mathcal{G}\mright|_{U_i \cap U_j} \mright)$ with $s_i = s_j \cdot \sigma_{ji}$ on $U_i \cap U_j$.

Then we have over $U_i \cap U_j$ that
\begin{equation*}
A_{s_i}
=
\mathrm{Ad}_{\sigma_{ji}^{-1}}\circ A_{s_j}
	+ \Delta\sigma_{ji}.
\end{equation*}
\end{propositions}

\begin{proof}
\leavevmode\newline
Define $H_{ji} \in \sAut\mleft( \mleft.\mathcal{P}\mright|_{U_i \cap U_j} \mright)$ as in \ref{prop:GaugeChangeAsBundleAutomorph} (also following its notation), then observe
\begin{equation*}
\mleft(H_{ji} \circ s_j\mright)_x
=
\mleft.s_i\mright|_x \cdot {\underbrace{\Phi^{-1}_{s_j}\mleft( \mleft.s_j\mright|_x \mright)}_{= e_x}}
=
\mleft.s_i\mright|_x
\end{equation*}
for all $x \in U_i \cap U_j$, and thus
\begin{align*}
s_j^! H_{ji}^!A
&=
\mleft( H_{ji} \circ s_j \mright)^!A
=
A_{s_i}.
\end{align*}
By \ref{prop:GaugeTrafoOfGaugeBoson} we also have
\begin{align*}
H_{ji}^!A
&=
{\sAd_{\mathrm{pr}_2\circ\mleft(\sigma^{H_{ji}}\mright)^{-1}}} \circ A
	+ \mleft(\pi^*\Delta\mright)\sigma^{H_{ji}}.
\end{align*}
We now want to apply $s_j^!$ on both hand sides.
Recall \ref{rem:PullBackGaugeFieldRemark} and \ref{rem:OtherNotationForLGBSectionsAsGaugeTrafo}, we will now use several natural isomorphisms without further mention, especially the base-point component will be treated in a "sloppy" way; the advantage will be a cleaner notation, and this is the reason why $\sAd$ becomes $\mathrm{Ad}$ by applying $s_j^!$ on the right hand side.
Using \ref{prop:GaugeChangeAsBundleAutomorph} and \ref{rem:PullBackDarboux} we can derive
\begin{equation*}
\Delta \sigma_{ji}
=
\bigl( \mleft(\pi \circ s_j\mright)^* \Delta \bigr)\mleft( s_j^*\sigma^{H_{ji}} \mright)
=
\bigl( s_j^*\pi^* \Delta \bigr)\mleft( s_j^*\sigma^{H_{ji}} \mright)
=
s_j^!\mleft(\bigl( \pi^* \Delta \bigr)\mleft( \sigma^{H_{ji}} \mright)\mright),
\end{equation*}
and
\begin{equation*}
s_j^!\mleft( {\sAd_{\mathrm{pr}_2\circ\mleft(\sigma^{H_{ji}}\mright)^{-1}}} \circ A \mright)
=
{\mathrm{Ad}_{\mathrm{pr}_2\circ\mleft(\sigma^{H_{ji}}\mright)^{-1}\circ s_j}} \circ s_j^!A
=
{\mathrm{Ad}_{\mleft(\mathrm{pr}_2\circ\sigma^{H_{ji}}\circ s_j\mright)^{-1}}} \circ s_j^!A
=
{\mathrm{Ad}_{\sigma_{ji}^{-1}}} \circ A_{s_j},
\end{equation*}
where we used again the trivial relation (alternatively recall \ref{cor:PullbackLGB})
\begin{equation*}
\mleft(\mathrm{pr}_2 \circ \sigma^{H_{ji}} \mright)^{-1}
=
\mathrm{pr}_2 \circ \mleft(\sigma^{H_{ji}}\mright)^{-1}~.
\end{equation*}
Altogether, concluding the proof with
\begin{equation*}
A_{s_i}
=
s_j^! H_{ji}^!A
=
{\mathrm{Ad}_{\sigma_{ji}^{-1}}} \circ A_{s_j}
	+ \Delta \sigma_{ji}~.
\end{equation*}
\end{proof}

This allows us to finally show the first relation to the infinitesimal gauge theory developed by Alexei Kotov and Thomas Strobl.

\begin{remarks}{Integrating curved Yang-Mills gauge theories, part I}{IntegratingKotovStrobl}
Let the situation be as in \ref{prop:LocalGaugeTrafoChangeGauge}, then set 
\begin{align*}
\sigma_{ji}
&\coloneqq
\e^{t \varepsilon}
\end{align*}
for a $\varepsilon \in \Gamma\mleft( \mleft.\mathcal{g}\mright|_{U_i \cap U_j} \mright)$ and $t \in \mathbb{R}$. Then we define the \textbf{infinitesimal gauge transformation $\delta_\varepsilon A_{s_j}$ of $A$ along $\varepsilon$ (w.r.t.\ $s_j$)} as an element of $\Omega^1\mleft( U_i \cap U_j; \mleft.\mathcal{g}\mright|_{U_i \cap U_j} \mright)$ by
\begin{align*}
\delta_\varepsilon A_{s_j}
&\coloneqq
\mleft.\frac{\mathrm{d}}{\mathrm{d}t}\mright|_{t=0} A_{s_i}.
\end{align*}
The corresponding derivative related to the adjoint representation can be calculated as usual, and for the corresponding term related to the Darboux derivative we use the vector bundle connection introduced in \ref{def:ConnectionOnLAB}. Thus, we get\footnote{Rigorously, one has to define $\delta_\varepsilon A_{s_j}$ via the natural projection $\mathrm{V}\mathcal{g} \to \mathcal{g}$ as we did in \ref{prop:FinallyTheNablaInduction}.}
\begin{align*}
\delta_\varepsilon A_{s_j}
&=
\nabla^{\mathcal{G}} \varepsilon 
	- \mleft[ \varepsilon, A_{s_j} \mright]_{\mathcal{g}}.
\end{align*}
This is \textit{precisely} the infinitesimal gauge transformation developed by Alexei Kotov and Thomas Strobl, see \cite{CurvedYMH} for a concise summary or \cite{MyThesis} for an extended introduction. These references are for the very general situation using Lie algebroids, hence see alternatively \cite{My1stpaper} for this type of gauge theory restricted to Lie algebra bundles.

Thus, we successfully integrated this definition of infinitesimal gauge transformations.
\end{remarks}

\begin{remarks}{Closure of gauge transformations}{ClosureOfGaugeTrafos}
The gauge transformations close simply by construction, that is,
\begin{equation*}
H'^!H^!A
=
(H \circ H')^!A
\end{equation*}
for all $H', H \in \sAut(\mathcal{P})$. The reader might wonder why, given that existing literature for the infinitesimal version (like \cite[Thm.\ 4.3.37 and Thm.\ 4.3.43]{MyThesis}, \cite[Prop.\ 8, Thm.\ 1]{EichtrafoKruemmungUrspruenglich} and \cite[Eq.\ 9, 10 and 11]{mayerlieAuchEichtrafoStuff}) prove that one needs a \textbf{multiplicative connection} on $\mathcal{G}$, that is, the parallel transport on $\mathcal{G}$ has to be a homomorphism of the $\mathcal{G}$-multiplication, while we did not assume this in this Subsection. However, the culprit is that $A$ as connection 1-form on $\mathcal{P}$ exists if and only if the connection on $\mathcal{G}$ is a multiplicative connection; recall \ref{cor:ExOfEhresmann}.
\end{remarks}

Alexei Kotov's and Thomas Strobl's theory also generalizes the curvature/field strength related to gauge theory. Hence, let us now turn to this notion and discuss why we need further generalisations.

\subsection{Generalized curvature/field strength}\label{CurvatureSubsection}

\subsubsection{Multiplicative Yang-Mills connections on LGBs}\label{MultiplicativeForms}

In the following we will introduce the field strength; in order to achieve gauge invariance later in the physical gauge theory, we will now need to assume what we will call \textbf{compatibility conditions} on the horizontal distribution of the structural LGB $\mathcal{G}$ of which we already encountered one in \ref{cor:ExOfEhresmann}. Here in this work we will just present these compatibility conditions, and you will be able to see in the proofs why this leads to gauge invariance later. However, if you want to understand the following conditions coming from a point of view not knowing the solution beforehand, then either see \cite{CurvedYMH} or \cite{MyThesis}; the latter also provides a physical motivation using what I called \textbf{field redefinitions} which naturally motivate the gauge theory presented here by transforming a classical gauge theory in such a way that the physics is preserved; now recall \ref{def:ConnectionOnLAB} for the following.

We start with the infinitesimal version of the compatibility conditions, already known and pointed out in the previously-mentioned references.

\begin{definitions}{Infinitesimal multiplicative Yang-Mills connection}{YangMillsConnection}
Let $\mathcal{G} \to M$ be an LGB over a smooth manifold $M$, and $\mathrm{H}\mathcal{G}$ be a horizontal distribution of $\mathcal{G}$. Then we say that $\mathrm{H}\mathcal{G}$ is an \textbf{infinitesimal multiplicative Yang-Mills connection (on $\mathcal{G}$)} if there is a \textbf{primitive} $\zeta \in \Omega^2(M; \mathcal{g})$ such that the $\mathcal{G}$-connection $\nabla^{\mathcal{G}}$ on the associated LAB $\mathcal{g}$ with bracket $\mleft[ \cdot, \cdot \mright]_{\mathcal{g}}$ satisfies the \textbf{infinitesimal compatibility conditions}
\begin{align}\label{CondSGleichNullLAB}
\nabla^{\mathcal{G}}\mleft( \mleft[ \mu, \nu \mright]_{\mathcal{g}} \mright)
&=
\mleft[ \nabla^{\mathcal{G}} \mu, \nu \mright]_{\mathcal{g}}
	+ \mleft[ \mu, \nabla^{\mathcal{G}} \nu \mright]_{\mathcal{g}},
\\
R_{\nabla^{\mathcal{G}}}(X, Y)\mu
&=
\mleft[ \zeta(X, Y), \mu \mright]_{\mathcal{g}}\label{CondKruemmungmitBLAB}
\end{align}
for all $\mu, \nu \in \Gamma(\mathcal{g})$ and $X, Y \in \mathfrak{X}(M)$, where $R_{\nabla^{\mathcal{G}}}$ is the curvature of $\nabla^{\mathcal{G}}$. We then also call $\nabla^{\mathcal{G}}$ an \textbf{(infinitesimal) multiplicative Yang-Mills connection} and denote it by $\nabla^\caG$.

In the following, $\zeta$ is always such a 2-form as given here, and even though it is not uniquely given we will treat it as if it is fixed for a given $\nabla^\caG$, though one may want to write that $\mleft( \nabla^\caG, \zeta \mright)$ is a multiplicative Yang-Mills connection, giving an emphasis on the non-uniqueness of $\zeta$. 
\end{definitions}

\begin{remarks}{Lie derivation laws covering a coupling}{MackenziesStuffRelation}
As illustrated in \cite[\S 5.1]{MyThesis} and \cite{My1stpaper} such connections also arise in \cite[\S 7.2, page 271ff.]{mackenzieGeneralTheory}: There these connections are called \textbf{Lie derivation laws covering a coupling} due to the fact that these are important in the construction of ``coupling'' an LAB to a Lie algebroid via an Atiyah sequence. There is a strong relationship of this study with the gauge theory we are going to define, but we will not need this so that we will not repeat what has been discussed in \cite[\S 5.1]{MyThesis} and \cite{My1stpaper}. However, see the classification results in the example section later.
%
\end{remarks}

We want to integrate these infinitesimal compatibility conditions to conditions on $\mathrm{H}\mathcal{G}$ directly.
Let us initially focus on the first infinitesmal compatibility condition \eqref{CondSGleichNullLAB}. In the following we will also view $\mathcal{G}$ as a principal $\mathcal{G}$-bundle which is equipped with the same horizontal distribution as $\mathcal{G}$ as LGB; related notions and notations carry over. One can show that such Ehresmann connections (modifying themselves)\footnote{In the LGB context such connections are often simply also called Ehresmann connections; however, note that our notion is more general.} satisfy \eqref{CondSGleichNullLAB}, and in fact Ehresmann connections on LGBs (and Lie groupoids) were already discussed in \cite{LAURENTGENGOUXStienonXuMultiplicativeForms}; and there is also a rather recent preprint related to a similar subject, see \cite{FernandesMarcutMultiplicativeForms}. Such Ehresmann connections are called \textbf{multiplicative}.

\textbf{Special thanks to Camille Laurent-Gengoux for pointing out these references and discussing related subjects, this helped me tremendously.}

\begin{lemmata}{Multiplicative connections induce Lie bracket derivations}{YangMillsConnAnEhresmann}
\tcbsubtitle[before skip=\baselineskip]{\cite[\S 4.5, Prop.\ 4.21]{LAURENTGENGOUXStienonXuMultiplicativeForms}}
Let $\mathcal{G} \to M$ be an LGB over a smooth manifold $M$, and $\mathrm{H}\mathcal{G}$ a horizontal distribution on $\mathcal{G}$. If $\mathrm{H}\mathcal{G}$ is an (Ehresmann) connection on $\mathcal{G}$ as principal bundle modifying itself, then the infinitesimal compatibility condition \eqref{CondSGleichNullLAB} is satisfied.
\end{lemmata}

\begin{proof}
\leavevmode\newline
This can be quickly answered by recalling how the parallel transport behaves; recall \ref{cor:OtherDefsForEhresmann}. This means that the corresponding parallel transport $\mathrm{PT}^{\mathcal{G}}$ of $\mathcal{G}$ (omitting the notation of the involved curve) is multiplicative, \textit{i.e.}\
\begin{equation*}
\mathrm{PT}^{\mathcal{G}}(gq)
=
\mathrm{PT}^{\mathcal{G}}(g) ~ \mathrm{PT}^{\mathcal{G}}(q)
\end{equation*}
for all $(g, q) \in \mathcal{G} * \mathcal{G}$. Thus, infinitesimally the associated parallel transport on $\mathcal{g}$ is a homomorphism w.r.t.\ the Lie bracket, and thus $\nabla^{\mathcal{G}}$ is a Lie bracket derivation.
\end{proof}

\begin{remarks}{Base manifold a horizontal leave}{baseManifoldHoriz}
Observe that the proof's argument implies that a parallel transported neutral element is again the neutral element of the corresponding final fibre, if $\mathrm{H}\mathcal{G}$ is such an Ehresmann connection. Thus, $M$ naturally embedded via the neutral section $e$ can then be viewed as a horizontal leave of $\mathrm{H}\mathcal{G}$.
\end{remarks}

Due to this, given compatibility condition \eqref{CondSGleichNullLAB}, the following is clear by definition.

\begin{corollaries}{$\mutot$ as Ehresmann connection}{TotMCFormIsConnectionForm}
Let $\mathcal{G} \to M$ be an LGB over a smooth manifold $M$, and $\mathrm{H}\mathcal{G}$ a multiplicative Ehresmann connection on $\mathcal{G}$. Then $\mutot$ is the connection 1-form corresponding to $\mathrm{H}\mathcal{G}$ in sense of \ref{thm:OurConnectionHasAUniqueoneForm}.
\end{corollaries}

\begin{proof}
\leavevmode\newline
This simply follows by \ref{def:TotMCFormOnLGB} and \ref{thm:OurConnectionHasAUniqueoneForm}, making use of $\rmH \caP = \rmH \caG$ (here essentially $\caP = \caG$).
\end{proof}

\begin{remarks}{Darboux derivative as local gauge field}{DavouxderivativeALocalGaugeField}
If $\mathrm{H}\mathcal{G}$ is a multiplicative Ehresmann connection on $\mathcal{G}$, we know by \ref{cor:TotMCFormIsConnectionForm} that the Darboux derivative $\Delta \sigma$ of a (local) section $\sigma$ of $\mathcal{G}$ (\ref{def:DarbouxDerivativeOnLGBs}) is a local gauge field (w.r.t.\ $\sigma$) corresponding to $\mutot$, recall \ref{def:LocalGaugeField}. That is,
\begin{equation*}
\Delta \sigma
=
\mleft(\mutot\mright)_\sigma.
\end{equation*}
\end{remarks}

Furthermore, following \cite{LAURENTGENGOUXStienonXuMultiplicativeForms}, we can derive that $\mutot$ is a multiplicative form if and only if $\mathrm{H}\mathcal{G}$ is a multiplicative Ehresmann connection; let us henceforth introduce multiplicative 1-forms. Recall \ref{lem:PullbackFibreBundleItsTangentSp}.

\begin{definitions}{Multiplicative forms}{MultiplicativeFormsDef}
\tcbsubtitle[before skip=\baselineskip]{\cite[\S 2.1, special situation of Def.\ 2.1]{crainic2015multiplicative}}
Let $\pi_{\mathcal{G}} \colon \mathcal{G} \to M$ be an LGB over a smooth manifold $M$. Then we call an $\omega \in \Omega^1\mleft( \mathcal{G}; \pi_{\mathcal{G}}^*\mathcal{g} \mright)$ a \textbf{multiplicative form} if
\begin{equation*}
\omega_{gq}\mleft( \mathrm{D}_{(g, q)}\Phi(X, Y)  \mright)
=
\sAd_{q^{-1}}\bigl( \omega_{g}(X) \bigr)
	+ \bigl( gq, \widehat{\omega}_{q}(Y) \bigr)
\end{equation*}
for all $(g, q) \in \mathcal{G}*\mathcal{G}$ and $(X, Y) \in \mathrm{T}_{(g, q)}(\mathcal{G}*\mathcal{G})$,
where $\Phi\colon \mathcal{G} * \mathcal{G} \to \mathcal{G}$ is the multiplication in $\mathcal{G}$, and $\widehat{\omega} \coloneqq \mathrm{pr}_2 \circ \omega$, where $\mathrm{pr}_2\colon \pi_{\mathcal{G}}^*\mathcal{g} \to \mathcal{g}$ is the natural projection onto the second component.
\end{definitions}

\begin{remarks}{Ah, those nasty base points}{sloppynotationformultiplicativity}
The shape of the second summand is again just because of technical reasons due to the definition of pullback bundles and their base points. If we do not care so much about the involved base points since these are clear by context, then we can also shortly write
\begin{equation*}
\mleft(\Phi^!\omega\mright)_{(g, q)}(X, Y)
=
\sAd_{q^{-1}}\bigl( \omega_{g}(X) \bigr)
	+ \omega_{q}(Y).
\end{equation*}
Also recall what we mentioned in \ref{BasicNotations} about pullback sections.
\end{remarks}

\begin{propositions}{Ehresmann connection 1-forms on LGBs are multiplicative}{TotMaurerIsMultiplicativeIfSEqual0}
\tcbsubtitle[before skip=\baselineskip]{\cite[\S 4.4, implication of Lemma 4.14]{LAURENTGENGOUXStienonXuMultiplicativeForms}}
Let $\pi_{\mathcal{G}} \colon \mathcal{G} \to M$ be an LGB over a smooth manifold $M$, and $\mathrm{H}\mathcal{G}$ a horizontal distribution on $\mathcal{G}$ with corresponding connection 1-form $\mutot$. Then $\mathrm{H}\mathcal{G}$ is a multiplicative (Ehresmann) connection on $\mathcal{G}$ as principal bundle if and only if $\mutot$ is multiplicative.
\end{propositions}

\begin{proof}
\leavevmode\newline
The general approach in the mentioned reference is different to ours; hence, let us provide a proof suitable to our constructions to highlight and provide further justification of the notion of Ehresmann connection provided here. We will view $\mathcal{G}$ as a principal $\mathcal{G}$-bundle whose horizontal distribution is also $\mathrm{H}\mathcal{G}$. We will also use the same notation as in \ref{def:MultiplicativeFormsDef}.

To answer whether or not $\mathrm{H}\mathcal{G}$ is a multiplicative Ehresmann connection we just need to discuss whether $\mutot$ is a connection 1-form on $\mathcal{G}$ due to the fact that $\mathrm{H}\mathcal{G}$ is the kernel of $\mutot$ by definition (see also \ref{cor:TotMCFormIsConnectionForm}), also using \ref{thm:OurConnectionHasAUniqueoneForm}. Also by definition, $\mutot$ already satisfies
\begin{equation*}
\mutot\mleft( \overline{\nu} \mright)
=
\pi_{\mathcal{G}}^*\nu.
\end{equation*}
By additionally using \ref{prop:DiffOfLGBAction} (especially the associated \ref{rem:DarbouxInActionInfinit}) we show
\begin{align*}
\mleft(\mutot\mright)_{gq}\mleft( \mathrm{D}_{(g,q)}\Phi(X, Y) \mright)
&=
\mleft(\mutot\mright)_{gq}\mleft(
	\mathrm{D}_gR_\sigma(X)
		- \mleft.{\overline{ \mleft.\mleft(\pi_{\mathcal{G}}^!\Delta \sigma\mright)\mright|_g (X)}}\mright|_{gq}
		+ \mleft.{\overline{\mleft( \mutot\mright)_{q} (Y)}}\mright|_{gq}
\mright)
\\
&=
\mleft(\mutot\mright)_{gq}\bigl(
	\mathcal{r}_{q*}(X)
\bigr)
		+ \mleft( gq, \mleft( \widehat{\mutot}\mright)_{q} (Y) \mright)
\\
&=
\mleft(\mathcal{r}_q^!\mutot\mright)_{g}(X)
		+ \mleft( gq, \mleft( \widehat{\mutot}\mright)_{q} (Y) \mright)
\end{align*} 
for all $(g, q) \in \mathcal{G}*\mathcal{G}$ and $(X, Y) \in \mathrm{T}_{(g, q)}(\mathcal{G}*\mathcal{G})$, where $\sigma$ is a (local) section of $\mathcal{G}$ such that $\sigma_{x} = q$. Thus, $\mutot$ is multiplicative if and only if
\begin{equation*}
\sAd_{q^{-1}}\mleft(\mleft(\mutot\mright)_{g}(X)\mright)
		+ \mleft( gq, \mleft( \widehat{\mutot}\mright)_{q} (Y) \mright)
=
\mleft(\mathcal{r}_q^!\mutot\mright)_{g}(X)
		+ \mleft( gq, \mleft( \widehat{\mutot}\mright)_{q} (Y) \mright)
\end{equation*}
if and only if
\begin{equation*}
\sAd_{q^{-1}}\mleft(\mleft(\mutot\mright)_{g}(X)\mright)
=
\mleft(\mathcal{r}_q^!\mutot\mright)_{g}(X)
\end{equation*}
if and only if $\mutot$ is a connection 1-form. This finishes the proof.
%
\end{proof}

This Theorem also implies the Leibniz rule of the Darboux derivative.

\begin{propositions}{Leibniz rule of the Darboux derivative}{DarbouxLeibnizRule}
Let $\pi_{\mathcal{G}} \colon \mathcal{G} \to M$ be an LGB over a smooth manifold $M$, and $\mathrm{H}\mathcal{G}$ a multiplicative Ehresmann connection on $\mathcal{G}$. Then we have
\begin{equation*}
\Delta(\sigma \tau)
=
\mathrm{Ad}_{\tau^{-1}} \circ \Delta \sigma
	+ \Delta\tau
\end{equation*}
and
\begin{equation*}
\Delta\mleft(\sigma^{-1}\mright)
=
- \mathrm{Ad}_{\sigma} \circ \Delta \sigma
\end{equation*}
for all $\sigma, \tau \in \Gamma(\mathcal{G})$.
\end{propositions}

\begin{remarks}{Old equations new}{OldIsNew}
These are precisely the formulas one expects, see \textit{e.g.}\ \cite[\S 5.1, Eq.\ (2) and (3), page 182]{mackenzieGeneralTheory} for the "classical" case; beware that the reference uses right- instead of left-invariant vector fields to describe the Lie algebra, so that the referenced formulas look explicitly a bit different. Nevertheless, the similarity of the structure should be obvious.

The mentioned reference for \ref{prop:TotMaurerIsMultiplicativeIfSEqual0}, \cite[\S 4.4, Lemma 4.14]{LAURENTGENGOUXStienonXuMultiplicativeForms}, actually uses the Leibniz rule of the Darboux derivative as an equivalent description of multiplicative connections, and essentially links multiplicativeness to closedness w.r.t.\ a natural simplicial differential; see also \ref{rem:SimplicialDifferentialStuff} later.
\end{remarks}

\begin{proof}[Proof of \ref{prop:DarbouxLeibnizRule}]
\leavevmode\newline
For this calculation we will take care about base points of pullback bundles w.r.t.\ the Darboux derivative, thus recall the last part of \ref{rem:RemarkABoutDarbouxNotationWRTPullback}. Using the same notation as in \ref{def:MultiplicativeFormsDef}, this proposition follows simply by \ref{prop:TotMaurerIsMultiplicativeIfSEqual0}, that is, set $Y \coloneqq \mathrm{D}_x\sigma(X)$ and $Z \coloneqq \mathrm{D}_x\tau(X)$ for $X \in \mathrm{T}_xM$ ($x \in M$), then
\begin{align*}
\bigl( \sigma_x\tau_x, \Delta(\sigma\tau)_x(X) \bigr)
&=
\mleft(\mutot\mright)_{\sigma_x\tau_x}\bigl( \mathrm{D}_x(\sigma \tau)(X) \bigr)
\\
&=
\mleft(\mutot\mright)_{\sigma_x\tau_x}\bigl( \mathrm{D}_{(\sigma_x, \tau_x)}\Phi(Y, Z) \bigr)
\\
&=
\sAd_{\tau_{x}^{-1}}\mleft(\mleft(\mutot\mright)_{\sigma_x}(Y)\mright)
		+ \mleft( \sigma_x\tau_x, \mleft( \widehat{\mutot}\mright)_{\tau_x} (Z) \mright)
\\
&=
\mleft( 
	\sigma_x\tau_x, 
	\mathrm{Ad}_{\tau_x^{-1}}\mleft( \mleft( \widehat{\mutot}\mright)_{\sigma_x} (Y) \mright) 
	+ \mleft( \widehat{\mutot}\mright)_{\tau_x} (Z) 
\mright)
\\
&=
\mleft( 
	\sigma_x\tau_x, 
	\mathrm{Ad}_{\tau_x^{-1}}\bigl( \mleft.\mleft( \Delta \sigma \mright)\mright|_x(X) \bigr) 
	+ \mleft.\mleft( \Delta \tau \mright)\mright|_x(X)
\mright).
\end{align*}
Hence,
\begin{equation*}
\Delta(\sigma \tau)
=
\mathrm{Ad}_{\tau^{-1}} \circ \Delta \sigma
	+ \Delta\tau.
\end{equation*}
Now recall \ref{rem:baseManifoldHoriz} which implies for the neutral element section $e$ that
\begin{equation*}
\Delta e
\equiv
0,
\end{equation*}
and therefore, using the previous result just derived,
\begin{equation*}
0
=
\Delta e
=
\Delta \mleft( \sigma \sigma^{-1} \mright)
=
\mathrm{Ad}_{\sigma} \circ \Delta \sigma
	+ \Delta\mleft(\sigma^{-1}\mright),
\end{equation*}
thus, concluding the proof with
\begin{equation*}
\Delta\mleft(\sigma^{-1}\mright)
=
- \mathrm{Ad}_{\sigma} \circ \Delta \sigma.
\end{equation*}
\end{proof}

The Leibniz rule of the Darboux derivative extends to $\nabla^{\mathcal{G}}$.

\begin{lemmata}{Conjugation preserving infinitesimal compatibility conditions}{FieldRedefViaLGBSection}
Let $\pi_{\mathcal{G}} \colon \mathcal{G} \to M$ be an LGB over a smooth manifold $M$, and $\mathrm{H}\mathcal{G}$ a multiplicative Ehresmann connection on $\mathcal{G}$. Then we have
\begin{equation}\label{SpecialFieldRedef}
\mathrm{Ad}_{\sigma^{-1}} \circ \nabla^{\mathcal{G}} \circ \mathrm{Ad}_\sigma
=
\nabla^{\mathcal{G}}
	+ \mathrm{ad}_{\Delta \sigma}
\end{equation}
for all $\sigma \in \Gamma(\mathcal{G})$, that is,
\begin{equation*}
\mleft(\mathrm{Ad}_{\sigma^{-1}} \circ \nabla^{\mathcal{G}}_X \circ \mathrm{Ad}_\sigma\mright)\nu
=
\nabla^{\mathcal{G}}_X\nu
	+ \mleft[ \Delta\sigma(X), \nu \mright]_{\mathcal{g}}
\end{equation*}
for all $\nu \in \Gamma(\mathcal{g})$ and $X \in \mathfrak{X}(M)$.
If $\mathrm{H}\mathcal{G}$ is an infinitesimal multiplicative Yang-Mills connection, in particular $R_{\nabla^\caG} = \mathrm{ad}_\zeta$ for $\zeta \in \Omega^2(M; \mathcal{g})$, then $\overline{\nabla}^\sigma \coloneqq \mathrm{Ad}_{\sigma^{-1}} \circ \nabla^\caG \circ \mathrm{Ad}_\sigma$ is also an infinitesimal multiplicative Yang-Mills connection with
\begin{equation*}
R_{\overline{\nabla}^\sigma}
=
\mathrm{ad}_{\mathrm{Ad}_{\sigma^{-1}}(\zeta)},
\end{equation*}
in particular we get
\begin{equation}\label{AlmostGeneralizedMaurerCartanEquation}
\mleft[ \mathrm{d}^{\nabla^\caG} \Delta \sigma + \frac{1}{2} \mleft[ \Delta \sigma \stackrel{\wedge}{,} \Delta\sigma \mright]_{\mathcal{g}} + \zeta - \mathrm{Ad}_{\sigma^{-1}} \circ \zeta, ~ \nu \mright]_{\mathcal{g}}
=
0
\end{equation}
for all $\nu \in \Gamma(\mathcal{g})$.
\end{lemmata}

\begin{proof}
\leavevmode\newline
By \ref{prop:DarbouxLeibnizRule} we get
\begin{align*}
\Delta\mleft( \e^{\mathrm{Ad}_\sigma (\nu)} \mright)
&=
\Delta\mleft( \sigma \e^{\nu} \sigma^{-1} \mright)
\\
&=
\mathrm{Ad}_{\sigma} \circ \Delta \mleft( \sigma \e^\nu \mright)
	+ \Delta \mleft( \sigma^{-1} \mright)
\\
&=
\mathrm{Ad}_{\sigma} \circ \mleft( \mathrm{Ad}_{\e^{-\nu}} \circ \Delta \sigma + \Delta e^\nu \mright)
	- \mathrm{Ad}_\sigma \circ \Delta \sigma
\\
&=
\mathrm{Ad}_{\sigma} \circ \mleft( \mathrm{Ad}_{\e^{-\nu}} \circ \Delta \sigma
	- \Delta \sigma + \Delta \e^\nu \mright)
\end{align*}
for all $\sigma \in \Gamma(\mathcal{G})$ and $\nu \in \Gamma(\mathcal{g})$, where the first equality is a well-known fact, see \textit{e.g.}\ \cite[\S 1.7, Thm.\ 1.7.16, page 59]{Hamilton}. Then
\begin{align*}
\nabla^{\mathcal{G}}\mleft( \mathrm{Ad}_\sigma(\nu) \mright)
&=
\mleft. \frac{\mathrm{d}}{\mathrm{d}t} \mright|_{t=0} \mleft(
	\Delta\mleft( \e^{\mathrm{Ad}_\sigma (t\nu)} \mright)
\mright)
\\
&=
\mleft. \frac{\mathrm{d}}{\mathrm{d}t} \mright|_{t=0} \Bigl(
	\mathrm{Ad}_{\sigma} \circ \mleft( \mathrm{Ad}_{\e^{-t\nu}} \circ \Delta \sigma
	- \Delta \sigma + \Delta \e^{t\nu} \mright)
\Bigr)
\\
&=
\mathrm{Ad}_{\sigma} \circ \mleft( \mathrm{ad}_{-\nu} \circ \Delta \sigma
	+ \nabla^{\mathcal{G}} \nu \mright)
\\
&=
\mathrm{Ad}_{\sigma} \circ \mleft( \mathrm{ad}_{\Delta \sigma}(\nu)
	+ \nabla^{\mathcal{G}} \nu \mright)
\end{align*}
making use of that this calculation is pointwise at $x \in M$ just a standard derivative in the vector space $\mathcal{g}_x$. Thus,
\begin{equation*}
\mathrm{Ad}_{\sigma^{-1}} \circ \nabla^{\mathcal{G}} \circ \mathrm{Ad}_\sigma
=
\nabla^{\mathcal{G}}
	+ \mathrm{ad}_{\Delta \sigma}.
\end{equation*}
In order to show $R_{\overline{\nabla}^\sigma} = \mathrm{ad}_{\mathrm{Ad}_{\sigma^{-1}}(\zeta)}$ one uses the definition $\overline{\nabla}^\sigma \coloneqq \mathrm{Ad}_{\sigma^{-1}} \circ \nabla^\caG \circ \mathrm{Ad}_\sigma$ to calculate the curvature, 
\begin{align*}
R_{\overline{\nabla}^\sigma} (X, Y) \nu
&=
\overline{\nabla}^\sigma_X \overline{\nabla}^\sigma_Y \nu
	- \overline{\nabla}^\sigma_Y \overline{\nabla}^\sigma_X \nu
	- \overline{\nabla}^\sigma_{[X, Y]} \nu
\\
&=
\mleft(\mathrm{Ad}_{\sigma^{-1}} \circ \mleft( \nabla^\caG_X \circ \nabla^\caG_Y 
	- \nabla^\caG_Y \circ \nabla^\caG_X
	- \nabla^\caG_{[X, Y]} \mright) \circ \mathrm{Ad}_\sigma\mright) (\nu)
\\
&=
\mleft( \mathrm{Ad}_{\sigma^{-1}} \circ
	R_{\nabla^{\mathcal{G}}}(X, Y) \circ \mathrm{Ad}_{\sigma}
\mright) (\nu)
\\
&=
\mathrm{Ad}_{\sigma^{-1}}\mleft( \mleft[ \zeta(X, Y), \mathrm{Ad}_{\sigma}(\nu) \mright]_{\mathcal{g}} \mright)
\\
&=
\mleft[ \mathrm{Ad}_{\sigma^{-1}}\bigl(\zeta(X, Y)\bigr), \nu \mright]_{\mathcal{g}}
\end{align*}
for all $X, Y \in \mathfrak{X}(M)$ and $\nu \in \Gamma(\mathcal{g})$. The remaining parts of the statement are trivial and straight-forward calculations, making use of \ref{def:YangMillsConnection}; in fact, we have proven all the remaining statements already in \cite[\S 3, part of Thm.\ 3.6]{My1stpaper} and in a more general manner in \cite[\S 4.6, Thm.\ 4.6.9]{MyThesis}: That $\overline{\nabla}^\sigma$ is a Lie bracket derivation is trivial to show with \ref{SpecialFieldRedef}, and one uses the right hand side of \ref{SpecialFieldRedef} to show another identity of its curvature,
\begin{equation*}
R_{\overline{\nabla}^\sigma} 
=
\mleft[ \mathrm{d}^{\nabla^\caG} \Delta \sigma + \frac{1}{2} \mleft[ \Delta \sigma \stackrel{\wedge}{,} \Delta\sigma \mright]_{\mathcal{g}} + \zeta, \cdot \mright]_{\mathcal{g}}.
\end{equation*}
Combining both identities for $R_{\overline{\nabla}^\sigma}$ we conclude with a proof of \ref{AlmostGeneralizedMaurerCartanEquation}. 
\end{proof}

We also get a generalization of statements like \cite[\S 5.5, Lemma 5.5.5, page 276]{Hamilton}:

\begin{lemmata}{Lie bracket of horizontal and vertical vector}{BracketVertHor}
Let $\mathcal{G} \to M$ be an LGB over a smooth manifold $M$ and $\pi\colon\mathcal{P} \to M$ a principal $\mathcal{G}$-bundle, also let $\mathrm{H}\mathcal{G}$ be a horizontal distribution on $\mathcal{G}$ and $A \in \Omega^1(\mathcal{P}; \pi^*\mathcal{g})$ be a connection 1-form on $\mathcal{P}$. Then we have
%
\begin{equation*}
A\bigl( \mleft[ X, \overline{\nu} \mright] \bigr)
=
\mleft( \pi^*\nabla^{\mathcal{G}} \mright)_{X}(\pi^*\nu)
\end{equation*}
for all $\nu \in \Gamma(\mathcal{g})$ and horizontal vector fields $X \in \mathfrak{X}(\mathcal{P})$.
\end{lemmata}

\begin{remarks}{Other references}{OtherRefsForAOnBracket}
For $\caP = \caG$ , this statement got also stated in \cite[Eq.\ (2.8)]{FernandesMarcutMultiplicativeForms}; we were unaware of this proof, which is why our proof is a bit different.
\end{remarks}

\begin{proof}[Proof of \ref{lem:BracketVertHor}]
\leavevmode\newline
First of all recall the very basic fact (see \textit{e.g.}\ \cite[\S A.1, Thm.\ A.1.46, page 615]{Hamilton}) that for two vector fields $X, Y \in \mathfrak{X}(\mathcal{P})$ we have\footnote{Rigorously, similarly to \ref{prop:FinallyTheNablaInduction} one has a projection involved, but as explained in \ref{rem:RemarkAboutPrTwoInDarbouxDerivative} we will omit notating this projection; in the context of this proof we can allow ourselves to take the typical ``less rigorous'' approach for simplicity.}
\begin{equation*}
\mleft[ X, Y \mright]_p
=
-\mleft.\frac{\mathrm{d}}{\mathrm{d}t}\mright|_{t=0}
	\mathrm{D}_{\phi_t(p)}\phi_{-t} \mleft(X_{\phi_t(p)}\mright)
\end{equation*}
for all $p \in \mathcal{P}$, where $\phi_t$ is the local flow of $Y$ (with parameter $t$ in an open interval of $\mathbb{R}$ containing 0). Now let $X$ be a horizontal vector field and $Y = \overline{\nu}$. For such a $Y$ we know by \ref{def:FundVecs} that
\begin{equation*}
\overline{\nu}_p
=
\mathrm{D}_{e_{x}}\Phi_p\mleft(\nu_{x}\mright)
=
\mleft. \frac{\mathrm{d}}{\mathrm{d}t} \mright|_{t=0}
\mleft( p \cdot \e^{t\nu_x} \mright)
\end{equation*}
for all $p \in \mathcal{P}$, where $\Phi_p$ is the orbit map through $p$, $x \coloneqq \pi(p)$, and $e_{x}$ the neutral element of $\mathcal{G}_x$. Thus, 
\begin{equation*}
\phi_t(p)
=
p \cdot \e^{t\nu_x}
=
r_{\e^{t\nu}}(p),
\end{equation*}
and by rewriting in sense of \ref{def:WeModTheRIghtPush} we achieve
\begin{align*}
\mleft[ X, \overline{\nu} \mright]_p
&=
-\mleft.\frac{\mathrm{d}}{\mathrm{d}t}\mright|_{t=0}
	\mathrm{D}_{p \cdot \e^{t\nu_x}}r_{\e^{-t\nu}} \mleft(X_{p \cdot \e^{t\nu_x}}\mright)
\\
&=
- \mleft.\frac{\mathrm{d}}{\mathrm{d}t}\mright|_{t=0} \mleft(
	\mathrm{D}_{p \cdot \e^{t\nu_x}}r_{\e^{-t\nu}} \mleft(X_{p \cdot \e^{t\nu_x}}\mright)
	- \mleft.{\overline{
		\mleft. \mleft( \pi^!\Delta \e^{-t\nu} \mright) \mright|_{p \cdot \e^{t\nu_x}}\mleft(X_{p \cdot \e^{t\nu_x}}\mright)
	}}\mright|_{p}
\mright)
\\
&\hspace{1cm}
- \mleft.\frac{\mathrm{d}}{\mathrm{d}t}\mright|_{t=0} \mleft(
	\mleft.{\overline{
		\mleft. \mleft( \pi^!\Delta \e^{-t\nu} \mright) \mright|_{p \cdot \e^{t\nu_x}}\mleft(X_{p \cdot \e^{t\nu_x}}\mright)
	}}\mright|_{p}
\mright)
\\
&=
- \mleft.\frac{\mathrm{d}}{\mathrm{d}t}\mright|_{t=0} \Bigl(
	\mathcal{r}_{\e^{-t\nu_x}*}\mleft(X_{p \cdot \e^{t\nu_x}}\mright)
\Bigr)
\\
&\hspace{1cm}
- \mleft.\frac{\mathrm{d}}{\mathrm{d}t}\mright|_{t=0} \mleft(
	\mleft.{\overline{
		\mleft. \mleft( \pi^!\Delta \e^{-t\nu} \mright) \mright|_{p \cdot \e^{t\nu_x}}\mleft(X_{p \cdot \e^{t\nu_x}}\mright)
	}}\mright|_{p}
\mright).
\end{align*}
The first summand, $t \mapsto \mathcal{r}_{\e^{-t\nu_x}*}\mleft(X_{p \cdot \e^{t\nu_x}}\mright)$, is a curve with values in $\mathrm{H}_{p}\mathcal{P}$ due to the fact that $X$ is a horizontal vector field. Thus, the velocity of this curve at $t=0$ will also be an element of $\mathrm{H}_p\mathcal{P}$, and the first summand is therefore in the kernel of $A$. Similarly,
\begin{equation*}
t \mapsto \eta(t) \coloneqq
\mleft. \mleft( \pi^!\Delta \e^{-t\nu} \mright) \mright|_{p \cdot \e^{t\nu_x}}\mleft(X_{p \cdot \e^{t\nu_x}}\mright)
=
\mleft. \mleft( \Delta \e^{-t\nu} \mright) \mright|_{x}\mleft(\mathrm{D}_{p \cdot \e^{t\nu_x}}\pi \mleft(X_{p \cdot \e^{t\nu_x}}\mright)\mright)
\end{equation*}
is a curve with values in $\mathcal{g}_x$, and therefore its fundamental vector field at $p$ is a curve with values in $\mathrm{V}_p\mathcal{P}$. Hence, we can calculate in the sense of vector spaces to derive
\begin{equation*}
- \mleft.\frac{\mathrm{d}}{\mathrm{d}t}\mright|_{t=0} \overline{\eta(t)}_p
=
\mleft.\frac{\mathrm{d}}{\mathrm{d}t}\mright|_{t=0} \Bigl( \mathrm{D}_{e_x}\Phi_p \bigl(\eta(-t)\bigr) \Bigr)
=
\mathrm{D}_{e_x}\Phi_p \mleft( \mleft.\frac{\mathrm{d}}{\mathrm{d}t}\mright|_{t=0} \eta(-t)\mright)
=
\mleft.\overline{\mleft.\frac{\mathrm{d}}{\mathrm{d}t}\mright|_{t=0} \eta(-t)}\mright|_p
\end{equation*}
where we substituted $t \leftrightarrow -t$ after the first equality sign. In total we achieve
\begin{equation*}
A_p\mleft( \mleft[ X, \overline{\nu} \mright]_p \mright)
=
A_p\mleft( \mleft.\overline{\mleft.\frac{\mathrm{d}}{\mathrm{d}t}\mright|_{t=0} \eta(-t)}\mright|_p \mright)
=
\mleft( 
	p,
	\mleft.\frac{\mathrm{d}}{\mathrm{d}t}\mright|_{t=0} \Bigl(
		\mleft. \mleft( \Delta \e^{t\nu} \mright) \mright|_{x}\mleft(\mathrm{D}_{p \cdot \e^{-t\nu_x}}\pi \mleft(X_{p \cdot \e^{-t\nu_x}}\mright)\mright)
	\Bigr)
\mright)~,
\end{equation*}
%
By \ref{cor:ExOfEhresmann}, $\rmH \caG$ is multiplicative, thus recall \ref{rem:baseManifoldHoriz}; that is, for the neutral element section $e$,
\begin{equation*}
\Delta e
\equiv
0~,
\end{equation*}
and, again, $t \mapsto \mleft( \Delta \e^{t\nu} \mright)_x$ and $t \mapsto \mathrm{D}_{p \cdot \e^{-t\nu_x}}\pi \mleft(X_{p \cdot \e^{-t\nu_x}}\mright)$ are curves with values in the vector spaces $\mathrm{T}^*_xM \otimes \mathcal{g}_x$ and $\mathrm{T}_xM$, respectively. Thus, we can just apply typical Leibniz rules to calculate the derivative of the contraction, also using \ref{def:ConnectionOnLAB},
\begin{align*}
&\mleft.\frac{\mathrm{d}}{\mathrm{d}t}\mright|_{t=0} \Bigl(
		\mleft. \mleft( \Delta \e^{t\nu} \mright) \mright|_{x}\mleft(\mathrm{D}_{p \cdot \e^{-t\nu_x}}\pi \mleft(X_{p \cdot \e^{-t\nu_x}}\mright)\mright)
	\Bigr)
\\
&\hspace{1cm}=
\mleft. \mleft( \mleft.\frac{\mathrm{d}}{\mathrm{d}t}\mright|_{t=0} \Delta \e^{t\nu} \mright) \mright|_{x}\bigl(\mathrm{D}_{p}\pi \mleft(X_{p}\mright)\bigr)
	+ {\underbrace{\mleft. \mleft( \Delta e \mright) \mright|_{x}}_{= 0}}\mleft(\mleft.\frac{\mathrm{d}}{\mathrm{d}t}\mright|_{t=0}\mathrm{D}_{p \cdot \e^{-t\nu_x}}\pi \mleft(X_{p \cdot \e^{-t\nu_x}}\mright)\mright)
\\
&\hspace{1cm}=
\nabla^{\mathcal{G}}_{\mathrm{D}_{p}\pi \mleft(X_{p}\mright)} \nu~,
\end{align*}
and by the definition of pullback connections we conclude
\begin{equation*}
A\bigl( \mleft[ X, \overline{\nu} \mright] \bigr)
=
\mleft( \pi^*\nabla^{\mathcal{G}} \mright)_{X}(\pi^*\nu)~.
\end{equation*}
\end{proof}

This concludes our discussion about the infinitesimal compatibility condition \eqref{CondSGleichNullLAB} and its integrated version. Via \ref{prop:TotMaurerIsMultiplicativeIfSEqual0} and \ref{lem:YangMillsConnAnEhresmann} we therefore concluded that compatibility condition \eqref{CondSGleichNullLAB} holds if the connection 1-form on $\cag$ is multiplicative. Let us therefore now turn to compatibility condition \eqref{CondKruemmungmitBLAB} and its integration.

\begin{theorems}{Generalized Maurer-Cartan equation}{GenMCEq}
Let $\pi_{\mathcal{G}} \colon \mathcal{G} \to M$ be an LGB over a smooth manifold $M$ with connected fibres, $\mathrm{H}\mathcal{G}$ a multiplicative Ehresmann connection on $\mathcal{G}$, and $\zeta \in \Omega^2\mleft( M; \mathcal{g} \mright)$. Then $\mathrm{H}\mathcal{G}$ satisfies the infinitesimal compatibility condition \eqref{CondKruemmungmitBLAB} w.r.t.\ $\zeta$ if and only if $\mutot$ satisfies
\begin{equation}\label{THEGeneralizedMCEq}
\mleft.\mleft(\mathrm{d}^{\pi_{\mathcal{G}}^*\nabla^{\mathcal{G}}} \mutot
	+ \frac{1}{2} \mleft[ \mutot \stackrel{\wedge}{,} \mutot \mright]_{\pi_{\mathcal{G}}^*\mathcal{g}}
	+ \pi_{\mathcal{G}}^! \zeta\mright)\mright|_g
=
\mleft( g, \mathrm{Ad}_{g^{-1}} \circ \zeta_x \circ \mleft(\mathrm{D}_g \pi_{\mathcal{G}}, \mathrm{D}_g \pi_{\mathcal{G}}\mright) \mright)
\end{equation}
for all $g \in \mathcal{G}_x$ ($x \in M$). We then say that $\mutot$ solves the \textbf{generalized Maurer-Cartan equation (w.r.t.\ $\zeta$)}.
\end{theorems}

\begin{remarks}{Dropping the base points, again...}{BasePointDifficultiesinGenMCEq}
The structure on the right hand side is again due to denoting the base point in pullback bundles; one may decide to drop the base point for simplicity in the notation. In that case, since the base point is clear by context, one may also just write
\begin{equation*}
\sAd_{g^{-1}} \circ \mleft.\pi_{\mathcal{G}}^!\zeta\mright|_g
~\text{ or }~
\mathrm{Ad}_{g^{-1}} \circ \mleft.\pi_{\mathcal{G}}^!\zeta\mright|_g
\end{equation*}
on the right hand side. As we will discuss in \ref{rem:SimplicialDifferentialStuff}, this equation says that the curvature of $\mutot$ is multiplicative-exact.
%
\end{remarks}

\begin{remarks}{Classical Maurer-Cartan equation recovered}{ClassicalMCEqinGeneralizedOne}
Recall \ref{rem:DarbouxOnCanonFlat}. If $\mathcal{G}$ is a trivial LGB, $\mathrm{H}\mathcal{G}$ its canonical flat connection and $\zeta \equiv 0$, then $\nabla^{\mathcal{G}}$ is the canonical flat connection on the trivial LAB $\mathcal{g}$ (recall \ref{ex:CanonicalFlatGConnection}) and it clearly satisfies the compatibility conditions in \ref{def:YangMillsConnection} w.r.t.\ $\zeta \equiv 0$. Furthermore, the generalized Maurer-Cartan equation then reduces to the classical Maurer-Cartan equation.
\end{remarks}

\begin{proof}[Proof of \ref{thm:GenMCEq}]
\leavevmode\newline
The idea of the proof is similar to the proof of a statement about when two multiplicative 2-forms\footnote{See \ref{rem:SimplicialDifferentialStuff} later; multiplicativity is related to closedness w.r.t.\ a differential. However, we will not need this general notion.} are equal, see \textit{e.g.}\ \cite[\S 3, Cor.\ 3.4]{bursztyn2004integration}. That is, we want to discuss under what conditions 
\begin{equation*}
\mathcal{G} \ni g \mapsto \omega_g \coloneqq
\mleft.\mleft(\mathrm{d}^{\pi_{\mathcal{G}}^*\nabla^{\mathcal{G}}} \mutot
	+ \frac{1}{2} \mleft[ \mutot \stackrel{\wedge}{,} \mutot \mright]_{\pi_{\mathcal{G}}^*\mathcal{g}}
	+ \pi_{\mathcal{G}}^! \zeta\mright)\mright|_g
	- \mleft( g, \mathrm{Ad}_{g^{-1}} \circ \zeta_x \circ \mleft(\mathrm{D}_g \pi_{\mathcal{G}}, \mathrm{D}_g \pi_{\mathcal{G}}\mright) \mright)
\end{equation*}
is constant along the fibres of $\mathcal{G}$, that is, constant along the flows of vertical tangent vectors, the left-invariant vector fields (recall \ref{cor:TLGBAsLGB}). 

$\bullet$ First of all observe that $\omega$ vanishes if it is contracted with at least one tangent vector is vertical. Let $Y \in \mathrm{V}_g\mathcal{G}$ ($g \in \mathcal{G}_x$, $x \in M$), and write $Y = \mleft.\overline{\nu_x}\mright|_g$ for a $\nu_x \in \mathcal{g}_x$. Due to $\mathrm{D}\pi_{\mathcal{G}} \mleft( Y \mright) = 0$ it is clear that the contraction of $\pi_{\mathcal{G}}^! \zeta$ with $Y$ is 0. Let $Z \in \mathrm{T}_g\mathcal{G}$ for which we write
\begin{equation*}
Z 
\coloneqq
\mleft.\overline{\mu_x}\mright|_g
	+ X_g
\end{equation*}
for a $\mu_x \in \mathcal{g}_x$ and $X_g \in \mathrm{H}_g\mathcal{G}$;
then by \ref{cor:TotMCFormIsConnectionForm}
\begin{equation*}
\frac{1}{2} \mleft.\mleft[ \mutot \stackrel{\wedge}{,} \mutot \mright]_{\pi_{\mathcal{G}}^*\mathcal{g}}\mright|_g (Y, Z)
=
\mleft[ \mleft(\mutot\mright)_g(Y), \mleft(\mutot\mright)_g(Z) \mright]_{\pi_{\mathcal{G}}^*\mathcal{g}}
=
\mleft( g,
\mleft[ \nu_x, \mu_x \mright]_{\mathcal{g}_x}
\mright).
\end{equation*}
Extend $Y$ and $Z$ naturally to vector fields on $\mathcal{G}$ (for simplicity also denoted by $Y$ and $Z$, respectively) by viewing $\nu_x$ and $\mu_x$ as values of sections $\nu$ and $\mu$, respectively, of $\mathcal{g}$ at $x$, and extending $X_g$ (locally) to a horizontal vector field $X$; thence,
\begin{align*}
\mleft(\mathrm{d}^{\pi_{\mathcal{G}}^*\nabla^{\mathcal{G}}} \mutot\mright)_g(Y, Z)
&=
\mleft( \pi_{\mathcal{G}}^*\nabla^{\mathcal{G}} \mright)_{Y_g} \mleft( \mutot(Z) \mright)
	- \mleft( \pi_{\mathcal{G}}^*\nabla^{\mathcal{G}} \mright)_{Z_g} \mleft( \mutot(Y) \mright)
	- \mleft(\mutot\mright)_g\mleft( [Y, Z] \mright)
\\
&=
{\underbrace{\mleft( \pi_{\mathcal{G}}^*\nabla^{\mathcal{G}} \mright)_{\overline{\nu}_g} \mleft( \pi^*_{\mathcal{G}}\mu \mright)}
	_{\mathclap{ = \mleft. \pi^*_{\mathcal{G}}\mleft( \nabla^{\mathcal{G}}_{\mathrm{D} \pi_{\mathcal{G}} \mleft( \overline{\nu} \mright)} \mu \mright) \mright|_g = 0 }}}
	- \mleft( \pi_{\mathcal{G}}^*\nabla^{\mathcal{G}} \mright)_{\overline{\mu}_g + X_g} \mleft( \pi^*_{\mathcal{G}}\nu \mright)
\\
&\hspace{1cm}
	- \mleft(\mutot\mright)_g\mleft( [\overline{\nu}, \overline{\mu}] \mright)
	- \mleft(\mutot\mright)_g\mleft( [\overline{\nu}, X] \mright)
\\
&=
- \mleft( \pi_{\mathcal{G}}^*\nabla^{\mathcal{G}} \mright)_{X_g} \mleft( \pi^*_{\mathcal{G}}\nu \mright)
	- \mleft( g, \mleft[ \nu_x, \mu_x \mright]_{\mathcal{g}_x} \mright)
	+ \mleft( \pi^*_{\mathcal{G}}\nabla^{\mathcal{G}} \mright)_{X_g}\mleft( \pi^*_{\mathcal{G}}\nu \mright)
\\
&=
- \mleft( g, \mleft[ \nu_x, \mu_x \mright]_{\mathcal{g}_x} \mright)
\end{align*}
using \ref{rem:FundVecsAreLABActions} (bracket of fundamental/left-invariant vector fields is fundamental/left-invariant) and \ref{lem:BracketVertHor}. Thus,
\begin{equation*}
\mleft.\mleft(\mathrm{d}^{\pi_{\mathcal{G}}^*\nabla^{\mathcal{G}}} \mutot
	+ \frac{1}{2} \mleft[ \mutot \stackrel{\wedge}{,} \mutot \mright]_{\pi_{\mathcal{G}}^*\mathcal{g}}\mright)\mright|_g(Y, Z)
=
0.
\end{equation*}
It follows that $\omega$ vanishes if contracted with a vertical tangent vector.

$\bullet$ Therefore we can study $\omega$ just w.r.t.\ tangent vectors complementary to the vertical structure, these do not necessarily need to be elements of $\mathrm{H}\mathcal{G}$. For a given $g \in \mathcal{G}_x$ ($x \in M$) define
\begin{align*}
\mathbb{R} &\to \Gamma(\mathcal{G}),\\
t &\mapsto \gamma(t) \coloneqq \sigma \e^{t\nu},
\end{align*}
where $\nu$ is any section of $\mathcal{g}$ and $\sigma \in \Gamma(\mathcal{G})$ (local) with $\sigma_x = g$; pointwise we denote $\gamma(t)$ as a map $M \ni x \mapsto \gamma_x(t)$. By \ref{cor:TLGBAsLGB} the flows of elements of $\mathrm{V}_g\mathcal{G}$ can be precisely described by curves like $\gamma$, so that the derivative w.r.t.\ $t$ of pullbacks of tensors with $\gamma$ characterizes any derivative along a fibre of $\mathcal{G}$. Furthermore, $\gamma(t)$ is by definition a section of $\mathcal{G}$ for all $t$ and thence describes a (local) embedding of $M$ into $\mathcal{G}$. Thus, the $t$-derivatives of terms like $\gamma(t)^!\omega$ describe the derivatives of $\omega$ along fibres of $\mathcal{G}$ contracted with tangent vectors complementary to the vertical structure. Therefore let us calculate $\gamma(t)^!\omega$: We have
\begin{equation*}
\gamma(t)^!\pi_{\mathcal{G}}^!\zeta
=
{\underbrace{\bigl( \pi_{\mathcal{G}} \circ \gamma(t) \bigr)^!}_{\equiv \mathrm{id}_M^!}}\zeta
=
\zeta
\end{equation*}
and under the identification $\gamma^*\pi_{\mathcal{G}}^*\mathcal{g} \cong (\pi_{\mathcal{G}} \circ \gamma)^*\mathcal{g} = \mathrm{id}_M^*\mathcal{g} \cong \mathcal{g}$ we get similarly
\begin{equation*}
\mleft( \gamma(t)^!\mleft( g, \mathrm{Ad}_{g^{-1}}\circ \zeta_{\pi_{\mathcal{G}}(g)} \circ \mathrm{D}_g\pi_{\mathcal{G}} \mright) \mright)_x
\cong
\mathrm{Ad}_{\gamma^{-1}_x(t)} \circ \zeta_{\pi_{\mathcal{G}}\mleft(\gamma_x(t)\mright)} \circ \mathrm{D}_x\bigl( \pi_{\mathcal{G}} \circ \gamma(t) \bigr)
=
\mathrm{Ad}_{\gamma^{-1}_x(t)} \circ \zeta_{x}
\end{equation*}
for all $x \in M$, where $\gamma^{-1}(t)$ is the $\mathcal{G}$-inverse of $\gamma(t)$. It is also a straight-forward calculation to show that 
\begin{equation*}
\gamma(t)^!\mathrm{d}^{\pi_{\mathcal{G}}^*\nabla^{\mathcal{G}}} \mutot
=
\mathrm{d}^{\gamma(t)^*\pi_{\mathcal{G}}^*\nabla^{\mathcal{G}}} \mleft( \gamma(t)^!\mutot\mright)
=
\mathrm{d}^{\mleft(\pi_{\mathcal{G}}\circ \gamma(t)\mright)^*\nabla^{\mathcal{G}}} \mleft( \gamma(t)^!\mutot\mright)
=
\mathrm{d}^{\nabla^{\mathcal{G}}} \mleft( \gamma(t)^!\mutot\mright),
\end{equation*}
using the definition of pullback vector bundle connections; alternatively see the explicit calulations in \cite[Appendix, Prop.\ A.1, Eq.\ (A.1)]{My1stpaper} or \cite[Appendix, Prop.\ A.1.1, Eq.\ (A.2)]{MyThesis}, and we have
\begin{align*}
&\mleft(\gamma(t)^!\mleft( \frac{1}{2} \mleft[ \mutot \stackrel{\wedge}{,} \mutot \mright]_{\pi_{\mathcal{G}}^*\mathcal{g}} \mright)\mright)_x(X,Y)
\\
&\hspace{1cm}=
\mleft[ \mleft(\mutot\mright)_{\gamma_x(t)}\mleft( \mathrm{D}_x\bigl(\gamma(t)\bigr)(X) \mright), 
\mleft(\mutot\mright)_{\gamma_x(t)}\mleft( \mathrm{D}_x\bigl(\gamma(t)\bigr)(Y) \mright) \mright]_{\mathcal{g}_{x}}
\\
&\hspace{1cm}=
\mleft[ \mleft(\gamma(t)^!\mutot\mright)_{x}(X), 
\mleft(\gamma(t)^!\mutot\mright)_{x}(Y) \mright]_{\mathcal{g}_{x}}
\\
&\hspace{1cm}=
\mleft( \frac{1}{2}
	\mleft[ \gamma(t)^!\mutot \stackrel{\wedge}{,}
	\gamma(t)^!\mutot \mright]_{\mathcal{g}}
\mright)_x(X,Y)
\end{align*}
for all $x \in M$ and $X, Y \in \mathrm{T}_xM$. Thus, we have so far
\begin{align}\label{PullbackofMCGeneralCurvPlusExtra}
\gamma(t)^!\omega
&=
\mathrm{d}^{\nabla^{\mathcal{G}}} \mleft( \gamma(t)^!\mutot\mright)
	+ \frac{1}{2}
	\mleft[ \gamma(t)^!\mutot \stackrel{\wedge}{,}
	\gamma(t)^!\mutot \mright]_{\mathcal{g}}
	+ \zeta
	- \mathrm{Ad}_{\gamma^{-1}(t)} \circ \zeta
\nonumber
\\
&=
\mathrm{d}^{\nabla^{\mathcal{G}}} \bigl( \Delta\gamma(t) \bigr)
	+ \frac{1}{2} \mleft[ \Delta\gamma(t) \stackrel{\wedge}{,} \Delta\gamma(t) \mright]_{\mathcal{g}}
	+ \zeta
	- \mathrm{Ad}_{\gamma^{-1}(t)} \circ \zeta.
\end{align}
By \ref{prop:DarbouxLeibnizRule} we get
\begin{equation*}
\Delta\gamma(t)
=
\mathrm{Ad}_{\e^{-t\nu}} \circ \Delta \sigma
	+ \Delta \e^{t\nu},
\end{equation*}
and via \ref{lem:FieldRedefViaLGBSection} we can derive
\begin{align*}
&\mleft(\mathrm{d}^{\nabla^{\mathcal{G}}} \mleft(\mathrm{Ad}_{\e^{-t\nu}} \circ \Delta \sigma \mright)\mright)(X, Y)
\\
&\hspace{1cm}=
\nabla^{\mathcal{G}}_X \bigl( \mathrm{Ad}_{\e^{-t\nu}} \mleft( \Delta \sigma(Y)\mright) \bigr)
	- \nabla^{\mathcal{G}}_Y \bigl( \mathrm{Ad}_{\e^{-t\nu}} \mleft( \Delta \sigma(X)\mright) \bigr)
	- \mathrm{Ad}_{\e^{-t\nu}} \bigl( \Delta \sigma([X, Y])\bigr)
\\
&\hspace{1cm}=
\mathrm{Ad}_{\e^{-t\nu}} \Bigl(
	\nabla^{\mathcal{G}}_X \bigl( \Delta \sigma(Y) \bigr)
	+ \mleft[\mleft(\Delta \e^{-t\nu}\mright)(X), \Delta \sigma(Y) \mright]_{\mathcal{g}}
\\
&\hspace{3.5cm}
	- \nabla^{\mathcal{G}}_Y \bigl( \Delta \sigma(X) \bigr)
	- \mleft[\mleft(\Delta \e^{-t\nu}\mright)(Y), \Delta \sigma(X) \mright]_{\mathcal{g}}
\\
&\hspace{3.5cm}
	-\Delta \sigma([X, Y])
\Bigr)
\\
&\hspace{1cm}=
\mathrm{Ad}_{\e^{-t\nu}} \mleft(
	\mleft(\mathrm{d}^{\nabla^{\mathcal{G}}} \Delta \sigma\mright)(X, Y)
	+ \mleft[\mleft(\Delta \e^{-t\nu}\mright)(X), \Delta \sigma(Y) \mright]_{\mathcal{g}}
	- \mleft[\mleft(\Delta \e^{-t\nu}\mright)(Y), \Delta \sigma(X) \mright]_{\mathcal{g}}
\mright)
\\
&\hspace{1cm}=
\mleft(\mathrm{Ad}_{\e^{-t\nu}} \circ \mleft(
	\mathrm{d}^{\nabla^{\mathcal{G}}} \Delta \sigma
	+ \mleft[\Delta \e^{-t\nu} \stackrel{\wedge}{,} \Delta \sigma \mright]_{\mathcal{g}}
\mright)\mright)(X, Y)
\end{align*}
for all $X, Y \in \mathfrak{X}(M)$, hence,
\begin{align*}
\mleft. \frac{\mathrm{d}}{\mathrm{d}t} \mright|_{t=0}\mleft(
	\mathrm{d}^{\nabla^{\mathcal{G}}} \Delta \gamma(t)
\mright)
&=
\mleft[ \mathrm{d}^{\nabla^{\mathcal{G}}} \Delta \sigma, \nu \mright]_{\mathcal{g}}
	- \mleft[\Delta \sigma \stackrel{\wedge}{,} \nabla^{\mathcal{G}}\nu \mright]_{\mathcal{g}}
	+ \mathrm{d}^{\nabla^{\mathcal{G}}} \nabla^{\mathcal{G}}\nu
\\
&=
\mleft[ \mathrm{d}^{\nabla^{\mathcal{G}}} \Delta \sigma, \nu \mright]_{\mathcal{g}}
	- \mleft[\Delta \sigma \stackrel{\wedge}{,} \nabla^{\mathcal{G}}\nu \mright]_{\mathcal{g}}
	+ R_{\nabla^{\mathcal{G}}} (\cdot, \cdot) \nu
\end{align*}
making use of that this calculation is pointwise at $x \in M$ just a standard derivative in the vector space $\mathcal{g}_x$ and that $\Delta e \equiv 0$ \ref{rem:baseManifoldHoriz}; we also used \ref{prop:GradedExtensionPlusAntiSymm}. In a similar fashion,
\begin{equation*}
\mleft.\frac{\mathrm{d}}{\mathrm{d}t}\mright|_{t=0}\mleft[ \Delta\gamma(t) \stackrel{\wedge}{,} \Delta\gamma(t) \mright]_{\mathcal{g}}
=
2\mleft[ \mleft[\Delta\sigma, \nu \mright]_{\mathcal{g}} \stackrel{\wedge}{,} \Delta\sigma \mright]_{\mathcal{g}}
	+ 2\mleft[ \Delta \sigma \stackrel{\wedge}{,} \nabla^{\mathcal{G}}\nu \mright]_{\mathcal{g}},
\end{equation*}
which we can rewrite by the Jacobi identity
\begin{align*}
\mleft[ \mleft[\Delta\sigma, \nu \mright]_{\mathcal{g}} \stackrel{\wedge}{,} \Delta\sigma \mright]_{\mathcal{g}}(X, Y)
&=
\mleft[ \mleft[\Delta\sigma(X), \nu \mright]_{\mathcal{g}} , \Delta\sigma(Y) \mright]_{\mathcal{g}}
	- \mleft[ \mleft[\Delta\sigma(Y), \nu \mright]_{\mathcal{g}} , \Delta\sigma(X) \mright]_{\mathcal{g}}
\\
&=
\mleft[ \mleft[\Delta\sigma(X), \Delta\sigma(Y) \mright]_{\mathcal{g}} , \nu \mright]_{\mathcal{g}}
\\
&=
\mleft[ \frac{1}{2} \mleft[\Delta\sigma \stackrel{\wedge}{,} \Delta\sigma \mright]_{\mathcal{g}} , \nu \mright]_{\mathcal{g}}(X, Y)
\end{align*}
for all $X, Y \in \mathfrak{X}(M)$, and we also trivially have
\begin{equation*}
\mleft.\frac{\mathrm{d}}{\mathrm{d}t}\mright|_{t=0} \mathrm{Ad}_{\gamma^{-1}(t)} \circ \zeta
=
\mleft.\frac{\mathrm{d}}{\mathrm{d}t}\mright|_{t=0} \mathrm{Ad}_{\e^{-t\nu}} \circ \mathrm{Ad}_{\sigma^{-1}} \circ \zeta
=
\mleft[ \mathrm{Ad}_{\sigma^{-1}} \circ \zeta, \nu \mright]_{\mathcal{g}}.
\end{equation*}
Collecting everything we derive
\begin{align}\label{PullbackofMCEqGivesCurv}
\mleft.\frac{\mathrm{d}}{\mathrm{d}t}\mright|_{t=0} \gamma(t)^!\omega
&=
\mleft[ \mathrm{d}^{\nabla^{\mathcal{G}}} \Delta \sigma, \nu \mright]_{\mathcal{g}}
	- \mleft[\Delta \sigma \stackrel{\wedge}{,} \nabla^{\mathcal{G}}\nu \mright]_{\mathcal{g}}
	+ R_{\nabla^{\mathcal{G}}} (\cdot, \cdot) \nu
\nonumber
\\
&\hspace{1cm}
	+ \mleft[ \frac{1}{2} \mleft[\Delta\sigma \stackrel{\wedge}{,} \Delta\sigma \mright]_{\mathcal{g}} , \nu \mright]_{\mathcal{g}}
	+ \mleft[ \Delta \sigma \stackrel{\wedge}{,} \nabla^{\mathcal{G}}\nu \mright]_{\mathcal{g}}
	- \mleft[ \mathrm{Ad}_{\sigma^{-1}} \circ \zeta, \nu \mright]_{\mathcal{g}}
\nonumber
\\
&=
R_{\nabla^{\mathcal{G}}} (\cdot, \cdot) \nu
	+ \mleft[ 
		\mathrm{d}^{\nabla^{\mathcal{G}}} \Delta \sigma
		+ \frac{1}{2} \mleft[\Delta\sigma \stackrel{\wedge}{,} \Delta\sigma \mright]_{\mathcal{g}}
		- \mathrm{Ad}_{\sigma^{-1}} \circ \zeta
		, \nu \mright]_{\mathcal{g}}
\end{align}
for all $\nu \in \Gamma(\mathcal{g})$ and $\sigma \in \Gamma(\mathcal{G})$.

$\bullet$ On the one hand, if $\mutot$ satisfies Eq.\ \eqref{THEGeneralizedMCEq}, then $\omega \equiv 0$. As previously mentioned, we have $\Delta e \equiv 0$, so that we then get in total for Eq.\ \eqref{PullbackofMCEqGivesCurv} w.r.t.\ $\sigma \coloneqq e$ that
\begin{equation*}
R_{\nabla^{\mathcal{G}}} (\cdot, \cdot) \nu
=
\mleft[ \zeta, \nu \mright]_{\mathcal{g}}
\end{equation*}
for all $\nu \in \Gamma(\mathcal{g})$, which is the infinitesimal compatibility condition \eqref{CondKruemmungmitBLAB}.

$\bullet$ On the other hand, if $\mathrm{H}\mathcal{G}$ satisfies the infinitesimal compatibility condition \eqref{CondKruemmungmitBLAB}, $R_{\nabla^{\mathcal{G}}} (\cdot, \cdot) \nu = \mleft[ \zeta, \cdot \mright]_{\mathcal{g}}$, then $\mathrm{H}\mathcal{G}$ is an infinitesimal multiplicative Yang-Mills connection by \ref{lem:YangMillsConnAnEhresmann}, and so we know by Eq.\ \eqref{AlmostGeneralizedMaurerCartanEquation} (\ref{lem:FieldRedefViaLGBSection}) that 
\begin{equation*}
\mleft[ \mathrm{d}^{\nabla^\caG} \Delta \sigma + \frac{1}{2} \mleft[ \Delta \sigma \stackrel{\wedge}{,} \Delta\sigma \mright]_{\mathcal{g}} - \mathrm{Ad}_{\sigma^{-1}} \circ \zeta, ~ \nu \mright]_{\mathcal{g}}
=
- \mleft[ \zeta, \nu \mright]_{\mathcal{g}},
\end{equation*}
so that Eq.\ \eqref{PullbackofMCEqGivesCurv} then has the form
\begin{equation*}
\mleft.\frac{\mathrm{d}}{\mathrm{d}t}\mright|_{t=0} \gamma(t)^!\omega
=
R_{\nabla^{\mathcal{G}}} (\cdot, \cdot) \nu
	- \mleft[ \zeta, \nu \mright]_{\mathcal{g}}
\stackrel{\eqref{CondKruemmungmitBLAB}}{=}
0
\end{equation*}
for all $\nu \in \Gamma(\mathcal{g})$ and $\sigma \in \Gamma(\mathcal{G})$.
As aforementioned in the discussion around the definition of $\gamma$, also recall that $\omega$ vanishes if contracted with a vertical vector, we can conclude that $\omega$ is constant along the fibres of $\mathcal{G}$ which are connected, so that this implies
\begin{equation*}
\omega
=
\pi_{\mathcal{G}}^!\xi
\end{equation*}
for a $\xi \in \Omega^2(M;\mathcal{g})$ defined by 
\begin{equation*}
e^!\omega 
=
\xi~.
\end{equation*}
But we already have calculated this in Eq.\ \eqref{PullbackofMCGeneralCurvPlusExtra}: Set $t=0$ and $\sigma \coloneqq e$, then also again by $\Delta e = 0$ we have
\begin{equation*}
\xi
=
e^!\omega
\stackrel{\eqref{PullbackofMCGeneralCurvPlusExtra}}{=}
\zeta
	- \zeta
=
0.
\end{equation*}
Finally we get
\begin{equation*}
\omega
=
\pi_{\mathcal{G}}^!\xi
=
0,
\end{equation*}
which finishes the proof.
\end{proof}

Trivially, this theorem is an extension of Eq.\ \eqref{AlmostGeneralizedMaurerCartanEquation} (\ref{lem:FieldRedefViaLGBSection}).

\begin{corollaries}{Pullback of generalized Maurer-Cartan equation}{PullbackOfMCSupperEquation}
Let $\pi_{\mathcal{G}} \colon \mathcal{G} \to M$ be an LGB over a smooth manifold $M$, $\mathrm{H}\mathcal{G}$ a multiplicative Ehresmann connection on $\mathcal{G}$ so that $\mutot$ satisfies the generalized Maurer-Cartan equation \eqref{THEGeneralizedMCEq} w.r.t.\ a $\zeta \in \Omega^2\mleft( M; \mathcal{g} \mright)$. Then
\begin{equation*}
\mathrm{d}^{\nabla^\caG} \Delta \sigma 
	+ \frac{1}{2} \mleft[ \Delta \sigma \stackrel{\wedge}{,} \Delta\sigma \mright]_{\mathcal{g}} 
	+ \zeta 
=
\mathrm{Ad}_{\sigma^{-1}} \circ \zeta
\end{equation*}
for all $\sigma \in \Gamma(\mathcal{G})$.
\end{corollaries}

\begin{proof}
\leavevmode\newline
Keeping the same notation as in the proof of \ref{thm:GenMCEq}, we have derived Eq.\ \eqref{PullbackofMCGeneralCurvPlusExtra}, and due to \ref{thm:GenMCEq} we get $\omega = 0$; altogether Eq.\ \eqref{PullbackofMCGeneralCurvPlusExtra} at $t=0$ gives then
\begin{equation*}
0
=
\mathrm{d}^{\nabla^\caG} \Delta\sigma
	+ \frac{1}{2} \mleft[ \Delta\sigma \stackrel{\wedge}{,} \Delta\sigma \mright]_{\mathcal{g}}
	+ \zeta
	- \mathrm{Ad}_{\sigma^{-1}} \circ \zeta~.
\end{equation*}
\end{proof}

Thus, we can finally define the connection we are going to need on the LGB.

\begin{definitions}{multiplicative Yang-Mills connection}{NowReallyYangMillsConnectio}
Let $\mathcal{G} \to M$ be an LGB over a smooth manifold $M$, and $\mathrm{H}\mathcal{G}$ be a horizontal distribution of $\mathcal{G}$. Then we say that $\mathrm{H}\mathcal{G}$ is a \textbf{multiplicative Yang-Mills (YM) connection (w.r.t.\ a $\zeta \in \Omega^2(M; \mathcal{g})$)}, if it satisfies the \textbf{compatibility conditions}:
\begin{enumerate}
	\item $\mathrm{H}\mathcal{G}$ is a multiplicative Ehresmann connection,
	\item The associated connection 1-form $\mutot$ satisfies the generalized Maurer-Cartan equation (w.r.t.\ a $\zeta \in \Omega^2(M; \mathcal{g})$).
\end{enumerate}
%
We then label $\mutot$ as a multiplicative YM connection, too.
\end{definitions}

\begin{remarks}{Integration of compatibilities}{YangMillsEqualsInfYM}
By \ref{lem:YangMillsConnAnEhresmann} and \ref{thm:GenMCEq}, every multiplicative Yang-Mills connection is also an infinitesimal multiplicative Yang-Mills connections, i.e.\ the infinitesimal compatibility conditions are naturally satisfied.
\end{remarks}

\begin{remarks}{Relation to simplicial differential}{SimplicialDifferentialStuff}
In fact, \ref{def:MultiplicativeFormsDef} comes from a simplicial differential (w.r.t.\ points on $\mathcal{G}$) defined in \cite[beginning of \S 1.2]{crainic2003differentiable}. We do not have time to introduce it completely, but we give a short sketch about its definitions, also following the style of \cite[appendix]{FernandesMarcutMultiplicativeForms}. On degree one the cited differential is a map
\begin{align*}
\Omega^\bullet\mleft(\mathcal{G}; \pi_{\mathcal{G}}^*\mathcal{g}\mright)
&\to
\Omega^\bullet\mleft(\mathcal{G} * \mathcal{G}; \pi_{\mathcal{G}}^*\mathcal{g}\mright),
\\
\omega
&\mapsto
\delta\omega,
\end{align*}
with
\begin{equation*}
\mleft(\delta\omega\mright)_{(g,q)}
\coloneqq
\sAd_{q^{-1}}\circ\mleft(\mathrm{pr}_1^!\omega\mright)_{(g,q)} 
	+ \mleft(\mathrm{pr}_2^!\omega\mright)_{(g, q)}
	- \mleft( \Phi^!\omega \mright)_{(g,q)}
\end{equation*}
for all $(g, q) \in \mathcal{G} * \mathcal{G}$, where $\Phi: \mathcal{G}*\mathcal{G}\to \mathcal{G}$ is the $\mathcal{G}$-multiplication, and $\mathrm{pr}_i\colon \mathcal{G}*\mathcal{G} \to \mathcal{G}$ are the canonical projections onto the $i$-th component ($i \in \{1,2\}$); $\pi_{\mathcal{G}}$ is canonically extended to $\mathcal{G} * \mathcal{G}$. Recall \ref{def:MultiplicativeFormsDef} and its \ref{rem:sloppynotationformultiplicativity}, $\omega$ can be defined to be multiplicative if $\delta\omega = 0$, i.e.\ $\omega$ is closed w.r.t.\ $\delta$. 

On degree 0 $\delta$ is a map $\Omega^\bullet(M; \mathcal{g}) \to \Omega^\bullet\mleft(\mathcal{G}; \pi^{*}_{\mathcal{g}}\mright)$ given for $\xi \in \Omega^\bullet(M; \mathcal{g})$ by
\begin{equation*}
\mleft(\delta\xi\mright)_g
\coloneqq
\bigl( g, \mathrm{Ad}_{g^{-1}} \circ \xi_x \circ {\underbrace{\mleft(\mathrm{D}_g \pi_{\mathcal{G}}, \dotsc, \mathrm{D}_g \pi_{\mathcal{G}}\mright)}_{ \text{times the degree of $\xi$} }} \bigr)
	- \mleft.\mleft(\pi_{\mathcal{G}}^! \xi\mright)\mright|_g
\end{equation*}
for all $g\in \mathcal{G}$. Then observe that we can rewrite the generalized Mauer-Cartan equation to
\begin{equation*}
F_{\mathcal{G}}
=
\delta \zeta,
\end{equation*}
where
\begin{equation*}
F_{\mathcal{G}}
=
\mathrm{d}^{\pi_{\mathcal{G}}^*\nabla^{\mathcal{G}}} \mutot
	+ \frac{1}{2} \mleft[ \mutot \stackrel{\wedge}{,} \mutot \mright]_{\pi_{\mathcal{G}}^*\mathcal{g}}
\in \Omega^2\mleft( \mathcal{G}; \pi_{\mathcal{G}}^*\mathcal{g} \mright)
\end{equation*}
is the ``ordinary'' curvature measuring the failure of $\rmH \caG$ being a regular foliation.
It is straight-forward to check that $\delta F_{\mathcal{G}} = 0$ (see e.g.\ \cite[\S 4.6, Thm.\ 4.27, Eq.\ (53)]{LAURENTGENGOUXStienonXuMultiplicativeForms} or \cite[\S 2.5, Prop.\ 2.22]{FernandesMarcutMultiplicativeForms}) by making use of that $\mathrm{H}\mathcal{G}$ is a multiplicative connection. In total we concluded that compatibility condition \eqref{CondSGleichNullLAB} is implied, if $\mutot$ is multiplicative, which also implies that $F_{\mathcal{G}}$ is multiplicative and thus $\delta$-closed; additionally with compatibility condition \eqref{CondKruemmungmitBLAB} we achieve $\delta$-exactness of $F_{\mathcal{G}}$. This is no wonder, the compatibility conditions already tell us by definition that the connection-1-form of $\nabla$ is closed w.r.t.\ the Chevalley-Eilenberg complex, which also implies the same for the curvature, however the curvature has to be exact by \eqref{CondKruemmungmitBLAB}. We depict that as $\delta^{\mathrm{CE}}\nabla^{\mathcal{G}} = 0$ and $R_{\nabla^{\mathcal{G}}} = \delta^{\mathrm{CE}}\zeta$; a summary is shown in \ref{tab:CompatibilityConditionsOnLGBAndItsLAB}.
\end{remarks}

\begin{table}[htbp]
	\centering
		\caption{Compatibility conditions on $\mathcal{g}$ and $\caG$} 
		\begin{tabular}{llll}
			& \multicolumn{1}{c}{$\mathcal{g}$} && \multicolumn{1}{c}{$\mathcal{G}$}
			\vspace{.2cm}\\
			\textbf{Closedness}
	& $\delta^{\mathrm{CE}}\nabla^{\mathcal{G}} = 0$ $\mleft(\Rightarrow \delta^{\mathrm{CE}}R_{\nabla^{\mathcal{G}}} = 0\mright)$
	&& $\delta\mutot = 0$ $\mleft(\Rightarrow \delta F_{\mathcal{G}} = 0\mright)$
	\\
	\textbf{Exactness}
		& $R_{\nabla^{\mathcal{G}}} = \delta^{\mathrm{CE}}\zeta$
		&& $F_{\mathcal{G}} = \delta \zeta$
		\\
		\end{tabular}
	\label{tab:CompatibilityConditionsOnLGBAndItsLAB}
\end{table}

To conclude this section, let us provide canonical examples of multiplicative Yang-Mills connections; we will make use of \ref{thm:GenMCEq} in order to prove that the generalized Maurer-Cartan equation is satisfied.

\begin{examples}{(Pre-)Classical multiplicative Yang-Mills connection}{ClassicalYangMillsConnection}
Let us look at a classical principal bundle $P$ with trivial LGB $\mathcal{G} = M \times G$, $G$ a Lie group; recall \ref{ex:TheCLassicalPrincAsEx} and \ref{ex:OurConnectionIsReallyMoreGeneral}. The canonical flat connection $\mathrm{H}\mathcal{G}$ on $\mathcal{G}$ induces the canonical flat connection $\nabla^{\mathcal{G}}$ on its LAB $\mathcal{g} = M \times \mathfrak{g}$, recall \ref{ex:CanonicalFlatGConnection}. Thus,
\begin{equation*}
R_{\nabla^{\mathcal{G}}} = 0
\end{equation*}
and so we can set $\zeta \equiv 0$. By \ref{ex:OurConnectionIsReallyMoreGeneral} (viewing $\mathcal{G}$ also as a principal bundle equipped with the same horizontal distribution) we know that $\mathrm{H}\mathcal{G}$ is a multiplicative connection, that is, the Maurer-Cartan form is (of course) multiplicative.

We therefore call flat $\nabla^{\mathcal{G}}$ a \textbf{classical multiplicative Yang-Mills connection}, regardless of whether or not $\mathcal{G}$ is trivial.

However, observe that $\zeta$ can be centre-valued and non-zero while $\nabla^{\mathcal{G}}$ is still flat. In such an occasion we speak of a \textbf{pre-classical multiplicative Yang-Mills connection}.
\end{examples}

\begin{examples}{Associated connections}{OurVeryImportantExample}
Recall \ref{ex:AssociatedLGBsAndTheirCanonicalConnection}, its setup and notation. We have discussed a canonical connection on associated LGBs $\mathcal{H} = P \times_\psi H$, called associated connection which is an Ehresmann connection. 

As it is well-known, the curvature measures the holonomy, that is, the parallel transport over closed (contractible) curves. Together with Eq.\ \eqref{CondKruemmungmitBLAB} and the Ambrose-Singer theorem, the parallel transport over a closed (contractible) curve $\alpha$ should act as conjugation with some group element. Let us see whether this holds here; there is a unique $g \in G$ for $p \in P$ such that
\begin{equation*}
\mathrm{PT}^{P}_\alpha(p)
=
p \cdot g,
\end{equation*}
and thus for $[p, h] \in \mathcal{H}$
\begin{equation*}
\mathrm{PT}_\alpha^\mathcal{H}\bigl( [p, h] \bigr)
=
\mleft[ p \cdot g, h \mright]
=
\mleft[ p, \psi_g(h) \mright].
\end{equation*}
Thus, if $\mathcal{H} = c_G(P)$, the inner group bundle of \ref{ex:InnerLGBs}, we get
\begin{equation*}
\mathrm{PT}_\alpha^\mathcal{H}\bigl( [p, h] \bigr)
=
\mleft[ p, ghg^{-1} \mright]
=
\mleft[p, g\mright] \cdot
\mleft[p, h\mright] \cdot
\mleft[p, g^{-1}\mright],
\end{equation*}
the conjugation of $[p, h]$ with $[p, g]$ in $c_G(P)$. It follows that there is a $\zeta \in \Omega^2(M; \mathcal{g})$ such that Eq.\ \eqref{CondKruemmungmitBLAB} is satisfied. This also implies that the associated connection of an inner group bundle is a multiplicative Yang-Mills connection. See the next approach for an explicit expression for $\zeta$.

Observe that $\alpha$ did not need to be contractible so that the inner group bundle may not describe the most general situation; indeed, one can extend this argument to $G \subset {\rm Aut}(H)$ containing $H$ as inner automorphisms and $\psi$ being the canonical action. See the final example section later.
\end{examples}

\begin{examples}{Alternative point of view of associated connections}{NablaYMAsAdjointConnection}
Alternatively one can prove that $\mutot$ solves the generalized Maurer-Cartan equation in \ref{ex:OurVeryImportantExample} by using the aforementioned relationship of the compatibility conditions with Mackenzie's study about extending Lie algebroids via LABs; recall \ref{rem:MackenziesStuffRelation}. The following is an information for the experienced reader: The LAB of $c_G(P)$ is given by the adjoint bundle $\mathrm{Ad}(P)$; recall \ref{ex:AssociatedLABsFromAssocLGBs}. $\mathrm{Ad}(P)$ is the kernel of the Atiyah sequence (see \textit{e.g.}\ \cite[\S 3.2, page 90ff.]{mackenzieGeneralTheory}), a short exact sequence of Lie algebroids over $M$
\begin{center}
	\begin{tikzcd}
		0 \arrow{r} & \mathrm{Ad}(P) \arrow{r} & \mathrm{At}(P) \arrow{r} & \mathrm{T}M \arrow{r} & 0,
	\end{tikzcd}
\end{center}
where $\mathrm{At}(P)$ is the Atiyah bundle of $P$. A connection on $P$ in the ``classical'' sense (that is, related to a canonical flat connection on $M \times G$ in our sense) has a 1:1 correspondence to a splitting $\chi\colon \mathrm{T}M \to \mathrm{AT}(P)$ of that sequence, \textit{i.e.}\ a section of the right non-trivial arrow. We can then construct a vector bundle connection $\nabla^\caG$ on the adjoint bundle by using the Lie algebroid bracket $\mleft[ \cdot, \cdot \mright]_{\mathrm{At}(P)}$ on $\mathrm{At}(P)$, that is,
\begin{align*}
\nabla^\caG_X \nu
&=
\mleft[ \chi(X), \nu \mright]_{\mathrm{At}(P)}
\end{align*}
for all $X \in \mathfrak{X}(M)$ and $\nu \in \Gamma(\mathrm{Ad}(P))$. It follows by straight-forward calculations as in \cite[\S 7.3, Prop.\ 7.3.2 and Lemma 7.3.3, page 278]{mackenzieGeneralTheory} that $\nabla^\caG$ is indeed a multiplicative Yang-Mills connection with $\zeta \in \Omega^2(M;\mathrm{Ad}(P))$ for example given by 
\begin{align*}
\zeta(Y, Z)
&\coloneqq
\mleft[ \chi(Y), \chi(Z) \mright]_{\mathrm{At}(P)} - \chi([Y, Z])
\end{align*}
for all $Y, Z \in \mathfrak{X}(M)$. In other words, $\nabla^\caG$ is the \textbf{adjoint connection} induced by a connection on $P$ (see for example \cite[\S 5.3, especially Prop.\ 5.3.13, page 199]{mackenzieGeneralTheory}). Thence, $\nabla^\caG$ corresponds to the natural parallel transport $\mathrm{PT}^{\mathrm{Ad}(P)}$ on $\mathrm{Ad}(P)$ (omitting the notation of the corresponding curve) given by
\begin{align*}
\mathrm{PT}^{\mathrm{Ad}(P)}\bigl( [p, v] \bigr)
&\coloneqq
\mleft[ \mathrm{PT}^{P}(p), v \mright]
\end{align*}
for all $[p, v] \in \mathrm{Ad}(P)$. This is clearly just the infinitesimal parallel transport inherited by $\mathrm{PT}^{c_G(P)}$, so that we can conclude that $\nabla^\caG$ corresponds to the multiplicative Yang-Mills connection on $c_G(P)$ in sense of \ref{def:ConnectionOnLAB} (see also \cite{LAURENTGENGOUXStienonXuMultiplicativeForms} for explicit proofs).

This proof gives a certain interpretation of $\nabla^\caG$ and $\zeta$: In the case of $\mathcal{G}$ being the inner group bundle of a typical gauge theory related to a principal $G$-bundle $P$, one can view $\nabla^\caG$ as the adjoint connection on the adjoint bundle, induced by $\chi$ a classical connection on $P$, this implies by Eq.\ \eqref{CondKruemmungmitBLAB} that $\zeta$ is the field strength related to $\chi$ (as a field strength with values in the adjoint bundle). Thus one may say in this case that the generalised sense of connection as in \ref{def:GaugeBosonsOnLGBPrincies} is modified by a classical connection from a classical principal bundle, thence also the label as multiplicative Yang-Mills connection to reflect its relation to the classical theory.
\end{examples}

\subsubsection{Field strength related to multiplicative Yang-Mills connections}

Let us now introduce and discuss the field strength.

\begin{definitions}{(Generalized) Field strength}{NewFieldStrength}
Let $\mathcal{G} \to M$ be an LGB over a smooth manifold $M$ and $\pi\colon\mathcal{P} \to M$ a principal $\mathcal{G}$-bundle, also let $\mathrm{H}\mathcal{G}$ be a multiplicative Yang-Mills connection on $\mathcal{G}$ (w.r.t.\ a $\zeta \in \Omega^2(M; \mathcal{g})$) and $A \in \Omega^1(\mathcal{P}; \pi^*\mathcal{g})$ be a connection 1-form on $\mathcal{P}$. Then we define the \textbf{(generalized) curvature} or \textbf{(generalized) field strength $F$ (of $A$)} as an element of $\Omega^2\mleft( \mathcal{P}; \pi^*\mathcal{g} \mright)$ by
\begin{equation*}
F
\coloneqq
\mathrm{d}^{\pi^*\nabla^\caG} A \circ \mleft( \pi^{\mathrm{H}\mathcal{P}}, \pi^{\mathrm{H}\mathcal{P}} \mright)
	+ \pi^!\zeta,
\end{equation*}
where $\pi^{\mathrm{H}\mathcal{P}}\colon \mathrm{T}\mathcal{P} \to \mathrm{H}\mathcal{P}$ is the canonical projection onto the associated Ehresmann connection $\mathrm{H}\mathcal{P}$ on $\mathcal{P}$; that is,
\begin{equation*}
F(X, Y)
=
\mathrm{d}^{\pi^*\nabla^\caG} A \mleft( \pi^{\mathrm{H}\mathcal{P}}(X), \pi^{\mathrm{H}\mathcal{P}}(Y) \mright)
	+ \mleft(\pi^*\zeta\mright) \bigl( \mathrm{D}\pi(X), \mathrm{D}\pi(Y) \bigr)
\end{equation*}
for all $X, Y \in \mathfrak{X}(\mathcal{P})$.
\end{definitions}

Of course, we expect certain properties, naturally generalized by our previous discussions:

\begin{propositions}{Properties of the generalized field strength}{NewFieldStrengthWithCoolProps}
Let $\mathcal{G} \to M$ be an LGB over a smooth manifold $M$ and $\pi\colon\mathcal{P} \to M$ a principal $\mathcal{G}$-bundle, also let $\mathrm{H}\mathcal{G}$ be a multiplicative Yang-Mills connection on $\mathcal{G}$ (w.r.t.\ a $\zeta \in \Omega^2(M; \mathcal{g})$) and $A \in \Omega^1(\mathcal{P}; \pi^*\mathcal{g})$ be a connection 1-form on $\mathcal{P}$. Then we have the following properties of the field strength:
\begin{itemize}
	\item \textbf{(Form of type $\sAd$)}
	\begin{equation*}
	\mathcal{r}_{\sigma}^!F
	=
	\sAd_{\sigma^{-1}} \circ F
	\end{equation*}
	for all (local) $\sigma \in \Gamma(\mathcal{G})$.
	\item \textbf{(Horizontal form)}
	\newline
	For $X, Y \in \mathrm{T}_p\mathcal{P}$ ($p \in \mathcal{P}$) we have
	\begin{equation*}
	F(X, Y)
	=
	0
	\end{equation*}
	if either of $X$ and $Y$ is vertical.
\end{itemize}
\end{propositions}

However, despite the easy proof in the classical theory, the proof of the first property in \ref{prop:NewFieldStrengthWithCoolProps} will be rather involved, but most of the proof is still straight-forward calculation, and then use our discussion about multiplicative Yang-Mills connections; the quintessence will be using \ref{lem:BracketVertHor} (for first order terms w.r.t.\ $\sigma$) and \ref{cor:PullbackOfMCSupperEquation} (for second order terms w.r.t.\ $\sigma$).

\begin{proof}[Proof of \ref{prop:NewFieldStrengthWithCoolProps}]
\leavevmode\newline
For this proof we will neglect caring for the base point information in the involved pullback bundles since the base point information is clear by context; recall \ref{BasicNotations}. In that sense we also have $\sAd_\sigma \cong \pi^*\mathrm{Ad}_\sigma \cong \mathrm{Ad}_\sigma$ for all $\sigma \in \Gamma(\mathcal{G})$ \textit{etc.}; this will simplify the following notation a bit.

$\bullet$ The second bullet point quickly follows by construction: The first summand of $F$ (recall \ref{def:NewFieldStrength}) is clearly zero if either of $X$ and $Y$ is vertical; similar holds for the second summand because the vertical bundle is the kernel of $\mathrm{D}\pi$.

$\bullet$ Thence, let us focus on proving the first bullet point, and denote with $\pi_h\colon \mathrm{T}\mathcal{P} \to \mathrm{H}\mathcal{P}$ the canonical projection onto the associated Ehresmann connection $\mathrm{H}\mathcal{P}$ on $\mathcal{P}$, so that we write
\begin{equation*}
F
=
\mathrm{d}^{\pi^*\nabla^\caG} A \circ \mleft( \pi_h, \pi_h \mright)
	+ \pi^!\zeta.
\end{equation*}
We have for $\sigma \in \Gamma(\mathcal{G})$
\begin{align*}
\mleft(\mathcal{r}_\sigma^!\pi^!\zeta\mright)_p(X, Y)
&=
\zeta_{x}\bigl(
	\mathrm{D}_{p \cdot \sigma_x}\pi\mleft( \mathcal{r}_{\sigma_x*}(X) \mright),
	\mathrm{D}_{p \cdot \sigma_x}\pi\mleft( \mathcal{r}_{\sigma_x*}(Y) \mright)
\bigr),
\\
&=
\zeta_{x}\bigl(
	{\underbrace{\mathrm{D}_{p \cdot \sigma_x}\pi\mleft( \mathrm{D}_pr_\sigma\mleft( X \mright) \mright)}_{\mathclap{ = \mathrm{D}_p\mleft( \pi \circ r_\sigma \mright) = \mathrm{D}_p\pi }}},
	\mathrm{D}_{p \cdot \sigma_x}\pi\mleft( \mathrm{D}_pr_\sigma\mleft( Y \mright) \mright)
\bigr)
\\
&=
\mleft(\pi^!\zeta\mright)_p(X, Y)
\end{align*}
for all $p \in \mathcal{P}_x$ ($x \in M$) and $X, Y \in \mathrm{T}_p\mathcal{P}$, making use of that fundamental vector fields are in the kernel of $\mathrm{D}\pi$ so that $\mathrm{D}\pi \circ \mathcal{r}_{\sigma_x*} = \mathrm{D}\pi \circ r_{\sigma_x*}$. For the first term in the field strength use \ref{cor:ModifiedRightPushyCommutesWithProj} and \ref{rem:RemOohThesePullbacksConfusOrNotToConfus} so that
\begin{align*}
\mathcal{r}_\sigma^!\mleft(\mathrm{d}^{\pi^*\nabla^\caG} A \circ \mleft( \pi_h, \pi_h \mright)\mright)
&=
\mleft(r_\sigma^*\mathrm{d}^{\pi^*\nabla^\caG} A\mright) \circ \mleft( r_\sigma^*\pi_h \circ \mathcal{r}_{\sigma*}, r_\sigma^*\pi_h \circ \mathcal{r}_{\sigma*} \mright)
\\
&=
\mleft(r_\sigma^*\mathrm{d}^{\pi^*\nabla^\caG} A\mright) \circ \mleft( \mathcal{r}_{\sigma*} \circ \pi_h, \mathcal{r}_{\sigma*} \circ \pi_h \mright)
\\
&=
\mathcal{r}_\sigma^!\mleft(\mathrm{d}^{\pi^*\nabla^\caG} A\mright) \circ (\pi_h, \pi_h).
\end{align*}
Let us denote $X^h \coloneqq \pi_h(X)$ and $Y^h \coloneqq \pi_h(Y)$, where $X$ and $Y$ are now elements of $\mathfrak{X}(\mathcal{P})$; since we are looking at tensorial equations we can w.l.o.g.\ assume that $X^h$ and $Y^h$ are the horizontal lifts of vector fields $\omega^X$ and $\omega^Y$, respectively, on $M$, especially $\mathrm{D}\pi\mleft( X^h \mright) = \mathrm{D}\pi\mleft( X \mright) = \pi^*\omega^X$ and $\mathrm{D}\pi\mleft( Y^h \mright) = \mathrm{D}\pi\mleft( Y \mright) = \pi^*\omega^Y$.
We also rewrite
\begin{equation*}
\mathcal{r}_{\sigma*}(X)
=
\mathrm{D}r_{\sigma}(X)
	- r_{\sigma}^*\mleft({\overline{ \mleft(\pi^! \Delta \sigma \mright) (X)}}\mright)
\in
\Gamma\mleft( r_\sigma^* \mathrm{T}\mathcal{P} \mright) \cong \mathfrak{X}(\mathcal{P}).
\end{equation*}
In the following we will make use of the canonical identification $\mathfrak{X}(\mathcal{P}) \cong \Gamma\mleft( r_\sigma^* \mathrm{T}\mathcal{P} \mright)$, $X \mapsto r_\sigma^*X$, especially in order to calculate Lie brackets.
%

Then make use of \ref{def:FinallyTheConnection} and \ref{thm:OurConnectionHasAUniqueoneForm} to show
\begin{align}\label{TransformedClassicStrengthTerm}
&
r_\sigma^*\mleft(
\mleft(\mathrm{d}^{\pi^*\nabla^\caG} A\mright) \mleft(\mathcal{r}_{\sigma*}\mleft(X^h\mright), \mathcal{r}_{\sigma*}\mleft(Y^h\mright)\mright)
\mright)
\nonumber
\\
&\hspace{1cm}=
r_\sigma^*\biggl(
\mleft( \pi^*\nabla^\caG \mright)_{\mathcal{r}_{\sigma*}\mleft(X^h\mright)} {\underbrace{\mleft( (A \circ \mathcal{r}_{\sigma*})\mleft(Y^h\mright) \mright)}_{= 0}}
\nonumber
\\
&\hspace{2.5cm}
	- \mleft( \pi^*\nabla^\caG \mright)_{\mathcal{r}_{\sigma*}\mleft(Y^h\mright)} \mleft( (A \circ \mathcal{r}_{\sigma*})\mleft(X^h\mright) \mright)
\nonumber
\\
&\hspace{2.5cm}
	- A\mleft( \mleft[ \mathcal{r}_{\sigma*}\mleft(X^h\mright), \mathcal{r}_{\sigma*}\mleft(Y^h\mright) \mright] \mright)
\biggr)
\nonumber
\\
&\hspace{1cm}=r_\sigma^*\biggl(
	- A\mleft( \mleft[ \mathcal{r}_{\sigma*}\mleft(X^h\mright), \mathcal{r}_{\sigma*}\mleft(Y^h\mright) \mright] \mright)
\biggr)
\nonumber
\\
&\hspace{1cm}=r_\sigma^*\Biggl(
- A\mleft(
	\mleft[\mathrm{D}r_\sigma\mleft( X^h \mright), \mathrm{D}r_\sigma\mleft(Y^h\mright)\mright]
\mright)
\nonumber
\\
&\hspace{2.5cm}
	+ A\mleft(
	\mleft[\mathrm{D}r_\sigma\mleft(X^h\mright), r_{\sigma}^*\mleft({\overline{ \mleft(\pi^! \Delta \sigma \mright) \mleft(Y^h\mright) }}\mright) \mright]
	\mright)
\nonumber
\\
&\hspace{2.5cm}
	+ A\mleft(
	\mleft[r_{\sigma}^*\mleft({\overline{ \mleft(\pi^! \Delta \sigma \mright) \mleft(X^h\mright) }}\mright), \mathrm{D}r_\sigma\mleft(Y^h\mright) \mright]
	\mright)
\nonumber
\\
&\hspace{2.5cm}
	- A\mleft(
	\mleft[r_{\sigma}^*\mleft({\overline{ \mleft(\pi^! \Delta \sigma \mright) \mleft(X^h \mright) }}\mright), r_{\sigma}^*\mleft({\overline{ \mleft(\pi^! \Delta \sigma \mright)\mleft(Y^h \mright) }}\mright) \mright]
	\mright)
\Biggr).
\end{align}
Let us look at each summand individually; for the first summand we use the very well-known fact (see \textit{e.g.}\ \cite[\S A.1.11, Cor.\ A.1.51, page 615]{Hamilton}) that
\begin{equation*}
\mleft[\mathrm{D}r_\sigma\mleft( X^h \mright), \mathrm{D}r_\sigma\mleft(Y^h\mright)\mright]
=
\mathrm{D}r_\sigma\mleft(\mleft[ X^h, Y^h\mright]\mright),
\end{equation*}
making use of that $r_\sigma$ is an isomorphism, and thus we get for the first summand by using \ref{def:GaugeBosonsOnLGBPrincies}
\begin{align*}
&r_\sigma^*\biggl(
- A\mleft(
	\mleft[\mathrm{D}r_\sigma\mleft( X^h \mright), \mathrm{D}r_\sigma\mleft(Y^h\mright)\mright]
\mright)
\biggr)
\\
&\hspace{1cm}=
-\mleft( r_\sigma^*A \mright)\mleft( \mathcal{r}_{\sigma*} \mleft( \mleft[ X^h, Y^h\mright] \mright)\mright)
	- \mleft( r_\sigma^* A \mright) \mleft( r_{\sigma}^*\mleft({\overline{ \mleft(\pi^! \Delta \sigma \mright) \mleft(\mleft[ X^h, Y^h\mright]\mright)}}\mright) \mright)
\\
&\hspace{1cm}=
-\mleft( \mathcal{r}_\sigma^!A \mright)\mleft( \mleft[ X^h, Y^h\mright] \mright)
	- r_{\sigma}^*\mleft(\mleft(\pi^! \Delta \sigma \mright) \mleft(\mleft[ X^h, Y^h\mright]\mright)\mright)
\\
&\hspace{1cm}=
-\mleft( \sAd_{\sigma^{-1}} \circ A \mright)\mleft( \mleft[ X^h, Y^h\mright] \mright)
	- r_{\sigma}^*\mleft(\mleft(\pi^! \Delta \sigma \mright) \mleft(\mleft[ X^h, Y^h\mright]\mright)\mright)
\\
&\hspace{1cm}=
\sAd_{\sigma^{-1}}\mleft(\mleft( \mathrm{d}^{\pi^*\nabla^\caG}A \mright)\mleft( X^h, Y^h \mright) \mright)
	- r_{\sigma}^*\mleft(\mleft(\pi^! \Delta \sigma \mright) \mleft(\mleft[ X^h, Y^h\mright]\mright)\mright),
\end{align*}
where we used a similar argument for the last equality as for the beginning of the calculation for Eq.\ \eqref{TransformedClassicStrengthTerm}.
We know by \ref{rem:FundVecsAreLABActions} that the map to fundamental vector fields is a homomorphism of Lie algebras, so that
\begin{equation*}
\mleft[r_{\sigma}^*\mleft({\overline{ \mleft(\pi^! \Delta \sigma \mright) \mleft(X^h \mright) }}\mright), r_{\sigma}^*\mleft({\overline{ \mleft(\pi^! \Delta \sigma \mright)\mleft(Y^h \mright) }}\mright) \mright]
=
r_\sigma^*\mleft(
{\overline{
	\mleft[ \mleft(\pi^! \Delta \sigma \mright) \mleft(X^h \mright), \mleft(\pi^! \Delta \sigma \mright)\mleft(Y^h \mright) \mright]_{\mathcal{\pi^*\mathcal{g}}}
}}
\mright),
\end{equation*}
and therefore we can derive for the fourth term in Eq.\ \eqref{TransformedClassicStrengthTerm} that
\begin{equation*}
\mleft( r_\sigma^*A \mright)\mleft(
	\mleft[r_{\sigma}^*\mleft({\overline{ \mleft(\pi^! \Delta \sigma \mright) \mleft(X^h \mright) }}\mright), r_{\sigma}^*\mleft({\overline{ \mleft(\pi^! \Delta \sigma \mright)\mleft(Y^h \mright) }}\mright) \mright]
	\mright)
=
r_\sigma^*\mleft(
	\mleft[ \mleft(\pi^! \Delta \sigma \mright) \mleft(X^h \mright), \mleft(\pi^! \Delta \sigma \mright)\mleft(Y^h \mright) \mright]_{\mathcal{\pi^*\mathcal{g}}}
\mright).
\end{equation*}
For the second and third summands in Eq.\ \eqref{TransformedClassicStrengthTerm} we want to use \ref{lem:BracketVertHor} and our result of the fourth summand to show
\begin{align*}
&\mleft( r_\sigma^*A \mright)\mleft(
	\mleft[\mathrm{D}r_\sigma\mleft(X^h\mright), r_{\sigma}^*\mleft({\overline{ \mleft(\pi^! \Delta \sigma \mright) \mleft(Y^h\mright) }}\mright) \mright]
	\mright)
\\ 
&\hspace{0.5cm}
=
\mleft( r_\sigma^*A \mright)\mleft(
	\mleft[\mathcal{r}_{\sigma*}\mleft(X^h\mright), r_{\sigma}^*\mleft({\overline{ \mleft(\pi^! \Delta \sigma \mright) \mleft(Y^h\mright) }}\mright) \mright]
	\mright)
\\&\hspace{1.5cm}
	+ \mleft( r_\sigma^*A \mright)\mleft(\mleft[
	r_{\sigma}^*\mleft({\overline{ \mleft(\pi^! \Delta \sigma \mright) \mleft(X^h\mright) }}\mright), r_{\sigma}^*\mleft({\overline{ \mleft(\pi^! \Delta \sigma \mright) \mleft(Y^h\mright) }}\mright) \mright]
	\mright)
\\ 
&\hspace{0.5cm}
=
r_\sigma^*\Biggl(
\mleft( \pi^*\nabla^\caG \mright)_{\mathcal{r}_{\sigma*}\mleft(X^h\mright)}\biggl( \mleft(\pi^! \Delta \sigma \mright) \mleft(Y^h\mright)\biggr)
\Biggr)
	+ r_\sigma^*\mleft(
	\mleft[ \mleft(\pi^! \Delta \sigma \mright) \mleft(X^h \mright), \mleft(\pi^! \Delta \sigma \mright)\mleft(Y^h \mright) \mright]_{\mathcal{\pi^*\mathcal{g}}}
\mright)
\end{align*}
using that $\mathcal{r}_{\sigma*}\mleft(X^h\mright)$ is horizontal due to the fact that $X^h$ is horizontal, and, last but not least, we can write for terms like
\begin{equation*}
\mleft(\pi^! \Delta \sigma \mright) \mleft(Y^h\mright)
=
\mleft(\pi^* \Delta \sigma \mright) \mleft(\mathrm{D}\pi\mleft(Y^h\mright)\mright)
=
\mleft(\pi^* \Delta \sigma \mright) \mleft(\pi^*\omega^Y\mright)
=
\pi^*\Bigl( \mleft(\Delta \sigma \mright) \mleft(\omega^Y\mright) \Bigr),
\end{equation*}
and
\begin{equation*}
\mathrm{D}\pi\mleft( \mathcal{r}_{\sigma*}\mleft(X^h\mright) \mright)
=
\mathrm{D}\pi\mleft( \mathrm{D}r_{\sigma}\mleft(X^h\mright)
	- r_{\sigma}^*\mleft({\overline{ \mleft(\pi^! \Delta \sigma \mright) \mleft(X^h\mright)}}\mright) \mright)
=
\mathrm{D}{\underbrace{\mleft( \pi \circ r_\sigma \mright)}_{= \pi}}\mleft(X^h\mright)
=
\pi^*\omega^X,
\end{equation*}
using that fundamental vector fields are in the kernel of $\mathrm{D}\pi$, 
and therefore
\begin{equation*}
\mleft( \pi^*\nabla^\caG \mright)_{\mathcal{r}_{\sigma*}\mleft(X^h\mright)}\biggl( \mleft(\pi^! \Delta \sigma \mright) \mleft(Y^h\mright)\biggr)
=
\pi^*\mleft(
	\nabla^\caG_{\omega^X}\Bigl( \mleft(\Delta \sigma \mright) \mleft(\omega^Y\mright)\Bigr)
\mright).
\end{equation*}

Collecting all these results and using \ref{cor:PullbackOfMCSupperEquation} we can write Eq.\ \eqref{TransformedClassicStrengthTerm} as
\begin{align*}
&
r_\sigma^*\mleft(
\mleft(\mathrm{d}^{\pi^*\nabla^\caG} A\mright) \mleft(\mathcal{r}_{\sigma*}\mleft(X^h\mright), \mathcal{r}_{\sigma*}\mleft(Y^h\mright)\mright)
\mright)
\\&\hspace{1cm}
=
\sAd_{\sigma^{-1}}\mleft(\mleft( \mathrm{d}^{\pi^*\nabla^\caG}A \mright)\mleft( X^h, Y^h \mright) \mright)
\\&\hspace{1.5cm}
	+ r_{\sigma}^*\Biggl(
		\pi^*\mleft(
	\nabla^\caG_{\omega^X}\Bigl( \mleft(\Delta \sigma \mright) \mleft(\omega^Y\mright)\Bigr)
\mright)
		- \pi^*\mleft(
	\nabla^\caG_{\omega^Y}\Bigl( \mleft(\Delta \sigma \mright) \mleft(\omega^X\mright)\Bigr)
\mright)
\\&\hspace{3cm}
		- {\underbrace{\mleft(\pi^! \Delta \sigma \mright) \mleft(\mleft[ X^h, Y^h\mright]\mright)}_{= \pi^* \mleft(\Delta \sigma \mleft(\mleft[ \omega^X, \omega^Y\mright]\mright)\mright)}}
		+ \mleft[ \pi^*\Bigl( \mleft(\Delta \sigma \mright) \mleft(\omega^X\mright) \Bigr), \pi^*\Bigl( \mleft(\Delta \sigma \mright) \mleft(\omega^Y\mright) \Bigr) \mright]_{\mathcal{\pi^*\mathcal{g}}}
\Biggr)
\\&\hspace{1cm}
=
\sAd_{\sigma^{-1}}\mleft(\mleft( \mathrm{d}^{\pi^*\nabla^\caG}A \mright)\mleft( X^h, Y^h \mright) \mright)
	+ {\underbrace{r_{\sigma}^* \pi^*}_{= \pi^*}}\mleft( 
		\mleft(
			\mathrm{d}^{\nabla^\caG} \Delta\sigma
			+ \frac{1}{2} \mleft[ \Delta\sigma \stackrel{\wedge}{,} \Delta\sigma \mright]_{\mathcal{g}}
		\mright)
	\mleft(\omega^X, \omega^Y\mright) \mright)
\\&\hspace{1cm}
=
\sAd_{\sigma^{-1}}\mleft(\mleft( \mathrm{d}^{\pi^*\nabla^\caG}A \mright)\mleft( X^h, Y^h \mright) \mright)
	+ \bigl( \pi^*
		\mleft( \mathrm{Ad}_{\sigma^{-1}} \circ \zeta 
		- \zeta \mright) \bigr)
	\bigl(\mathrm{D}\pi(X), \mathrm{D}\pi(Y)\bigr)
\\&\hspace{1cm}
=
\sAd_{\sigma^{-1}}\mleft(\mleft( \mathrm{d}^{\pi^*\nabla^\caG}A \mright)\mleft( X^h, Y^h \mright) \mright)
	+ \mleft(\mleft(\pi^!\mleft(  
		\mathrm{Ad}_{\sigma^{-1}} \circ \zeta 
		- \zeta 
	\mright)\mright)(X, Y) \mright)
\\&\hspace{1cm}
=
\mleft(
	\sAd_{\sigma^{-1}}\circ \mleft( \mathrm{d}^{\pi^*\nabla^\caG}A \circ ( \pi_h, \pi_h) + \pi^!\zeta \mright) 
		- \pi^!\zeta 
\mright)(X, Y)
\end{align*}
where we used that $\mathrm{D}\pi (X) = \pi^*\omega^X$ and $\mathrm{D}\pi (Y) = \pi^*\omega^Y$, which immediately implies
\begin{equation*}
\mathrm{D}\pi\mleft( [X, Y] \mright)
=
\pi^*\mleft( \mleft[\omega^X, \omega^Y\mright] \mright),
\end{equation*}
which is a well-known fact, as also given in \cite[Proposition A.1.49; page 615]{Hamilton}, but also straight-forward to check. 

Finally, we can conclude this proof by combining the previous result with our first calculation of this proof, that is,
\begin{equation*}
\mathcal{r}_{\sigma}^!F
=
\sAd_{\sigma^{-1}}\circ \mleft( \mathrm{d}^{\pi^*\nabla^\caG}A \circ ( \pi_h, \pi_h) + \pi^!\zeta \mright) 
		- \pi^!\zeta
		+ \pi^!\zeta
=
\sAd_{\sigma^{-1}} \circ F~.
\end{equation*}
\end{proof}

We achieve of course also a structure equation.

\begin{propositions}{Structure equation of the generalized field strength}{StructureEq}
Let $\mathcal{G} \to M$ be an LGB over a smooth manifold $M$ and $\pi\colon\mathcal{P} \to M$ a principal $\mathcal{G}$-bundle, also let $\mathrm{H}\mathcal{G}$ be a multiplicative Yang-Mills connection on $\mathcal{G}$ (w.r.t.\ a $\zeta \in \Omega^2(M; \mathcal{g})$) and $A \in \Omega^1(\mathcal{P}; \pi^*\mathcal{g})$ be a connection 1-form on $\mathcal{P}$. Then we have the \textbf{structure equation}
\begin{equation*}
F
=
\mathrm{d}^{\pi^*\nabla^\caG} A
	+ \frac{1}{2} \mleft[ A \stackrel{\wedge}{,} A \mright]_{\pi^*\mathcal{g}}
	+ \pi^!\zeta~.
\end{equation*}
\end{propositions}

\begin{remarks}{$\rmH \caP$ and foliations}{CurvoftotMCForm}
Again viewing $\pi_{\mathcal{G}} \colon \mathcal{G} \to M$ as the principal bundle itself with $\mathrm{H}\mathcal{P} = \mathrm{H}\mathcal{G}$, we know by \ref{cor:TotMCFormIsConnectionForm} that $\mutot$ is the connection 1-form corresponding to $\mathrm{H}\mathcal{G}$. Denoting the associated generalized field strength by $F\mleft( \mutot \mright)$, the structure equation and the generalized Maurer-Cartan equation, \ref{thm:GenMCEq} (also recall \ref{rem:BasePointDifficultiesinGenMCEq}), imply
\begin{equation*}
F_g\mleft( \mutot \mright)
=
\sAd_{g^{-1}} \circ \mleft. \pi_{\mathcal{G}}^!\zeta \mright|_g
\end{equation*}
for all $g \in \mathcal{G}_x$ ($x \in M$). This implies that $F\mleft( \mutot \mright)$ is fully encoded by $\zeta$, especially we have
\begin{equation}\label{ZetaIsTheCurvatureOnG}
e^!\Bigl(F\mleft( \mutot \mright)\Bigr)
=
\zeta,
\end{equation}
where $e$ is the neutral section of $\mathcal{G}$. Thus, $\zeta$ is the (generalised) curvature of $\mutot$ restricted on $M$, and 
\begin{equation*}
F
=
\mathrm{d}^{\pi^*\nabla^\caG} A
	+ \frac{1}{2} \mleft[ A \stackrel{\wedge}{,} A \mright]_{\pi^*\mathcal{g}}
	+ (e \circ \pi)^!\Bigl(F\mleft( \mutot \mright)\Bigr).
\end{equation*}
Following similar calculations as in the proof of \ref{prop:NewFieldStrengthWithCoolProps}, it is trivial to show that the first two summands measure the failure of $\mathrm{H}\mathcal{P}$ being a foliation, hence this meaning carries over from the classical theory as also argued in \cite[\S 2.5, Cor.\ 2.23]{FernandesMarcutMultiplicativeForms} for LGBs. Due to the third summand, $\mathrm{H}\mathcal{P}$ is involutive if and only if
\begin{equation*}
F
=
(e \circ \pi)^!\Bigl(F\mleft( \mutot \mright)\Bigr).
\end{equation*}
In that sense Eq.\ \eqref{ZetaIsTheCurvatureOnG} reflects the fact that $M$ is a horizontal leave of $\mathrm{H}\mathcal{G}$. Furthermore, $F$ can be non-zero while $\mathrm{H}\mathcal{P}$ is involutive, if $F\mleft( \mutot \mright)$ is non-zero along $M$.
\end{remarks}

\begin{proof}[Proof of \ref{prop:StructureEq}]
\leavevmode\newline
The idea of the proof is similar to the proof of the "classical" statement; following the structure of \cite[\S 5.5, Lemma 5.5.5, page 276]{Hamilton}, see also works like \cite[\S 4.6, Lemma 4.25]{LAURENTGENGOUXStienonXuMultiplicativeForms} and \cite[\S 2.5, Prop.\ 2.24]{FernandesMarcutMultiplicativeForms} where something similar is shown on Lie groupoids instead of principal bundles, but with no $\zeta$. That is, we will look at vertical and horizontal tangent vectors and their mixed terms; recall \ref{def:GaugeBosonsOnLGBPrincies} and \ref{thm:OurConnectionHasAUniqueoneForm}.

$\bullet$ We have by \ref{prop:NewFieldStrengthWithCoolProps}
\begin{equation*}
F\mleft( \overline{\nu}, \overline{\mu} \mright)
=
0
\end{equation*}
for all $\mu, \nu \in \Gamma(\mathcal{g})$. Regarding the right hand side of the structure equation, this trivially also holds for the third summand $\mleft(\pi^!\zeta\mright)\mleft( \overline{\nu}, \overline{\mu} \mright) = 0$. We also have
\begin{equation*}
\mleft(\frac{1}{2} \mleft[ A \stackrel{\wedge}{,} A \mright]_{\pi^*\mathcal{g}}\mright)\mleft( \overline{\nu}, \overline{\mu} \mright)
=
\mleft[ A\mleft( \overline{\nu} \mright), A\mleft( \overline{\mu} \mright) \mright]_{\pi^*\mathcal{g}}
=
\mleft[ \pi^*\nu, \pi^*\mu \mright]_{\pi^*\mathcal{g}}
=
\pi^*\mleft( \mleft[ \nu, \mu \mright]_{\mathcal{g}} \mright),
\end{equation*}
and
\begin{equation*}
\mleft(\mathrm{d}^{\pi^*\nabla^\caG} A\mright)\mleft( \overline{\nu}, \overline{\mu} \mright)
=
{\underbrace{\mleft(\pi^*\nabla^\caG\mright)_{\overline{\nu}} \bigl(A\mleft( \overline{\mu} \mright) \bigr)}_{\mathclap{ = \pi^*\mleft( \nabla^\caG_{\mathrm{D}\pi\mleft( \overline{\nu} \mright)} \mu \mright) = 0 }}}
	- \mleft(\pi^*\nabla^\caG\mright)_{\overline{\mu}} \bigl(A\mleft( \overline{\nu} \mright) \bigr)
	- A {\underbrace{\bigl( \mleft[ \overline{\nu}, \overline{\mu} \mright] \bigr)}_{= \overline{[\nu, \mu]}}}
=
- \pi^*\mleft( \mleft[ \nu, \mu \mright]_{\mathcal{g}} \mright).
\end{equation*}
Thus,
\begin{equation*}
\mleft(
	\mathrm{d}^{\pi^*\nabla^\caG} A
	+ \frac{1}{2} \mleft[ A \stackrel{\wedge}{,} A \mright]_{\pi^*\mathcal{g}}
	+ \pi^!\zeta
\mright)\mleft( \overline{\nu}, \overline{\mu} \mright)
=
0
=
F\mleft( \overline{\nu}, \overline{\mu} \mright).
\end{equation*}

$\bullet$ Let $X$ and $Y$ be horizontal vector fields of $\mathcal{P}$, then clearly
\begin{equation*}
\mleft(\frac{1}{2} \mleft[ A \stackrel{\wedge}{,} A \mright]_{\pi^*\mathcal{g}}\mright)(X, Y)
=
\bigl[ A\mleft( X \mright), A\mleft( Y \mright) \bigr]_{\pi^*\mathcal{g}}
=
0,
\end{equation*}
then clearly
\begin{equation*}
\mleft(
	\mathrm{d}^{\pi^*\nabla^\caG} A
	+ \frac{1}{2} \mleft[ A \stackrel{\wedge}{,} A \mright]_{\pi^*\mathcal{g}}
	+ \pi^!\zeta
\mright)(X, Y)
=
\mleft(
	\mathrm{d}^{\pi^*\nabla^\caG} A \circ \mleft( \pi_h, \pi_h \mright)
	+ \pi^!\zeta
\mright)(X, Y)
=
F(X, Y),
\end{equation*}
where $\pi_h\colon \mathrm{T}\mathcal{P} \to \mathrm{H}\mathcal{P}$ denotes the canonical projection onto the horizontal bundle $\mathrm{H}\mathcal{P}$ (especially, $\pi_h(X) = X$, $\pi_h(Y)= Y$).

$\bullet$ Now let $X \in \mathfrak{X}(\mathcal{P})$ be again horizontal and $\nu \in \Gamma(\mathcal{g})$; then again by \ref{prop:NewFieldStrengthWithCoolProps}
\begin{equation*}
F\mleft(X, \overline{\nu}\mright)
=
0,
\end{equation*}
as also clearly $\mleft( \pi^!\zeta \mright)\mleft(X, \overline{\nu}\mright) = 0$,
furthermore
\begin{equation*}
\mleft(\frac{1}{2} \mleft[ A \stackrel{\wedge}{,} A \mright]_{\pi^*\mathcal{g}}\mright)\mleft( X, \overline{\nu} \mright)
=
\mleft[ A(X), A\mleft( \overline{\nu} \mright) \mright]_{\pi^*\mathcal{g}}
=
0,
\end{equation*}
and by \ref{lem:BracketVertHor}
\begin{equation*}
\mleft(\mathrm{d}^{\pi^*\nabla^\caG} A\mright)\mleft( X, \overline{\nu} \mright)
=
\mleft( \pi^*\nabla^\caG \mright)_X \mleft( \pi^*\nu \mright)
	- A\bigl( \mleft[ X, \overline{\nu} \mright] \bigr)
=
0,
\end{equation*}
which concludes the proof with
\begin{equation*}
F\mleft(X, \overline{\nu}\mright)
=
\mleft(
	\mathrm{d}^{\pi^*\nabla^\caG} A
	+ \frac{1}{2} \mleft[ A \stackrel{\wedge}{,} A \mright]_{\pi^*\mathcal{g}}
	+ \pi^!\zeta
\mright)\mleft(X, \overline{\nu}\mright).
\end{equation*}
\end{proof}

Concluding this subsubsection, we also achieve a generalized Bianchi identity, which we already have proven in \cite[\S 7, Thm.\ 7.3]{My1stpaper} and \cite[\S 5, Thm.\ 5.1.42]{MyThesis}; the proof is straightforward to show, making use of the infinitesimal compatibility conditions, \ref{def:YangMillsConnection}.

\begin{lemmata}{Generalized Bianchi identity}{GenBianchi}
\tcbsubtitle[before skip=\baselineskip]{\cite[\S 5, Thm.\ 5.1.42]{MyThesis}}
Let $\mathcal{G} \to M$ be an LGB over a smooth manifold $M$ and $\pi\colon\mathcal{P} \to M$ a principal $\mathcal{G}$-bundle, also let $\mathrm{H}\mathcal{G}$ be a multiplicative Yang-Mills connection on $\mathcal{G}$ (w.r.t.\ a $\zeta \in \Omega^2(M; \mathcal{g})$) and $A \in \Omega^1(\mathcal{P}; \pi^*\mathcal{g})$ be a connection 1-form on $\mathcal{P}$. Then we have the \textbf{(generalized) Bianchi identity}
\begin{equation*}
\mathrm{d}^{\pi^*\nabla^\caG}F
	+ \mleft[ A \stackrel{\wedge}{,} F \mright]_{\pi^*\mathcal{g}}
=
\pi^! \mathrm{d}^{\nabla^\caG} \zeta~.
\end{equation*}
\end{lemmata}

\subsubsection{Gauge transformation of the generalized field strength}

Similar to \ref{GaugeTrafoForA} we will now discuss the gauge transformation of the generalized field strength $F$; recall \ref{prop:GaugeTrafoOfGaugeBoson}.

\begin{propositions}{Gauge transformation of the generalized field strength}{GaugeTrafoOfCurv}
Let $\mathcal{G} \to M$ be an LGB over a smooth manifold $M$ and $\pi\colon\mathcal{P} \to M$ a principal $\mathcal{G}$-bundle, also let $\mathrm{H}\mathcal{G}$ be a multiplicative Yang-Mills connection on $\mathcal{G}$ and $A \in \Omega^1(\mathcal{P}; \pi^*\mathcal{g})$ be a connection 1-form on $\mathcal{P}$. Furthermore, let $H \in \sAut(\mathcal{P})$. We then have that $H^!F$ is the field strength related to $H^!A$ and
\begin{equation*}
H^!F
=
{\sAd_{\mathrm{pr}_2\circ\mleft(\sigma^H\mright)^{-1}}} \circ F,
\end{equation*}
where $\sigma^H \in C^\infty(\mathcal{P}; \mathcal{G})^{\mathcal{G}}$ is defined as in \ref{prop:GaugeTrafoAsBundleIsomIsASectionOfConjugationMaps} and $\mathrm{pr}_2\colon \pi^*\mathcal{G} \to \mathcal{G}$ is the projection onto the second component. Similar to Definition \eqref{MultiWithPulli} we may shortly just write 
\begin{equation*}
H^!F
=
{\sAd_{\mleft(\sigma^H\mright)^{-1}}} \circ F.
\end{equation*}
\end{propositions}

\begin{proof}
\leavevmode\newline
That $H^!F$ is the field strength related to $H^!A$ follows quickly by the same calculation as for proving Eq.\ \eqref{PullbackofMCGeneralCurvPlusExtra}, using $\pi \circ H = \pi$,
\begin{align*}
H^!F
&=
\mathrm{d}^{(\pi \circ H)^*\nabla^\caG} \mleft(H^!A\mright)
	+ \frac{1}{2} \mleft[ H^!A \stackrel{\wedge}{,} H^!A \mright]_{\pi^*\mathcal{g}}
	+ (\pi \circ H)^!\zeta
\\
&=
\mathrm{d}^{\pi^*\nabla^\caG} \mleft(H^!A\mright)
	+ \frac{1}{2} \mleft[ H^!A \stackrel{\wedge}{,} H^!A \mright]_{\pi^*\mathcal{g}}
	+ \pi^!\zeta,
\end{align*}
which is the field strength of $H^!A$. Now recall Eq.\ \eqref{DiffOfPrincAutom}, that is,
\begin{equation*}
\mathrm{D}_p H(X)
=
\mathcal{r}_{\overline{\sigma}_p*}(X)
	+ \mleft.{\overline{ \mleft(\mleft(\pi^*\Delta\mright)\sigma^H\mright)_p(X) }}\mright|_{H(p)}
\end{equation*}
for all $X \in \mathrm{T}_p\mathcal{P}$ ($p \in \mathcal{P}$), where $\overline{\sigma}_p \coloneqq \mathrm{pr}_2\mleft( \sigma^H_p \mright)$. Thus, by \ref{prop:NewFieldStrengthWithCoolProps},
\begin{align*}
\mleft( H^!F \mright)_p(X, Y)
&=
F_{H(p)}\bigl( \mathrm{D}_pH(X), \mathrm{D}_pH(Y) \bigr)
\\
&=
F_{p \cdot \overline{\sigma}_p}\bigl( \mathcal{r}_{\overline{\sigma}_p*}(X), \mathcal{r}_{\overline{\sigma}_p*}(Y) \bigr)
\\
&=
\mleft(\mathcal{r}_{\overline{\sigma}}^!F\mright)_p(X, Y)
\\
&=
\mleft(\sAd_{\overline{\sigma}_p^{-1}} \circ F\mright)_p(X, Y)
\end{align*}
for all $X, Y \in \mathrm{T}_p \mathcal{P}$; this finishes the proof.
\end{proof}

We will now extend this again to a local change of gauges (sections of $\mathcal{P}$).

\begin{definitions}{Local field strength}{LocalFieldStrength}
Let $\mathcal{G} \to M$ be an LGB over a smooth manifold $M$ and $\pi\colon\mathcal{P} \to M$ a principal $\mathcal{G}$-bundle, also let $\mathrm{H}\mathcal{G}$ be a multiplicative Yang-Mills connection on $\mathcal{G}$ and $A \in \Omega^1(\mathcal{P}; \pi^*\mathcal{g})$ be a connection 1-form on $\mathcal{P}$. Furthermore, let $s \in \Gamma(\mathcal{P}|_U)$ be a (local) gauge over an open subset $U \subset M$. Then we define the \textbf{local curvature} of \textbf{local field strength $F_s \in \Omega^2\mleft(U; \mleft.\mathcal{g}\mright|_U\mright)$ (w.r.t.\ $s$)} by
\begin{equation*}
F_s
\coloneqq
s^!F.
\end{equation*}
\end{definitions}

In previous calculations regarding $\mutot$ and the generalized Maurer-Cartan equation we have basically already shown the pullback of the structure equation.

\begin{corollaries}{Pullback of the structure equation}{PullbackOfStructureEq}
Let $\mathcal{G} \to M$ be an LGB over a smooth manifold $M$ and $\pi\colon\mathcal{P} \to M$ a principal $\mathcal{G}$-bundle, also let $\mathrm{H}\mathcal{G}$ be a multiplicative Yang-Mills connection on $\mathcal{G}$ and $A \in \Omega^1(\mathcal{P}; \pi^*\mathcal{g})$ be a connection 1-form on $\mathcal{P}$. Furthermore, let $s \in \Gamma(\mathcal{P}|_U)$ be a (local) gauge over an open subset $U \subset M$. Then we can express the local field strength as the pullback of the structure equation, that is,
\begin{equation*}
F_s
=
\mathrm{d}^{\nabla^\caG}A_s
	+ \frac{1}{2} \mleft[ A_s \stackrel{\wedge}{,} A_s \mright]_{\mathcal{g}}
	+ \zeta.
\end{equation*}
\end{corollaries}

\begin{proof}
\leavevmode\newline
This follows quickly by the same calculation as for proving Eq.\ \eqref{PullbackofMCGeneralCurvPlusExtra}.
\end{proof}

The gauge transformation of the field strength $F$ describes the behaviour of $F$ under a change of gauge.

\begin{propositions}{Gauge transformations again as a change of gauge}{LocalGaugeTrafoChangeGaugeFieldStrength}
Let $\mathcal{G} \to M$ be an LGB over a smooth manifold $M$ and $\pi\colon\mathcal{P} \to M$ a principal $\mathcal{G}$-bundle, also let $\mathrm{H}\mathcal{G}$ be a multiplicative Yang-Mills connection on $\mathcal{G}$ and $A \in \Omega^1(\mathcal{P}; \pi^*\mathcal{g})$ be a connection 1-form on $\mathcal{P}$. Also let $U_i$ and $U_j$ be two open subsets of $M$ so that $U_i \cap U_j \neq \emptyset$, two gauges $s_i \in \Gamma\mleft(\mathcal{P}|_{U_i}\mright)$ and $s_j \in \Gamma\mleft(\mathcal{P}|_{U_j}\mright)$, and the unique $\sigma_{ji} \in \Gamma\mleft( \mleft.\mathcal{G}\mright|_{U_i \cap U_j} \mright)$ with $s_i = s_j \cdot \sigma_{ji}$ on $U_i \cap U_j$.

Then we have for the fields strength of $A$ over $U_i \cap U_j$ that
\begin{equation*}
F_{s_i}
=
\mathrm{Ad}_{\sigma_{ji}^{-1}}\circ F_{s_j}.
\end{equation*}
\end{propositions}

\begin{proof}
\leavevmode\newline
The proof is precisely as for \ref{prop:LocalGaugeTrafoChangeGauge}, just without catering for the Darboux derivative in the gauge transformation due to that one uses \ref{prop:GaugeTrafoOfCurv} instead of \ref{prop:GaugeTrafoOfGaugeBoson}.
\end{proof}

\begin{remarks}{Integrating curved Yang-Mills gauge theories, part II}{IntegratingKotovStroblFieldStrength}
Similar to \ref{rem:IntegratingKotovStrobl}, \ref{cor:PullbackOfStructureEq} and \ref{prop:LocalGaugeTrafoChangeGaugeFieldStrength} show that we recover the field strength and its gauge transformation as developed by Alexei Kotov and Thomas Strobl, see \cite{CurvedYMH} for a concise summary or \cite{MyThesis} for an extended introduction. These references are for the very general situation using Lie algebroids, hence see alternatively \cite{My1stpaper} for this type of gauge theory restricted to Lie algebra bundles.
\end{remarks}


\section{Curved Yang-Mills gauge theory}\label{CYMSection}

\subsection{Definition and gauge invariance}\label{CYMDefGaugeInv}

We eventually are able to conclude this paper with the definition of what we will call \textbf{curved Yang-Mills gauge theory}, highlighting that this theory is an integral of the infinitesimal gauge theory originally developed by Alexei Kotov and Thomas Strobl. For the following we say that a fibre metric $\kappa$ of $\mathcal{g}$ is \textbf{$\mathrm{Ad}$-invariant} if
\begin{equation*}
\kappa\bigl( \mathrm{Ad}_g(\nu), \mathrm{Ad}_g(\mu) \bigr)
=
\kappa (\nu, \mu)
\end{equation*}
for all $g \in \mathcal{G}_x$ ($x \in M$) and $\mu, \nu \in \mathcal{g}_x$.

\begin{corollaries}{The Lagrangian}{ContractionOfLocalFIeldWellDefined}
Let $\mathcal{G} \to M$ be an LGB over a spacetime $M$ so that its LAB $\mathcal{g}$ admits an $\mathrm{Ad}$-invariant fibre metric $\kappa$, and let $\pi\colon\mathcal{P} \to M$ be a principal $\mathcal{G}$-bundle; also let $\mathrm{H}\mathcal{G}$ be a multiplicative Yang-Mills connection on $\mathcal{G}$ and $A \in \Omega^1(\mathcal{P}; \pi^*\mathcal{g})$ be a connection 1-form on $\mathcal{P}$. Furthermore, let $\mleft( U_i \mright)_i$ be an open covering of $M$ so that there are subordinate gauges $s_i \in \Gamma\mleft(\mathcal{P}|_{U_i}\mright)$.

Then the top-degree form $\mathfrak{L}_{\mathrm{CYM}}[A] \in \Omega^{\mathrm{dim}(M)}(M; \mathbb{R})$, defined locally by
\begin{equation*}
\mleft.\bigl(\mathfrak{L}_{\mathrm{CYM}}[A]\bigr)\mright|_{U_i}
\coloneqq 
- \frac{1}{2} \kappa \mleft( F_{s_i} \stackrel{\wedge}{,} *F_{s_i} \mright),
\end{equation*}
is well-defined and independent of the choice of gauge, where $*$ is the Hodge star operator w.r.t.\ the spacetime metric of $M$.
\end{corollaries}

\begin{proof}
\leavevmode\newline
Let $s_j$ be another gauge corresponding to $U_j$ so that $U_i \cap U_j \neq \emptyset$, and we have a unique $\sigma_{ji} \in \Gamma\mleft( \mleft.\mathcal{G}\mright|_{U_i \cap U_j} \mright)$ with $s_i = s_j \cdot \sigma_{ji}$ on $U_i \cap U_j$. Then it follows by \ref{prop:LocalGaugeTrafoChangeGaugeFieldStrength} and the definition of the Hodge star operator (which is just a certain contraction w.r.t.\ $M$) that
\begin{equation*}
\kappa \mleft( F_{s_i} \stackrel{\wedge}{,} *F_{s_i} \mright)
=
\kappa \mleft( \mleft(\mathrm{Ad}_{\sigma_{ji}^{-1}}\circ F_{s_j} \mright) \stackrel{\wedge}{,} *\mleft( \mathrm{Ad}_{\sigma_{ji}^{-1}}\circ F_{s_j} \mright) \mright)
=
\kappa \mleft( F_{s_j} \stackrel{\wedge}{,} *F_{s_j} \mright),
\end{equation*}
using the $\mathrm{Ad}$-invariance of $\kappa$. This concludes the proof.
\end{proof}

\begin{definitions}{Curved Yang-Mills gauge theory}{CYMGTFinally}
Let $\mathcal{G} \to M$ be an LGB over a spacetime $M$ so that its LAB $\mathcal{g}$ admits an $\mathrm{Ad}$-invariant fibre metric $\kappa$, and let $\pi\colon\mathcal{P} \to M$ be a principal $\mathcal{G}$-bundle; furthermore let $\mathrm{H}\mathcal{G}$ be a multiplicative Yang-Mills connection on $\mathcal{G}$ and $A \in \Omega^1(\mathcal{P}; \pi^*\mathcal{g})$ be a connection 1-form on $\mathcal{P}$.

Then the functional $A \mapsto \mathfrak{L}_{\mathrm{CYM}}[A]$ is called the \textbf{curved Yang-Mills Lagrangian}, where $\mathfrak{L}_{\mathrm{CYM}}[A]$ is defined in \ref{cor:ContractionOfLocalFIeldWellDefined}.
\end{definitions}

By construction, this Lagrangian is gauge-invariant.

\begin{theorems}{Gauge invariance of the curved Yang-Mills Lagrangian}{GaugeInvarianceOfLagrangian}
Let $\mathcal{G} \to M$ be an LGB over a spacetime $M$ so that its LAB $\mathcal{g}$ admits an $\mathrm{Ad}$-invariant fibre metric $\kappa$, and let $\pi\colon\mathcal{P} \to M$ be a principal $\mathcal{G}$-bundle; also let $\mathrm{H}\mathcal{G}$ be a multiplicative Yang-Mills connection on $\mathcal{G}$ and $A \in \Omega^1(\mathcal{P}; \pi^*\mathcal{g})$ be a connection 1-form on $\mathcal{P}$.

Then we have
\begin{equation*}
\mathfrak{L}_{\mathrm{CYM}}\mleft[ H^!A \mright]
=
\mathfrak{L}_{\mathrm{CYM}}[A]
\end{equation*}
for all $H \in \sAut(\mathcal{P})$.
\end{theorems}

\begin{proof}
\leavevmode\newline
By \ref{prop:GaugeTrafoOfCurv} we have 
\begin{equation*}
H^!F
=
{\sAd_{\mathrm{pr}_2\circ\mleft(\sigma^H\mright)^{-1}}} \circ F,
\end{equation*}
where $\sigma^H \in C^\infty(\mathcal{P}; \mathcal{G})^{\mathcal{G}}$ is defined as in \ref{prop:GaugeTrafoAsBundleIsomIsASectionOfConjugationMaps} and $\mathrm{pr}_2: \pi^*\mathcal{G} \to \mathcal{G}$ is the projection onto the second component. Similar to the argument in the proof of \ref{prop:LocalGaugeTrafoChangeGauge} we have
\begin{equation*}
\mleft( H^!F \mright)_s
=
\mathrm{Ad}_{\mathrm{pr}_2\circ\mleft(\sigma^H\mright)^{-1} \circ s} \circ F_s
\end{equation*}
for a given gauge $s \in \Gamma(\mathcal{P}|_U)$, where $U$ is an open subset of $M$. With the same argument as in \ref{cor:ContractionOfLocalFIeldWellDefined} it follows that
\begin{equation*}
\mathfrak{L}_{\mathrm{CYM}}\mleft[ H^!A \mright]
=
\mathfrak{L}_{\mathrm{CYM}}[A].
\end{equation*}
\end{proof}

\begin{remarks}{Infinitesimal gauge invariance}{InfinitesimalGaugeInv}
As we have seen previously, the compatibility conditions (\ref{def:NowReallyYangMillsConnectio}) were important to derive how $F$ behaves under modified right push-forwards. Recall \ref{rem:IntegratingKotovStrobl}, we have a corresponding formulation of infinitesimal gauge transformation and it is straight-forward to show that we achieve \textbf{infinitesimal gauge invariance} as in the classical theory. However, as argued by Alexei Kotov and Thomas Strobl in \cite{CurvedYMH} (for a concise summary; \cite{MyThesis} for an extended introduction, or alternatively \cite{My1stpaper} for a simplified language) the corresponding infinitesimal gauge invariance just needs the infinitesimal compatibility conditions (\ref{def:YangMillsConnection}).
\end{remarks}

\subsection{Field redefinitions}

As mentioned in the introduction and developed in \cite[the latter reference developed the following for Lie algebroids and groupoids as structure]{My1stpaper, MyThesis}: There is an equivalence relation of curved Yang-Mills gauge theories, preserving the dynamics and kinematics of this theory. Let us recapitulate and integrate these:

\begin{definitions}{Field redefinitions on the LGB}{FieldRedefsAsDefIntegrated}
Let $\pi_{\mathcal{G}} \colon \mathcal{G} \to M$ be an LGB over a smooth manifold $M$ with LAB $\mathcal{g}$; furthermore let $\mutot$ be a multiplicative Yang-Mills connection on $\mathcal{G}$ (w.r.t.\ a $\zeta \in \Omega^2(M; \cag)$). We then define \textbf{field redefinitions w.r.t.\ $\lambda \in \Omega^1(M; \mathcal{g})$} of $\mutot$ by
\begin{equation*}
\mleft.\mutot^{\lambda}\mright|_g
\coloneqq
\mleft.\mutot\mright|_g
	+ \sAd_{g^{-1}} \circ \mleft. \pi_{\mathcal{G}}^!\lambda\mright|_g
	- \mleft.\pi_{\mathcal{G}}^!\lambda\mright|_g
\end{equation*}
for all $g \in \caG$
and of $\zeta$ by
\begin{equation*}
\zeta^\lambda
\coloneqq
\zeta
	+ \mathrm{d}^{\nabla^\caG}\lambda + \frac{1}{2} \mleft[ \lambda \stackrel{\wedge}{,} \lambda \mright]_{\mathcal{g}}.
\end{equation*}
\end{definitions}

\begin{remark}
Equivalently, since we introduced $\mutot$ as a composition of the vertical Maurer-Cartan form with the projection $\pi^{\mathrm{vert}}$ onto the vertical part w.r.t.\ the horizontal distribution on $\mathcal{G}$, we can define the field redefinition of $\pi^{\mathrm{vert}}$ by
\begin{equation*}
\pi^{\mathrm{vert}, \lambda}
\coloneqq
\pi^{\mathrm{vert}}
	+ \overline{\pi_{\mathcal{G}}^! \lambda}^l
	+ \overline{\pi_{\mathcal{G}}^! \lambda}^r,
\end{equation*}
where $\overline{\pi_{\mathcal{G}^!} \lambda}^r$ denotes the fundamental vector field of $\pi_{\mathcal{G}}^! \lambda$ via right-multiplication, that is,
\begin{equation*}
\mleft.\overline{\pi_{\mathcal{G}^!} \lambda}^r\mright|_g
=
\mathrm{D}_{e_x}R_g \circ \lambda_{x} \circ \mathrm{D}_x \pi_{\mathcal{G}}
\end{equation*}
for all $g \in \mathcal{G}_x$, and $\overline{\pi_{\mathcal{G}}^! \lambda}^l$ is just the typical definition for fundamental vector fields, with an emphasis on that it is via left-multiplication. 
\end{remark}

\begin{propositions}{Field redefinitions preserving multiplicative YM connections}{YMStaysYM}
Given the context of \ref{def:FieldRedefsAsDefIntegrated}, the field redefinition of $\mutot$ is a multiplicative Yang-Mills connection w.r.t.\ $\zeta^\lambda$.
\end{propositions}

\begin{proof}
\leavevmode\newline
It is a straight-forward check to see that $\mutot^{\lambda}$ is still multiplicative, and well-known in the mentioned literature regarding multiplicative connections; a quick way to see this is by following \ref{rem:SimplicialDifferentialStuff}: We actually add an exact term,
\begin{equation*}
\mutot^{\lambda}
=
\mutot
	+ \delta \lambda ~.
\end{equation*}
Thus, $\mutot^{\lambda}$ stays multiplicatively closed, and hence it is multiplicative. That it satisfies the generalised Maurer-Cartan equation w.r.t.\ $\zeta^\lambda$ is a straight-forward but a bit lengthy calculation, which is why we keep this as an exercise for the reader; for the infinitesimal, i.e.\ in the connected case by \ref{thm:GenMCEq}, see \cite{My1stpaper, MyThesis}.
\end{proof}

In fact, a simple field redefinition of an Ehresmann connection preserves the compatibilities with the new multiplicative connection, and this is precisely one of the ideas of the field redefinitions. As usual, the difference of two connection 1-forms $A - A^\prime$ should be a one-form which is horizontal and transforms via the Adjoint representations under the modified push-forward. However, we have seen here that the modified push-forward depends on a connection on the group bundle. Thus, is it possible to vary $A$ with a 1-form which is horizontal \textit{and} does not transform via the Adjoint transformation such that $A^\prime$ is a connection 1-form but w.r.t.\ a different connection on the group bundle? The answer to this are the field redefinitions; for the following observe that the field redefinition of $\mutot$ induces a field redefinition of $\nabla^\caG$ given by
\begin{equation*}
\nabla^{\caG, \lambda}
=
\nabla^\caG
	+ \mathrm{ad} \circ \lambda ~ .
\end{equation*}

\begin{theorems}{Field redefinitions preserving the physics}{EhresmannTransformedEhresmann}
Let $\mathcal{G} \to M$ be an LGB over a smooth manifold $M$ so that its LAB $\mathcal{g}$ admits an $\mathrm{Ad}$-invariant fibre metric $\kappa$, and let $\pi\colon\mathcal{P} \to M$ be a principal $\mathcal{G}$-bundle; furthermore let $\mutot$ be a multiplicative Yang-Mills connection on $\mathcal{G}$ (w.r.t.\ a $\zeta \in \Omega^2(M; \cag)$) and $A \in \Omega^1(\mathcal{P}; \pi^*\mathcal{g})$ be a connection 1-form on $\mathcal{P}$ w.r.t.\ $\mutot$. We define the \textbf{field redefinitions w.r.t.\ $\lambda \in \Omega^1(M; \mathcal{g})$ of $A$} by
\begin{equation*}
A^\lambda
\coloneqq
A - \pi^!\lambda ~.
\end{equation*}

Then $A^\lambda$ is a connection 1-form w.r.t.\ $\mu^{\lambda}$ and
\begin{equation*}
F^\lambda
\coloneqq
\mathrm{d}^{\pi^*\nabla^{\mathrm{YM},\lambda}} A^\lambda
	+ \frac{1}{2} \mleft[ A^\lambda \stackrel{\wedge}{,} A^\lambda \mright]_{\pi^*\mathcal{g}}
	+ \pi^!\zeta^\lambda
\equiv
F
\end{equation*}
for all $\lambda \in \Omega^1(M; \mathcal{G})$. 
\end{theorems}

\begin{proof}
\leavevmode\newline
That $F^\lambda = F$ follows by similar calculations as in \cite{My1stpaper, MyThesis}. Therefore let us only prove that $A^\lambda$ is still a connection 1-form. $A^\lambda$ is clearly still the identity on the vertical structure, and henceforth let us check its behaviour w.r.t.\ the modified pushforward induced by $\mutot^{\lambda}$, denoted by $\mathcal{r}_\sigma^\lambda$ for $\sigma \in \Gamma(\mathcal{G})$: We have to study
\begin{equation*}
\mleft(\mleft(\mathcal{r}_\sigma^\lambda\mright)^!A^\lambda\mright)_p(Y)
\end{equation*}
for all $p \in \mathcal{P}_x$ and $Y \in \mathrm{T}_p\mathcal{P}$. First of all, similar to the beginning of the proof of \ref{prop:NewFieldStrengthWithCoolProps} we have
\begin{equation*}
\mleft(\mathcal{r}_\sigma^\lambda\mright)^! \mleft(\pi^! \lambda\mright)
=
\pi^! \lambda ~ ,
\end{equation*}
and thus we essentially only need to study the field redefined Darboux derivative:
\begin{equation*}
\Delta^\lambda \sigma
\coloneqq
\sigma^! \mutot^{\lambda}
=
\Delta \sigma 
	+ \mathrm{Ad}_{\sigma^{-1}} \circ \lambda
	- \lambda ~ .
\end{equation*}
In total,
\begin{align*}
\mleft(\mleft(\mathcal{r}_\sigma^\lambda\mright)^!A^\lambda\mright)_p(Y)
&=
A_{p \cdot \sigma_x}
	\mleft( 
		\mathcal{r}_{\sigma*}(Y)
	\mright)
	- \mleft(\pi^!\mleft( \mathrm{Ad}_{\sigma^{-1}} \circ \lambda \mright)\mright)_p(Y)
	+ \lambda_x(X)
	- \lambda_x(X)
\\
&=
\mleft( \sAd_{\sigma^{-1}} \circ A - \sAd_{\sigma^{-1}} \circ \pi^!\lambda \mright)_p(Y)
\\
&=
\mleft(\sAd_{\sigma^{-1}} \circ A^\lambda\mright)_p(Y) ~ ,
\end{align*}
where $X \coloneqq \mathrm{D}_p\pi(Y)$, $\mathcal{r}_\sigma$ denotes the modified pushforward via $\mutot$ as usual, and we made use of $A$ being a connection 1-form w.r.t.\ $\mutot$.
\end{proof}

We see there is a transformation keeping the Lagrangian invariant such that the dynamics are preserved; $\mathcal{P}$ itself does not change and thus the gauge transformations in form of (local) principal bundle automorphisms also stay the same, and so the kinematics are preserved, too. However, field redefinitions non-trivially transform data like $\mutot$ so that it may become flat, and we may end up at a pre-classical theory describing the same physics. $\zeta$ itself might survive as shown in \cite{My1stpaper, MyThesis}, but in general the field redefinitions make it difficult to find ``truly new'' examples, especially from a local point of view. Before we discuss this, let us summarize these results from the point of view as in \ref{rem:SimplicialDifferentialStuff}.

\begin{remarks}{Meaning of Field redefinitions}{FieldRedefIntegrated}
Recall \ref{rem:SimplicialDifferentialStuff} and its notation, \textit{i.e.}\ there is a simplicial differential $\delta$ so that
\begin{align*}
\delta \mutot &= 0~,\\
F_{\mathcal{G}} &= \delta\zeta~,
\end{align*}
where $F_{\mathcal{G}}$ denotes the curvature of $\mutot$ in an ordinary way, that is,
\begin{equation*}
F_{\mathcal{G}}
=
\mathrm{d}^{\pi_{\mathcal{G}}^*\nabla^\caG} \mutot
	+ \frac{1}{2} \mleft[ \mutot \stackrel{\wedge}{,} \mutot \mright]_{\pi_{\mathcal{G}}^*\mathcal{g}}~,
\end{equation*}
where $\pi_{\mathcal{G}}$ is the projection of $\mathcal{G}$.
The field redefinition of $\mutot$ can be written as
\begin{equation*}
\mutot^{\lambda}
=
\mutot
	+ \delta \lambda ~ .
\end{equation*}
\ref{prop:YMStaysYM} and \ref{thm:EhresmannTransformedEhresmann} have now several interpretations and implications, where we denote with $\mleft[\mutot\mright]_{\delta}$ the equivalence class of $\mutot$ inherited by $\delta$:
\begin{enumerate}
	\item The field redefinition of $\mutot$ is about picking any other representative for $\mleft[\mutot\mright]_{\delta}$.
	\item Every other representative of $\mleft[\mutot\mright]_{\delta}$ is not just multiplicative, but also a multiplicative Yang-Mills connection, that is,
	\begin{align*}
	\delta \mutot^{\lambda} &= 0,\\
	F_{\mathcal{G}}^{\lambda} &= \delta\zeta^{\lambda},
	\end{align*}
	where $F_{\mathcal{G}}^{\lambda}$ is defined like $F_{\mathcal{G}}$ but using $\mutot^{\lambda}$ instead of $\mutot$.
	\item Regarding the choice of $\mutot$, the Lagrangian only depends on the choice of class $\mleft[\mutot\mright]_{\delta}$. However, the explicit choice of $\zeta$ may still matter: One could vary $\zeta$ with centre-valued forms while keeping $\mutot$ fixed, this is in general not totally covered by the field redefinitions and may affect the Lagrangian as discussed in \cite{MyThesis} (or \cite{My1stpaper}).
\end{enumerate}
\end{remarks}

Last but not least, as also mentioned in the introduction, these field redefinitions are not just any transformation preserving the physics. Curved Yang-Mills gauge theory is a reformulation of classical gauge theory in such a way that gauge theory is form-invariant w.r.t.\ field redefinitions. This reformulation introduces extra terms, $\zeta$ and connection 1-forms of $\mutot$ and $\nabla^\caG$, that is, every classical gauge theory is equivalent to a possibly curved theory with a non-zero $\zeta$ contributing to the field strength. From that point of view it is tempting to ask whether there are examples which cannot be flattened by field redefinitions:

\subsection{Examples}\label{CYMExamples}

First examples which cannot be flattened by field redefinitions were found in \cite{My1stpaper, MyThesis}; but many more were found after the initial upload of this work: It turned out that connections like $\mutot$ help classifying singular foliations (\cite{Fischer:2401.05966}), and in return singular foliation theory provides a full theory of examples and even a classification of such examples as long as the group bundle is centerless (\cite[the appendix]{Fischer:2401.05966}); motivated by \ref{ex:NablaYMAsAdjointConnection} (recall also \ref{ex:OurVeryImportantExample}) such a classification is via viewing the LGB and its connection as associated to a ``suitable'' principal bundle.

Let us first point out the obvious:

\begin{examples}{Classical examples}{ClassicalYMTheoriesRecovered}
By \ref{ex:ClassicalYangMillsConnection}, we just recover the classical formalism of gauge theory if having a classical principal bundle $P$ on which a trivial LGB $\mathcal{G} = M \times G$ acts on the right, where $G$ is a Lie group with Lie algebra $\mathfrak{g}$. $\mathrm{H}\mathcal{G}$ is the canonical flat multiplicative Yang-Mills connection, $\nabla^\caG$ is the canonical flat connection on $\mathcal{g} = M \times \mathfrak{g}$, and $\zeta \equiv 0$. Then every example known in the classical formalism carries over.

Furthermore, starting with such a classical example, by \cite[\S 6, Cor.\ 6.2]{My1stpaper} we know that if $\mathfrak{g}$ has a non-zero centre, then one can always add a centre-valued $\zeta$ to this gauge theory, transforming it into a flat pre-classical gauge theory with non-trivial $\zeta$. If $M$ is at least three-dimensional, then it is possible to have $\mathrm{d}^{\nabla^\caG} \zeta \neq 0$, and thus there is no $\lambda \in \Omega^1(M; \mathcal{g})$ with $\zeta^\lambda = 0$.
\end{examples}

This shows that there are examples with non-trivial $\zeta$, stable w.r.t.\ field redefinitions. However, as shown in \cite[\S 5.2, Thm.\ 5.16]{My1stpaper} (or alternatively \cite[\S 5.1.5, Thm.\ 5.1.33]{MyThesis}), if $M$ is contractible, then there is $\lambda \in \Omega^1(M; \mathcal{g})$ so that $\nabla^{\caG, \lambda}$ is flat. Thus, if $M$ is contractible, then every curved Yang-Mills Lagrangian is equivalent to a Lagrangian of a pre-classical gauge theory ($\zeta$ may be non-vanishing). This may be due to that $\mathcal{G}$ is a trivial fibre bundle, and then the LGB action breaks down to Lie group action; recall all our discussions related to classical principal bundles equipped with an LGB action of a trivial LGB as in the previous example. 
Hence we conjecture:

\begin{conjectures}{Trivial LGBs are associated with pre-classical theories}{TrivialLGBIsPreclassical}
Let $\mathcal{G} \to M$ be a trivial LGB over a smooth manifold $M$; furthermore let $\mutot$ be a multiplicative Yang-Mills connection on $\mathcal{G}$. Then there is a $\lambda \in \Omega^1(M; \mathcal{g})$ so that $\mutot^{\lambda}$ is flat.
\end{conjectures}

For centerless simple Lie group bundles over connected $M$ this turns out to be true, see \cite{Fischer:2024vak}. Obviously, non-trivial LGBs are not used in the classical formalism, and by the previous discussion one may ask if there is a (global) example with a curved multiplicative Yang-Mills connection starting with a non-trivial LGB. Let us describe these examples, already presented in \cite{MyThesis, Fischer:2024vak}.

There is a natural class of multiplicative Yang-Mills connections which cannot be flattened via field redefinitions. In order to understand this, we follow and use results of~\cite[in particular the appendix]{Fischer:2401.05966}: Assume an ordinary principal $G$-bundle $P$, where $G$ is a Lie group, and assume that $M$ is connected. Also assume that we have a manifold $\sfT$ on which $G$ acts (from the left).\footnote{$\sfT$ and its related constructions and assumptions are only needed to provide an extra geometric interpretation as in \ref{tab:ComparingTheGeometryGaugeTheoryWithSingularFoliations}.} Then we have two associated bundles: 
    \begin{equation}
        \begin{aligned}
            c_G(P) &\coloneqq (P \times G) \Big/ G~,
            \\
            \scT &\coloneqq (P \times \sfT) \Big/ G~.
        \end{aligned}
    \end{equation}
		As discussed, we have a canonical left-action, $c_G(P) \curvearrowright \scT$, and an Ehresmann connection on $P$ (in the usual sense) induces associated connections on $c_G(P)$ and $\scT$. The one on $c_G(P)$ is a multiplicative Yang-Mills connection on $c_G(P)$ whose primitive ($= \zeta$) is given by the corresponding curvature on $P$, and the one on $\scT$ is a connection that has been called \emph{compatible Yang--Mills connection}~\cite{Fischer:2401.05966}. That is, this is an Ehresmann connection on $\scT$ w.r.t.~the $c_G(P)$-action and the multiplicative Yang-Mills connection on $c_G(P)$, which satisfies a certain curvature equation.
    
    The following classification of \cite{Fischer:2401.05966} provides further motivation for considering Lie group bundles like $c_G(P)$ beyond the existence of natural multiplicative Yang-Mills connections. The following correspondence is via associated bundle constructions, more general than $\caG$ being an inner group bundle; field redefinitions of an associated connection then just corresponds changing the corresponding Ehresmann connection on $P$.
		
		\begin{propositions}{Classification of Yang-Mills bundles with trivial center}
		{classYMcenterless}
		\tcbsubtitle[before skip=\baselineskip]{\cite[Appendix]{Fischer:2401.05966}}
		Let $G$ be a Lie group with trivial center. The identity component of $G$ we denote by $G_0$, and $M$ is a connected manifold with point $\ell \in M$. We have a natural 1:1 correspondence between:

		\begin{itemize}
			\item Pairs $(\caG, [\mutot])$ made of Lie group bundles $\caG \to M$ with structural Lie group $G = \mathcal G_\ell$ and a class, up to field redefinition, of multiplicative Yang-Mills connections $\mutot$.
			\item Equivalence classes of pairs made of:
				\begin{enumerate}
				\item a group morphism\footnote{Here $G_0$ is seen as a sub-group of $ \mathrm{Aut}(G)$ through conjugation. Notice that if $G$ is connected, $ \mathrm{Aut}(G)/G_0= \mathrm{Out}(G)$.} $\Xi \colon \pi_1(M,\ell) \longrightarrow \mathrm{Aut}(G)/G_0$, that we call the outer holonomy, where $\pi_1(M,\ell)$ is the first fundamental group based at $\ell$,
				\item a principal $H$-bundle $P \to M $ with $H \subset {\mathrm{Aut}}(G)$ the inverse image through ${\mathrm{Aut}}(G) \to {\mathrm{Aut}}(G)/G_0$ of the image of $\Xi$, naturally inducing that the quotient $P/G_0$ is the Galois bundle $ \tilde M \to M$ associated to the kernel of $ \Xi$. 
				\end{enumerate}
				We identify to such pairs $(\Xi_1,P_1) $ and $ (\Xi_2,P_2)$, with $P_1$ a principal $ H_1$-bundle and $ P_2$ a principal $ H_2 $-bundle, if there exists $ \Psi \in {\rm Aut}(G)$ such that $ \Xi_1= \overline{\Psi}\circ \Xi_2 \circ \overline{\Psi}^{-1}$, where $\overline \Psi$ is the class of $\Psi$ in $\mathrm{Aut}(G)/G_0$, and if there exists an fiber preserving diffeomorphism $ \chi \colon P_1 \simeq P_2 $ such that $ \chi (p \cdot g) = \chi(p) \cdot \Psi(g)$ for all $g \in H_1$.
				Notice that $ \Psi(g)$ automatically belongs to $ H_2$, so that this condition makes sense.
				\end{itemize} 
		\end{propositions}
    
    
    In particular \cite{Fischer:2401.05966} provides the following (a modification of \cite{blazquez2022group}):
    
    \begin{lemmata}{Simple Lie group bundles}{SemiSimpleLGBAdjust}
		\tcbsubtitle[before skip=\baselineskip]{\cite[Appendix]{Fischer:2401.05966}}
        If $G$ is simple and centerless, then all multiplicative Yang-Mills connections on $\scG$ are associated to an Ehresmann connection of a unique principal $G$-bundle $P$, and their primitive is the corresponding curvature on $P$.
    \end{lemmata}
		
		Thus, assuming that $G$ has a trivial center implies that the associated connection on $c_G(P)$ is in one-to-one correspondence with the Ehresmann connection on $P$. Furthermore assume that $\sfT = \mathbb{R}^d$, and that $G$ acts faithfully on $\sfT$ with 0 as fixed point. Then there is a canonical singular foliation $\scF$ on $\scT$ with $M$ as a leaf generated by vector fields of the form
    \begin{equation}\label{anchor}
        \chi_\scT(X) + \overline{\nu}
    \end{equation}
    for all $X \in \mathfrak{X}(M)$ and $\nu \in \Gamma(\mathrm{ad}(P))$, where $\chi_\scT$ is the projectable horizontal lift of the associated connection on $\scT$, and $\nu \mapsto \overline{\nu}$ denotes the inherited Lie algebra bundle action. The connection on $\scT$ is then also what one calls an $\scF$-connection.
    
    Because the action is faithful, there is the following one-to-one correspondence between:
    \begin{itemize}
        \item the multiplicative Yang-Mills connection on $c_G(P)$ modelled as associated to $P$;
        \item the Ehresmann connection on $P$~;
        \item the $\scF$-connection on $\scT$~.
    \end{itemize}
    The major result of~\cite{Fischer:2401.05966} implies that (in a formal setting, and in a neighborhood around a leaf) every foliation can be modeled like this; however $P$ is then in general infinite-dimensional.\footnote{In our context we simply assume a finite-dimensional $P$ so that the arguments of~\cite{Fischer:2401.05966} carry over to a smooth setting.} Altogether, we can study multiplicative Yang-Mills connections on such $c_G(P)$ as structure group bundle by looking at $\scF$ and $P$. In particular, $c_G(P)$ admits a flat multiplicative Yang-Mills connection (up to field redefinition) associated\footnote{If $G$ is simple, then all multiplicative YM connections are associated to the same $P$ by \ref{lem:SemiSimpleLGBAdjust}.} to $P$ if and only if $\scF$ admits a flat $\scF$-connection, which is the case if and only if $P$ is flat. Furthermore,~\cite{Fischer:2401.05966} provides a clear geometric interpretation of the field redefinitions, extending the discussion of~\cite{My1stpaper} and~\cite[\S 5.1]{MyThesis}.
    
    \begin{table*}[htbp]
        \caption{Comparing the geometry of multiplicative Yang-Mills connections (YM conn.) with the geometry of singular foliations. Here, $\scF$-connections only serve to highlight a geometric interpretation, but we actually provided a notion only defined as associated to $P$ as defined above; however, the general notion of $\scF$-connections allows for more than such connections if $G$ is not simple, needing a ``bigger'' principal bundle as in \ref{prop:classYMcenterless}; this will not matter here, see~\cite{Fischer:2401.05966} for the general classification of $\scF$-connections and their relation to multiplicative Yang-Mills connections.}
        \label{tab:ComparingTheGeometryGaugeTheoryWithSingularFoliations}
        \centering
        \begin{tabular}{l|l|l}
            Lie group bundle $c_G(P)$ & Singular foliations $\scF$ & principal $G$-bundle $P$\\ \hline
            multiplicative YM conn.\ $\mu$ & $\scF$-connection & Ehresmann connection \\
            $c_G(P)$ admits flat $\mu$ & $\scF$ is flat & $P$ is flat \\
            field redefinition of $\mu$ & change of $\scF$-connection & change of Ehresmann connection
        \end{tabular}
    \end{table*}
    
    Thus, in order to find non-trivial multiplicative Yang-Mills connections which cannot be flattened by field redefinitions one only needs to apply the previous construction on such principal bundles $P$ which do not admit flat connections; alternatively search for foliations $\scF$ which do not admit flat $\scF$-connections. There is plenty of literature on both cases.
    
    \begin{examples}{Cooking recipe}{Hopf}
        Using non-flat principal bundles with a simple and centerless structural group, for example a non-trivial principal bundle $P$ over a simply connected base manifold with such a structure, such as the Hopf fibration $\mathbb{S}^7 \to \mathbb{S}^4$, leads to curved multiplicative Yang-Mills connections on $c_G(P)$ in the above construction which cannot be flattened by field redefinitions. The general examples with centerless Lie groups are associated to principal bundles $P$ as defined in \ref{prop:classYMcenterless}.
    \end{examples}
		
		What about the full theory? For this we need to require the existence of an $\mathrm{Ad}$-invariant connection on $c_G(P)$ which is simply given by assuming that $G$ is simple which is covered by the previous discussion. Thus, it is only left to find a principal $c_G(P)$-bundle $\scP$; for this we have several canonical choices, which are of course not the only ones:
		\begin{itemize}
			\item $\scP = c_G(P)$ itself;
			\item $\scP = P$, where $c_G(P)$ acts via gauge transformations ($\sAut(\caP) \cong \Gamma(c_G(P))$);
			\item $\scP = (P \times P) \Big/ G$, the Atiyah groupoid of $P$, where $G$ acts diagonally on $P \times P$.
		\end{itemize}
		
		About the last bullet point, for the experienced reader: We have a short exact sequence of Lie groupoids
		\begin{equation*}
        \begin{tikzcd}
            c_G(P) \arrow[hook]{r} &
            (P \times P) \Big/ G \arrow[two heads]{r} &
            \scP\mathrm{air}(M)~,
        \end{tikzcd}
    \end{equation*}
		where $\scP\mathrm{air}(M)$ is the pair groupoid of $M$. It is a straightforward task to show that the Atiyah groupoid is a principal $c_G(P)$-bundle,
\begin{equation*}
	\begin{tikzcd}
		\mathcal{P} = (P \times P) \Big/ G \arrow{d}{\pi_{\mathcal{P}}} \arrow{dr}{s_{\scP}} & \arrow[bend right]{l} c_G(P) \arrow{d} \\
		\scP\mathrm{air}(M) \arrow[swap]{r}{s_{\scP\mathrm{air}(M)}} & M
	\end{tikzcd}
\end{equation*}
where $s_\scP$ and $s_{\scP\mathrm{air}(M)}$ are the source arrows of $(P \times P) \Big/ G$ and $\scP\mathrm{air}(M)$, respectively. Here one needs the notion of LGB-actions along a moment map, here $s_\scP$, which we did not discuss here, but it is straight-forward to extend the previous constructions to such setups; see \cite{Fischer:2024vak}. Observe however, that the base manifold of the Atiyah groupoid as such a principal bundle is the pair groupoid.

Using such principal bundles thence lead to a curved Yang-Mills gauge theory which cannot be described in a classical way.

\section{Future prospects}\label{conclusions}

\subsection{Associated bundles and connections}

As a next step one should classify all possible setups, extending \ref{prop:classYMcenterless}, which is already in progress, and one could now define minimal couplings. Given a vector bundle $V \to M$, equipped with a LGB representation $\rho \colon \mathcal{G} \to \mathrm{Aut}(V)$, we should be able to define associated vector bundles $\mathcal{V} \coloneqq ( \mathcal{P} \times_M V ) \Big/ \mathcal{G}$, as we already mentioned before. W.r.t.\ a local gauge $s$ of $\mathcal{P}$ every section $\Psi \in \Gamma(\mathcal{V})$ has locally the form
\begin{align*}
[s, \psi_s],
\end{align*}
where $\psi$ is a local section of $V$ and $[ \cdot, \cdot ]$ denotes the equivalence class related to the construction in $\mathcal{V}$. Then one can define the \textbf{minimal coupling $\nabla^A \Psi$} locally by
\begin{align*}
\mleft[  s,
	\nabla \psi_s + \rho_*(A_s)(\psi_s)
\mright]
\end{align*} 
where $\rho_*$ is the LAB representation inherited by $\rho$, and $\nabla$ is a vector bundle connection with
\begin{align*}
\rho\mleft( \sigma^{-1} \mright) \circ \nabla \circ \rho(\sigma)
&=
\nabla
    + \rho_*(\Delta \sigma),
\\
R_\nabla
&=
\rho_* \circ \zeta
\end{align*}
for all $\sigma \in \Gamma(\mathcal{G})$.
Observe the similarity with the infinitesimal compatibility conditions \ref{def:YangMillsConnection} and \ref{lem:FieldRedefViaLGBSection}; the first condition should lead to that $\nabla^A \Psi$ is well-defined and independent of the choice of $s$. The second condition should imply that the curvature of $\nabla^A$ is given by the generalised field strength of $A$. In sense of parallel transports, denoted as usual with $\rm PT$ and a superscript denoting the corresponding bundle, one will have
\begin{equation*}
{\rm PT}^V(v\cdot g)
=
{\rm PT}^V(v) \cdot
{\rm PT}^\caG(g)
\end{equation*}
for all $(v, g) \in V*\caG$ (as for $\caF$-connections and ``compatible pairs of Yang-Mills connections'' in \cite{Fischer:2401.05966}) and a corresponding curvature equation. The minimal coupling is then defined as
\begin{equation*}
{\rm PT}^\caV([p, v])
=
\mleft[ {\rm PT}^\caP(p), {\rm PT}^V(v) \mright]
\end{equation*}
for all $[p, v] \in \caV$. After showing all of this one could try to construct all the other obvious analogues of the classical formalism and discuss their applications. Essentially, one achieves a special type of curved Yang-Mills-Higgs theory as introduced in \cite{Fischer:2024vak}, related to the corresponding action groupoid. Which leads us to:

\subsection{Principal Lie groupoid bundles}

As mentioned, there is also the notion of curved Yang-Mills-Higgs theories, where the structure on the principal bundle is given by a Lie groupoid; infinitesimally introduced in \cite{CurvedYMH}, and summarized in \cite{MyThesis}. Essentially, this theory got already integrated: \cite{Fischer:2024vak} integrated the gauge transformations, Ehresmann connections and generalised field strength which is the most difficult part; the Lagrangian is already trivially motivated by the infinitesimal picture. A summary of all of that is work in progress by the authors of \cite{Fischer:2024vak}, let us therefore just give a quick summary of the mentioned references:

As we have seen, there are two steps: Generalising the notion of Ehresmann connection by fixing a multiplicative connection on the structure, and generalising the notion of field strength making use of suitable curvature equations on the structure. Thus, if $\caG$ is a Lie groupoid over a manifold $N$, then the proper generalisation of multiplicative Yang-Mills connections is called \textbf{covariant adjustment}: That is, a \textbf{Cartan connection} $\mutot$ which is a horizontal and multiplicative distribution complementary to the target arrow,\footnote{We define Lie algebroids as left-invariant vector fields, i.e.\ via the vertical structure of the target arrow.} and we are interested into those Cartan connections which satisfy the proper generalisation of the Maurer-Cartan equation given by saying that the curvature of $\mutot$ is multiplicative-exact, i.e.\ there is a $\zeta \in \Omega^2(N; s^*\cag)$
\begin{equation*}
		\left.\left(\mathrm{d}^{s^*\nabla^\caG} \mutot
		- \frac{1}{2} \left(s^*t_{\nabla^\caG_\rho}\right)\left( \mutot \stackrel{\wedge}{,} \mutot \right) \right) \right|_{g}
		=
		\mathrm{Ad}_{g^{-1}}\circ \left.t^! \zeta\right|_g 
		- \left.s^! \zeta\right|_g~,
\end{equation*}
for all $g \in \caG$, where $s$ and $t$ is the source and target arrow, respectively, $\cag$ the Lie algebroid of $\caG$ with anchor $\rho$ and bracket $[\cdot, \cdot]_\cag$, $\mathrm{Ad}$ is the adjoint representation induced by the choice of Cartan connection (\cite{crainic2015multiplicative}), and $t_{\nabla^\caG_\rho}$ the torsion of the induced connection $\nabla^\caG$ on $\cag$ where $\nabla^\caG$ is contracted with $\rho$ in the lower argument:
\begin{equation*}
t_{\nabla^\caG_\rho}(\mu, \nu)
\coloneqq
\nabla^\caG_{\rho(\mu)}\nu
	- \nabla^\caG_{\rho(\nu)}\mu
	- \mleft[ \mu, \nu \mright]_\cag
\end{equation*}
for all $\mu, \nu \in \Gamma(\cag)$.
A Cartan connection satisfying such a curvature condition is a covariant adjustment. 

Now assume a principal $\caG$-bundle $\caP$ over a possible different manifold $M$ on which $\caG$ acts along a map $\phi \colon \caP \to N$, the \textbf{lifted Higgs field}. An Ehresmann connection on $\caP$ in the sense of \cite{Fischer:2024vak} is a horizontal distribution on $\caP$ invariant under the modified push-forward which is defined as here, e.g.\ as in the first bullet point of \ref{rem:AlternModPush} which also has a similar expression via the corresponding Darboux derivative. As here, the existence and notion of an Ehresmann connection does not require the curvature condition. However, the curvature condition on $\mutot$ appears in defining the generalised field strength of a lifted Higgs field $\phi$ and a connection 1-form $A \in \Omega^1(\caP; \phi^*\cag)$, given by:
\begin{equation*}
F
=
\mathrm{d}^{\phi^*\nabla^\caG} A
	- \frac{1}{2} \mleft(\phi^*t_{\nabla^\caG_\rho}\mright)\mleft( A \stackrel{\wedge}{,} A \mright)
	+ \frac12 \mleft(\phi^*\zeta\mright)\bigl( \rmD\phi - (\phi^*\rho)(A) \stackrel{\wedge}{,} \rmD\phi - (\phi^*\rho)(A) \bigr)~.
\end{equation*}
Note the appearance of the minimal coupling in the last term (\cite{CurvedYMH, MyThesis, Fischer:2024vak}) and that we also vary w.r.t.\ $\phi$. The corresponding Lagrangian is now contracting not just $F$, but also contracting the minimal coupling, and adding a potential, similar to how we usually construct Yang-Mills-Higgs theories; in order to contract those terms one needs metrics invariant under the adjoint representations in the context of Lie groupoids (which in general depend on the choice of $\mutot$!). See the mentioned references for this construction and a proof for gauge invariance, e.g.\ \cite{CurvedYMH} or \cite{MyThesis}.

As before there are field redefinitions, an equivalence relation in this context, \cite[Def.\ 4.5.1 and 4.7.10]{MyThesis} which are now heavily more complicated due to the simple fact that $\rm Ad$ now depends in general on $\mutot$. Curved examples stable under field redefinitions were provided in \cite{Fischer:2024vak} and (partial) classifications are work in progress. Sadly, the field redefinitions may imply that locally such theories are not that new, and even globally may be reconstructed by LGBs equipped with an action, so that the language of groupoids may be nice language-wise but may otherwise not add many new examples. This could also be seen as the possibly restrictive nature of choosing a Cartan connection (with curvature conditions) on the structure. See future works regarding these phenomena, in particular see \cite{grad2026yangmillstheorymultiplicativeehresmann} for a new approach on defining Lagrangians with multiplicative Ehresmann connections; this reference also includes a conclusion discussing a possible umbrella framework for defining Ehresmann connections.

\textbf{Acknowledgements:} I want to thank Siye Wu, Camille Laurent-Gengoux, Mark John David Hamilton and Alessandra Frabetti for their great help and support in making this paper. Also big thanks to my wife Rachelle, my mother, father, Dennis, Gregor, Marco, Nico, Jakob, Kathi, Konstantin, Lukas, Locki, Gareth, Philipp, Ramona, Annerose, Michael, Maxim, and Anna for all their love and support. Special thanks also to Žan Grad for very fruitful discussions, and to my referees for much appreciated advises!






\appendix
\setcounter{equation}{0}
\renewcommand{\theequation}{\Alph{section}.\arabic{equation}} 

\renewcommand\refname{List of References}


\bibliography{Literatur}
\bibliographystyle{unsrt}

\end{document}